\documentclass[manuscript,screen]{acmart}
\AtBeginDocument{%
  \providecommand\BibTeX{{%
    \normalfont B\kern-0.5em{\scshape i\kern-0.25em b}\kern-0.8em\TeX}}}

\setcopyright{acmcopyright}
\copyrightyear{2026}
\acmYear{2026}

\usepackage{wrapfig}
\usepackage{amsmath,amsfonts}
\usepackage{algorithmic}
\usepackage{graphicx}
\usepackage{xcolor}
\usepackage{tikz}
\usetikzlibrary{arrows.meta,positioning,calc}
\usepackage{enumitem}
\usepackage[normalem]{ulem}
\usepackage{textcomp}
\usepackage{multirow}
\usepackage{indentfirst}
\usepackage{adjustbox}
\usepackage{hyperref}
\hypersetup{
    colorlinks=true,
    linkcolor=black,
    filecolor=black,
    urlcolor=black
}
\usepackage{color}
\usepackage{longtable}
\usepackage{makecell}

\definecolor{darkgrn}{rgb}{0, 0.8, 0}

\usepackage{pifont}
\newcommand{\cmark}{\ding{51}}%
\usepackage{gensymb}
\usepackage{soul}
\usepackage{float}
\usepackage{multirow}
\usepackage{makecell}
\usepackage[table]{xcolor}   
\usepackage{colortbl}
\usepackage{adjustbox}
\usepackage{pifont}
\usepackage{rotating}
\usepackage{wasysym}

\usepackage{paracol}

\usepackage{caption}
\begin{document}

\title{LLM-Based Agents for Software and Systems Security: Approaches, Applications, and Assessment}


\author{Jingjing Nie}
\email{jnie5@buffalo.edu}
\orcid{0009-0005-4829-660X}
\affiliation{%
  \institution{University at Buffalo, SUNY}
  \streetaddress{105 White Road}
  \city{Buffalo}
  \state{New York}
  \country{USA}
  \postcode{14260-2500}
}

\author{Jiawei Guo}
\authornote{Equal contribution to the first author.}
\email{jiaweigu@buffalo.edu}
\orcid{0009-0009-9134-9307}
\affiliation{%
  \institution{University at Buffalo, SUNY}
  \streetaddress{105 White Road}
  \city{Buffalo}
  \state{New York}
  \country{USA}
  \postcode{14260-2500}
}

\author{Krishna Meda}
\email{krishnamohanmeda@gmail.com}
\orcid{0009-0007-9368-4470}
\affiliation{%
  \institution{University at Buffalo, SUNY}
  \streetaddress{105 White Road}
  \city{Buffalo}
  \state{New York}
  \country{USA}
  \postcode{14260-2500}
}
\author{Haipeng Cai}
\email{haipengc@buffalo.edu}
\orcid{0000-0002-5224-9970}
\affiliation{%
  \institution{University at Buffalo, SUNY}
  \streetaddress{105 White Road}
  \city{Buffalo}
  \state{New York}
  \country{USA}
  \postcode{14260-2500}
}


\renewcommand{\shortauthors}{Nie et al.}

\begin{abstract}

Software and systems security workflows are typically procedural: analysts inspect
heterogeneous artifacts, form hypotheses, invoke tools, interpret outputs, and
revise plans. Large language model (LLM)-based agents, which can plan, use tools,
retain state, and revise actions across multi-step workflows, are being rapidly
adopted to automate this work. Given the consequences of delegating security
decisions to autonomous systems, understanding how such agents are built, used, and
assessed is crucial. Yet to this date, there remains a lack of systematic
understanding of what has been done and how far we are in this field: the term
``agent'' is applied inconsistently, applications differ sharply in risk, and
assessment protocols are often incomparable. To gain a comprehensive and coherent
view of this area hence inform relevant future research, this paper provides a
systematic literature review of the (1) technical approaches, including agent
architecture, perception, memory, reasoning and planning, action space,
orchestration, and self-improvement, (2) applications, with respect to the security
tasks served, and (3) assessment, including the datasets, outcome and trajectory
metrics, safety measures, and baselines considered, 
over the peer-reviewed literature spanning the emergence of this area (2023--2026). 
Our synthesis reveals a field that has built agents able
to act but not yet agents whose authority is bounded or whose behavior is auditable.
In addition to knowledge systematization, we also extend our insights into the
limitations of and challenges faced by current approach, application, and assessment
designs, which shed light on potentially promising future research directions.

\end{abstract}


\maketitle

\section{Introduction}\label{sec:intro}

Software and systems security increasingly depends on reasoning across heterogeneous artifacts: source code, binaries, configurations, logs, vulnerability reports, execution traces, network services, tickets, and human-written policies. Security work is also intrinsically procedural. Analysts rarely solve a task by producing one answer in isolation; they inspect evidence, form hypotheses, call tools, interpret outputs, revise plans, and communicate findings. This makes security a natural but demanding setting for large language models (LLMs). Modern LLMs can interpret code and natural language, but many security tasks require more than text generation. They require systems that can plan, use tools, maintain task state, and act over multi-step workflows.

LLM-based agents provide one response to this need. In the broader AI literature, agentic LLM systems have been developed to interleave reasoning and action, invoke external tools, and operate over interactive environments~\cite{yao2023react,schick2023toolformer}. In software engineering, benchmarks such as SWE-bench show that realistic tasks often require repository-level context, environment interaction, and iterative repair rather than isolated code completion~\cite{jimenez2024swebench}. These characteristics are even more notable in security. A vulnerability-analysis agent may need to inspect a repository, trace a data flow, query a vulnerability database, run a proof of concept, and explain exploitability. A penetration-testing agent may need to perform reconnaissance, choose tools, interpret command output, and adapt to failed attempts. A repair agent may need to localize a defect, synthesize a patch, run tests, and preserve functionality. A Security Operations Center (SOC) agent may need to correlate alerts, enrich evidence, reduce hallucination, and produce analyst-facing narratives.

Recent research reflects this shift from LLMs as answer generators to LLMs as operational security agents. Systems such as \textsc{PentestGPT}~\cite{79} and BountyBench~\cite{25zhang2026bountybench} explore agentic workflows for penetration testing and offensive security. CVE-Bench~\cite{38zhu2025cvebenchbenchmarkaiagents} assesses agents in executable vulnerability environments. \textsc{RepoAudit}~\cite{89guo2025repoauditautonomousllmagentrepositorylevel} studies repository-level vulnerability auditing, while \textsc{PatchAgent}~\cite{75} targets vulnerability repair. \textsc{Cognitive SOC}~\cite{32} illustrates how multi-agent designs can support evidence-grounded security operations. These examples show the breadth of current work: LLM-based agents are being applied to vulnerability detection, exploit development, red teaming, fuzzing, malware analysis, reverse engineering, incident response, and access-control assessment, among many others. 

At the same time, the rise of LLM-based security agents introduces new scientific and practical challenges. First, the term ``agent'' is used inconsistently. Some papers describe a single tool-using LLM as an agent; others build multi-agent systems with planners, executors, critics, memory, retrieval, and policy layers. Second, security applications differ sharply in risk. An agent that summarizes alerts has a different safety profile from one that executes exploits, scans internet-facing assets, patches production code, or changes access-control policies. Third, assessment is difficult. Final task success is important, but it does not show whether an agent used evidence faithfully, stayed within scope, avoided hallucinations, respected tool permissions, or consumed a practical amount of time and money. 
%


These issues make the literature hard to interpret without systematic organization.
Existing papers vary in their architectures, autonomy levels, memory designs, tool
interfaces, security targets, datasets, baselines, and reported metrics. Some focus
on end-to-end offensive workflows; others on repository auditing, patching, SOC
support, cyber-defense simulation, or benchmark construction. Some assess agents
against real-world systems or executable environments, while others use synthetic
tasks, curated datasets, or restricted artifacts. Yet to this date, there remains a
lack of systematic understanding of what has been done and how far we are in this
field. Without a shared taxonomy, it is difficult to compare how agents are built,
determine which security tasks they serve and which remain underexplored, or judge
whether reported results are reproducible, safe, and comparable across studies.



To address this gap, we systematically review 100 peer-reviewed papers on LLM-based
agents for software and systems security, published from January 2023 through March
2026---a window that spans the entire history of this emerging area, 
since the capable LLMs that serve as these agents' underlying models only emerged at the end of 2022.
We organize the literature along three dimensions:
\emph{Approach} captures agent architecture, perception, memory, reasoning and
planning, action space, orchestration, and self-improvement; \emph{Application}
captures the security tasks served; and \emph{Assessment} captures datasets, outcome
and trajectory metrics, safety measures, baselines, and experimental protocols.

This survey makes three main contributions:
\begin{itemize}[leftmargin=18pt,topsep=0pt,itemsep=-0.5ex,partopsep=0.5ex,parsep=0.5ex]
\item We provide an extensive literature search and selection protocol that yields
a corpus of 100 peer-reviewed papers spanning the emergence of this area.
\item We derive and apply a taxonomy characterizing the surveyed literature across
Approach, Application, and Assessment, enabling structured comparison of agent
designs, security tasks, datasets, metrics, and baselines.
\item We synthesize the findings into a central observation---that the field has
built agents able to act, but not yet agents whose authority is bounded or whose
behavior is auditable---and extend it into the limitations of and challenges facing
current designs, pointing to future directions in evidence-centered architecture,
risk-aware autonomy, realistic and reproducible benchmarks, trajectory-level
assessment, and accountable human-agent collaboration.
\end{itemize}

\noindent
\textbf{Paper organization}.
The rest of the paper is organized as follows. Section~\ref{sec:background-relwork} provides background concepts used in the survey. Section~\ref{sec:osm} summarizes the survey methodology, and Section~\ref{sec:ls} describes the literature search and selection process. Section~\ref{sec:std} presents the taxonomy derivation. Section~\ref{sec:res} maps the surveyed papers to the taxonomy and analyzes the results for Approach, Application, and Assessment. Section~\ref{sec:sra} discusses limitations, challenges, and future research directions. Section~\ref{sec:threats} discusses threats to validity, and Section~\ref{sec:con} concludes the paper.

\section{Background and Related Work}\label{sec:background-relwork}

This section provides essential background on LLM-based agents and software/systems security, followed by a discussion of related surveys.

\vspace{-4pt}
\subsection{LLM-Based Agents}\label{subsec:llm-agents}

A \ul{large language model} (LLM) is a neural network trained on large-scale text corpora that can generate, interpret, and reason over natural language and code~\cite{zhao2023llmsurvey}. Foundation-scale LLMs---including GPT~\cite{openai2023gpt4}, Claude~\cite{anthropic2024claude}, Gemini~\cite{google2023gemini}, and open-weight models such as LLaMA~\cite{touvron2023llama} and DeepSeek~\cite{deepseek2024}---have demonstrated broad capabilities in code understanding, program repair, and security reasoning when used with appropriate prompting strategies.

An \ul{LLM-based agent} extends a standalone LLM into a task-oriented system that can perceive its environment, reason about observations, plan multi-step actions, invoke external tools, and learn from feedback~\cite{wang2024autonomousagents}. Whereas a bare LLM processes a single prompt and returns a single response, an agent operates in a loop: it observes the current state (e.g., tool output, code context, alert data), selects an action (e.g., run a scanner, query a knowledge base, generate a patch), executes that action, and then reasons over the result to decide its next step. This observe-reason-act cycle may repeat over many iterations before the agent produces a final output.

Several components distinguish agents from standalone LLM usage:

\vspace{3pt}\noindent
{\bf Agent architecture.}
An agent may be a single autonomous entity or a coordinated team of multiple agents. Single-agent designs handle the entire task loop in one model, while multi-agent designs decompose the task across specialised agents that communicate through message passing, shared memory, or hierarchical delegation.

\vspace{3pt}\noindent
{\bf Memory.}
Agents maintain state across steps through in-context working memory (the current prompt window), external short-term memory (e.g., an exploration trace stored outside the context), and long-term or persistent memory (e.g., a vector database of prior findings, a knowledge graph of discovered assets). Memory enables continuity: an agent can remember what it has already scanned, which hypotheses failed, and what evidence supports a current finding.

\vspace{3pt}\noindent
{\bf Perception.}
Perception concerns how raw artifacts---source code, binary disassembly, execution traces, log entries, network packets, alert records---are transformed into reasoning-ready context for the LLM. Common mechanisms include direct ingestion into the prompt, retrieval-augmented generation (RAG) from external corpora, summarisation of verbose tool output, and embedding-based semantic matching.

\vspace{3pt}\noindent
{\bf Reasoning and planning.}
Agents employ various reasoning paradigms to decide what to do next. ReAct~\cite{yao2023react} interleaves reasoning, tool action, and observation. Chain-of-thought (CoT)~\cite{wei2022cot} makes intermediate reasoning explicit. Reflection and self-critique allow the agent to evaluate its own intermediate products before continuing. Planning strategies include dynamic/reactive planning (revising steps based on new observations), hierarchical task decomposition (separating high-level goals from low-level execution), and backtracking on failure.

\vspace{3pt}\noindent
{\bf Action space and tools.}
The action space defines what operations the agent can invoke: static analysis tools (linters, taint analysers, formal verifiers), dynamic analysis tools (fuzzers, debuggers, execution sandboxes), code execution environments (shells, compilers, containers), information retrieval tools (code search, documentation retrieval, RAG), and security-specific platforms (CTF environments, penetration-testing frameworks, SIEM/SOAR systems, MITRE ATT\&CK integration). The breadth of the action space largely determines what security tasks the agent can address.

\vspace{3pt}\noindent
{\bf Self-improvement and learning.}
Some agents adapt their behaviour over time through in-context learning from session feedback, cross-session trajectory storage, fine-tuning on solved tasks, or reinforcement learning with environment-based reward signals. These mechanisms allow agents to improve on repeated tasks without manual prompt engineering.

\vspace{-4pt}
\subsection{Software and Systems Security}\label{subsec:softsec}

Software and systems security encompasses the protection of software artifacts, computing infrastructure, and networked environments against threats that compromise their confidentiality, integrity, or availability~\cite{anderson2020security}. The security lifecycle spans a broad set of tasks that range from proactive analysis through reactive response:

\vspace{3pt}\noindent
{\bf Vulnerability discovery and management.}
This includes the detection, localization, classification, prioritization, and tracking of software weaknesses in source code, compiled binaries, smart contracts, APIs, and infrastructure configurations. Traditional approaches rely on static analysis, dynamic testing, and manual code review; LLM-based agents augment these approaches with natural-language reasoning over code context, automated hypothesis generation, and iterative refinement.

\vspace{3pt}\noindent
{\bf Offensive security.}
Penetration testing, red-teaming, exploit development, and adversarial simulation assess the resilience of deployed systems by actively attempting to compromise them. These tasks are inherently interactive and multi-step, making them a natural fit for agentic workflows that can plan, execute, and adapt offensive strategies.

\vspace{3pt}\noindent
{\bf Defensive monitoring and response.}
Intrusion detection, incident response, log analysis, anomaly detection, threat intelligence, and 
SOC workflows monitor deployed systems for active or recent compromises. Agents in this space reason over heterogeneous runtime telemetry---logs, alerts, network flows---and must correlate signals into actionable defensive decisions.

\vspace{3pt}\noindent
{\bf Low-level artifact analysis.}
Binary analysis, reverse engineering, protocol inference, and malware analysis reason over artifacts below the level of 
source code. These tasks require agents to compensate for missing structure---no source, no specification, or only externally-observable behavior---by combining LLM reasoning with established analysis tools.

\vspace{3pt}\noindent
{\bf Repair and hardening.}
Automated program repair (APR), patch generation, code hardening, and security policy enforcement produce defensive modifications that close known weaknesses or prevent classes of future vulnerabilities. Agentic approaches extend traditional APR by integrating fault localization, patch synthesis, and validation into a single iterative loop.

\setlength{\tabcolsep}{3pt}
\begin{table}[t]
\caption{Comparison with closely related surveys. \emph{Agents}: requires an agentic
architecture rather than LLM usage alone. \emph{Instrument}: agents as security
practitioners (vs.\ agents as the object of attack). \textit{Approach/Application} (to security)\textit{/Assessment}: aspects
covered ($\CIRCLE$ full, $\LEFTcircle$ partial, $\Circle$ absent). \textit{Year}: year of surveyed studies covered up to}
\label{tab:relatedsurveys}
\vspace{-10pt}
\small
\begin{tabular}{llcccccl}
\toprule
Survey & Scope & Agents & Instrument & Approach & Application & Assessment & Year \\
\midrule
Xu et al.~\cite{xu2025large} & LLMs for cybersecurity & \Circle & \LEFTcircle & \LEFTcircle & \CIRCLE & \LEFTcircle & 2024 \\
Zhou et al.~\cite{zhou2024vulndetectrepair} & LLMs for vuln.\ detect./repair & \Circle & \LEFTcircle & \LEFTcircle & \LEFTcircle & \LEFTcircle & 2024 \\
Fan et al.~\cite{fan2023llmse} & LLMs for SE & \Circle & \Circle & \LEFTcircle & \Circle & \Circle & 2023 \\
Liu et al.~\cite{liu2024llmagentsse} & LLM agents for SE & \CIRCLE & \Circle & \CIRCLE & \Circle & \LEFTcircle & 2024 \\
Xu et al.~\cite{xu2025autonomouscyberattacks} & Agents for cyberattacks & \CIRCLE & \CIRCLE & \LEFTcircle & \LEFTcircle & \Circle & 2025 \\
Hassanin et al.~\cite{hassanin2024cyberdefences} & LLMs for cyber defence & \Circle & \LEFTcircle & \Circle & \LEFTcircle & \Circle & 2024 \\
Tang et al.~\cite{tang2025security} & Attacks/defenses on agents & \CIRCLE & \Circle & \LEFTcircle & \LEFTcircle & \LEFTcircle & 2025 \\
Shahriar et al.~\cite{shahriar2025survey} & Agentic security landscape & \CIRCLE & \LEFTcircle & \Circle & \CIRCLE & \Circle & 2025 \\
Wang et al.~\cite{wang2024autonomousagents} & LLM agents (domain-agnostic) & \CIRCLE & \Circle & \CIRCLE & \Circle & \Circle & 2024 \\
\midrule
\textbf{This survey} & \textbf{LLM agents for SW/sys.\ security} & \CIRCLE & \CIRCLE & \CIRCLE & \CIRCLE & \CIRCLE & \textbf{2026} \\
\bottomrule
\end{tabular}
\vspace{-8pt}
\end{table}

\vspace{-4pt}
\subsection{Related Work}\label{subsec:related-work}

Table~\ref{tab:relatedsurveys} contrasts our survey with the most closely related
prior work. No existing survey combines our focus (LLM-based \emph{agents}), domain
(software and systems security), and three-aspect taxonomy (approach, application,
and assessment). We group and contrast these surveys below.

\vspace{1pt}\noindent
{\bf LLMs for cybersecurity.}
Xu et al.~\cite{xu2025large} conduct a systematic literature review of LLMs
for cybersecurity, analyzing 185 papers across vulnerability detection, malware
analysis, and intrusion detection, and characterizing how LLM architectures and
adaptation techniques map onto security tasks. Zhou et al.~\cite{zhou2024vulndetectrepair}
narrow the scope to vulnerability detection and repair, comparing prompting,
fine-tuning, and retrieval-augmented strategies. Both are closest to our application
categories, but both treat the LLM as a model applied to a task rather than as an
operational system: neither characterizes the agentic execution
model---planning, tool invocation, memory, orchestration, and iterative
refinement---that determines what an agent can attempt and how its behavior must be
assessed. Our corpus is disjoint from theirs by construction, since we require an
agentic architecture rather than LLM usage alone.

\vspace{1pt}\noindent
{\bf LLM-based agents for software engineering.}
Fan et al.~\cite{fan2023llmse} survey LLMs for software engineering and identify open
problems across code generation, program repair, testing, and maintenance, treating
the LLM largely as a single-shot predictor. Liu et al.~\cite{liu2024llmagentsse}
narrow this to LLM-based \emph{agents} for software engineering, organizing the
literature by agent roles, planning, memory, and multi-agent collaboration. Both
establish that agentic designs improve robustness on complex SE tasks, but neither
focuses on security: they do not treat the offensive/defensive duality, adversarial
robustness, dual-use risk, or the security-specific benchmarks and safety
constraints that define our targeted domain.

\vspace{1pt}\noindent
{\bf LLM-based agents for offense or defense.}
Xu et al.~\cite{xu2025autonomouscyberattacks} survey LLM-based agents in autonomous
cyberattacks, organizing offensive capabilities along the cyber kill chain, while
Hassanin et al.~\cite{hassanin2024cyberdefences} provide a complementary
overview of LLMs for cyber defense. Domain-specific surveys narrow further: 
\cite{srinivas2025ai} reviews LLMs and agents across SOC functions.
Each addresses one side of the offensive/defensive divide or one application area;
our work unifies both orientations and eleven task categories under a single
application taxonomy, and additionally characterizes how these agents are
constructed and assessed.

\vspace{1pt}\noindent
{\bf Security \emph{of} LLM-based agents.}
A distinct line of surveys treats the agent as the object, not the instrument
of security. Tang et al.~\cite{tang2025security} systematize attacks and
defenses against LLM-based agents and propose evaluation criteria for each, and 
Shahriar et al.~\cite{shahriar2025survey} survey 
around applications, threats, and defenses. These surveys are
complementary rather than competing: they focus on the agent's own attack surface,
whereas we address the agent as a security \emph{practitioner} and treat attacks on
agents as one application category among eleven. Critically, none of them
characterizes agent construction at the granularity we require---perception,
memory operations, action space, orchestration, and self-improvement---or
systematizes assessment practice across datasets, outcome and trajectory metrics,
safety measures, and baselines.

\vspace{1pt}\noindent
{\bf General surveys of LLM-based autonomous agents.}
Wang et al.~\cite{wang2024autonomousagents} survey LLM-based autonomous agents across
domains, providing a widely-adopted taxonomy of agent construction (profiling,
memory, planning, and action). We adopt a compatible view of agent architecture
(Section~\ref{ssec:resultapproach}) but specialize it to security: our categories
capture sandboxed execution, integration with scanners and CTF environments,
scope containment, and robustness against prompt injection---adaptations that
domain-agnostic agent surveys do not address.


In contrast to all of the above, our work is, to our knowledge, the first systematic
survey to focus specifically on LLM-based \emph{agents}---rather than standalone
LLMs---as instruments of software and systems security, with a unified taxonomy
spanning how such agents are built (\textit{approach}), what security problems/tasks they address
(\textit{application}), and how they are assessed (\textit{assessment}). This three-aspect coupling allows us to observe that autonomy has outpaced governance and that assessment
practice has not kept pace with either---a finding no single-aspect survey would make.



\begin{figure}[t]
  \centering
  \includegraphics[width=\linewidth]{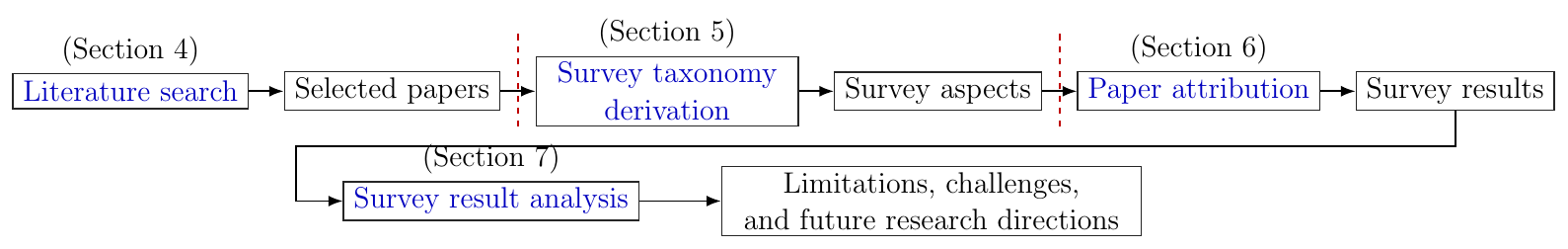}
  \vspace{-20pt}
  \caption{Overview of our survey methodology and workflow.}
  \label{fig:survey-workflow}
  \vspace{-12pt}
\end{figure}

\section{Overview of Survey Methodology}\label{sec:osm}
To gain a systematic understanding of LLM-based agents for software and system
security, our survey is organized around three research questions (\textbf{RQ}s), to answer which 
we follow a structured methodology shown in Figure~\ref{fig:survey-workflow}.

\begin{itemize}[leftmargin=*,itemsep=0pt]
\item \textbf{RQ1 (Approach).} How are LLM-based agents for software and
systems security constructed and operated, in terms of architecture,
perception, memory, reasoning and planning, action space, orchestration,
and self-improvement?
\item \textbf{RQ2 (Application).} Which software and systems security tasks
do these agents address, and how is effort distributed across the security
lifecycle?
\item \textbf{RQ3 (Assessment).} How is agent behavior assessed, in terms of
datasets, outcome and trajectory metrics, safety measures, and baselines?
\end{itemize}
%
%

We first performed a systematic \emph{literature search}
($\S$\ref{sec:ls}) to collect peer-reviewed studies published
between January 2023 and March 2026. This time window captures the period in
which LLM-based agent systems became a visible research direction while keeping
the scope focused on recent work. The search process combined keyword-based
searching and forward/backward snowballing, followed by manual eligibility
filtering based on explicit inclusion and exclusion criteria. This process
produced the final set of 100 papers used in the survey.

Based on the selected papers, we then derived the \emph{survey taxonomy}
($\S$\ref{sec:std}). The taxonomy characterizes the literature from
three complementary aspects: \emph{approach}, which captures how LLM-based
security agents are designed and operated; \emph{application}, which captures
the security tasks addressed by the agents; and \emph{assessment}, which
captures the datasets, metrics, and experimental protocols used to assess
agentic security systems.

After deriving the taxonomy, we conducted \emph{paper attribution}
($\S$\ref{sec:res}) by mapping each selected paper to the
taxonomy items. This attribution process allowed us to summarize cross-paper
patterns, compare agent designs and applications, and identify how the
surveyed works assess agent behavior. Finally, we analyzed the attributed
results ($\S$\ref{sec:sra}) to identify limitations, open challenges, and future research directions for LLM-based agents in
software and systems security.

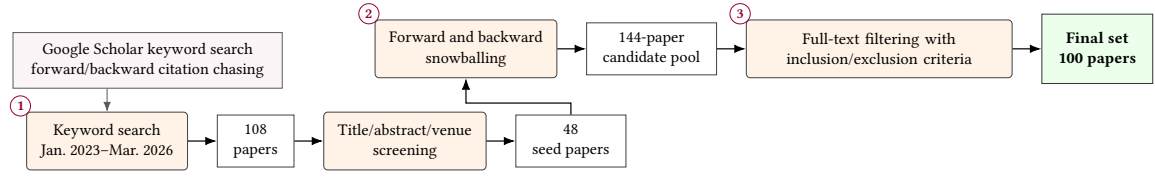
\begin{figure*}[tp]
  \centering
  \resizebox{\textwidth}{!}{%
  \begin{tikzpicture}[
    font=\small,
    >=Latex,
    node distance=5mm,
    source/.style={draw=black!65, fill=purple!4, minimum height=6mm, align=center, inner xsep=9pt,inner ysep=5pt},
    process/.style={draw=black!75, fill=orange!10, rounded corners=2pt, minimum height=10mm, align=center, inner xsep=7pt, inner ysep=3pt},
    count/.style={draw=black!75, fill=white, minimum height=9mm, align=center, inner xsep=8pt},
    final/.style={draw=black!80, fill=green!8, minimum height=12mm, align=center, inner xsep=8pt, font=\bfseries\small},
    phase/.style={circle, draw=purple!80!black, fill=white, text=purple!80!black, font=\bfseries\footnotesize, inner sep=1.5pt},
    flow/.style={->, line width=0.6pt},
    thinflow/.style={->, line width=0.4pt, draw=black!65}
  ]
    \node[source,xshift=-30mm] (src) {Google Scholar keyword search\\forward/backward citation chasing};
    \node[process, xshift=25mm,yshift=2mm] (snow) {Forward and backward\\snowballing};
    \node[count, right=of snow] (c144) {144-paper\\candidate pool};
    \node[process, right=of c144, text width=41mm] (elig) {Full-text filtering with\\inclusion/exclusion criteria};
    \node[final, right=of elig] (final) {Final set\\100 papers};
    \node[process, below=4mm of src, xshift=-7mm] (p1) {Keyword search\\Jan. 2023--Mar. 2026};
    \node[count, right=of p1] (c108) {108\\papers};
    \node[process, right=of c108] (screen) {Title/abstract/venue\\screening};
    \node[count, right=of screen] (c48) {48\\seed papers};

    \draw[thinflow] (src.south -| p1.north) -- (p1.north);
    \draw[flow] (p1) -- (c108);
    \draw[flow] (c108) -- (screen);
    \draw[flow] (screen) -- (c48);
    \draw[flow] (c48.north) -- ++(0,2mm) -| (snow.south);
    \draw[flow] (snow.east) -- (c144.west);
    \draw[flow] (c144) -- (elig);
    \draw[flow] (elig) -- (final);
    \node[phase] at ([xshift=-1mm,yshift=1mm]p1.north west) {1};
    \node[phase] at ([xshift=-1mm,yshift=1mm]snow.north west) {2};
    \node[phase] at ([xshift=-1mm,yshift=1mm]elig.north west) {3};
  \end{tikzpicture}%
  }
  \vspace{-12pt}
  \caption{Overview of our literature search process.}
  \label{fig:literature-search}
  \vspace{-12pt}
\end{figure*}

\section{Literature Search} \label{sec:ls}
We searched, identified, and filtered studies on LLM agents for software
and system security in three phases, as summarized in
Figure~\ref{fig:literature-search}. The goal of the search was to collect
papers that explicitly use or develop LLM-based agents for security-related
software or system tasks, while excluding work that primarily studies the
security of agents or LLMs themselves in general.


\vspace{-4pt}
\subsection{Phase 1}~\label{subsubsec:p1}
We search broadly for all potentially relevant research papers on on Google Scholar
and main academic publication portals/digital libraries such as ACM DL, IEEE Xplore, Scopus, and DBLP. 
We used the following two groups of topic-relevant keywords and their (pair-wise) combinations
as queries: 
Group-1: "AI agent", "LLM agent", "LLM-based agent", "agentic", 
"multi-agent", "autonomous agent"; and 
Group-2: "vulnerability detection", 
"penetration testing", "exploit", "fuzzing", "security testing", 
"vulnerability assessment", "intrusion detection", "malware detection", 
"cyber attack". 
We use these queries as it combines agent-related terms with common software/system security
tasks, including vulnerability detection, penetration testing, exploit
analysis, fuzzing, security testing, vulnerability assessment, intrusion
detection, malware detection, and cyber attacks. 
%
%
This initial search returned 108 candidate papers. We then manually inspected 
each paper and its 
metadata in detail. Papers that were clearly outside the target scope were removed at
this stage. After this screening, we retained 48 seed papers for the next
phase.


\vspace{-4pt}
\subsection{Phase 2} \label{subsubsec:p2}
To improve coverage beyond keyword matching, we performed manual forward and
backward snowballing from the 48 seed papers. Backward snowballing identified
relevant papers cited by the seed papers, while forward snowballing identified
later papers that cited them. During this process, we considered whether each
candidate paper addressed an LLM-based agent and whether its task concerned
software or system security.

The snowballing phase expanded the candidate pool to 144 papers. 
This pool was then passed to the 
next phase.


\vspace{-4pt}
\subsection{Phase 3}  \label{subsubsec:p3}
In the third phase, we manually inspected each candidate paper in the expanded
pool and applied the following inclusion and exclusion criteria.

\noindent
We used three \ul{inclusion criteria}:
\begin{itemize}[leftmargin=18pt,topsep=0pt,itemsep=-1ex,partopsep=1ex,parsep=1ex]
\item The paper must use or develop an AI agent whose core intelligence is LLM-based.
\item The paper must use the agent for software/system security-related studies.
\item The paper must have been published between January 2023 (early enough date of published work on LLM-based agent)
and March 2026 (our survey start date).
\end{itemize}

\noindent
We also used the following \ul{exclusion criteria}:
\begin{itemize}[leftmargin=18pt,topsep=0pt,itemsep=-1ex,partopsep=1ex,parsep=1ex]
\item Papers that do not address both an LLM-based agent and a software/system security task.
\item Short articles, abstracts, posters (i.e., shorter than four pages) or non-English publications. 
\item Studies exclusively available as preprints (e.g., on arXiv) only, which were omitted to keep the survey focused on peer-reviewed work.
\end{itemize}

\noindent
For each candidate paper, we examined the title, abstract, 
keywords, and main text. 
A paper was included only if it
satisfied all inclusion criteria and none of the exclusion criteria. At the end
of this phase, we were left with \textbf{100} papers as the final corpus for our
survey. For each included paper, we recorded its research problem, agent design,
security application, datasets or experimental subjects, assessment metrics,
main findings, limitations, and future-work discussions when available.

\section{Survey Taxonomy Derivation} \label{sec:std}
In this section, we describe how we derived the taxonomy used to characterize the surveyed literature on LLM-based agents for software and systems security, 
to answer the three RQs ($\S$\ref{sec:osm}. Accordingly, we organize the taxonomy around three high-level aspects: \emph{Approach}, \emph{Application}, and \emph{Assessment}. 

We derived the taxonomy in two steps. First, we identified candidate attributes and items by reading and coding the selected papers (\S\ref{subsubsec:ai}). Second, we generalized and consolidated these candidates into a taxonomy (\S\ref{subsubsec:ag}). The resulting taxonomy is summarized in Table~\ref{tab:overall-survey-taxonomy} and then elaborated for each aspect in Section~\ref{ss:taxonomy-approach}, Section~\ref{ss:taxonomy-application}, and Section~\ref{ss:taxonomy-assessment}.

\vspace{-4pt}
\subsection{Attribute/Item Identification} \label{subsubsec:ai}
Our attribute/item identification step aims to refine the three survey aspects
(\emph{Approach}, \emph{Application}, and \emph{Assessment}) into concrete
taxonomy dimensions. We first reviewed the selected papers and recorded recurring
terms, concepts, and design choices that describe LLM-based agents for software
and system security. For the \emph{Approach} aspect, we identified attributes
such as agent architecture, memory, perception, reasoning and planning, action
space, workflow and orchestration, and self-improvement. For the
\emph{Application} aspect, we recorded the security tasks addressed by the
agents, such as vulnerability detection, penetration testing, exploit analysis,
repair, fuzzing, intrusion detection, and security operations. For the
\emph{Assessment} aspect, we collected dataset characteristics, assessment
metrics, baselines, safety measurements, and process-level indicators used to
assess agent behavior and performance for the application.


\vspace{-3pt}
\subsection{Attribute/Item Generalization} \label{subsubsec:ag}
After the initial attributes and items were identified, we generalized them
through several iterations to keep the taxonomy manageable and reusable across
papers. We merged semantically similar terms, resolved overlapping categories,
and normalized paper-specific wording into broader taxonomy items. For example,
under the \emph{Approach} aspect, terms such as tool invocation, shell execution,
sandbox execution, and browser automation were organized under the broader
attribute of agent action space. Under the \emph{Application} aspect, related
tasks such as bug finding, vulnerability localization, 
and vulnerability triage were generalized into vulnerability detection and
analysis. Similarly, under the \emph{Assessment} aspect, metrics such as task
completion, exploit success, patch validity, precision, recall, hallucination rate, 
and tool-call correctness were grouped into broader metric families. After
this generalization step, the resulting attributes and items were documented as
the survey taxonomy used for paper attribution.

\newcommand{\myrotcell}[1]{\rotcell{\makebox[0pt][l]{#1}}}
\begingroup
\scriptsize
\setlength{\tabcolsep}{3pt}
\renewcommand{\arraystretch}{0.72}
\begin{longtable}{|>{\raggedright\arraybackslash}p{0.08\textwidth}|>{\raggedright\arraybackslash}p{0.12\textwidth}|>{\raggedright\arraybackslash}p{0.20\textwidth}|>{\raggedright\arraybackslash}p{0.54\textwidth}|}
\caption{Overview of the derived survey taxonomy}
\label{tab:overall-survey-taxonomy}\\
\hline
\textbf{Aspect} & \textbf{Attribute} & \textbf{Item} & \textbf{Description} \\
\hline
\endfirsthead
\hline
\textbf{Aspect} & \textbf{Attribute} & \textbf{Item} & \textbf{Description} \\
\hline
\endhead
\hline
\endfoot
\hline
\endlastfoot
\multirow{24}{*}{Approach} & \multirow{3}{0.12\textwidth}{Agent architecture} & Agent multiplicity & Single vs. multi-agent organization. \\
\cline{3-4}
 &  & Internal structure & Module, pipeline, guardrail, fine-tuning, or hybrid layout. \\
\cline{3-4}
 &  & Backbone LLM role & LLM as actor, planner, verifier, critic, or controller. \\
\cline{2-4}
 & \multirow{2}{0.12\textwidth}{Agent perception} & Input sources & Code, binaries, traces, logs, environments, images, or standards. \\
\cline{3-4}
 &  & Perception mechanisms & Ingest, retrieve, summarize, embed, or observe via tools. \\
\cline{2-4}
 & \multirow{3}{0.12\textwidth}{Agent memory} & Memory types & In-context, external, long-term, graph, or parametric memory. \\
\cline{3-4}
 &  & Memory operations & Read, write, update, and revise memory. \\
\cline{3-4}
 &  & Security-specific memory & Vulnerability knowledge, repos, traces, CTI, and tool docs. \\
\cline{2-4}
 & \multirow{3}{0.12\textwidth}{Agent reasoning \\\& planning} & Reasoning paradigms & Patterns for analysis, decomposition, critique, and hypotheses. \\
\cline{3-4}
 &  & Planning strategies & Sequencing, decomposition, revision, and recovery. \\
\cline{3-4}
 &  & Decision-making under uncertainty & Branching, voting, checkpoints, and uncertainty handling. \\
\cline{2-4}
 & \multirow{7}{0.12\textwidth}{Agent\\action space} & Static analysis tools & Linters, taint analysis, symbolic execution, and verification. \\
\cline{3-4}
 &  & Dynamic analysis tools & Fuzzers, debuggers, sandboxes, and traffic analysis. \\
\cline{3-4}
 &  & Code execution \& environment & Shells, interpreters, containers, browsers, and automation. \\
\cline{3-4}
 &  & Information retrieval tools & Search, documentation lookup, vector stores, and RAG. \\
\cline{3-4}
 &  & Security-specific platforms & CTFs, pentest tools, scanners, SIEM/SOAR, ATT\&CK, cloud, OSINT. \\
\cline{3-4}
 &  & Inter-agent communication & Messaging, aggregation, consensus, and audit trails. \\
\cline{3-4}
 &  & Human interaction & Human queries, approvals, reviews, and feedback. \\
\cline{2-4}
 & \multirow{3}{0.12\textwidth}{Agent workflow \\\& orchestration} & Workflow topology & Seq., parallel, iterative, hierarchical, debate, adversarial. \\
\cline{3-4}
 &  & Agent roles & Planner, analyst, executor, verifier, reporter, critic, coord. \\
\cline{3-4}
 &  & Human-in-the-loop integration & Autonomous, human-on-loop, human-in-loop, or collaborative modes. \\
\cline{2-4}
 & \multirow{3}{0.12\textwidth}{Agent learning \\\& self-improvement} & In-context adaptation & Within-session feedback, examples, and corrections. \\
\cline{3-4}
 &  & Cross-session learning & Persistent trajectories, experience, or fine-tuning data. \\
\cline{3-4}
 &  & RL-based improvement & Reward- and environment-driven improvement. \\
\hline
\multirow{11}{*}{Application} & \multicolumn{2}{>{\raggedright\arraybackslash}p{0.36\textwidth}|}{Vulnerability detection \& analysis} & Detect, localize, classify, analyze, triage, and correlate vulnerabilities. \\
\cline{2-4}
 & \multicolumn{2}{>{\raggedright\arraybackslash}p{0.36\textwidth}|}{Exploit development \& analysis} & Generate, chain, craft, detect, and analyze exploits or attack paths. \\
\cline{2-4}
 & \multicolumn{2}{>{\raggedright\arraybackslash}p{0.36\textwidth}|}{Penetration testing \& red-teaming} & Support reconnaissance, access, escalation, movement, CTFs, and red-team automation. \\
\cline{2-4}
 & \multicolumn{2}{>{\raggedright\arraybackslash}p{0.36\textwidth}|}{Vulnerability repair \& hardening} & Generate, validate, prioritize, review, and harden vulnerable code or dependencies. \\
\cline{2-4}
 & \multicolumn{2}{>{\raggedright\arraybackslash}p{0.36\textwidth}|}{Fuzzing \& test generation} & Generate harnesses, seeds, mutations, security tests, and crash triage workflows. \\
\cline{2-4}
 & \multicolumn{2}{>{\raggedright\arraybackslash}p{0.36\textwidth}|}{Malware analysis \& detection} & Detect, classify, attribute, deobfuscate, and analyze malware behavior. \\
\cline{2-4}
 & \multicolumn{2}{>{\raggedright\arraybackslash}p{0.36\textwidth}|}{Reverse engineering} & Analyze disassembly, decompiled binaries, functions, types, symbols, similarity, and protocols. \\
\cline{2-4}
 & \multicolumn{2}{>{\raggedright\arraybackslash}p{0.36\textwidth}|}{Intrusion \& threat detection, incident response} & Analyze logs or traffic, triage alerts, hunt threats, investigate incidents, and automate response. \\
\cline{2-4}
 & \multicolumn{2}{>{\raggedright\arraybackslash}p{0.36\textwidth}|}{Access control \& authentication assessment} & Test authentication, authorization, firewall, ACL, and access-control policies. \\
\cline{2-4}
 & \multicolumn{2}{>{\raggedright\arraybackslash}p{0.36\textwidth}|}{Threat intelligence \& security operations (SecOps)} & Synthesize threat intelligence, reports, advisories, policies, Q\&A, and CTI analysis. \\
\cline{2-4}
 & \multicolumn{2}{>{\raggedright\arraybackslash}p{0.36\textwidth}|}{Model attack generation \& robustness assessment} & Assess jailbreaks, prompt injection, agent guardrails, and adversarial ML/LLM robustness. \\
\hline
\multirow{51}{*}{Assessment} & \multirow{2}{0.12\textwidth}{Dataset origin} & Real-world & Production, competition, threat-feed, or curated public artifacts. \\
\cline{3-4}
 &  & Synthetic / controlled & Generated, mutated, or purpose-built controlled datasets. \\
\cline{2-4}
 & \multirow{6}{0.12\textwidth}{Dataset\\characterisation\\axes} & Realism & Synthetic, semi-synthetic, or real-world fidelity. \\
\cline{3-4}
 &  & Label quality & Automatic, competition-verified, expert/reference-verified, weak, or noisy labels; unspecified. \\
\cline{3-4}
 &  & Scale & Small, medium, or large dataset size. \\
\cline{3-4}
 &  & Language / artifact & Source, binary, network, log, or other artifact modality. \\
\cline{3-4}
 &  & Vulnerability type coverage & Breadth of vulnerability classes or task types. \\
\cline{3-4}
 &  & Accessibility & Public, registered, restricted, proprietary, or internal access. \\
\cline{2-4}
 & \multirow{12}{0.12\textwidth}{Dataset by\\security task} & Vuln. detection \& classification & Named corpora, vulnerability databases, and other audit datasets. \\
\cline{3-4}
 &  & Pentest \& CTF & Penetration-testing, CTF, and attack-simulation tasks. \\
\cline{3-4}
 &  & Smart contracts & Smart-contract vulnerability detection, auditing, and verification. \\
\cline{3-4}
 &  & Web/phishing/scam & Web threats, phishing, scams, and app-layer abuse. \\
\cline{3-4}
 &  & Repair/bug-fix & Patch generation, bug fixing, and vulnerability repair. \\
\cline{3-4}
 &  & Cyber defense / simulation & Defensive simulations, incidents, and cyber ranges. \\
\cline{3-4}
 &  & Security testing / fuzzing & API, program, and protocol testing or fuzzing. \\
\cline{3-4}
 &  & AI / agent security & Security risks and defenses for learned systems, LLMs, or agents. \\
\cline{3-4}
 &  & Asset discovery / OSINT & Asset, exposed-service, and OSINT discovery. \\
\cline{3-4}
 &  & Malware / Android apps & Malware, mobile apps, and Android artifacts. \\
\cline{3-4}
 &  & Intrusion / SOC / SOAR & Intrusion data, security alerts, triage records, and SOAR workflows. \\
\cline{3-4}
 &  & Other security task & Security-task datasets outside the listed families. \\
\cline{2-4}
 & \multirow{10}{0.12\textwidth}{Task effectiveness\\metrics} & Accuracy/classification & Classification accuracy or labeled-output correctness. \\
\cline{3-4}
 &  & Precision/recall/F1 & Retrieval or detection quality from TP/FP/FN counts. \\
\cline{3-4}
 &  & Error rates & Incorrect outputs, failures, false alarms, and error modes. \\
\cline{3-4}
 &  & Success/completion & Task completion or objective achievement. \\
\cline{3-4}
 &  & Findings/yield & Discovered vulnerabilities, bugs, exploits, alerts, or findings. \\
\cline{3-4}
 &  & Correctness/validity & Validity of reports, patches, exploits, or artifacts. \\
\cline{3-4}
 &  & Coverage & Codebase, task-space, dataset, or attack-surface coverage. \\
\cline{3-4}
 &  & Time/responsiveness & Elapsed time, latency, speed, or responsiveness. \\
\cline{3-4}
 &  & Availability/reward & Benchmark availability, rewards, bounty, or payoff. \\
\cline{3-4}
 &  & Qualitative/user assess. & Human-rated usefulness, clarity, satisfaction, or value. \\
\cline{2-4}
 & \multirow{7}{0.12\textwidth}{Safety \& reliability\\metrics} & Scope containment & Authorized-scope and sandbox adherence. \\
\cline{3-4}
 &  & Prompt injection robustness & Resistance to malicious, conflicting, or tool-mediated prompts. \\
\cline{3-4}
 &  & Consistency & Stability across runs, prompts, tasks, or environments. \\
\cline{3-4}
 &  & ASR & Attack success rate or adversarial outcome success. \\
\cline{3-4}
 &  & Defense success & Prevention, detection, or mitigation success. \\
\cline{3-4}
 &  & Utility-security tradeoff & Balance between utility and security constraints. \\
\cline{3-4}
 &  & Overhead & Added cost, latency, compute, or operational burden. \\
\cline{2-4}
 & \multirow{8}{0.12\textwidth}{Agentic process\\/ trajectory metrics} & Steps to completion & Reasoning, action, or workflow steps to finish. \\
\cline{3-4}
 &  & Tool call accuracy & Correct tool selection, parameters, and interpretation. \\
\cline{3-4}
 &  & Hallucination rate & Unsupported, fabricated, or misleading claims/actions. \\
\cline{3-4}
 &  & Cost / token efficiency & Token, API, compute, or per-task efficiency. \\
\cline{3-4}
 &  & Iteration count & Retries, planning loops, repair cycles, or attempts. \\
\cline{3-4}
 &  & Artifact correctness & Correctness of files, patches, reports, exploits, deliverables. \\
\cline{3-4}
 &  & Report / explanation & Quality, faithfulness, and usefulness of reports. \\
\cline{3-4}
 &  & Economic / bounty value & Monetary value, bounty size, or economic significance. \\
\cline{2-4}
 & \multirow{6}{0.12\textwidth}{Comparative\\\& baseline metrics} & vs. non-AI tools & Comparison with traditional tools or non-AI automation. \\
\cline{3-4}
 &  & vs. prior ML / DL & Comparison with earlier ML or DL approaches. \\
\cline{3-4}
 &  & vs. plain LLMs & Comparison with a non-agent LLM under a comparable task setup. \\
\cline{3-4}
 &  & vs. humans & Comparison with analysts, developers, experts, or students. \\
\cline{3-4}
 &  & vs. agent systems & Comparison with other agents, frameworks, or baselines. \\
\cline{3-4}
 &  & Ablation studies & Component removal or variation to estimate contribution. \\
\end{longtable}
\endgroup

\vspace{-4pt}
\subsection{Resulting Taxonomy}\label{ssec:attributionresults}
Following the attribute/item identification and generalization steps, we derived our taxonomy 
(as outlined in Table~\ref{tab:overall-survey-taxonomy}), as elaborated below 
for each of the three high-level aspects separately. 

\vspace{-4pt}
\subsubsection{Approach}\label{ss:taxonomy-approach}

In this aspect, we characterize how LLM-based agents for software and system security are technically constructed and operated. As summarized in Table~\ref{tab:overall-survey-taxonomy}, we identify seven attributes. The first three attributes, {\bf agent architecture}, {\bf agent perception}, and {\bf agent memory}, describe the agent's structural foundation. Agent architecture captures whether the system is single- or multi-agent, how its internal components are organized, and what role the backbone LLM plays. Agent perception captures the inputs and observation mechanisms through which the agent understands code, logs, binaries, traces, environments, or natural-language artifacts. Agent memory captures how security-relevant information is retained, retrieved, updated, and reused during or across tasks.

The remaining attributes describe how the agent acts over time. {\bf Agent reasoning \& planning} records the reasoning paradigms, planning strategies, and uncertainty-handling mechanisms used to derive and revise decisions. {\bf Agent action space} specifies the tools and external operations available to the agent, including analysis tools, execution environments, retrieval systems, security platforms, inter-agent communication, and human interaction. {\bf Agent workflow \& orchestration} captures the process structure, such as sequential, iterative, parallel, hierarchical, or human-involved workflows. Finally, {\bf agent self-improvement \& learning} records whether the agent adapts from feedback, stores successful trajectories, or improves through reward-driven interaction. Together, these attributes treat an LLM security agent as an operational system rather than a single prompt-response model.

\vspace{-4pt}
\subsubsection{Application}\label{ss:taxonomy-application}
In this aspect, we examine the concrete security tasks to which LLM-based agents are applied. The resulting taxonomy contains eleven application domains. Several domains follow the vulnerability lifecycle: {\bf vulnerability detection \& analysis} concerns discovering, localizing, classifying, triaging, or explaining weaknesses; {\bf exploit development \& analysis} concerns constructing, chaining, detecting, or analyzing exploits; {\bf penetration testing \& red-teaming} covers staged or end-to-end offensive assessment; and {\bf vulnerability repair \& hardening} addresses patching, validation, secure coding, code review, and defensive hardening.

Other domains capture specialized security-analysis tasks and operational defense. {\bf Fuzzing \& test generation}, {\bf malware analysis \& detection}, and {\bf reverse engineering} cover testing, malicious-code analysis, and low-level software or protocol understanding. {\bf Intrusion \& threat detection, incident response}, {\bf access control \& authentication assessment}, and {\bf threat intelligence \& security operations (SecOps)} capture defensive monitoring, alert triage, access-policy assessment, threat-intelligence synthesis, and analyst-support workflows. Finally, {\bf model attack generation \& robustness assessment} accounts for work where the LLM, ML component, or agentic workflow itself becomes the target of security assessment. These domains are organized by task objective, so they remain applicable even when different papers use different architectures, datasets, or tools.

\vspace{-4pt}
\subsubsection{Assessment}\label{ss:taxonomy-assessment}
In this aspect, we examine how surveyed works assess LLM-based security agents. Since these agents often act through multi-step trajectories, assessment cannot be limited to final-answer correctness. We therefore organize assessment around both datasets and metrics. For datasets, {\bf dataset origin} distinguishes real-world subjects from synthetic or controlled environments, while {\bf dataset characterisation axes} record properties such as realism, label quality, scale, artifact modality, vulnerability coverage, and accessibility. {\bf Dataset by security task} further indicates which application area the dataset supports, such as vulnerability detection, penetration testing, repair, cyber-defense simulation, malware analysis, or SOC/SOAR workflows.

For metrics, we distinguish outcome-oriented, safety-oriented, process-oriented, and comparative measurements. {\bf Task effectiveness metrics} capture whether the agent achieves the intended security outcome, such as successful detection, exploitation, patching, coverage, or report generation. {\bf Safety \& reliability metrics} capture whether the agent remains robust, bounded, and consistent under adversarial or high-impact conditions. {\bf Agentic process / trajectory metrics} assess how the agent reaches its result, including tool calls, hallucinations, cost, iterations, artifacts, and explanations. Finally, {\bf comparative \& baseline metrics} record whether results are compared against non-AI tools, prior ML/DL methods, plain LLMs without agentic scaffolding, humans, other agent systems, or ablated variants.

\vspace{-6pt}
\section{Paper Attribution} \label{sec:res}
The next phase of our survey process is to apply the attribution framework 
derived and survey aspects/attributes (in $\S$\ref{sec:std}) to the relevant literature identified (in $\S$\ref{sec:ls}). 
In this section, we focus on summarizing the surveyed papers as per (i.e., mapping them to) our taxonomy of LLM-based agents for software and systems security. 
We start with statistics on the collected papers based on their venues ($\S$\ref{subsec:asr}), followed by the summary of the research body in each of the three aspects, Approach ($\S$\ref{ssec:resultapproach}), Application ($\S$\ref{ssec:resultapplication}), and Assessment ($\S$\ref{ssec:resultassessment}), to answer the three RQs that guide our survey. 

Note that our attribution records only what a paper explicitly
describes. A paper is left unmarked under an attribute when the corresponding
capability is optional in, and absent from, its agent design. 

\vspace{-4pt}
\subsection{Paper Distribution by Venues}\label{subsec:asr}
\renewcommand{\arraystretch}{0.85}

\begin{table}[tp]
  \centering
  \caption{Distribution of Publication Venues of Surveyed Papers}
  \vspace{-10pt}
  \resizebox{0.95\columnwidth}{!}{%
    \begin{tabular}{rr|l|r}
    \hline
    \multicolumn{1}{|c|}{\textbf{Type}} & \multicolumn{1}{c|}{\textbf{Acronym}} & \multicolumn{1}{c|}{\textbf{Description}} & \multicolumn{1}{c|}{\textbf{\# Papers}} \\
    \hline
    \multicolumn{1}{|c|}{\multirow{13}[2]{*}{Conference}}
      & \multicolumn{1}{l|}{TrustCom} & IEEE International Conference on Trust, Security and Privacy in Computing and Communications & \multicolumn{1}{r|}{5} \\
    \multicolumn{1}{|c|}{} & \multicolumn{1}{l|}{USENIX Security} & USENIX Security Symposium & \multicolumn{1}{r|}{4} \\
    \multicolumn{1}{|c|}{} & \multicolumn{1}{l|}{ICSE} & IEEE/ACM International Conference on Software Engineering & \multicolumn{1}{r|}{3} \\
    \multicolumn{1}{|c|}{} & \multicolumn{1}{l|}{ICAIC} & IEEE International Conference on AI in Cybersecurity & \multicolumn{1}{r|}{3} \\
    \multicolumn{1}{|c|}{} & \multicolumn{1}{l|}{BigData} & IEEE International Conference on Big Data & \multicolumn{1}{r|}{3} \\
    \multicolumn{1}{|c|}{} & \multicolumn{1}{l|}{MILCOM} & IEEE Military Communications Conference & \multicolumn{1}{r|}{3} \\
    \multicolumn{1}{|c|}{} & \multicolumn{1}{l|}{NeurIPS} & Conference on Neural Information Processing Systems & \multicolumn{1}{r|}{3} \\
    \multicolumn{1}{|c|}{} & \multicolumn{1}{l|}{ICML} & International Conference on Machine Learning & \multicolumn{1}{r|}{3} \\
    \multicolumn{1}{|c|}{} & \multicolumn{1}{l|}{PST} & Annual International Conference on Privacy, Security, and Trust & \multicolumn{1}{r|}{2} \\
    \multicolumn{1}{|c|}{} & \multicolumn{1}{l|}{ICLR} & International Conference on Learning Representations & \multicolumn{1}{r|}{2} \\
    \multicolumn{1}{|c|}{} & \multicolumn{1}{l|}{ICMCIS} & International Conference on Military Communication and Information Systems & \multicolumn{1}{r|}{2} \\
    \multicolumn{1}{|c|}{} & \multicolumn{1}{l|}{AsiaCCS} & ACM Asia Conference on Computer and Communications Security & \multicolumn{1}{r|}{2} \\
    \multicolumn{1}{|c|}{} & \multicolumn{1}{l|}{ICAART} & International Conference on Agents and Artificial Intelligence & \multicolumn{1}{r|}{2} \\
    \hline
    \multicolumn{1}{|l|}{\multirow{2}[2]{*}{Journal}}
      & \multicolumn{1}{l|}{IEEE Access} & IEEE Access & \multicolumn{1}{r|}{3} \\
    \multicolumn{1}{|l|}{} & \multicolumn{1}{l|}{IEEE Commun. Mag.} & IEEE Communications Magazine & \multicolumn{1}{r|}{2} \\
    \hline
    \multicolumn{2}{|l|}{Various other venues} & Venues represented by one surveyed paper each & \multicolumn{1}{r|}{58} \\
    \hline
    \multicolumn{2}{|l|}{All venues} & Total & \multicolumn{1}{r|}{100} \\
    \hline
    \end{tabular}%
    
  }
  \label{tab:vs}
  \vspace{-18pt}
\end{table}%
\renewcommand{\arraystretch}{0.80}
The 100 papers in our dataset were published between 2023 and March 2026 across 73 distinct publication venues. For this analysis, annual editions and minor naming variants of the same venue were consolidated under a single venue name. Table~\ref{tab:vs} lists the 15 venues represented by at least two papers: 13 conferences and two journals. Together, these venues represented by at least two papers account for 42 papers, whereas the remaining 58 venues each contribute one paper and are therefore summarized rather than listed individually.

The distribution reflects the interdisciplinary impact of LLM-based agents for software and systems security. Security- and privacy-focused conferences are the largest category among the recurring venues. TrustCom contributes 5 papers, the largest number from any single venue, followed by USENIX Security with 4. ICAIC contributes 3 papers, while PST and AsiaCCS contribute 2 papers each. Collectively, these 5 security venues account for 16\% of the surveyed papers.

General AI and machine-learning venues are also well represented, contributing 10 papers across four recurring venues. NeurIPS and ICML each contribute 3 papers, while ICLR and ICAART each contribute 2, for a combined total of 10 papers. Software engineering and data-oriented research are represented by ICSE and IEEE BigData, with 3 papers each. MILCOM, ICMCIS, and IEEE Communications Magazine together contribute 7 papers.


\vspace{-4pt}
\subsection{Survey Result: Approach}
\label{ssec:resultapproach}

Next, we present the results of our survey, with one subsection devoted to each of the three highest-level survey aspects: approach ($\S$\ref{ssec:resultapproach}), application ($\S$\ref{ssec:resultapplication}), and assessment ($\S$\ref{ssec:resultassessment}). 
Under each subsection (e.g., $\S$\ref{ssec:resultapproach}: \textbf{Survey Result: Approach}), we discuss the surveyed works along the respective taxonomies (e.g., the taxonomies of agentic approaches as shown in Table~\ref{tab:approach-agent-architecture} and Table~\ref{tab:approach-agent-memory}), using
each sub-subsection (e.g., $\S$\ref{sssec:agentarch}: \textit{Agent architecture}), heading (e.g., \textbf{Agent multiplicity}), underline (e.g., \ul{Single-agent}), italics (e.g., \textit{Collaborative}) as structural anchors for the different (from high to low) levels of the respective categorization (taxonomy). 
Due to space constraints, the lowest-level categories are not always listed in respective taxonomy tables.

\vspace{-4pt}
\subsubsection{Agent architecture}\label{sssec:agentarch}

\begin{table*}[t]
  \centering
  \vspace{-10pt}
  \caption{Paper attribution for Approach: Agent architecture}
  \vspace{-10pt}
  \scriptsize
  \setlength{\tabcolsep}{6.0pt}
  \renewcommand{\arraystretch}{0.72}
\begin{adjustbox}{width=\textwidth,max totalheight=0.88\textheight,center}
    \begin{tabular}{|c|c|c|c|c|c|c|c|c|c|c|c|c|c|c|}

\hline
    & & \multicolumn{2}{c|}{\makecell[c]{Agent multiplicity}} & \multicolumn{7}{c|}{\makecell[c]{Internal structure}} & \multicolumn{4}{c|}{\makecell[c]{Backbone LLM role}} \\
    \cline{3-4}\cline{5-11}\cline{12-15}
    \multicolumn{1}{|c|}{\multirow{-2}{*}{\makecell[c]{Year}}} & 
    \multicolumn{1}{c|}{\multirow{-2}{*}{\makecell[c]{Paper}}} & 
    \multicolumn{1}{c|}{\cellcolor{blue!36}\makecell[c]{Single-\\agent \\ (54\%)}} & 
    \multicolumn{1}{c|}{\cellcolor{blue!31}\makecell[c]{Multi-agent\\systems \\ (46\%)}} & 
    \multicolumn{1}{c|}{\cellcolor{blue!61}\makecell[c]{Modular\\ pipeline \\ (91\%)}} & 
    \multicolumn{1}{c|}{\cellcolor{blue!5}\makecell[c]{Mixture-of-\\agents (MoA) \\ (4\%)}} & 
    \multicolumn{1}{c|}{\cellcolor{blue!5}\makecell[c]{Fine-tuned\\LLM \\ (7\%)}} & 
    \multicolumn{1}{c|}{\cellcolor{blue!13}\makecell[c]{Hybrid\\ML+LLM\\ (20\%)}} & 
    \multicolumn{1}{c|}{\cellcolor{blue!9}\makecell[c]{Guardrail/policy\\enforcement layer \\ (13\%)}} & 
    \multicolumn{1}{c|}{\cellcolor{blue!12}\makecell[c]{Neuro-symbolic\\hybrid \\ (18\%)}} & 
    \multicolumn{1}{c|}{\cellcolor{blue!5}\makecell[c]{Unsp-\\ecified \\ (3\%)}} & 
    \multicolumn{1}{c|}{\cellcolor{blue!5}\makecell[c]{Actor\\only \\ (8\%)}} & 
    \multicolumn{1}{c|}{\cellcolor{blue!59}\makecell[c]{Planner\\+actor\\ (88\%)}} & 
    \multicolumn{1}{c|}{\cellcolor{blue!10}\makecell[c]{Critic/verifier\\/judge\\(15\%)}} & 
    \multicolumn{1}{c|}{\cellcolor{blue!5}\makecell[c]{All-\\in-one \\ (4\%)}} \\
    \hline

\cellcolor{white}2026 & \cite{4} &  & \cmark & \cmark &  &  &  &  & \cmark &  &  & \cmark &  &  \\
\rowcolor{gray!15} \cellcolor{white} & \cite{5} &  & \cmark &  & \cmark &  &  & \cmark &  &  &  & \cmark &  &  \\
\cellcolor{white} & \cite{11} &  & \cmark & \cmark &  &  &  &  &  &  &  & \cmark & \cmark &  \\
\rowcolor{gray!15} \cellcolor{white} & \cite{19} & \cmark &  & \cmark &  &  & \cmark &  &  &  &  & \cmark &  & \cmark \\
\cellcolor{white} & \cite{23lbathaviator} &  & \cmark & \cmark &  & \cmark &  &  & \cmark &  &  & \cmark &  &  \\
\rowcolor{gray!15} \cellcolor{white} & \cite{25zhang2026bountybench} & \cmark &  & \cmark &  &  &  &  &  &  &  & \cmark &  & \cmark \\
\cellcolor{white} & \cite{26RIGAKI2026129987} &  & \cmark & \cmark &  & \cmark & \cmark &  &  &  &  & \cmark &  &  \\
\rowcolor{gray!15} \cellcolor{white} & \cite{36ZOU2026104305} & \cmark &  & \cmark &  &  &  &  &  &  &  & \cmark &  &  \\
\cellcolor{white} & \cite{40zhuo2026cyberzero} & \cmark &  & \cmark &  & \cmark &  &  &  &  &  & \cmark &  &  \\
\rowcolor{gray!15} \cellcolor{white} & \cite{57} & \cmark &  & \cmark &  &  &  &  &  &  &  & \cmark &  &  \\
\cellcolor{white} & \cite{64happe2026llms} & \cmark &  & \cmark &  &  &  &  &  &  &  & \cmark &  &  \\
\rowcolor{gray!15} \cellcolor{white} & \cite{84WANG2026103731} &  & \cmark & \cmark &  &  &  &  & \cmark &  &  & \cmark &  &  \\
\cellcolor{white} & \cite{86} & \cmark &  & \cmark &  &  &  &  &  &  &  & \cmark &  &  \\
\rowcolor{gray!15} \cellcolor{white} & \cite{91zhu-etal-2026-teams} &  & \cmark & \cmark &  &  &  &  &  &  &  & \cmark &  &  \\
\cellcolor{white} & \cite{92} & \cmark &  & \cmark &  &  &  &  &  &  &  & \cmark &  &  \\
\rowcolor{gray!15} \cellcolor{white} & \cite{96} &  & \cmark & \cmark &  &  & \cmark &  &  &  &  & \cmark &  &  \\
\cellcolor{white} & \cite{97} &  & \cmark & \cmark &  &  &  & \cmark & \cmark &  &  & \cmark & \cmark &  \\
\rowcolor{gray!15} \cellcolor{white} & \cite{101} & \cmark &  & \cmark &  &  &  & \cmark &  &  &  & \cmark &  &  \\
\hline
\cellcolor{white}2025 & \cite{2} &  & \cmark &  &  &  & \cmark &  &  &  & \cmark &  &  &  \\
\rowcolor{gray!15} \cellcolor{white} & \cite{3} &  & \cmark & \cmark &  &  & \cmark &  &  &  & \cmark &  &  &  \\
\cellcolor{white} & \cite{6} &  & \cmark & \cmark &  &  &  &  &  &  &  &  & \cmark &  \\
\rowcolor{gray!15} \cellcolor{white} & \cite{7} &  & \cmark & \cmark &  &  &  &  &  &  &  & \cmark & \cmark &  \\
\cellcolor{white} & \cite{8zhang2025agent} & \cmark &  & \cmark &  &  &  & \cmark &  &  &  & \cmark &  & \cmark \\
\rowcolor{gray!15} \cellcolor{white} & \cite{9jie2025agent4vul} &  & \cmark &  &  &  & \cmark &  &  &  & \cmark &  &  &  \\
\cellcolor{white} & \cite{12} &  & \cmark & \cmark &  &  &  &  &  &  &  & \cmark &  &  \\
\rowcolor{gray!15} \cellcolor{white} & \cite{13} & \cmark &  & \cmark &  &  &  &  &  &  &  & \cmark &  &  \\
\cellcolor{white} & \cite{15} & \cmark &  & \cmark &  &  &  &  &  &  & \cmark &  &  &  \\
\rowcolor{gray!15} \cellcolor{white} & \cite{16} &  & \cmark & \cmark &  &  &  &  &  &  &  & \cmark &  &  \\
\cellcolor{white} & \cite{17} &  & \cmark & \cmark &  &  &  &  &  &  &  & \cmark &  & \cmark \\
\rowcolor{gray!15} \cellcolor{white} & \cite{18} & \cmark &  & \cmark &  &  &  &  &  &  &  & \cmark &  &  \\
\cellcolor{white} & \cite{20} &  & \cmark & \cmark &  &  &  &  & \cmark &  &  & \cmark &  &  \\
\rowcolor{gray!15} \cellcolor{white} & \cite{21} & \cmark &  & \cmark &  &  &  &  &  &  &  & \cmark & \cmark &  \\
\cellcolor{white} & \cite{22} & \cmark &  & \cmark &  &  &  &  & \cmark &  &  & \cmark &  &  \\
\rowcolor{gray!15} \cellcolor{white} & \cite{24yildiz-etal-2025-benchmarking} & \cmark &  & \cmark &  &  &  &  &  &  &  & \cmark &  &  \\
\cellcolor{white} & \cite{27} & \cmark &  & \cmark &  &  &  &  &  &  &  & \cmark &  &  \\
\rowcolor{gray!15} \cellcolor{white} & \cite{29} & \cmark &  & \cmark &  &  &  &  & \cmark &  &  & \cmark & \cmark &  \\
\cellcolor{white} & \cite{30} & \cmark &  &  &  &  &  & \cmark &  &  &  &  &  &  \\
\rowcolor{gray!15} \cellcolor{white} & \cite{31} &  & \cmark & \cmark &  &  &  &  &  &  &  & \cmark &  &  \\
\cellcolor{white} & \cite{32} &  & \cmark & \cmark &  &  &  & \cmark &  &  &  & \cmark &  &  \\
\rowcolor{gray!15} \cellcolor{white} & \cite{33} &  & \cmark &  & \cmark &  &  &  &  &  &  &  & \cmark &  \\
\cellcolor{white} & \cite{34} &  & \cmark & \cmark &  & \cmark &  &  &  &  &  &  & \cmark &  \\
\rowcolor{gray!15} \cellcolor{white} & \cite{35} &  & \cmark & \cmark &  &  &  &  &  &  &  & \cmark &  &  \\
\cellcolor{white} & \cite{37app15169096} &  & \cmark & \cmark &  &  &  &  &  &  &  & \cmark &  &  \\
\rowcolor{gray!15} \cellcolor{white} & \cite{38zhu2025cvebenchbenchmarkaiagents} & \cmark &  & \cmark &  &  &  &  &  &  &  & \cmark &  &  \\
\cellcolor{white} & \cite{39} & \cmark &  & \cmark &  &  & \cmark &  &  &  &  & \cmark &  &  \\
\rowcolor{gray!15} \cellcolor{white} & \cite{41jiao2025deepvulhunter} & \cmark &  & \cmark &  &  &  &  &  &  & \cmark &  &  &  \\
\cellcolor{white} & \cite{42LOEVENICH2025111162} & \cmark &  & \cmark &  &  & \cmark &  &  &  &  & \cmark &  &  \\
\rowcolor{gray!15} \cellcolor{white} & \cite{45} &  & \cmark & \cmark & \cmark &  &  & \cmark &  &  &  & \cmark &  &  \\
\cellcolor{white} & \cite{46abramovich2025enigma} & \cmark &  & \cmark &  &  &  &  &  &  &  & \cmark &  &  \\
\rowcolor{gray!15} \cellcolor{white} & \cite{47chowdhry2025evaluating} & \cmark &  &  &  &  & \cmark &  &  &  &  & \cmark &  &  \\
\cellcolor{white} & \cite{48} & \cmark &  & \cmark &  &  & \cmark &  &  &  &  & \cmark &  &  \\
\rowcolor{gray!15} \cellcolor{white} & \cite{49} &  & \cmark & \cmark &  &  &  & \cmark &  &  &  & \cmark &  &  \\
\cellcolor{white} & \cite{51} &  & \cmark & \cmark &  &  & \cmark & \cmark &  &  &  & \cmark &  &  \\
\rowcolor{gray!15} \cellcolor{white} & \cite{53} & \cmark &  & \cmark &  &  &  &  &  &  &  & \cmark &  &  \\
\cellcolor{white} & \cite{55} &  & \cmark & \cmark &  &  &  &  &  &  &  & \cmark &  &  \\
\rowcolor{gray!15} \cellcolor{white} & \cite{56} & \cmark &  & \cmark &  &  &  & \cmark &  &  &  & \cmark &  &  \\
\cellcolor{white} & \cite{58} &  & \cmark & \cmark &  &  &  &  & \cmark &  &  & \cmark &  &  \\
\rowcolor{gray!15} \cellcolor{white} & \cite{59} & \cmark &  & \cmark &  &  &  & \cmark &  &  &  & \cmark &  &  \\
\cellcolor{white} & \cite{60} &  & \cmark & \cmark &  &  & \cmark &  &  &  &  & \cmark &  &  \\
\rowcolor{gray!15} \cellcolor{white} & \cite{61} &  & \cmark & \cmark &  &  &  &  &  &  &  & \cmark &  &  \\
\cellcolor{white} & \cite{62} & \cmark &  &  &  &  &  &  &  & \cmark & \cmark &  &  &  \\
\rowcolor{gray!15} \cellcolor{white} & \cite{65} &  & \cmark & \cmark &  & \cmark &  &  &  &  &  & \cmark &  &  \\
\cellcolor{white} & \cite{67} & \cmark &  & \cmark &  &  &  & \cmark & \cmark &  &  & \cmark &  &  \\
\rowcolor{gray!15} \cellcolor{white} & \cite{68} &  & \cmark & \cmark &  &  &  &  &  &  &  & \cmark &  &  \\
\cellcolor{white} & \cite{69} & \cmark &  & \cmark &  &  &  &  & \cmark &  &  & \cmark &  &  \\
\rowcolor{gray!15} \cellcolor{white} & \cite{70} & \cmark &  & \cmark &  &  &  &  &  &  &  & \cmark &  &  \\
\cellcolor{white} & \cite{71} &  & \cmark & \cmark &  &  &  &  & \cmark &  &  & \cmark &  &  \\
\rowcolor{gray!15} \cellcolor{white} & \cite{72} &  & \cmark & \cmark &  &  &  &  &  &  &  & \cmark &  &  \\
\cellcolor{white} & \cite{73} &  & \cmark & \cmark &  &  & \cmark &  &  &  &  & \cmark &  &  \\
\rowcolor{gray!15} \cellcolor{white} & \cite{75} & \cmark &  & \cmark &  &  &  &  & \cmark &  &  & \cmark & \cmark &  \\
\cellcolor{white} & \cite{78} &  & \cmark & \cmark &  &  &  &  &  &  &  & \cmark &  &  \\
\rowcolor{gray!15} \cellcolor{white} & \cite{80} &  & \cmark & \cmark & \cmark &  & \cmark &  &  &  &  & \cmark &  &  \\
\cellcolor{white} & \cite{81Cao_Huang_Li_Huilin_He_Oo_Hooi_2025} & \cmark &  & \cmark &  &  &  &  &  &  &  & \cmark &  &  \\
\rowcolor{gray!15} \cellcolor{white} & \cite{83} & \cmark &  & \cmark &  &  &  &  & \cmark &  &  & \cmark &  &  \\
\cellcolor{white} & \cite{87} & \cmark &  & \cmark &  &  & \cmark &  &  &  &  & \cmark &  &  \\
\rowcolor{gray!15} \cellcolor{white} & \cite{88} & \cmark &  & \cmark &  &  &  &  &  &  &  & \cmark &  &  \\
\cellcolor{white} & \cite{89guo2025repoauditautonomousllmagentrepositorylevel} & \cmark &  & \cmark &  &  &  &  & \cmark &  &  & \cmark & \cmark &  \\
\rowcolor{gray!15} \cellcolor{white} & \cite{90} & \cmark &  &  &  &  &  &  &  & \cmark &  & \cmark &  &  \\
\cellcolor{white} & \cite{93info16050365} & \cmark &  & \cmark &  &  &  &  &  &  &  & \cmark &  &  \\
\rowcolor{gray!15} \cellcolor{white} & \cite{94} & \cmark &  & \cmark &  &  &  &  &  &  &  & \cmark &  &  \\
\cellcolor{white} & \cite{99} &  & \cmark & \cmark &  & \cmark &  &  & \cmark &  &  & \cmark & \cmark &  \\
\rowcolor{gray!15} \cellcolor{white} & \cite{100} & \cmark &  & \cmark &  &  &  &  & \cmark &  &  & \cmark & \cmark &  \\
\hline
\cellcolor{white}2024 & \cite{1} & \cmark &  & \cmark &  &  &  &  &  &  & \cmark &  &  &  \\
\rowcolor{gray!15} \cellcolor{white} & \cite{10debenedetti2024agentdojo} & \cmark &  & \cmark &  &  &  &  &  &  &  & \cmark &  &  \\
\cellcolor{white} & \cite{14chen2024agentpoison} & \cmark &  &  &  &  &  &  &  & \cmark &  & \cmark &  &  \\
\rowcolor{gray!15} \cellcolor{white} & \cite{28} & \cmark &  & \cmark &  &  & \cmark &  &  &  &  & \cmark &  &  \\
\cellcolor{white} & \cite{43} & \cmark &  & \cmark &  &  & \cmark &  &  &  &  & \cmark &  &  \\
\rowcolor{gray!15} \cellcolor{white} & \cite{50} & \cmark &  & \cmark &  & \cmark &  &  &  &  & \cmark &  &  &  \\
\cellcolor{white} & \cite{54} & \cmark &  & \cmark &  &  &  &  &  &  &  & \cmark &  &  \\
\rowcolor{gray!15} \cellcolor{white} & \cite{63zemicheal2024llm} &  & \cmark & \cmark &  &  &  &  &  &  &  & \cmark &  &  \\
\cellcolor{white} & \cite{66} &  & \cmark & \cmark &  &  &  &  &  &  &  & \cmark &  &  \\
\rowcolor{gray!15} \cellcolor{white} & \cite{74icaart24} & \cmark &  & \cmark &  &  &  &  &  &  &  & \cmark &  &  \\
\cellcolor{white} & \cite{76} &  & \cmark & \cmark &  &  & \cmark &  &  &  &  & \cmark &  &  \\
\rowcolor{gray!15} \cellcolor{white} & \cite{77} &  & \cmark & \cmark &  &  &  &  &  &  &  & \cmark &  &  \\
\cellcolor{white} & \cite{79} & \cmark &  & \cmark &  &  &  &  &  &  &  & \cmark &  &  \\
\rowcolor{gray!15} \cellcolor{white} & \cite{82rigaki2024prompt} & \cmark &  & \cmark &  &  &  &  &  &  &  & \cmark &  &  \\
\cellcolor{white} & \cite{85} &  & \cmark & \cmark &  &  &  &  &  &  &  & \cmark & \cmark &  \\
\rowcolor{gray!15} \cellcolor{white} & \cite{95} &  & \cmark & \cmark &  &  & \cmark &  &  &  &  & \cmark &  &  \\
\cellcolor{white} & \cite{98} &  & \cmark & \cmark &  &  &  & \cmark & \cmark &  &  & \cmark & \cmark &  \\
\hline
\rowcolor{gray!15} \cellcolor{white}2023 & \cite{52} & \cmark &  & \cmark &  &  & \cmark &  & \cmark &  &  & \cmark & \cmark &  \\
\hline
    \end{tabular}%
\end{adjustbox}
  \label{tab:approach-agent-architecture}%
  \vspace{-18pt}
\end{table*}%

Agents designed for software/system security have adopted various architectural designs in terms of agent multiplicity (i.e., single- or multi-agent), internal structure, and the role of the backbone LLM, as summarized in Table~\ref{tab:approach-agent-architecture}.

\noindent
{\bf Agent multiplicity.} 
Agent multiplicity describes whether a system employs a single autonomous agent or coordinates multiple interacting agents. 
In this dimension, architectures are classified as either single-agent or multi-agent.

\ul{Single-agent} designs remain the most common architectural choice (e.g., \cite{1,8zhang2025agent,10debenedetti2024agentdojo,38zhu2025cvebenchbenchmarkaiagents,39,40zhuo2026cyberzero,41jiao2025deepvulhunter,42LOEVENICH2025111162,43,46abramovich2025enigma,48,54,56,57,59,62,64happe2026llms,67,86,87,88}). In these systems, one agent typically handles the end-to-end loop of interpreting the task, invoking tools, and producing the final security outcome. For example, \textsc{RepoAudit} uses one autonomous repository-auditing agent that incrementally explores relevant functions, stores intermediate findings in memory, and validates candidate bug reports without delegating the task to peer agents \cite{89guo2025repoauditautonomousllmagentrepositorylevel}. Similarly, VVF-AI adopts a single-agent workflow in which an LLM-based agent generates and executes vulnerability-verification logic to filter false positives from detection results \cite{100}. 
PentestGPT~\cite{79}
adopts a stratified variant in which a Reasoning Module maintains a high-level
Pentesting Task Tree while a Generation Module translates strategic goals into
concrete commands, addressing context-loss by preventing the planner from being
overwhelmed by execution detail.

\ul{Multi-agent} design is also a major architectural pattern across the surveyed
papers~\cite{2,3,4,5,6,7,9jie2025agent4vul,11,12,16,17,20,23lbathaviator,31,32,
33,34,35,37app15169096,45,49,51,55,58,60,61,63zemicheal2024llm,65,66,68,71,72,
73,76,77,78,80,84WANG2026103731,91zhu-etal-2026-teams,95,96,97,98,99},
reflecting the increasing adoption of collaborative agent designs for complex security tasks.

In particular, the surveyed studies of multi-agent architecture fall into three types of structures:

\textit{Collaborative.} A frequently used coordination strategy is task-specialised
collaboration, where agents each own a distinct sub-task and their outputs are
integrated to drive the overall workflow. For example, PentestAgent~\cite{78}
deploys separate agents for intelligence gathering, vulnerability analysis, and
exploitation, each equipped with a customised toolset tied together through a
RAG-augmented knowledge pipeline. Similarly, AutoRestTest~\cite{2} assigns four
agents --- API, dependency, parameter, and value --- to optimise orthogonal
dimensions of REST API testing via multi-agent reinforcement learning, achieving
coverage gains of 12--27\% over single-agent baselines.


\textit{Hierarchical.} Several studies structure collaboration vertically,
with a high-level planning agent delegating to specialised
sub-agents~\cite{4,12,17,20,31,35,37app15169096,61,71,73,84WANG2026103731,
91zhu-etal-2026-teams,95,97,98}. HPTSA~\cite{91zhu-etal-2026-teams} is a
representative example: a planning agent analyses the target and autonomously
launches specialised subagents for zero-day exploitation subtasks. 

\textit{Peer-to-peer.} Some studies adopt a flat topology in which agents of
equal standing exchange information and resolve disagreements through debate or
voting~\cite{5,17,33,97}.

\noindent
{\bf Internal structure.}
The internal structure of an LLM-based security agent describes how the system decomposes the agent into functional components, and how it combines LLM reasoning with other computational mechanisms.

\ul{Modular pipeline} is the dominant internal organization across the architecture taxonomy. Within this broad category, \textit{planner/executor modules} separate strategic decision-making from concrete action. For example, \textsc{PentestAgent} uses a planning agent to consume reconnaissance results and generate exploitation plans, while other agents perform intelligence gathering, vulnerability analysis, and exploitation~\cite{78}. PTFusion follows a related pattern in web penetration testing: a MasterAgent maintains strategic planning, while ReconAgent and AttackAgent perform tactical tool-mediated execution through MCP-enabled tool interfaces~\cite{84WANG2026103731}.

\textit{Verifier/critic modules} are another recurring modular pattern. They are introduced because security agents frequently generate uncertain intermediate claims: a suspected vulnerability, a candidate exploit, a patch, or a data-flow fact may look plausible but still be wrong. \textsc{RepoAudit}, for instance, consists of initiator, explorer, and validator components; the validator checks generated data-flow facts and path-condition satisfiability before accepting candidate bug reports~\cite{89guo2025repoauditautonomousllmagentrepositorylevel}. \textsc{PatchAgent} similarly couples fault localization and patch generation with a verifier that tests candidate patches and feeds counterexamples back into the repair loop~\cite{75}.

\textit{Summariser/reporter modules} appear when the agent must transform low-level observations into analyst-facing outputs~\cite{3,11,17,21,32,36ZOU2026104305,51,55,60,63zemicheal2024llm,68,99}. For example, \textsc{Cognitive SOC} uses specialised Triage, Enrichment, and StoryWeaver agents: after triage classifies an alert and enrichment gathers contextual evidence, StoryWeaver turns the evidence into an incident narrative with citation chains~\cite{32}.

\ul{Mixture-of-agents (MoA)} structures combine multiple LLM agents or model instances as a deliberate ensemble rather than as a simple linear pipeline~\cite{5,33,45,80}. In these systems, \textit{model diversity and response aggregation} are part of the architecture. The Agentic Multi-LLM Network is a representative example: multiple LLM-equipped network nodes propose network-optimization strategies, and the framework uses collaboration plus a consensus-oriented trust layer to select reliable responses~\cite{5}. Perses also reflects this logic by assigning different small local models to distinct privilege-escalation roles, relying on heterogeneity rather than a single large model to carry the task~\cite{80}.

\ul{Fine-tuned LLM component} architectures include systems that adapt model behavior through domain-specific training data or specialised model components~\cite{23lbathaviator,26RIGAKI2026129987,40zhuo2026cyberzero,34,65,99,50}. In this category, \textit{task-specific model adaptation} supplements prompting and tool use. \textsc{Cyber-Zero}, for example, synthesizes cybersecurity trajectories from public CTF writeups and persona-driven simulation, then uses those trajectories to train open-weight cybersecurity agents~\cite{40zhuo2026cyberzero}. AVIATOR similarly targets vulnerability-injection workflows and dataset generation, using agentic processing to create high-fidelity security data for downstream training and assessment~\cite{23lbathaviator}.

\ul{Hybrid ML + LLM architecture} combines LLM reasoning with other machine-learning, reinforcement-learning, or learned prediction components~\cite{19,26RIGAKI2026129987,96,2,3,9jie2025agent4vul,39,42LOEVENICH2025111162,47chowdhry2025evaluating,48,51,60,73,80,87,28,43,76,95,52}. Here, \textit{statistical learning and language reasoning} play complementary roles. AutoRestTest is illustrative: multi-agent reinforcement learning coordinates API, dependency, parameter, and value agents, while the LLM is used to generate domain-specific parameter values for REST API testing~\cite{2}. In autonomous cyber-defense work, hybrid architectures combine DRL policies, knowledge graphs, and LLM interfaces so that learned policies can act in cyber ranges while LLMs provide explanation, retrieval, or policy guidance~\cite{42LOEVENICH2025111162,43,95}.

\ul{Guardrail/policy enforcement layer} architectures add explicit mechanisms that constrain, audit, or filter agent behavior~\cite{5,97,101,8zhang2025agent,30,32,45,49,51,56,59,67,98}. In these papers, \textit{trust and safety control} is part of the internal design rather than an external assessment concern. The Agentic Multi-LLM Network uses blockchain-based audit trails and Byzantine Fault Tolerance consensus to reject malicious or unreliable proposals from distributed LLM agents~\cite{5}. Agent Security Bench formalizes attacks and defenses against agent stages such as prompt handling, tool use, and memory retrieval, making guardrail design a central architectural dimension~\cite{8zhang2025agent}.

\ul{Neuro-symbolic hybrid} architectures combine neural LLM components with symbolic analysis, formal reasoning, static analysis, or verification tools~\cite{4,23lbathaviator,84WANG2026103731,97,20,22,29,58,67,69,71,75,83,89guo2025repoauditautonomousllmagentrepositorylevel,99,100,98,52}. The typical grouping is \textit{LLM generation plus symbolic checking}. \textsc{PropertyGPT} is a clear example: the LLM generates smart-contract properties, while compilation, static analysis, and formal verification components check whether the generated properties are usable and meaningful~\cite{83}. \textsc{RepoAudit} also fits this category because its LLM-guided repository exploration is constrained by data-flow facts and satisfiability checks~\cite{89guo2025repoauditautonomousllmagentrepositorylevel}.

\ul{Unspecified internal structure} is used only when a paper establishes that an agent is present but does not describe its internal organization precisely enough to distinguish a modular, ensemble, guardrailed, fine-tuned, or hybrid design~\cite{62,90,14chen2024agentpoison}. ``Unspecified'' records insufficient evidence, not meaning that the system has no internal structure.

\noindent
{\bf Backbone LLM role.}
The backbone LLM role captures what functional responsibility the LLM has inside the agent. Some systems use the LLM as a generator inside a larger architecture; others place it at the center of planning and tool use; still others use LLMs as judges, verifiers, or all-purpose agent backbones.

\ul{Actor only} roles use the LLM mainly to produce a bounded artifact or local action, while another mechanism handles the broader workflow~\cite{2,3,9jie2025agent4vul,15,41jiao2025deepvulhunter,62,1,50}. The common grouping is \textit{LLM as generator}. AutoRestTest is a representative case: the multi-agent RL system and semantic dependency graph coordinate REST API testing, while the LLM is used primarily by the value agent to generate realistic parameter values satisfying API constraints~\cite{2}. Similarly, IDS-rule generation uses the LLM to produce and generalize detection rules from vulnerability reports, payloads, and existing rules, but the surrounding IDS assessment and rule-deployment logic remain outside the LLM itself~\cite{1}. In this role, the LLM improves expressiveness and domain adaptation without becoming the main planner.

\ul{Planner + actor} is the most expansive role: the LLM plans steps, selects tools, issues commands, interprets observations, and revises subsequent actions. In this grouping, \textit{LLM as autonomous operator} is the central design idea. \textsc{RepairAgent}, for example, treats the LLM as an autonomous repair agent that plans tool use, retrieves repair ingredients, generates patches, and validates candidate repairs~\cite{88}. PTFusion similarly uses LLM agents to plan and execute web-penetration-testing tasks while maintaining a dynamic knowledge graph over the target environment~\cite{84WANG2026103731}.

\ul{Critic / verifier / judge} roles use the LLM, or an LLM-supported module, to assess another component's output rather than to produce the primary action~\cite{11,97,6,7,21,29,33,34,75,89guo2025repoauditautonomousllmagentrepositorylevel,99,100,85,98,52}. The narrative grouping here is \textit{LLM as quality-control layer}. In \textsc{RepoAudit}, the validator critiques the explorer's generated data-flow facts and checks path-condition satisfiability for bug candidates, reducing false-positive reports~\cite{89guo2025repoauditautonomousllmagentrepositorylevel}. In smart-contract and patching systems, critic or verifier roles similarly help decide whether a vulnerability claim, generated mitigation, or patch should be trusted~\cite{6,33,34,75,99,100}.

\ul{All-in-one} backbones absorb most roles into a single agent loop rather than clearly separating planner, executor, and verifier responsibilities~\cite{19,25zhang2026bountybench,8zhang2025agent,17}. The grouping can be described as \textit{LLM as unified task engine}. BountyBench exemplifies this style by assessing agents on real-world cybersecurity systems where the same agent loop must inspect code or services, reason about vulnerabilities, execute actions, and submit detection, exploitation, or patching outputs~\cite{25zhang2026bountybench}. Agent Security Bench similarly treats the agent as a broad system whose prompt handling, tool use, and memory stages can all become attack surfaces~\cite{8zhang2025agent}.

\enlargethispage{\baselineskip}
\vspace{-8pt}
\subsubsection{Agent memory}
Agents designed for software/system security also differ in how they store,
retrieve, and revise information, in terms of memory types, memory operations,
and security-specific memory, as summarized in
Table~\ref{tab:approach-agent-memory}.
\begin{table*}[t]
  \centering
  \vspace{-12pt}
  \caption{Paper attribution for Approach: Agent memory}
  \vspace{-12pt}
  \scriptsize
  \setlength{\tabcolsep}{1.5pt}
  \renewcommand{\arraystretch}{0.70}
\begin{adjustbox}{width=\textwidth,max totalheight=0.88\textheight,center}
    \begin{tabular}{|c|c|c|c|c|c|c|c|c|c|c|c|c|c|c|}
    \hline
\rowcolor{white} && 
\multicolumn{5}{c|}{\makecell[c]{Memory types}} & \multicolumn{3}{c|}{\makecell[c]{Memory operations}} & \multicolumn{5}{c|}{\makecell[c]{Security-specific memory}} \\
\cline{3-7}\cline{8-10}\cline{11-15}

\rowcolor{white} 
\multicolumn{1}{|c|}{\multirow{-4}{*}[-0.2ex]{\makecell[c]{Year}}} & 
\multicolumn{1}{c|}{\multirow{-4}{*}[-0.2ex]{\makecell[c]{Paper}}} & 
\multicolumn{1}{c|}{\cellcolor{blue!40}\makecell[c]{In-context working\\memory \\ (60\%)}} & \multicolumn{1}{c|}{\cellcolor{blue!5}\makecell[c]{External\\short-term memory \\ (7\%)}} & \multicolumn{1}{c|}{\cellcolor{blue!38}\makecell[c]{Long-term /\\persistent memory \\ (57\%)}} & \multicolumn{1}{c|}{\cellcolor{blue!7}\makecell[c]{Knowledge graph\\memory \\ (10\%)}} & \multicolumn{1}{c|}{\cellcolor{blue!67}\makecell[c]{Parametric\\memory \\ (100\%)}} & \multicolumn{1}{c|}{\cellcolor{blue!41}\makecell[c]{Read \\ (61\%)}} & \multicolumn{1}{c|}{\cellcolor{blue!5}\makecell[c]{Write \\ (7\%)}} & \multicolumn{1}{c|}{\cellcolor{blue!28}\makecell[c]{Update /\\reflection \\ (42\%)}} & \multicolumn{1}{c|}{\cellcolor{blue!17}\makecell[c]{Vulnerability\\knowledge bases \\ (25\%)}} & \multicolumn{1}{c|}{\cellcolor{blue!15}\makecell[c]{Code repositories\\and patch histories \\ (23\%)}} & \multicolumn{1}{c|}{\cellcolor{blue!22}\makecell[c]{Execution traces\\and fuzzing logs \\ (33\%)}} & \multicolumn{1}{c|}{\cellcolor{blue!8}\makecell[c]{Threat intelligence\\feeds \\ (12\%)}} & \multicolumn{1}{c|}{\cellcolor{blue!33}\makecell[c]{Tool\\documentation \\ (49\%)}} \\
    \hline

\cellcolor{white}2026 & \cite{4} &  &  & \cmark & \cmark & \cmark & \cmark &  & \cmark & \cmark &  &  &  & \cmark \\
\rowcolor{gray!15} \cellcolor{white} & \cite{5} & \cmark &  & \cmark &  & \cmark &  &  & \cmark &  &  &  &  & \cmark \\
\cellcolor{white} & \cite{11} &  &  &  &  & \cmark & \cmark &  &  &  & \cmark & \cmark &  &  \\
\rowcolor{gray!15} \cellcolor{white} & \cite{19} & \cmark &  & \cmark &  & \cmark & \cmark &  & \cmark &  &  & \cmark &  &  \\
\cellcolor{white} & \cite{23lbathaviator} &  &  & \cmark &  & \cmark & \cmark &  &  & \cmark & \cmark &  &  & \cmark \\
\rowcolor{gray!15} \cellcolor{white} & \cite{25zhang2026bountybench} &  &  &  &  & \cmark &  & \cmark &  &  & \cmark & \cmark &  &  \\
\cellcolor{white} & \cite{26RIGAKI2026129987} & \cmark &  & \cmark &  & \cmark & \cmark &  &  &  &  &  &  &  \\
\rowcolor{gray!15} \cellcolor{white} & \cite{36ZOU2026104305} & \cmark &  & \cmark &  & \cmark & \cmark &  &  &  & \cmark & \cmark &  & \cmark \\
\cellcolor{white} & \cite{40zhuo2026cyberzero} & \cmark &  & \cmark &  & \cmark & \cmark &  &  &  & \cmark &  &  &  \\
\rowcolor{gray!15} \cellcolor{white} & \cite{57} &  &  & \cmark &  & \cmark & \cmark &  &  &  &  &  &  &  \\
\cellcolor{white} & \cite{64happe2026llms} & \cmark &  & \cmark &  & \cmark &  &  & \cmark &  & \cmark &  &  &  \\
\rowcolor{gray!15} \cellcolor{white} & \cite{84WANG2026103731} & \cmark &  & \cmark & \cmark & \cmark &  &  & \cmark &  &  & \cmark &  & \cmark \\
\cellcolor{white} & \cite{86} & \cmark &  & \cmark &  & \cmark & \cmark &  & \cmark &  &  &  &  &  \\
\rowcolor{gray!15} \cellcolor{white} & \cite{91zhu-etal-2026-teams} &  &  & \cmark &  & \cmark & \cmark &  &  &  & \cmark & \cmark &  &  \\
\cellcolor{white} & \cite{92} & \cmark &  & \cmark &  & \cmark &  &  & \cmark &  &  &  &  &  \\
\rowcolor{gray!15} \cellcolor{white} & \cite{96} &  &  &  &  & \cmark & \cmark &  &  &  &  &  &  &  \\
\cellcolor{white} & \cite{97} & \cmark &  & \cmark &  & \cmark &  &  & \cmark &  &  &  &  & \cmark \\
\rowcolor{gray!15} \cellcolor{white} & \cite{101} & \cmark &  & \cmark &  & \cmark &  &  & \cmark &  & \cmark & \cmark &  & \cmark \\
\hline
\cellcolor{white}2025 & \cite{2} & \cmark &  &  & \cmark & \cmark & \cmark &  &  &  &  &  &  & \cmark \\
\rowcolor{gray!15} \cellcolor{white} & \cite{3} & \cmark &  &  &  & \cmark & \cmark &  &  &  &  &  &  &  \\
\cellcolor{white} & \cite{6} &  & \cmark & \cmark &  & \cmark &  &  & \cmark & \cmark &  &  &  &  \\
\rowcolor{gray!15} \cellcolor{white} & \cite{7} &  &  & \cmark &  & \cmark &  &  & \cmark &  &  &  &  &  \\
\cellcolor{white} & \cite{8zhang2025agent} & \cmark & \cmark & \cmark &  & \cmark & \cmark &  &  &  & \cmark & \cmark &  &  \\
\rowcolor{gray!15} \cellcolor{white} & \cite{9jie2025agent4vul} & \cmark &  &  &  & \cmark & \cmark &  &  &  &  &  &  &  \\
\cellcolor{white} & \cite{12} &  &  &  & \cmark & \cmark & \cmark &  &  &  &  & \cmark & \cmark & \cmark \\
\rowcolor{gray!15} \cellcolor{white} & \cite{13} &  &  & \cmark &  & \cmark & \cmark &  &  & \cmark &  &  &  & \cmark \\
\cellcolor{white} & \cite{15} & \cmark &  &  &  & \cmark & \cmark &  &  &  &  &  &  & \cmark \\
\rowcolor{gray!15} \cellcolor{white} & \cite{16} & \cmark &  &  &  & \cmark &  &  & \cmark &  &  &  & \cmark &  \\
\cellcolor{white} & \cite{17} &  &  &  &  & \cmark & \cmark &  &  &  &  &  &  &  \\
\rowcolor{gray!15} \cellcolor{white} & \cite{18} &  &  & \cmark & \cmark & \cmark & \cmark &  &  & \cmark &  &  & \cmark & \cmark \\
\cellcolor{white} & \cite{20} & \cmark &  & \cmark & \cmark & \cmark & \cmark &  &  &  &  &  &  & \cmark \\
\rowcolor{gray!15} \cellcolor{white} & \cite{21} & \cmark &  & \cmark &  & \cmark & \cmark &  &  & \cmark &  &  &  & \cmark \\
\cellcolor{white} & \cite{22} & \cmark &  &  &  & \cmark & \cmark &  &  & \cmark &  & \cmark &  & \cmark \\
\rowcolor{gray!15} \cellcolor{white} & \cite{24yildiz-etal-2025-benchmarking} & \cmark &  & \cmark &  & \cmark & \cmark &  &  & \cmark & \cmark &  &  &  \\
\cellcolor{white} & \cite{27} & \cmark &  & \cmark &  & \cmark & \cmark &  &  & \cmark &  & \cmark &  &  \\
\rowcolor{gray!15} \cellcolor{white} & \cite{29} & \cmark & \cmark &  &  & \cmark &  &  & \cmark &  &  & \cmark &  & \cmark \\
\cellcolor{white} & \cite{30} & \cmark &  &  &  & \cmark & \cmark &  &  &  &  & \cmark &  & \cmark \\
\rowcolor{gray!15} \cellcolor{white} & \cite{31} &  &  & \cmark &  & \cmark &  &  & \cmark &  &  &  & \cmark & \cmark \\
\cellcolor{white} & \cite{32} & \cmark &  & \cmark &  & \cmark &  & \cmark &  & \cmark &  &  & \cmark & \cmark \\
\rowcolor{gray!15} \cellcolor{white} & \cite{33} & \cmark &  &  &  & \cmark & \cmark &  &  &  &  &  &  &  \\
\cellcolor{white} & \cite{34} & \cmark &  &  &  & \cmark & \cmark &  &  & \cmark &  &  &  &  \\
\rowcolor{gray!15} \cellcolor{white} & \cite{35} & \cmark &  &  &  & \cmark & \cmark &  &  &  &  & \cmark &  &  \\
\cellcolor{white} & \cite{37app15169096} &  &  & \cmark &  & \cmark & \cmark &  & \cmark & \cmark &  & \cmark &  & \cmark \\
\rowcolor{gray!15} \cellcolor{white} & \cite{38zhu2025cvebenchbenchmarkaiagents} &  &  & \cmark &  & \cmark & \cmark &  &  & \cmark & \cmark &  &  & \cmark \\
\cellcolor{white} & \cite{39} &  &  & \cmark & \cmark & \cmark & \cmark &  & \cmark &  &  &  &  &  \\
\rowcolor{gray!15} \cellcolor{white} & \cite{41jiao2025deepvulhunter} & \cmark &  & \cmark &  & \cmark & \cmark &  &  & \cmark & \cmark &  &  &  \\
\cellcolor{white} & \cite{42LOEVENICH2025111162} & \cmark &  & \cmark & \cmark & \cmark &  & \cmark & \cmark & \cmark &  &  & \cmark & \cmark \\
\rowcolor{gray!15} \cellcolor{white} & \cite{45} & \cmark &  & \cmark &  & \cmark &  &  & \cmark &  &  & \cmark &  & \cmark \\
\cellcolor{white} & \cite{46abramovich2025enigma} & \cmark &  &  &  & \cmark & \cmark &  &  &  & \cmark &  &  & \cmark \\
\rowcolor{gray!15} \cellcolor{white} & \cite{47chowdhry2025evaluating} & \cmark &  &  &  & \cmark & \cmark &  &  &  &  &  &  &  \\
\cellcolor{white} & \cite{48} &  &  &  &  & \cmark & \cmark &  &  &  &  &  &  &  \\
\rowcolor{gray!15} \cellcolor{white} & \cite{49} & \cmark &  & \cmark &  & \cmark &  &  & \cmark &  &  &  &  &  \\
\cellcolor{white} & \cite{51} &  &  &  &  & \cmark &  &  & \cmark &  & \cmark & \cmark &  &  \\
\rowcolor{gray!15} \cellcolor{white} & \cite{53} & \cmark &  &  &  & \cmark & \cmark &  &  &  &  &  &  &  \\
\cellcolor{white} & \cite{55} &  &  &  &  & \cmark & \cmark &  &  &  &  & \cmark &  & \cmark \\
\rowcolor{gray!15} \cellcolor{white} & \cite{56} & \cmark &  &  &  & \cmark &  &  & \cmark &  &  & \cmark &  &  \\
\cellcolor{white} & \cite{58} &  &  & \cmark &  & \cmark &  &  & \cmark &  &  &  &  & \cmark \\
\rowcolor{gray!15} \cellcolor{white} & \cite{59} & \cmark &  & \cmark &  & \cmark &  &  & \cmark &  &  &  &  &  \\
\cellcolor{white} & \cite{60} & \cmark &  &  &  & \cmark &  &  & \cmark &  &  & \cmark &  &  \\
\rowcolor{gray!15} \cellcolor{white} & \cite{61} &  &  &  &  & \cmark &  &  & \cmark &  &  & \cmark &  &  \\
\cellcolor{white} & \cite{62} &  & \cmark & \cmark &  & \cmark & \cmark &  &  &  &  &  &  &  \\
\rowcolor{gray!15} \cellcolor{white} & \cite{65} & \cmark &  &  &  & \cmark & \cmark &  &  &  &  &  & \cmark & \cmark \\
\cellcolor{white} & \cite{67} &  &  &  &  & \cmark & \cmark &  &  &  &  &  & \cmark &  \\
\rowcolor{gray!15} \cellcolor{white} & \cite{68} &  &  &  &  & \cmark & \cmark &  &  &  &  &  & \cmark &  \\
\cellcolor{white} & \cite{69} & \cmark &  & \cmark &  & \cmark &  & \cmark &  & \cmark & \cmark & \cmark &  &  \\
\rowcolor{gray!15} \cellcolor{white} & \cite{70} & \cmark &  & \cmark &  & \cmark &  &  & \cmark &  &  & \cmark &  & \cmark \\
\cellcolor{white} & \cite{71} &  &  & \cmark &  & \cmark &  &  & \cmark &  &  & \cmark &  &  \\
\rowcolor{gray!15} \cellcolor{white} & \cite{72} & \cmark &  & \cmark &  & \cmark & \cmark &  & \cmark & \cmark &  &  &  & \cmark \\
\cellcolor{white} & \cite{73} &  &  &  &  & \cmark &  &  & \cmark &  &  &  &  & \cmark \\
\rowcolor{gray!15} \cellcolor{white} & \cite{75} & \cmark &  & \cmark &  & \cmark &  & \cmark & \cmark & \cmark &  & \cmark &  & \cmark \\
\cellcolor{white} & \cite{78} &  &  & \cmark & \cmark & \cmark &  &  & \cmark &  &  &  &  & \cmark \\
\rowcolor{gray!15} \cellcolor{white} & \cite{80} &  &  & \cmark &  & \cmark & \cmark &  &  &  &  &  &  & \cmark \\
\cellcolor{white} & \cite{81Cao_Huang_Li_Huilin_He_Oo_Hooi_2025} &  &  & \cmark &  & \cmark & \cmark &  &  &  &  & \cmark &  & \cmark \\
\rowcolor{gray!15} \cellcolor{white} & \cite{83} & \cmark &  & \cmark &  & \cmark & \cmark & \cmark & \cmark & \cmark & \cmark & \cmark &  & \cmark \\
\cellcolor{white} & \cite{87} & \cmark & \cmark & \cmark &  & \cmark &  &  & \cmark &  &  & \cmark &  & \cmark \\
\rowcolor{gray!15} \cellcolor{white} & \cite{88} & \cmark &  &  &  & \cmark &  &  & \cmark &  & \cmark & \cmark &  & \cmark \\
\cellcolor{white} & \cite{89guo2025repoauditautonomousllmagentrepositorylevel} & \cmark &  & \cmark &  & \cmark & \cmark &  &  &  & \cmark & \cmark &  &  \\
\rowcolor{gray!15} \cellcolor{white} & \cite{90} &  &  & \cmark &  & \cmark &  &  & \cmark &  &  &  &  & \cmark \\
\cellcolor{white} & \cite{93info16050365} &  &  & \cmark &  & \cmark &  &  & \cmark &  &  &  &  &  \\
\rowcolor{gray!15} \cellcolor{white} & \cite{94} &  &  &  &  & \cmark & \cmark &  &  &  &  &  &  & \cmark \\
\cellcolor{white} & \cite{99} & \cmark &  &  &  & \cmark & \cmark &  &  &  &  &  &  & \cmark \\
\rowcolor{gray!15} \cellcolor{white} & \cite{100} & \cmark &  &  &  & \cmark &  &  & \cmark & \cmark &  &  & \cmark & \cmark \\
\hline
\cellcolor{white}2024 & \cite{1} &  &  & \cmark &  & \cmark & \cmark &  &  & \cmark &  &  &  &  \\
\rowcolor{gray!15} \cellcolor{white} & \cite{10debenedetti2024agentdojo} & \cmark &  &  &  & \cmark & \cmark &  &  &  & \cmark &  &  & \cmark \\
\cellcolor{white} & \cite{14chen2024agentpoison} &  &  & \cmark &  & \cmark & \cmark &  &  &  &  &  &  &  \\
\rowcolor{gray!15} \cellcolor{white} & \cite{28} & \cmark &  &  &  & \cmark &  &  & \cmark &  &  &  &  &  \\
\cellcolor{white} & \cite{43} & \cmark &  & \cmark & \cmark & \cmark & \cmark &  &  &  &  &  & \cmark &  \\
\rowcolor{gray!15} \cellcolor{white} & \cite{50} & \cmark &  & \cmark &  & \cmark & \cmark &  &  & \cmark &  &  &  &  \\
\cellcolor{white} & \cite{54} &  &  &  &  & \cmark & \cmark &  &  &  & \cmark &  &  &  \\
\rowcolor{gray!15} \cellcolor{white} & \cite{63zemicheal2024llm} &  &  & \cmark &  & \cmark &  &  & \cmark & \cmark & \cmark &  &  & \cmark \\
\cellcolor{white} & \cite{66} &  &  &  &  & \cmark & \cmark &  &  &  &  &  &  &  \\
\rowcolor{gray!15} \cellcolor{white} & \cite{74icaart24} &  &  &  &  & \cmark & \cmark &  &  &  &  &  &  &  \\
\cellcolor{white} & \cite{76} & \cmark &  &  &  & \cmark &  & \cmark &  &  &  &  & \cmark & \cmark \\
\rowcolor{gray!15} \cellcolor{white} & \cite{77} & \cmark &  & \cmark &  & \cmark &  &  & \cmark &  &  & \cmark &  & \cmark \\
\cellcolor{white} & \cite{79} & \cmark &  & \cmark &  & \cmark & \cmark &  &  & \cmark & \cmark & \cmark &  & \cmark \\
\rowcolor{gray!15} \cellcolor{white} & \cite{82rigaki2024prompt} & \cmark & \cmark &  &  & \cmark & \cmark &  &  &  &  &  &  &  \\
\cellcolor{white} & \cite{85} & \cmark & \cmark &  &  & \cmark & \cmark &  &  &  &  & \cmark &  & \cmark \\
\rowcolor{gray!15} \cellcolor{white} & \cite{95} &  &  &  &  & \cmark &  &  & \cmark &  &  &  &  &  \\
\cellcolor{white} & \cite{98} & \cmark &  &  &  & \cmark &  &  & \cmark &  &  &  &  & \cmark \\
\hline
\rowcolor{gray!15} \cellcolor{white}2023 & \cite{52} & \cmark &  & \cmark &  & \cmark & \cmark &  &  & \cmark & \cmark &  &  & \cmark \\
\hline
    \end{tabular}%
\end{adjustbox}
  \label{tab:approach-agent-memory}%
  \vspace{-18pt}
\end{table*}%

\noindent
{\bf Memory types.}
\ul{In-context working memory} refers to task state carried inside the
agent's prompt or context window, including recent observations, intermediate
decisions, tool outputs, and partially constructed findings
\cite{5,19,26RIGAKI2026129987,36ZOU2026104305,40zhuo2026cyberzero,
64happe2026llms,84WANG2026103731,86,92,97,101,2,3,8zhang2025agent,
9jie2025agent4vul,15,16,20,21,22,24yildiz-etal-2025-benchmarking,27,29,
30,32,33,34,35,41jiao2025deepvulhunter,42LOEVENICH2025111162,
45,46abramovich2025enigma,47chowdhry2025evaluating,
49,53,56,59,60,65,69,70,72,75,83,87,88,
89guo2025repoauditautonomousllmagentrepositorylevel,99,100,
10debenedetti2024agentdojo,28,43,50,76,77,79,82rigaki2024prompt,85,98,52}.
\textit{Task-progress memory} appears when the agent must remember what has already been tried. Ginige et al.'s \textsc{AutoPentester}~\cite{21}, for instance, feeds previous testing results back into later decision steps so that the agent can continue a multi-step penetration-testing process rather than repeatedly restart from the target description. \textit{Evidence-carrying context} is similarly important in auditing and vulnerability detection. Guo et al.'s \textsc{RepoAudit}~\cite{89guo2025repoauditautonomousllmagentrepositorylevel} uses the current repository exploration path and accumulated data-flow evidence as local context while deciding which program locations deserve further inspection.

\ul{External short-term memory} stores temporary task state outside the
immediate LLM context
\cite{6,8zhang2025agent,29,62,87,82rigaki2024prompt,85}. This pattern is useful
when the agent must traverse a large state space, such as a binary, web target, or multi-step attack path, while keeping the prompt compact. Chen et al.'s \textsc{ClearAgent}~\cite{29} illustrates this design in binary vulnerability detection: the agent records an exploration trace over call-graph functions and control-flow basic blocks, then consults that trace when deciding where to inspect next and whether a candidate bug should be 
validated.

\ul{Long-term / persistent memory} covers reusable information stored across
sessions or beyond the current context window, including vector databases,
knowledge bases, validated facts, past reports, examples, or learned task
trajectories
\cite{4,5,19,23lbathaviator,26RIGAKI2026129987,36ZOU2026104305,
40zhuo2026cyberzero,57,64happe2026llms,84WANG2026103731,86,
91zhu-etal-2026-teams,92,97,101,6,7,8zhang2025agent,13,18,20,21,
24yildiz-etal-2025-benchmarking,27,31,32,37app15169096,
38zhu2025cvebenchbenchmarkaiagents,39,41jiao2025deepvulhunter,
42LOEVENICH2025111162,45,49,58,59,62,69,70,71,72,75,78,80,
81Cao_Huang_Li_Huilin_He_Oo_Hooi_2025,83,87,
89guo2025repoauditautonomousllmagentrepositorylevel,90,93info16050365,1,
14chen2024agentpoison,43,50,63zemicheal2024llm,77,79,52}.
\textit{Reusable domain exemplars} appear when the agent retrieves prior
security artifacts to guide a new case. Liu et al.'s
\textsc{PropertyGPT}~\cite{83} stores human-written smart-contract properties
in a vector database; for a new contract, it retrieves a similar property as an
in-context reference and then iteratively revises generated properties with
compiler, static-analysis, and prover feedback. \textit{Persistent task
experience} appears in reflective penetration-testing systems such as Dai et
al.'s \textsc{RefPentester}~\cite{87}, where experience from failed actions and
stage transitions is retained to reduce repeated mistakes in later testing
steps.

\ul{Knowledge graph memory} represents security-relevant entities and relations
explicitly, such as hosts, services, vulnerabilities, attack techniques,
dependencies, or environment states
\cite{4,84WANG2026103731,2,12,18,20,39,42LOEVENICH2025111162,78,43}. Compared
with unstructured retrieval, this memory type gives agents a structured view of
what is known and how facts are connected. Wang et al.'s
\textsc{PTFusion}~\cite{84WANG2026103731} uses a dynamic knowledge graph to
fuse web-penetration-testing context, executed actions, and discovered target
facts, enabling the planner to choose later commands with awareness of the
current attack surface. Loevenich et al.~\cite{42LOEVENICH2025111162} similarly
construct cybersecurity knowledge graphs from network logs, CTI reports, and
vulnerability frameworks, then use these structures to support autonomous cyber
defense and analyst-facing retrieval.

\ul{Parametric memory} denotes the knowledge implicitly stored in the backbone LLM's trained parameters. It appears throughout the surveyed agent systems. For example, in Kim et al.'s \textsc{AutoRestTest}~\cite{2}, the value agent uses the LLM's learned knowledge of API naming conventions, parameter formats, and common web-service semantics to generate realistic REST input values from specifications and constraints.

\noindent
{\bf Memory operations.}
\ul{Read} operations retrieve stored information and inject it into the next reasoning or action step \cite{4,19,23lbathaviator,86,13,18,20,21,37app15169096,39,72,83,1,14chen2024agentpoison}. For example, Liu et al.'s \textsc{PropertyGPT}~\cite{83} reads from a vector database of human-written properties to obtain a close reference before generating a new smart-contract property. Ginige et al.'s \textsc{AutoPentester}~\cite{21} reads retrieved penetration-testing knowledge and prior findings to produce more targeted commands for the next stage.

\ul{Write} operations store newly produced artifacts, including generated
reports, candidate patches, incident narratives, formal properties, or
intermediate analysis results \cite{25zhang2026bountybench,32,
42LOEVENICH2025111162,69,75,83,76}. The important distinction is that the agent
does not merely consume memory; it also turns its observations into reusable
records. Sheikhi et al.'s \textsc{Cognitive SOC}~\cite{32}, for example,
converts alert evidence and enrichment results into an evidence-backed incident
narrative for analyst review. In \textsc{PropertyGPT}, Liu et al.~\cite{83}
also write generated properties and retain selected top-ranked candidates for
later verification.

\ul{Update / reflection} operations revise memory content, future plans, or
action preferences based on execution feedback, failed attempts, self-critique,
or validation results
\cite{4,5,19,64happe2026llms,84WANG2026103731,86,92,97,101,6,7,16,29,31,
37app15169096,39,42LOEVENICH2025111162,45,49,51,56,58,59,60,61,70,71,72,
73,75,78,83,87,88,90,93info16050365,100,28,63zemicheal2024llm,77,95,98}.
\textit{Exploration reflection} appears in systems that must decide where to
search next, such as Chen et al.'s \textsc{ClearAgent}~\cite{29}, which reflects
on visited binary locations and candidate bug evidence before continuing the
analysis. \textit{Failure-driven reflection} appears in Dai et al.'s
\textsc{RefPentester}~\cite{87}, where unsuccessful penetration-testing actions
and their rewards are used to adjust later behavior.

\noindent
{\bf Security-specific memory.}
\ul{Vulnerability knowledge bases} store or retrieve information about known
vulnerabilities, CVEs, CWEs, attack patterns, exploit preconditions, security
properties, and bug classes
\cite{4,23lbathaviator,6,13,18,21,22,24yildiz-etal-2025-benchmarking,27,32,
34,37app15169096,38zhu2025cvebenchbenchmarkaiagents,41jiao2025deepvulhunter,
42LOEVENICH2025111162,69,72,75,83,100,1,50,63zemicheal2024llm,79,52}.
\textsc{PropertyGPT}~\cite{83}, for example, uses
vulnerability-related smart-contract properties as reusable references for
formal verification, while \textsc{AutoPentester}~\cite{21} uses
penetration-testing knowledge to guide scanning, assessment, and exploitation.

\ul{Code repositories and patch histories} capture source files, vulnerable and
fixed versions, commits, code slices, repository-level facts, and previous
repair decisions
\cite{11,23lbathaviator,25zhang2026bountybench,36ZOU2026104305,
40zhuo2026cyberzero,64happe2026llms,91zhu-etal-2026-teams,101,
8zhang2025agent,24yildiz-etal-2025-benchmarking,
38zhu2025cvebenchbenchmarkaiagents,41jiao2025deepvulhunter,
46abramovich2025enigma,51,69,83,88,
89guo2025repoauditautonomousllmagentrepositorylevel,10debenedetti2024agentdojo,
54,63zemicheal2024llm,79,52}. Guo et al.'s
\textsc{RepoAudit}~\cite{89guo2025repoauditautonomousllmagentrepositorylevel}
uses repository-level code facts and data-flow evidence to audit large code
bases, whereas Bouzenia et al.'s \textsc{RepairAgent}~\cite{88} searches the
codebase for repair ingredients and uses validation feedback to refine candidate
patches.

\ul{Execution traces and fuzzing logs} include runtime outputs, command logs,
failed exploit attempts, test results, path conditions, coverage observations,
and fuzzer-produced evidence
\cite{11,19,25zhang2026bountybench,36ZOU2026104305,84WANG2026103731,
91zhu-etal-2026-teams,101,8zhang2025agent,12,22,27,29,30,35,37app15169096,
45,51,55,56,60,61,69,70,71,75,81Cao_Huang_Li_Huilin_He_Oo_Hooi_2025,83,87,
88,89guo2025repoauditautonomousllmagentrepositorylevel,77,79,85}. In
penetration testing, Dai et al.'s \textsc{RefPentester}~\cite{87} uses
execution results and failure feedback as evidence for reflection. In binary
analysis, Chen et al.'s \textsc{ClearAgent}~\cite{29} uses exploration traces
and concrete-input validation to check whether a suspected vulnerability is
actually triggerable.

\ul{Threat intelligence feeds} refer to CTI reports, indicators of compromise,
MITRE ATT\&CK-style tactics and techniques, security-operation evidence, and
other external intelligence used to contextualize alerts or defensive actions
\cite{12,16,18,31,32,42LOEVENICH2025111162,65,67,68,100,43,76}. Sheikhi et
al.'s \textsc{Cognitive SOC}~\cite{32} uses alert evidence and threat context to
produce evidence-backed narratives for security analysts. Loevenich et
al.~\cite{42LOEVENICH2025111162} build cyber-defense knowledge graphs from
network logs, open-source CTI reports, and vulnerability frameworks, allowing
the agent to connect defensive observations with known adversary tactics.

\ul{Tool documentation} covers command syntax, API specifications, tool manuals,
usage constraints, expected outputs, and environment-specific operating
instructions
\cite{4,5,23lbathaviator,36ZOU2026104305,84WANG2026103731,97,101,2,12,13,
15,18,20,21,22,29,30,31,32,37app15169096,
38zhu2025cvebenchbenchmarkaiagents,42LOEVENICH2025111162,45,
46abramovich2025enigma,55,58,65,70,72,73,75,78,80,
81Cao_Huang_Li_Huilin_He_Oo_Hooi_2025,83,87,88,90,94,99,100,
10debenedetti2024agentdojo,63zemicheal2024llm,76,77,79,85,98,52}. Kim et al.'s
\textsc{AutoRestTest}~\cite{2} uses API specifications and parameter
constraints to generate valid REST inputs, while \textsc{AutoPentester}~\cite{21}
uses penetration-testing knowledge to turn high-level goals into commands
that can be executed against the target environment.

\vspace{-4pt}
\subsubsection{Agent perception}

Agent perception concerns the observations that enter the agent loop and the mechanisms used to turn raw security artifacts into reasoning-ready context.
Table~\ref{tab:approach-agent-perception} summarizes both the input sources perceived by security agents and the mechanisms they use to transform those observations into reasoning-ready context.
\begin{table*}[t]
  \centering
  \vspace{-10pt}
  \caption{Paper attribution for Approach: Agent perception}
  \vspace{-10pt}
  \scriptsize
  \setlength{\tabcolsep}{2.5pt}
  \renewcommand{\arraystretch}{0.70}
\begin{adjustbox}{width=\textwidth,max totalheight=0.88\textheight,center}
    \begin{tabular}{|c|c|c|c|c|c|c|c|c|c|c|c|c|c|c|c|}
    \hline
\rowcolor{white} & & 
\multicolumn{9}{c|}{\makecell[c]{Input sources}} & \multicolumn{5}{c|}{\makecell[c]{Perception mechanisms}} \\
\cline{3-11}\cline{12-16}
\rowcolor{white} 
\multicolumn{1}{|c|}{\multirow{-4}{*}[-0.2ex]{\makecell[c]{Year}}} & 
\multicolumn{1}{c|}{\multirow{-4}{*}[-0.2ex]{\makecell[c]{Paper}}} & 
\multicolumn{1}{c|}{\cellcolor{blue!22}\makecell[c]{Source\\code \\ (33\%)}} & \multicolumn{1}{c|}{\cellcolor{blue!5}\makecell[c]{Compiled binaries\\disassembly\\decompiled output \\ (5\%)}} & \multicolumn{1}{c|}{\cellcolor{blue!17}\makecell[c]{Execution\\traces \\ (25\%)}} & \multicolumn{1}{c|}{\cellcolor{blue!37}\makecell[c]{System\\ logs \\ (55\%)}} & \multicolumn{1}{c|}{\cellcolor{blue!28}\makecell[c]{Natural\\language \\ (42\%)}} & \multicolumn{1}{c|}{\cellcolor{blue!40}\makecell[c]{Structured\\ data \\ (60\%)}} & \multicolumn{1}{c|}{\cellcolor{blue!44}\makecell[c]{Environment\\observations \\ (66\%)}} & \multicolumn{1}{c|}{\cellcolor{blue!5}\makecell[c]{Images /\\screenshots \\ (3\%)}} & \multicolumn{1}{c|}{\cellcolor{blue!5}\makecell[c]{Protocol\\specifications\\ /standards \\ (3\%)}} & \multicolumn{1}{c|}{\cellcolor{blue!67}\makecell[c]{Direct\\LLM\\ingestion \\ (100\%)}} & \multicolumn{1}{c|}{\cellcolor{blue!20}\makecell[c]{Retrieval-\\augmented\\perception \\ (30\%)}} & \multicolumn{1}{c|}{\cellcolor{blue!23}\makecell[c]{Summarisation /\\compression \\ (35\%)}} & \multicolumn{1}{c|}{\cellcolor{blue!7}\makecell[c]{Embedding \\/semantic\\similarity \\ (10\%)}} & \multicolumn{1}{c|}{\cellcolor{blue!45}\makecell[c]{Tool-\\mediated\\observation \\ (68\%)}} \\
    \hline

\cellcolor{white}2026 & \cite{4} &  &  &  & \cmark & \cmark & \cmark & \cmark &  &  & \cmark & \cmark &  &  & \cmark \\
\rowcolor{gray!15} \cellcolor{white} & \cite{5} &  &  &  & \cmark & \cmark & \cmark & \cmark &  &  & \cmark &  &  &  & \cmark \\
\cellcolor{white} & \cite{11} & \cmark &  & \cmark &  & \cmark &  & \cmark &  &  & \cmark &  & \cmark &  & \cmark \\
\rowcolor{gray!15} \cellcolor{white} & \cite{19} &  &  & \cmark & \cmark &  & \cmark & \cmark &  &  & \cmark & \cmark &  &  & \cmark \\
\cellcolor{white} & \cite{23lbathaviator} & \cmark &  &  &  &  & \cmark &  &  &  & \cmark & \cmark &  & \cmark & \cmark \\
\rowcolor{gray!15} \cellcolor{white} & \cite{25zhang2026bountybench} & \cmark &  &  & \cmark & \cmark & \cmark & \cmark &  &  & \cmark &  &  &  & \cmark \\
\cellcolor{white} & \cite{26RIGAKI2026129987} &  &  & \cmark & \cmark &  & \cmark & \cmark &  &  & \cmark &  &  &  &  \\
\rowcolor{gray!15} \cellcolor{white} & \cite{36ZOU2026104305} & \cmark & \cmark &  &  & \cmark & \cmark & \cmark & \cmark &  & \cmark &  & \cmark &  & \cmark \\
\cellcolor{white} & \cite{40zhuo2026cyberzero} &  &  &  &  & \cmark & \cmark &  &  &  & \cmark &  &  &  &  \\
\rowcolor{gray!15} \cellcolor{white} & \cite{57} &  &  &  & \cmark & \cmark &  &  &  &  & \cmark & \cmark & \cmark &  &  \\
\cellcolor{white} & \cite{64happe2026llms} & \cmark &  &  &  &  & \cmark &  &  &  & \cmark &  &  &  & \cmark \\
\rowcolor{gray!15} \cellcolor{white} & \cite{84WANG2026103731} &  &  &  & \cmark &  & \cmark & \cmark &  &  & \cmark &  & \cmark &  & \cmark \\
\cellcolor{white} & \cite{86} &  &  &  &  &  & \cmark &  &  &  & \cmark & \cmark & \cmark &  & \cmark \\
\rowcolor{gray!15} \cellcolor{white} & \cite{91zhu-etal-2026-teams} &  &  &  &  & \cmark &  &  &  &  & \cmark &  &  &  & \cmark \\
\cellcolor{white} & \cite{92} &  &  &  & \cmark &  &  & \cmark &  &  & \cmark &  &  &  &  \\
\rowcolor{gray!15} \cellcolor{white} & \cite{96} &  &  &  &  & \cmark & \cmark &  &  &  & \cmark &  &  &  &  \\
\cellcolor{white} & \cite{97} &  &  &  & \cmark & \cmark & \cmark & \cmark &  & \cmark & \cmark &  &  & \cmark & \cmark \\
\rowcolor{gray!15} \cellcolor{white} & \cite{101} & \cmark &  &  &  &  &  &  &  &  & \cmark &  & \cmark & \cmark & \cmark \\
\hline
\cellcolor{white}2025 & \cite{2} &  &  &  &  &  & \cmark & \cmark &  &  & \cmark &  &  & \cmark & \cmark \\
\rowcolor{gray!15} \cellcolor{white} & \cite{3} &  &  &  & \cmark & \cmark & \cmark & \cmark &  &  & \cmark &  &  &  & \cmark \\
\cellcolor{white} & \cite{6} & \cmark &  &  &  & \cmark &  &  &  &  & \cmark &  &  &  &  \\
\rowcolor{gray!15} \cellcolor{white} & \cite{7} & \cmark & \cmark &  &  &  &  & \cmark &  &  & \cmark & \cmark &  &  &  \\
\cellcolor{white} & \cite{8zhang2025agent} &  &  & \cmark &  & \cmark &  & \cmark &  &  & \cmark & \cmark & \cmark &  & \cmark \\
\rowcolor{gray!15} \cellcolor{white} & \cite{9jie2025agent4vul} & \cmark & \cmark &  &  &  &  &  &  &  & \cmark &  &  & \cmark &  \\
\cellcolor{white} & \cite{12} &  &  & \cmark &  &  & \cmark & \cmark &  &  & \cmark &  &  &  & \cmark \\
\rowcolor{gray!15} \cellcolor{white} & \cite{13} &  &  &  &  &  & \cmark & \cmark &  &  & \cmark & \cmark &  &  & \cmark \\
\cellcolor{white} & \cite{15} &  &  &  &  & \cmark & \cmark &  &  & \cmark & \cmark &  &  &  &  \\
\rowcolor{gray!15} \cellcolor{white} & \cite{16} &  &  &  & \cmark & \cmark & \cmark & \cmark &  &  & \cmark &  & \cmark &  &  \\
\cellcolor{white} & \cite{17} &  &  & \cmark &  & \cmark & \cmark & \cmark &  &  & \cmark &  &  &  &  \\
\rowcolor{gray!15} \cellcolor{white} & \cite{18} &  &  &  &  & \cmark & \cmark & \cmark &  &  & \cmark & \cmark &  &  & \cmark \\
\cellcolor{white} & \cite{20} &  &  &  & \cmark &  & \cmark & \cmark &  &  & \cmark & \cmark &  &  & \cmark \\
\rowcolor{gray!15} \cellcolor{white} & \cite{21} &  &  &  & \cmark & \cmark &  & \cmark &  &  & \cmark & \cmark &  &  & \cmark \\
\cellcolor{white} & \cite{22} &  &  &  & \cmark &  & \cmark & \cmark &  &  & \cmark &  &  &  & \cmark \\
\rowcolor{gray!15} \cellcolor{white} & \cite{24yildiz-etal-2025-benchmarking} & \cmark &  &  &  &  & \cmark &  &  &  & \cmark & \cmark &  &  & \cmark \\
\cellcolor{white} & \cite{27} &  &  &  & \cmark & \cmark & \cmark &  &  &  & \cmark & \cmark & \cmark &  & \cmark \\
\rowcolor{gray!15} \cellcolor{white} & \cite{29} &  & \cmark & \cmark &  &  &  & \cmark &  &  & \cmark &  & \cmark &  & \cmark \\
\cellcolor{white} & \cite{30} &  &  &  & \cmark & \cmark & \cmark & \cmark &  &  & \cmark &  & \cmark &  & \cmark \\
\rowcolor{gray!15} \cellcolor{white} & \cite{31} &  &  & \cmark & \cmark &  &  & \cmark &  &  & \cmark &  &  &  &  \\
\cellcolor{white} & \cite{32} &  &  &  & \cmark &  &  & \cmark &  &  & \cmark & \cmark & \cmark &  & \cmark \\
\rowcolor{gray!15} \cellcolor{white} & \cite{33} & \cmark &  &  &  &  & \cmark &  &  &  & \cmark &  &  & \cmark &  \\
\cellcolor{white} & \cite{34} & \cmark &  &  &  &  &  &  &  &  & \cmark &  &  &  &  \\
\rowcolor{gray!15} \cellcolor{white} & \cite{35} &  &  &  & \cmark &  &  &  &  &  & \cmark &  &  &  & \cmark \\
\cellcolor{white} & \cite{37app15169096} & \cmark &  &  & \cmark &  & \cmark & \cmark &  &  & \cmark & \cmark & \cmark &  & \cmark \\
\rowcolor{gray!15} \cellcolor{white} & \cite{38zhu2025cvebenchbenchmarkaiagents} & \cmark &  &  & \cmark &  &  & \cmark &  &  & \cmark &  &  &  & \cmark \\
\cellcolor{white} & \cite{39} &  &  &  &  &  & \cmark & \cmark &  &  & \cmark &  &  &  &  \\
\rowcolor{gray!15} \cellcolor{white} & \cite{41jiao2025deepvulhunter} & \cmark &  &  &  & \cmark & \cmark &  &  &  & \cmark & \cmark &  &  &  \\
\cellcolor{white} & \cite{42LOEVENICH2025111162} &  &  & \cmark & \cmark &  & \cmark & \cmark &  &  & \cmark & \cmark &  & \cmark & \cmark \\
\rowcolor{gray!15} \cellcolor{white} & \cite{45} &  &  & \cmark & \cmark &  & \cmark & \cmark &  &  & \cmark &  &  &  & \cmark \\
\cellcolor{white} & \cite{46abramovich2025enigma} & \cmark &  &  &  &  &  & \cmark &  &  & \cmark &  & \cmark &  & \cmark \\
\rowcolor{gray!15} \cellcolor{white} & \cite{47chowdhry2025evaluating} &  &  & \cmark & \cmark &  & \cmark & \cmark &  &  & \cmark &  & \cmark &  &  \\
\cellcolor{white} & \cite{48} &  &  & \cmark &  &  &  & \cmark &  &  & \cmark &  &  &  &  \\
\rowcolor{gray!15} \cellcolor{white} & \cite{49} &  &  & \cmark & \cmark &  & \cmark & \cmark &  &  & \cmark &  & \cmark &  &  \\
\cellcolor{white} & \cite{51} &  &  & \cmark & \cmark & \cmark & \cmark & \cmark &  &  & \cmark &  & \cmark &  &  \\
\rowcolor{gray!15} \cellcolor{white} & \cite{53} &  &  & \cmark &  &  &  & \cmark &  &  & \cmark &  &  &  &  \\
\cellcolor{white} & \cite{55} &  &  &  &  & \cmark & \cmark & \cmark &  &  & \cmark &  &  &  & \cmark \\
\rowcolor{gray!15} \cellcolor{white} & \cite{56} &  &  & \cmark & \cmark & \cmark & \cmark & \cmark &  &  & \cmark &  &  &  & \cmark \\
\cellcolor{white} & \cite{58} & \cmark &  &  &  &  & \cmark & \cmark &  &  & \cmark & \cmark & \cmark &  & \cmark \\
\rowcolor{gray!15} \cellcolor{white} & \cite{59} &  &  &  & \cmark &  & \cmark & \cmark &  &  & \cmark &  &  &  & \cmark \\
\cellcolor{white} & \cite{60} &  &  & \cmark & \cmark &  & \cmark & \cmark &  &  & \cmark &  & \cmark &  &  \\
\rowcolor{gray!15} \cellcolor{white} & \cite{61} &  &  &  & \cmark & \cmark & \cmark & \cmark &  &  & \cmark &  & \cmark &  &  \\
\cellcolor{white} & \cite{62} &  &  &  & \cmark & \cmark & \cmark & \cmark &  &  & \cmark & \cmark &  &  &  \\
\rowcolor{gray!15} \cellcolor{white} & \cite{65} & \cmark & \cmark &  & \cmark &  &  & \cmark &  &  & \cmark &  &  &  & \cmark \\
\cellcolor{white} & \cite{67} &  &  & \cmark & \cmark &  &  & \cmark &  &  & \cmark &  &  &  & \cmark \\
\rowcolor{gray!15} \cellcolor{white} & \cite{68} &  &  &  &  & \cmark &  &  &  &  & \cmark &  & \cmark &  & \cmark \\
\cellcolor{white} & \cite{69} & \cmark &  & \cmark &  & \cmark & \cmark & \cmark &  &  & \cmark &  &  &  & \cmark \\
\rowcolor{gray!15} \cellcolor{white} & \cite{70} & \cmark &  &  & \cmark &  &  & \cmark &  &  & \cmark & \cmark &  &  & \cmark \\
\cellcolor{white} & \cite{71} & \cmark &  &  & \cmark &  & \cmark & \cmark &  &  & \cmark & \cmark & \cmark &  & \cmark \\
\rowcolor{gray!15} \cellcolor{white} & \cite{72} & \cmark &  &  &  &  & \cmark &  &  & \cmark & \cmark & \cmark &  & \cmark & \cmark \\
\cellcolor{white} & \cite{73} &  &  & \cmark & \cmark &  & \cmark & \cmark &  &  & \cmark &  & \cmark &  &  \\
\rowcolor{gray!15} \cellcolor{white} & \cite{75} & \cmark &  &  & \cmark &  &  &  &  &  & \cmark &  & \cmark &  & \cmark \\
\cellcolor{white} & \cite{78} &  &  &  & \cmark &  &  & \cmark &  &  & \cmark & \cmark & \cmark &  & \cmark \\
\rowcolor{gray!15} \cellcolor{white} & \cite{80} &  &  & \cmark & \cmark &  & \cmark & \cmark &  &  & \cmark &  & \cmark &  &  \\
\cellcolor{white} & \cite{81Cao_Huang_Li_Huilin_He_Oo_Hooi_2025} & \cmark &  &  & \cmark &  & \cmark &  & \cmark &  & \cmark & \cmark &  &  & \cmark \\
\rowcolor{gray!15} \cellcolor{white} & \cite{83} & \cmark &  &  & \cmark &  &  &  &  &  & \cmark & \cmark & \cmark &  & \cmark \\
\cellcolor{white} & \cite{87} &  &  &  & \cmark &  & \cmark & \cmark &  &  & \cmark & \cmark & \cmark &  & \cmark \\
\rowcolor{gray!15} \cellcolor{white} & \cite{88} & \cmark &  &  &  &  &  &  &  &  & \cmark &  &  &  & \cmark \\
\cellcolor{white} & \cite{89guo2025repoauditautonomousllmagentrepositorylevel} & \cmark &  & \cmark &  & \cmark &  &  &  &  & \cmark &  &  &  & \cmark \\
\rowcolor{gray!15} \cellcolor{white} & \cite{90} &  &  &  & \cmark & \cmark &  & \cmark &  &  & \cmark &  & \cmark &  & \cmark \\
\cellcolor{white} & \cite{93info16050365} &  &  &  & \cmark &  & \cmark &  &  &  & \cmark &  &  &  & \cmark \\
\rowcolor{gray!15} \cellcolor{white} & \cite{94} &  &  &  & \cmark &  & \cmark &  &  &  & \cmark &  &  &  & \cmark \\
\cellcolor{white} & \cite{99} & \cmark &  &  &  & \cmark &  &  &  &  & \cmark &  &  &  & \cmark \\
\rowcolor{gray!15} \cellcolor{white} & \cite{100} & \cmark &  &  &  &  &  & \cmark &  &  & \cmark &  &  &  & \cmark \\
\hline
\cellcolor{white}2024 & \cite{1} &  &  &  &  & \cmark & \cmark &  &  &  & \cmark & \cmark &  & \cmark &  \\
\rowcolor{gray!15} \cellcolor{white} & \cite{10debenedetti2024agentdojo} &  &  &  &  & \cmark &  & \cmark &  &  & \cmark &  &  &  & \cmark \\
\cellcolor{white} & \cite{14chen2024agentpoison} &  &  &  &  & \cmark & \cmark & \cmark &  &  & \cmark & \cmark &  & \cmark & \cmark \\
\rowcolor{gray!15} \cellcolor{white} & \cite{28} &  &  &  & \cmark & \cmark & \cmark & \cmark &  &  & \cmark &  &  &  &  \\
\cellcolor{white} & \cite{43} &  &  &  & \cmark &  &  & \cmark &  &  & \cmark & \cmark &  &  & \cmark \\
\rowcolor{gray!15} \cellcolor{white} & \cite{50} &  &  &  & \cmark & \cmark &  &  &  &  & \cmark &  &  &  &  \\
\cellcolor{white} & \cite{54} & \cmark &  &  &  &  &  & \cmark & \cmark &  & \cmark &  & \cmark &  & \cmark \\
\rowcolor{gray!15} \cellcolor{white} & \cite{63zemicheal2024llm} & \cmark &  &  & \cmark &  & \cmark &  &  &  & \cmark & \cmark & \cmark &  & \cmark \\
\cellcolor{white} & \cite{66} &  &  &  &  &  &  &  &  &  & \cmark &  &  &  & \cmark \\
\rowcolor{gray!15} \cellcolor{white} & \cite{74icaart24} &  &  & \cmark & \cmark & \cmark & \cmark & \cmark &  &  & \cmark &  & \cmark &  &  \\
\cellcolor{white} & \cite{76} &  &  & \cmark & \cmark & \cmark &  & \cmark &  &  & \cmark &  & \cmark &  &  \\
\rowcolor{gray!15} \cellcolor{white} & \cite{77} &  &  &  & \cmark &  & \cmark &  &  &  & \cmark &  &  &  & \cmark \\
\cellcolor{white} & \cite{79} &  &  & \cmark & \cmark & \cmark &  & \cmark &  &  & \cmark &  & \cmark &  & \cmark \\
\rowcolor{gray!15} \cellcolor{white} & \cite{82rigaki2024prompt} &  &  &  &  &  & \cmark & \cmark &  &  & \cmark &  &  &  & \cmark \\
\cellcolor{white} & \cite{85} &  &  &  &  & \cmark & \cmark & \cmark &  &  & \cmark &  & \cmark &  & \cmark \\
\rowcolor{gray!15} \cellcolor{white} & \cite{95} &  &  &  & \cmark &  & \cmark & \cmark &  &  & \cmark &  &  &  & \cmark \\
\cellcolor{white} & \cite{98} & \cmark &  &  & \cmark & \cmark &  &  &  &  & \cmark &  &  &  & \cmark \\
\hline
\rowcolor{gray!15} \cellcolor{white}2023 & \cite{52} & \cmark &  &  &  & \cmark &  &  &  &  & \cmark &  &  &  &  \\
\hline
    \end{tabular}%
\end{adjustbox}
  \label{tab:approach-agent-perception}%
  \vspace{-18pt}
\end{table*}%

\noindent

{\bf Input sources.}
\ul{Source code} is the dominant code-centric input source for security tasks such as
vulnerability detection, repository auditing, program repair, vulnerability
injection, and smart-contract verification
\cite{11,23lbathaviator,25zhang2026bountybench,36ZOU2026104305,
64happe2026llms,101,6,7,9jie2025agent4vul,
24yildiz-etal-2025-benchmarking,33,34,37app15169096,
38zhu2025cvebenchbenchmarkaiagents,41jiao2025deepvulhunter,
46abramovich2025enigma,58,65,69,70,71,72,75,
81Cao_Huang_Li_Huilin_He_Oo_Hooi_2025,83,88,
89guo2025repoauditautonomousllmagentrepositorylevel,99,100,54,
63zemicheal2024llm,98,52}. Guo et al.'s
\textsc{RepoAudit}~\cite{89guo2025repoauditautonomousllmagentrepositorylevel}
is an example: the agent perceives source files together with
data-flow evidence so that later reasoning is grounded in concrete paths
through the code. Liu et al.'s \textsc{PropertyGPT}~\cite{83} uses a different
code-facing view. It reads smart-contract code as the object for which formal
properties must be generated and verified.

\ul{Compiled binaries, disassembly, and decompiled output} are used when source
code is unavailable or when the security task is inherently binary-facing
\cite{36ZOU2026104305,7,9jie2025agent4vul,29,65}. Here, perception must bridge
the semantic gap between low-level representations and vulnerability concepts.
Chen et al.'s \textsc{ClearAgent}~\cite{29} shows this contrast clearly: rather
than asking the LLM to reason from source, the framework exposes binary
representations through an analysis interface and lets the agent iteratively
inspect functions, control-flow blocks, and candidate vulnerability locations.

\ul{Execution traces} and \ul{system logs} move perception from static artifacts
to runtime evidence. Execution traces capture observed behavior, command
results, paths, tests, or failed attempts,
whereas system logs provide operational evidence from hosts, networks, services,
or SOC pipelines. 
In penetration testing, Ginige
et al.'s \textsc{AutoPentester}~\cite{21} observes tool outputs after each
action and uses them as the basis for the next command. In SOC-style analysis,
Sheikhi et al.'s \textsc{Cognitive SOC}~\cite{32} perceives alert and
enrichment evidence so that the final narrative can cite concrete operational
signals rather than free-form speculation.

\ul{Natural language} inputs include user goals, challenge descriptions,
security policies, vulnerability reports, incident narratives, threat
descriptions, and human-authored task instructions. \ul{Structured data} covers API specifications, JSON-like tool output,
CVE metadata, graph data, formal properties, tables, and benchmark records. These two sources often appear together: the
natural-language part gives the task intent, while the structured part gives
machine-checkable constraints. Kim et al.'s \textsc{AutoRestTest}~\cite{2}
illustrates the structured side especially well, since the agent consumes REST
API specifications, parameter types, and constraints before generating test
inputs.

\ul{Environment observations} appear when the agent is embedded in an
interactive security environment rather than a one-shot analysis setting. In such systems, the agent sees a
changing world: open ports, service banners, exploit feedback, browser states,
or cyber-range state. Wang et al.'s \textsc{PTFusion}~\cite{84WANG2026103731}
uses this kind of evolving context in web penetration testing, combining
executed actions and discovered target facts before planning later steps.

\ul{Images / screenshots} and \ul{protocol specifications / standards} are
more specialized perceptual sources. Images and screenshots appear in
multimodal security settings, including phishing or interface-oriented analysis
\cite{36ZOU2026104305,81Cao_Huang_Li_Huilin_He_Oo_Hooi_2025,54}. Protocol
specifications and standards support agents that reason about protocol fields,
message formats, or standards-derived constraints \cite{97,15,72}. These
categories show that perception is not
limited to text and code; security agents may need to read visual evidence or
normative protocol knowledge before they can act correctly.

\noindent
{\bf Perception mechanisms.}
\ul{Direct LLM ingestion} places perceived artifacts directly into the prompt or conversation state for interpretation. For example, ~\cite{89guo2025repoauditautonomousllmagentrepositorylevel} provides selected code facts and path evidence so that the LLM reasons over a narrower audit context.

\ul{Retrieval-augmented perception} first searches an external corpus,
knowledge base, vector store, or graph before the LLM interprets the current
case
\cite{4,19,23lbathaviator,57,86,7,8zhang2025agent,13,18,20,21,
24yildiz-etal-2025-benchmarking,27,32,37app15169096,
41jiao2025deepvulhunter,42LOEVENICH2025111162,58,62,70,71,72,78,
81Cao_Huang_Li_Huilin_He_Oo_Hooi_2025,83,87,1,14chen2024agentpoison,43,
63zemicheal2024llm}. For example, Liu et al.'s
\textsc{PropertyGPT}~\cite{83} retrieves similar human-written smart-contract
properties before generating a new one. Shen et al.'s
\textsc{PentestAgent}~\cite{78} uses retrieval to compensate for limited
penetration-testing knowledge, bringing relevant domain knowledge into the
multi-agent testing workflow.

\ul{Summarisation / compression} is used when the observation is too large,
verbose, or noisy to pass forward unchanged
\cite{11,36ZOU2026104305,57,84WANG2026103731,86,101,8zhang2025agent,16,27,
29,30,32,37app15169096,46abramovich2025enigma,
47chowdhry2025evaluating,49,51,58,60,61,68,71,73,75,78,80,83,87,90,54,
63zemicheal2024llm,74icaart24,76,79,85}. For example, in
\textsc{Cognitive SOC}, Sheikhi et al.~\cite{32} use this kind of evidence
distillation to transform alert details and enrichment artifacts into an
analyst-facing narrative with clearer attribution.

\ul{Embedding / semantic similarity} perception maps artifacts into a semantic
space so that the agent can match related code, properties, vulnerabilities,
API elements, or threat descriptions
\cite{23lbathaviator,97,101,2,9jie2025agent4vul,33,
42LOEVENICH2025111162,72,1,14chen2024agentpoison}. In \textsc{AutoRestTest}, Kim et al.~\cite{2} use semantic information
from API specifications and dependency graphs to guide test generation. In the
autonomous cyber-defense setting, Loevenich et
al.~\cite{42LOEVENICH2025111162} use learned representations to connect CTI,
CVEs, and ATT\&CK-style tactics inside knowledge-graph-supported perception.

\ul{Tool-mediated observation} makes perception active. The agent does not only
read a provided artifact; it invokes scanners, debuggers, fuzzers, browsers,
code-search tools, cyber-range interfaces, or verification tools, and then
observes their outputs. For example, \textsc{ClearAgent}~\cite{29} obtains new binary-analysis
context by interacting with analysis tools, while Bouzenia et al.'s
\textsc{RepairAgent}~\cite{88} observes code-search and validation feedback
during autonomous repair. In both cases, perception and action form a loop:
what the agent sees next is partly determined by what it decided to inspect or
execute now.

\vspace{-4pt}
\subsubsection{Agent reasoning \& planning}
This attribute is how an agent turns perceived security evidence into ordered decisions. Among the surveyed papers, agents have adopted various types of reasoning and planning design, in terms of their
reasoning paradigm, in how they plan across multi-step tasks, and in how they handle uncertainty when evidence is incomplete or actions may fail, as shown in Table~\ref{tab:approach-agent-reasoning-planning}.
\begin{table*}[t]
  \centering
  \vspace{-10pt}
  \caption{Paper attribution for Approach: Agent reasoning \& planning}
  \vspace{-10pt}
  \scriptsize
  \setlength{\tabcolsep}{1.5pt}
  \renewcommand{\arraystretch}{0.70}
\begin{adjustbox}{width=\textwidth,max totalheight=0.88\textheight,center}
    \begin{tabular}{|c|c|c|c|c|c|c|c|c|c|c|c|c|c|c|c|c|c|c|c|}
    \hline
\rowcolor{white} & & 
\multicolumn{11}{c|}{\makecell[c]{Reasoning paradigms}} & \multicolumn{4}{c|}{\makecell[c]{Planning strategies}} & \multicolumn{3}{c|}{\makecell[c]{Decision-making under uncertainty}} \\
\cline{3-13}\cline{14-17}\cline{18-20}
\rowcolor{white} 
\multicolumn{1}{|c|}{\multirow{-4}{*}[-0.2ex]{\makecell[c]{Year}}} & 
\multicolumn{1}{c|}{\multirow{-4}{*}[-0.2ex]{\makecell[c]{Paper}}} & 
\multicolumn{1}{c|}{\cellcolor{blue!10}\makecell[c]{ReAct \\ (15\%)}} & \multicolumn{1}{c|}{\cellcolor{blue!10}\makecell[c]{CoT\\ (15\%)}} & \multicolumn{1}{c|}{\cellcolor{blue!7}\makecell[c]{Reflection\\ /self-\\critique \\ (10\%)}} & \multicolumn{1}{c|}{\cellcolor{blue!5}\makecell[c]{Ref-\\lexion \\ (2\%)}} & \multicolumn{1}{c|}{\cellcolor{blue!5}\makecell[c]{Prompt\\chaining \\ (2\%)}} & \multicolumn{1}{c|}{\cellcolor{blue!7}\makecell[c]{Few-shot\\promp-\\ting \\ (10\%)}} & \multicolumn{1}{c|}{\cellcolor{blue!10}\makecell[c]{Self-debug\\/self-repair \\ (15\%)}} & \multicolumn{1}{c|}{\cellcolor{blue!31}\makecell[c]{Least-to\\-most task\\decomposition \\ (46\%)}} & \multicolumn{1}{c|}{\cellcolor{blue!19}\makecell[c]{Hypothesis\\-driven\\reasoning \\ (29\%)}} & \multicolumn{1}{c|}{\cellcolor{blue!5}\makecell[c]{Graph-based\\reasoning / attack\\graph construction \\ (2\%)}} & \multicolumn{1}{c|}{\cellcolor{blue!12}\makecell[c]{Unsp-\\ecified \\ (18\%)}} & \multicolumn{1}{c|}{\cellcolor{blue!54}\makecell[c]{Dynamic /\\reactive\\planning \\ (80\%)}} & \multicolumn{1}{c|}{\cellcolor{blue!15}\makecell[c]{Hierarchical\\task\\decomposition \\ (22\%)}} & \multicolumn{1}{c|}{\cellcolor{blue!5}\makecell[c]{Backtracking\\on failure \\ (4\%)}} & \multicolumn{1}{c|}{\cellcolor{blue!11}\makecell[c]{Unsp-\\ecified \\ (16\%)}} & \multicolumn{1}{c|}{\cellcolor{blue!15}\makecell[c]{Confidence\\-based\\branching \\ (23\%)}} & \multicolumn{1}{c|}{\cellcolor{blue!5}\makecell[c]{Majority\\voting \\ (7\%)}} & \multicolumn{1}{c|}{\cellcolor{blue!5}\makecell[c]{Human-in-\\the-loop\\checkpoints \\ (4\%)}} \\
    \hline

\cellcolor{white}2026 & \cite{4} &  &  &  &  &  &  &  & \cmark &  & \cmark &  & \cmark &  &  &  &  &  &  \\
\rowcolor{gray!15} \cellcolor{white} & \cite{5} &  &  &  &  &  &  &  &  &  &  & \cmark & \cmark &  &  &  &  & \cmark &  \\
\cellcolor{white} & \cite{11} &  &  &  &  &  &  &  & \cmark &  &  &  & \cmark &  &  &  &  &  &  \\
\rowcolor{gray!15} \cellcolor{white} & \cite{19} &  &  &  &  &  &  &  &  & \cmark &  &  & \cmark &  &  &  &  &  &  \\
\cellcolor{white} & \cite{23lbathaviator} &  &  & \cmark &  &  & \cmark & \cmark & \cmark &  &  &  & \cmark &  &  &  &  &  &  \\
\rowcolor{gray!15} \cellcolor{white} & \cite{25zhang2026bountybench} & \cmark &  &  &  &  &  & \cmark &  &  &  &  & \cmark &  &  &  &  &  &  \\
\cellcolor{white} & \cite{26RIGAKI2026129987} &  &  &  &  &  &  &  & \cmark &  &  &  & \cmark &  &  &  & \cmark &  &  \\
\rowcolor{gray!15} \cellcolor{white} & \cite{36ZOU2026104305} &  &  &  &  &  &  & \cmark & \cmark &  &  &  & \cmark & \cmark &  &  & \cmark &  & \cmark \\
\cellcolor{white} & \cite{40zhuo2026cyberzero} &  &  &  &  &  &  & \cmark & \cmark &  &  &  & \cmark &  &  &  &  &  &  \\
\rowcolor{gray!15} \cellcolor{white} & \cite{57} &  &  & \cmark &  &  &  &  &  &  &  &  & \cmark &  &  &  &  &  &  \\
\cellcolor{white} & \cite{64happe2026llms} &  & \cmark & \cmark &  &  &  &  & \cmark & \cmark &  &  & \cmark &  &  &  &  &  &  \\
\rowcolor{gray!15} \cellcolor{white} & \cite{84WANG2026103731} &  & \cmark &  &  &  &  &  &  &  &  &  & \cmark & \cmark &  &  & \cmark &  &  \\
\cellcolor{white} & \cite{86} & \cmark &  &  &  &  &  &  & \cmark &  &  &  & \cmark &  &  &  & \cmark &  &  \\
\rowcolor{gray!15} \cellcolor{white} & \cite{91zhu-etal-2026-teams} &  &  &  &  &  &  &  &  &  &  & \cmark & \cmark & \cmark &  &  &  &  &  \\
\cellcolor{white} & \cite{92} &  &  &  &  &  &  &  &  &  &  & \cmark & \cmark &  &  &  &  &  &  \\
\rowcolor{gray!15} \cellcolor{white} & \cite{96} &  &  &  &  &  &  &  &  &  &  & \cmark &  &  &  & \cmark & \cmark &  &  \\
\cellcolor{white} & \cite{97} &  & \cmark & \cmark &  &  &  &  & \cmark & \cmark &  &  & \cmark & \cmark &  &  & \cmark & \cmark &  \\
\rowcolor{gray!15} \cellcolor{white} & \cite{101} &  &  &  &  &  &  & \cmark & \cmark &  &  &  & \cmark &  &  &  &  &  &  \\
\hline
\cellcolor{white}2025 & \cite{2} &  &  &  &  &  & \cmark &  & \cmark &  &  &  & \cmark & \cmark &  &  &  &  &  \\
\rowcolor{gray!15} \cellcolor{white} & \cite{3} &  & \cmark &  &  &  &  &  & \cmark &  &  &  &  &  &  & \cmark &  &  &  \\
\cellcolor{white} & \cite{6} &  &  & \cmark &  &  &  &  &  &  &  &  & \cmark &  &  &  &  &  &  \\
\rowcolor{gray!15} \cellcolor{white} & \cite{7} &  &  &  &  &  &  &  & \cmark &  &  &  & \cmark &  &  &  &  &  &  \\
\cellcolor{white} & \cite{8zhang2025agent} & \cmark &  &  &  &  &  &  &  &  &  &  & \cmark &  &  &  &  &  &  \\
\rowcolor{gray!15} \cellcolor{white} & \cite{9jie2025agent4vul} &  &  &  &  &  &  &  & \cmark &  &  &  &  &  &  & \cmark &  &  &  \\
\cellcolor{white} & \cite{12} &  &  &  &  &  &  &  & \cmark &  &  &  & \cmark &  &  &  &  &  &  \\
\rowcolor{gray!15} \cellcolor{white} & \cite{13} & \cmark &  &  &  &  &  &  & \cmark &  &  &  & \cmark &  &  &  & \cmark &  &  \\
\cellcolor{white} & \cite{15} &  & \cmark &  &  & \cmark &  &  & \cmark &  &  &  &  &  &  & \cmark &  &  &  \\
\rowcolor{gray!15} \cellcolor{white} & \cite{16} &  & \cmark &  &  &  &  &  &  &  &  &  & \cmark &  &  &  &  &  &  \\
\cellcolor{white} & \cite{17} &  &  &  &  &  &  &  &  & \cmark &  &  & \cmark &  &  &  &  & \cmark &  \\
\rowcolor{gray!15} \cellcolor{white} & \cite{18} &  &  &  &  &  &  &  & \cmark & \cmark & \cmark &  & \cmark &  &  &  &  &  &  \\
\cellcolor{white} & \cite{20} &  & \cmark &  &  &  &  &  & \cmark &  &  &  &  & \cmark &  &  &  &  &  \\
\rowcolor{gray!15} \cellcolor{white} & \cite{21} &  & \cmark &  &  &  &  &  & \cmark &  &  &  & \cmark &  &  &  & \cmark &  &  \\
\cellcolor{white} & \cite{22} & \cmark &  &  &  &  &  &  & \cmark &  &  &  & \cmark & \cmark &  &  & \cmark &  &  \\
\rowcolor{gray!15} \cellcolor{white} & \cite{24yildiz-etal-2025-benchmarking} & \cmark & \cmark &  &  &  & \cmark &  &  &  &  &  & \cmark &  &  &  &  &  &  \\
\cellcolor{white} & \cite{27} & \cmark &  &  &  &  &  &  &  & \cmark &  &  & \cmark &  &  &  & \cmark &  &  \\
\rowcolor{gray!15} \cellcolor{white} & \cite{29} &  & \cmark & \cmark &  &  &  &  &  & \cmark &  &  & \cmark &  & \cmark &  &  &  &  \\
\cellcolor{white} & \cite{30} &  &  &  &  &  &  &  &  &  &  & \cmark &  &  &  & \cmark &  &  &  \\
\rowcolor{gray!15} \cellcolor{white} & \cite{31} &  &  &  &  &  &  &  & \cmark & \cmark &  &  & \cmark & \cmark &  &  & \cmark &  &  \\
\cellcolor{white} & \cite{32} &  &  &  &  &  &  &  & \cmark &  &  &  &  &  &  & \cmark &  &  & \cmark \\
\rowcolor{gray!15} \cellcolor{white} & \cite{33} &  &  &  &  &  & \cmark &  &  &  &  &  &  &  &  & \cmark &  & \cmark &  \\
\cellcolor{white} & \cite{34} &  &  &  &  &  &  &  &  & \cmark &  &  &  &  &  & \cmark & \cmark & \cmark &  \\
\rowcolor{gray!15} \cellcolor{white} & \cite{35} &  &  &  &  &  &  &  & \cmark &  &  &  &  & \cmark &  &  &  &  &  \\
\cellcolor{white} & \cite{37app15169096} &  &  &  &  &  &  &  & \cmark & \cmark &  &  & \cmark & \cmark &  &  & \cmark &  &  \\
\rowcolor{gray!15} \cellcolor{white} & \cite{38zhu2025cvebenchbenchmarkaiagents} &  &  &  &  &  &  &  &  &  &  & \cmark & \cmark & \cmark &  &  &  &  &  \\
\cellcolor{white} & \cite{39} &  &  &  &  &  &  &  & \cmark &  &  &  & \cmark &  &  &  &  &  &  \\
\rowcolor{gray!15} \cellcolor{white} & \cite{41jiao2025deepvulhunter} &  & \cmark &  &  &  &  &  &  &  &  &  &  &  &  & \cmark &  &  &  \\
\cellcolor{white} & \cite{42LOEVENICH2025111162} & \cmark &  &  &  &  &  &  &  & \cmark &  &  & \cmark & \cmark & \cmark &  & \cmark & \cmark &  \\
\rowcolor{gray!15} \cellcolor{white} & \cite{45} &  &  &  &  &  &  &  & \cmark &  &  &  & \cmark &  &  &  &  &  &  \\
\cellcolor{white} & \cite{46abramovich2025enigma} &  &  &  &  &  &  & \cmark &  &  &  &  & \cmark &  &  &  &  &  &  \\
\rowcolor{gray!15} \cellcolor{white} & \cite{47chowdhry2025evaluating} &  &  &  &  &  &  &  &  &  &  & \cmark & \cmark &  &  &  &  &  &  \\
\cellcolor{white} & \cite{48} &  &  &  &  &  &  &  &  &  &  & \cmark & \cmark &  &  &  &  &  &  \\
\rowcolor{gray!15} \cellcolor{white} & \cite{49} &  & \cmark &  &  &  &  &  &  & \cmark &  &  & \cmark &  & \cmark &  & \cmark &  &  \\
\cellcolor{white} & \cite{51} &  &  &  &  &  & \cmark &  & \cmark & \cmark &  &  & \cmark &  &  &  &  &  &  \\
\rowcolor{gray!15} \cellcolor{white} & \cite{53} &  &  &  &  &  &  &  &  &  &  & \cmark & \cmark &  &  &  &  &  &  \\
\cellcolor{white} & \cite{55} & \cmark &  &  &  &  &  &  & \cmark &  &  &  & \cmark &  &  &  &  &  &  \\
\rowcolor{gray!15} \cellcolor{white} & \cite{56} & \cmark &  &  &  &  & \cmark &  & \cmark &  &  &  & \cmark &  &  &  &  &  & \cmark \\
\cellcolor{white} & \cite{58} &  &  &  &  &  &  &  & \cmark &  &  &  & \cmark &  &  &  &  &  &  \\
\rowcolor{gray!15} \cellcolor{white} & \cite{59} &  &  &  &  &  &  &  & \cmark &  &  &  & \cmark &  & \cmark &  &  &  &  \\
\cellcolor{white} & \cite{60} &  &  &  &  &  &  &  & \cmark & \cmark &  &  & \cmark & \cmark &  &  &  &  &  \\
\rowcolor{gray!15} \cellcolor{white} & \cite{61} &  &  &  &  &  &  &  & \cmark &  &  &  & \cmark & \cmark &  &  &  & \cmark &  \\
\cellcolor{white} & \cite{62} &  &  &  &  &  &  &  &  &  &  & \cmark & \cmark &  &  &  &  &  &  \\
\rowcolor{gray!15} \cellcolor{white} & \cite{65} &  &  &  &  &  &  &  &  & \cmark &  &  & \cmark &  &  &  &  &  &  \\
\cellcolor{white} & \cite{67} &  &  &  &  &  &  &  &  &  &  & \cmark & \cmark &  &  &  &  &  &  \\
\rowcolor{gray!15} \cellcolor{white} & \cite{68} &  &  &  &  &  &  &  &  &  &  & \cmark &  &  &  & \cmark & \cmark &  &  \\
\cellcolor{white} & \cite{69} &  &  &  &  &  &  &  &  & \cmark &  &  & \cmark &  &  &  &  &  &  \\
\rowcolor{gray!15} \cellcolor{white} & \cite{70} &  &  &  &  &  &  & \cmark &  &  &  &  & \cmark &  &  &  & \cmark &  &  \\
\cellcolor{white} & \cite{71} &  &  &  &  &  &  & \cmark & \cmark &  &  &  & \cmark & \cmark &  &  &  &  &  \\
\rowcolor{gray!15} \cellcolor{white} & \cite{72} & \cmark & \cmark &  &  &  &  &  & \cmark &  &  &  & \cmark &  &  &  &  &  &  \\
\cellcolor{white} & \cite{73} &  &  &  &  &  & \cmark &  &  & \cmark &  &  & \cmark & \cmark &  &  &  &  &  \\
\rowcolor{gray!15} \cellcolor{white} & \cite{75} &  &  & \cmark &  &  &  & \cmark &  &  &  &  & \cmark &  &  &  &  &  &  \\
\cellcolor{white} & \cite{78} &  &  &  &  &  &  & \cmark &  &  &  &  & \cmark &  &  &  & \cmark &  &  \\
\rowcolor{gray!15} \cellcolor{white} & \cite{80} &  &  &  &  &  &  &  & \cmark &  &  &  &  & \cmark &  &  &  &  &  \\
\cellcolor{white} & \cite{81Cao_Huang_Li_Huilin_He_Oo_Hooi_2025} &  &  &  &  &  &  &  &  & \cmark &  &  &  &  &  & \cmark &  &  &  \\
\rowcolor{gray!15} \cellcolor{white} & \cite{83} &  &  &  &  &  &  & \cmark &  & \cmark &  &  & \cmark &  &  &  &  &  &  \\
\cellcolor{white} & \cite{87} &  &  & \cmark & \cmark &  &  &  & \cmark &  &  &  & \cmark & \cmark &  &  &  &  &  \\
\rowcolor{gray!15} \cellcolor{white} & \cite{88} &  &  &  &  &  &  & \cmark &  &  &  &  & \cmark &  &  &  &  &  &  \\
\cellcolor{white} & \cite{89guo2025repoauditautonomousllmagentrepositorylevel} &  &  &  &  &  & \cmark & \cmark &  & \cmark &  &  & \cmark &  &  &  &  &  &  \\
\rowcolor{gray!15} \cellcolor{white} & \cite{90} &  &  &  &  &  &  &  &  &  &  & \cmark & \cmark &  &  &  &  &  &  \\
\cellcolor{white} & \cite{93info16050365} &  &  &  &  &  &  &  & \cmark & \cmark &  &  & \cmark &  &  &  & \cmark &  &  \\
\rowcolor{gray!15} \cellcolor{white} & \cite{94} &  &  &  &  &  &  &  & \cmark & \cmark &  &  & \cmark &  &  &  &  &  &  \\
\cellcolor{white} & \cite{99} &  &  &  &  &  &  &  & \cmark & \cmark &  &  &  &  &  & \cmark & \cmark &  &  \\
\rowcolor{gray!15} \cellcolor{white} & \cite{100} &  &  &  &  &  &  & \cmark &  & \cmark &  &  & \cmark &  &  &  &  &  &  \\
\hline
\cellcolor{white}2024 & \cite{1} &  & \cmark &  &  &  &  &  &  &  &  &  &  &  &  & \cmark &  &  &  \\
\rowcolor{gray!15} \cellcolor{white} & \cite{10debenedetti2024agentdojo} & \cmark &  &  &  &  &  &  &  &  &  &  & \cmark &  &  &  &  &  &  \\
\cellcolor{white} & \cite{14chen2024agentpoison} &  &  &  &  &  &  &  &  &  &  & \cmark &  &  &  & \cmark &  &  &  \\
\rowcolor{gray!15} \cellcolor{white} & \cite{28} &  &  & \cmark &  &  &  &  &  &  &  &  & \cmark &  &  &  & \cmark &  &  \\
\cellcolor{white} & \cite{43} & \cmark &  &  &  &  &  &  &  & \cmark &  &  & \cmark & \cmark &  &  &  &  &  \\
\rowcolor{gray!15} \cellcolor{white} & \cite{50} &  &  &  &  &  &  &  &  & \cmark &  &  &  &  &  & \cmark &  &  &  \\
\cellcolor{white} & \cite{54} &  &  &  &  &  &  &  &  &  &  & \cmark & \cmark &  &  &  &  &  &  \\
\rowcolor{gray!15} \cellcolor{white} & \cite{63zemicheal2024llm} &  &  &  &  &  &  &  &  & \cmark &  &  &  & \cmark &  &  &  &  & \cmark \\
\cellcolor{white} & \cite{66} &  &  &  &  &  &  &  & \cmark &  &  &  & \cmark &  &  &  &  &  &  \\
\rowcolor{gray!15} \cellcolor{white} & \cite{74icaart24} &  &  &  &  &  &  &  &  &  &  & \cmark & \cmark &  &  &  &  &  &  \\
\cellcolor{white} & \cite{76} &  &  &  &  &  &  &  & \cmark &  &  &  & \cmark &  &  &  &  &  &  \\
\rowcolor{gray!15} \cellcolor{white} & \cite{77} &  &  &  &  &  &  &  &  &  &  & \cmark & \cmark &  &  &  &  &  &  \\
\cellcolor{white} & \cite{79} &  &  &  &  &  &  &  & \cmark & \cmark &  &  & \cmark &  &  &  & \cmark &  &  \\
\rowcolor{gray!15} \cellcolor{white} & \cite{82rigaki2024prompt} & \cmark &  &  &  &  & \cmark &  &  &  &  &  & \cmark &  &  &  &  &  &  \\
\cellcolor{white} & \cite{85} & \cmark &  &  &  & \cmark &  &  & \cmark &  &  &  & \cmark &  &  &  &  &  &  \\
\rowcolor{gray!15} \cellcolor{white} & \cite{95} &  &  &  &  &  &  &  &  &  &  & \cmark & \cmark & \cmark &  &  &  &  &  \\
\cellcolor{white} & \cite{98} &  &  & \cmark & \cmark &  &  &  & \cmark & \cmark &  &  & \cmark & \cmark &  &  &  &  &  \\
\hline
\rowcolor{gray!15} \cellcolor{white}2023 & \cite{52} &  & \cmark &  &  &  & \cmark & \cmark &  &  &  &  &  &  &  & \cmark & \cmark &  &  \\
\hline
    \end{tabular}%
\end{adjustbox}
  \label{tab:approach-agent-reasoning-planning}%
  \vspace{-18pt}
\end{table*}%

\noindent
{\bf Reasoning paradigms.}
\ul{ReAct} reasoning interleaves reasoning, tool action, and observation, making
it a natural fit for interactive security tasks where each command changes what
the agent knows next
\cite{25zhang2026bountybench,86,8zhang2025agent,13,22,
24yildiz-etal-2025-benchmarking,27,42LOEVENICH2025111162,55,56,72,
10debenedetti2024agentdojo,43,82rigaki2024prompt,85}. Rather than producing a
single final answer from a static prompt, these agents repeatedly decide what to
inspect, execute, or query. Wu et al.'s \textsc{AutoPT}~\cite{22}, for example,
frames web penetration testing as an iterative process in which the agent
observes tool feedback and then chooses the next testing action. The same
pattern appears in benchmark-style agent settings such as BountyBench
\cite{25zhang2026bountybench}, where code inspection, command execution, and
final vulnerability handling are all part of the action-observation loop.

\ul{Chain-of-Thought (CoT)} makes intermediate reasoning explicit, usually by
asking the LLM to decompose the problem, compare evidence, or justify a next
step before acting
\cite{64happe2026llms,84WANG2026103731,97,3,15,16,20,21,
24yildiz-etal-2025-benchmarking,29,41jiao2025deepvulhunter,49,72,1,52}.
This paradigm is especially visible when the agent must translate noisy
observations into a plausible security interpretation. Ginige et al.'s
\textsc{AutoPentester}~\cite{21} uses CoT-style reasoning in its strategy
analysis stage to interpret previous penetration-testing results and choose
subsequent attack strategies. Wang et al.'s
\textsc{PTFusion}~\cite{84WANG2026103731} uses preference-based CoT prompting
to align heterogeneous penetration-testing tool outputs with the current
strategic plan.

\ul{Reflection / self-critique} gives the agent an explicit opportunity to
assess its own intermediate products, failed attempts, or candidate findings
before continuing
\cite{23lbathaviator,57,64happe2026llms,97,6,29,75,87,28,98}. The role of
reflection varies by task. In Chen et al.'s
\textsc{ClearAgent}~\cite{29}, reflection is tied to exploration: the agent
reconsiders visited binary locations and candidate bug evidence before deciding
where to inspect next. In Dai et al.'s
\textsc{RefPentester}~\cite{87}, reflection has a more operational role,
turning failed penetration-testing attempts into guidance for later actions.

\ul{Reflexion} is a narrower pattern in which feedback from previous attempts is
converted into reusable verbal guidance for future trials \cite{87,98}. Its
main value is continuity across attempts: the agent does not merely observe
failure, but writes down a lesson that changes later behavior. This is closely
aligned with self-reflective penetration testing, where the same class of
mistake can otherwise recur across multiple stages.

\ul{Prompt chaining} separates a complex task into a sequence of prompts, each
responsible for a narrower subproblem \cite{15,85}. \ul{Few-shot prompting}
instead provides examples inside the prompt so that the model can imitate the
desired analysis pattern or output form
\cite{23lbathaviator,2,24yildiz-etal-2025-benchmarking,33,51,56,73,
89guo2025repoauditautonomousllmagentrepositorylevel,82rigaki2024prompt,52}.
These two techniques address different forms of task complexity: prompt chains
control workflow order, while few-shot examples calibrate local reasoning
behavior. Kim et al.'s \textsc{AutoRestTest}~\cite{2}, for instance, uses
example-guided generation to help produce API inputs that respect parameter
semantics and constraints.

\ul{Self-debug / self-repair} appears when the agent generates an artifact,
tests or checks it, and then revises the artifact based on feedback
\cite{23lbathaviator,25zhang2026bountybench,36ZOU2026104305,
40zhuo2026cyberzero,101,46abramovich2025enigma,70,71,75,78,83,88,
89guo2025repoauditautonomousllmagentrepositorylevel,100,52}. Bouzenia et al.'s \textsc{RepairAgent}~\cite{88} illustrates the
repair side: the agent searches for repair ingredients, proposes a fix, and
uses validation feedback to revise its candidate patch. Liu et al.'s
\textsc{PropertyGPT}~\cite{83} follows a similar loop for formal properties,
using compiler, static-analysis, and prover feedback to revise generated
specifications.

\ul{Least-to-most task decomposition} breaks a large security objective into
smaller subtasks or stages before solving them
\cite{4,11,23lbathaviator,26RIGAKI2026129987,36ZOU2026104305,
40zhuo2026cyberzero,64happe2026llms,86,97,101,2,3,7,12,13,15,18,20,21,22,
31,32,35,37app15169096,39,45,51,55,56,58,59,60,61,71,72,80,87,
93info16050365,94,99,66,76,79,85,98}. For example, Deng et al.'s
\textsc{PentestGPT}~\cite{79} captures this idea through a pentesting task tree
that keeps high-level strategy separate from command-level execution.

\ul{Hypothesis-driven reasoning} starts from possible vulnerabilities, attack
paths, causes, or explanations and then seeks evidence to support or reject
them
\cite{19,64happe2026llms,97,17,18,27,29,31,34,37app15169096,
42LOEVENICH2025111162,49,51,60,65,69,73,
81Cao_Huang_Li_Huilin_He_Oo_Hooi_2025,83,
89guo2025repoauditautonomousllmagentrepositorylevel,93info16050365,94,99,100,
43,50,63zemicheal2024llm,79,98}. In \textsc{RepoAudit}, Guo et
al.~\cite{89guo2025repoauditautonomousllmagentrepositorylevel} use repository
exploration and validation to move from suspected data-flow facts toward
actionable bug reports. In \textsc{PropertyGPT}, Liu et al.~\cite{83} use
formal checking to determine whether generated properties are compilable,
appropriate, and verifiable.

\ul{Graph-based reasoning / attack graph construction} is important for tasks whose logic is relational, multi-hop, or path-dependent \cite{4,18}. Graph structures can represent preconditions, dependencies,
attack steps, or environment states, giving the agent a more explicit reasoning
substrate than a flat prompt.

\ul{Unspecified reasoning paradigm} denotes papers that do not name or provide enough procedural evidence for a particular reasoning paradigm~\cite{5,91zhu-etal-2026-teams,92,96,30,38zhu2025cvebenchbenchmarkaiagents,47chowdhry2025evaluating,48,53,62,67,68,90,14chen2024agentpoison,54,74icaart24,77,95}. We do not infer ReAct or chain-of-thought from generic iteration, tool use, or multi-step execution alone; accordingly, ``Unspecified'' means that the paradigm is under-reported, not that the agent performs no reasoning.

\noindent
{\bf Planning strategies.} Planning strategies capture how an agent plans. A paper can assign a
named Planner agent without describing a specific planning strategy (e.g., a
brief one-line mention of ``planning'' inside a larger pipeline), and a paper
can plan dynamically without ever naming a distinct Planner role (e.g., a
single ReAct-style loop that revises its next action after every
observation). The two tags therefore overlap substantially but are not
interchangeable, and their totals should not be read as describing the same
set of papers.
\ul{Dynamic / reactive planning} is the dominant planning style: agents revise
their next steps according to new observations, failed attempts, tool output, or
environment changes
\cite{4,5,8zhang2025agent,11,19,23lbathaviator,24yildiz-etal-2025-benchmarking,25zhang2026bountybench,
26RIGAKI2026129987,36ZOU2026104305,40zhuo2026cyberzero,57,
64happe2026llms,84WANG2026103731,86,91zhu-etal-2026-teams,92,94,97,101,2,6,7,12,
13,16,17,18,21,22,27,29,31,37app15169096,
38zhu2025cvebenchbenchmarkaiagents,39,42LOEVENICH2025111162,45,
46abramovich2025enigma,47chowdhry2025evaluating,48,49,51,53,55,56,58,59,60,
61,62,65,67,69,70,71,72,73,75,78,83,87,88,
89guo2025repoauditautonomousllmagentrepositorylevel,90,93info16050365,100,
10debenedetti2024agentdojo,28,43,54,66,74icaart24,76,77,79,82rigaki2024prompt,85,95,98}.
The important feature is not merely iteration, but contingency: the next plan
depends on what the agent just learned. \textsc{ClearAgent}~\cite{29} chooses
new binary-analysis targets based on its current exploration context, whereas
\textsc{RepairAgent}~\cite{88} changes repair behavior after tests or
validation expose problems with a candidate patch. \textsc{RedAgent}~\cite{86}
dynamically selects refinement actions based on real-time responses from the
target model, and the ReAct agents benchmarked in
\textsc{JITVUL}~\cite{24yildiz-etal-2025-benchmarking} and Agent Security
Bench~\cite{8zhang2025agent} inherit the same thought-action-observation
contingency from their underlying ReAct paradigm.

\ul{Hierarchical task decomposition} separates high-level objectives from
lower-level execution choices
\cite{36ZOU2026104305,63zemicheal2024llm,84WANG2026103731,91zhu-etal-2026-teams,97,2,20,22,31,
35,37app15169096,38zhu2025cvebenchbenchmarkaiagents,42LOEVENICH2025111162,
60,61,71,73,80,87,43,95,98}.
Zhu et al.'s HPTSA~\cite{91zhu-etal-2026-teams} is a clear example: a planning
agent analyzes the target and launches specialized subagents for exploitation
subtasks. \textsc{RefPentester}~\cite{87} uses a staged penetration-testing
structure, letting the current stage constrain which tactical actions are
reasonable. The NVIDIA CVE-triage system of ZeMicheal et
al.~\cite{63zemicheal2024llm} follows the same pattern at the task level: a
Plan-and-Execute-style LLM planner first decomposes each CVE into a
context-sensitive checklist, which is then executed step by step.

\ul{Backtracking on failure} appears when the agent returns to an earlier
choice, abandons a weak path, or retries from a different branch after evidence
does not support the current plan \cite{29,42LOEVENICH2025111162,49,59}. \textsc{ClearAgent}~\cite{29} provides a compact
example at the analysis level, revising its exploration direction when the
current binary path does not substantiate a candidate vulnerability.

\ul{Unspecified planning strategy} denotes papers that mention planning only
in passing, or not at all, without enough procedural detail to classify the
planning behavior as dynamic/reactive, hierarchical, or backtracking
\cite{1,3,9jie2025agent4vul,14chen2024agentpoison,15,30,32,33,34,
41jiao2025deepvulhunter,50,52,68,81Cao_Huang_Li_Huilin_He_Oo_Hooi_2025,96,99}.
As with unspecified reasoning paradigms, this label reflects under-reporting
rather than the absence of any planning; we introduce it as a fourth planning
strategy so that under-specified papers are marked explicitly instead of
silently contributing to the denominator without a tag.

\noindent
{\bf Decision-making uncertainty.}
\ul{Confidence-based branching} lets the agent choose different actions
depending on the strength of evidence, estimated risk, or confidence in an
intermediate finding
\cite{26RIGAKI2026129987,36ZOU2026104305,84WANG2026103731,86,96,97,13,21,22,
27,31,34,37app15169096,42LOEVENICH2025111162,49,68,70,78,93info16050365,99,
28,79,52}. \textsc{PTFusion}~\cite{84WANG2026103731} represents
this idea at the planning level by using context-aware knowledge fusion and
executed actions to guide more precise next commands.

\ul{Majority voting} handles uncertainty by comparing outputs from multiple
agents and selecting the most supported answer
\cite{5,97,17,33,34,42LOEVENICH2025111162,61}. This pattern is most natural
when the task admits plausible disagreement, such as smart-contract
vulnerability classification or ambiguous defensive decisions. In these
settings, the system treats disagreement itself as useful evidence rather than
assuming that a single generation is authoritative.

\ul{Human-in-the-loop checkpoints} reserve key decisions for human approval,
verification, or guidance \cite{36ZOU2026104305,32,56,63zemicheal2024llm}. This
is a practical response to uncertainty and risk: the agent may gather evidence,
rank hypotheses, or draft a recommendation, while a human analyst validates the
finding before a sensitive action is taken. In Sheikhi et al.'s
\textsc{Cognitive SOC}~\cite{32}, this boundary is visible in the
analyst-facing narrative: the agent organizes evidence and explanation, but the
human remains the decision maker for security-operation response.

\vspace{-4pt}
\subsubsection{Agent action space}

The security capability of an agent is often determined less by the language model alone than by what the model is allowed to inspect, execute, query, and validate through tools. The agent action space describes the external operations that an LLM-based security agent can invoke after reasoning about a task. In this survey, actions range from conventional software-analysis utilities to security platforms, multi-agent coordination mechanisms, and human-facing review steps, as summarized in Table~\ref{tab:approach-agent-action-space}.

\begin{table*}[t]
  \centering
  \vspace{-10pt}
  \caption{Paper attribution for Approach: Agent action space}
  \vspace{-10pt}
  \scriptsize
  \setlength{\tabcolsep}{1.5pt}
  \renewcommand{\arraystretch}{0.70}
\begin{adjustbox}{width=\textwidth,max totalheight=0.88\textheight,center}
    \begin{tabular}{|c|c|c|c|c|c|c|c|c|c|c|c|c|c|c|c|c|c|c|c|c|c|c|c|c|c|c|c|c|c|c|}
    \hline
\rowcolor{white}  & & 
\multicolumn{5}{c|}{\makecell[c]{Static analysis tools}} & \multicolumn{4}{c|}{\makecell[c]{Dynamic analysis tools}} & \multicolumn{4}{c|}{\makecell[c]{Code execution \& environment}} & \multicolumn{3}{c|}{\makecell[c]{Information retrieval tools}} & \multicolumn{8}{c|}{\makecell[c]{Security-specific platforms}} & \multicolumn{3}{c|}{\makecell[c]{Inter-agent communication}} & \multicolumn{2}{c|}{\makecell[c]{Human interaction}} \\
\cline{3-7}\cline{8-11}\cline{12-15}\cline{16-18}\cline{19-26}\cline{27-29}\cline{30-31}
\rowcolor{white} 
\multicolumn{1}{|c|}{\multirow{-4}{*}[-0.2ex]{\makecell[c]{Year}}} & 
\multicolumn{1}{c|}{\multirow{-4}{*}[-0.2ex]{\makecell[c]{Paper}}} & 
\multicolumn{1}{c|}{\cellcolor{blue!10}\makecell[c]{Linters\\and\\pattern\\matchers \\ (15\%)}} & \multicolumn{1}{c|}{\cellcolor{blue!5}\makecell[c]{Data-\\flow/\\taint\\analyzers\\ (5\%)}} & \multicolumn{1}{c|}{\cellcolor{blue!5}\makecell[c]{Symbolic\\execution\\engines \\ (2\%)}} & \multicolumn{1}{c|}{\cellcolor{blue!5}\makecell[c]{Formal\\verification\\tools \\ (5\%)}} & \multicolumn{1}{c|}{\cellcolor{blue!5}\makecell[c]{Binary\\analysis\\frameworks \\ (6\%)}} & \multicolumn{1}{c|}{\cellcolor{blue!5}\makecell[c]{Fuzzers \\ (4\%)}} & \multicolumn{1}{c|}{\cellcolor{blue!5}\makecell[c]{Debuggers \\ (4\%)}} & \multicolumn{1}{c|}{\cellcolor{blue!11}\makecell[c]{Execution\\sandboxes \\ (17\%)}} & \multicolumn{1}{c|}{\cellcolor{blue!6}\makecell[c]{Network\\traffic\\analysers \\ (9\%)}} & \multicolumn{1}{c|}{\cellcolor{blue!5}\makecell[c]{Compilers /\\interpreters \\ (5\%)}} & \multicolumn{1}{c|}{\cellcolor{blue!13}\makecell[c]{REPL/\\shell\\execution \\ (19\%)}} & \multicolumn{1}{c|}{\cellcolor{blue!7}\makecell[c]{Containerised\\environments \\ (11\%)}} & \multicolumn{1}{c|}{\cellcolor{blue!5}\makecell[c]{Browser\\/web\\automation \\ (7\%)}} & \multicolumn{1}{c|}{\cellcolor{blue!11}\makecell[c]{Code\\search \\ (17\%)}} & \multicolumn{1}{c|}{\cellcolor{blue!25}\makecell[c]{Documentation\\retrieval \\ (38\%)}} & \multicolumn{1}{c|}{\cellcolor{blue!19}\makecell[c]{Vector\\store\\/RAG\\retrieval \\ (29\%)}} & \multicolumn{1}{c|}{\cellcolor{blue!9}\makecell[c]{CTF\\platforms \\ (14\%)}} & \multicolumn{1}{c|}{\cellcolor{blue!17}\makecell[c]{Penetration\\testing\\frameworks \\ (25\%)}} & \multicolumn{1}{c|}{\cellcolor{blue!5}\makecell[c]{Vulnerability\\scanners \\ (6\%)}} & \multicolumn{1}{c|}{\cellcolor{blue!5}\makecell[c]{SIEM\\/ SOAR\\integration \\ (6\%)}} & \multicolumn{1}{c|}{\cellcolor{blue!6}\makecell[c]{MITRE\\ATT\&CK\\integration \\ (9\%)}} & \multicolumn{1}{c|}{\cellcolor{blue!6}\makecell[c]{Cloud\\/IaC\\platform\\tools \\ (9\%)}} & \multicolumn{1}{c|}{\cellcolor{blue!8}\makecell[c]{Cyber-range\\/network\\simulation\\environments \\ (12\%)}} & \multicolumn{1}{c|}{\cellcolor{blue!5}\makecell[c]{Asset\\search/\\OSINT\\platforms \\ (1\%)}} & \multicolumn{1}{c|}{\cellcolor{blue!13}\makecell[c]{Message\\passing\\between\\agents \\ (20\%)}} & \multicolumn{1}{c|}{\cellcolor{blue!5}\makecell[c]{Voting/\\aggregation\\protocols \\ (6\%)}} & \multicolumn{1}{c|}{\cellcolor{blue!5}\makecell[c]{Blockchain\\consensus\\/audit\\trail \\ (1\%)}} & \multicolumn{1}{c|}{\cellcolor{blue!5}\makecell[c]{Querying\\human\\expert \\ (1\%)}} & \multicolumn{1}{c|}{\cellcolor{blue!8}\makecell[c]{Findings\\ for\\human\\review \\ (12\%)}} \\
    \hline

\cellcolor{white}2026 & \cite{4} & \cmark &  &  &  &  &  &  & \cmark &  &  & \cmark & \cmark &  &  & \cmark & \cmark &  & \cmark & \cmark &  &  &  &  &  &  &  &  &  &  \\
\rowcolor{gray!15} \cellcolor{white} & \cite{5} &  &  &  &  &  &  &  &  &  &  &  &  &  &  &  &  &  &  &  &  &  &  &  &  & \cmark & \cmark & \cmark &  &  \\
\cellcolor{white} & \cite{11} &  &  & \cmark & \cmark &  &  &  &  &  & \cmark &  &  &  & \cmark &  &  &  &  &  &  &  &  &  &  &  &  &  &  &  \\
\rowcolor{gray!15} \cellcolor{white} & \cite{19} &  &  &  &  &  &  &  &  & \cmark &  &  &  &  &  &  & \cmark &  &  &  &  &  &  &  &  &  &  &  &  & \cmark \\
\cellcolor{white} & \cite{23lbathaviator} & \cmark &  &  & \cmark &  &  &  &  &  &  &  &  &  &  & \cmark & \cmark &  &  &  &  &  &  &  &  &  &  &  &  &  \\
\rowcolor{gray!15} \cellcolor{white} & \cite{25zhang2026bountybench} &  &  &  &  &  &  &  & \cmark &  &  & \cmark & \cmark &  & \cmark &  &  &  &  &  &  &  &  &  &  &  &  &  &  &  \\
\cellcolor{white} & \cite{26RIGAKI2026129987} &  &  &  &  &  &  &  & \cmark &  &  &  &  &  &  &  &  &  & \cmark &  &  & \cmark &  & \cmark &  &  &  &  &  &  \\
\rowcolor{gray!15} \cellcolor{white} & \cite{36ZOU2026104305} &  &  &  &  &  &  & \cmark &  &  &  & \cmark &  &  & \cmark & \cmark &  & \cmark &  &  &  &  &  &  &  &  &  &  &  & \cmark \\
\cellcolor{white} & \cite{40zhuo2026cyberzero} &  &  &  &  &  &  &  &  &  &  &  &  &  &  &  &  & \cmark &  &  &  &  &  &  &  &  &  &  &  &  \\
\rowcolor{gray!15} \cellcolor{white} & \cite{57} &  &  &  &  &  &  &  &  &  &  &  &  &  &  &  & \cmark &  &  &  &  &  &  &  &  &  &  &  &  &  \\
\cellcolor{white} & \cite{64happe2026llms} &  &  &  &  &  &  &  &  &  &  & \cmark &  &  & \cmark &  &  &  & \cmark &  &  &  &  &  &  &  &  &  &  &  \\
\rowcolor{gray!15} \cellcolor{white} & \cite{84WANG2026103731} & \cmark & \cmark &  &  &  &  &  &  &  &  & \cmark &  & \cmark &  & \cmark &  &  & \cmark &  &  &  &  &  &  &  &  &  &  &  \\
\cellcolor{white} & \cite{86} &  &  &  &  &  &  &  &  &  &  &  &  &  &  &  & \cmark &  &  &  &  &  &  &  &  &  &  &  &  &  \\
\rowcolor{gray!15} \cellcolor{white} & \cite{91zhu-etal-2026-teams} &  &  &  &  &  &  &  &  &  &  &  &  &  & \cmark &  &  &  &  &  &  &  &  &  &  &  &  &  &  &  \\
\cellcolor{white} & \cite{92} &  &  &  &  &  &  &  &  &  &  &  &  &  &  &  &  &  &  &  &  &  & \cmark &  &  &  &  &  &  &  \\
\rowcolor{gray!15} \cellcolor{white} & \cite{96} &  &  &  &  &  &  &  &  &  &  &  &  &  &  &  &  &  &  &  &  &  &  &  &  &  &  &  &  & \cmark \\
\cellcolor{white} & \cite{97} &  &  &  & \cmark &  &  &  &  &  &  &  &  &  &  & \cmark &  &  &  &  &  &  &  &  &  & \cmark & \cmark &  &  &  \\
\rowcolor{gray!15} \cellcolor{white} & \cite{101} &  & \cmark &  &  &  &  &  & \cmark &  &  &  &  &  &  & \cmark &  &  &  & \cmark &  &  &  &  &  &  &  &  &  &  \\
\hline
\cellcolor{white}2025 & \cite{2} &  &  &  &  &  &  &  &  &  &  &  &  & \cmark &  &  &  &  &  &  &  &  &  &  &  &  &  &  &  &  \\
\rowcolor{gray!15} \cellcolor{white} & \cite{3} &  &  &  &  &  &  &  &  &  &  &  &  &  &  &  &  &  &  &  &  &  &  &  &  & \cmark &  &  &  & \cmark \\
\cellcolor{white} & \cite{6} &  &  &  &  &  &  &  &  &  &  &  &  &  &  &  &  &  &  &  &  &  &  &  &  & \cmark &  &  &  &  \\
\rowcolor{gray!15} \cellcolor{white} & \cite{7} &  &  &  &  &  &  &  &  &  &  &  &  &  &  &  & \cmark &  &  &  &  &  &  &  &  &  &  &  &  &  \\
\cellcolor{white} & \cite{8zhang2025agent} &  &  &  &  &  &  &  &  &  &  &  &  &  & \cmark &  & \cmark &  &  &  &  &  &  &  &  &  &  &  &  &  \\
\rowcolor{gray!15} \cellcolor{white} & \cite{9jie2025agent4vul} & \cmark &  &  &  &  &  &  &  &  &  &  &  &  &  &  &  &  &  &  &  &  &  &  &  &  &  &  &  &  \\
\cellcolor{white} & \cite{12} &  &  &  &  &  &  &  & \cmark &  &  & \cmark &  &  &  & \cmark &  &  & \cmark &  &  &  &  &  &  & \cmark &  &  &  &  \\
\rowcolor{gray!15} \cellcolor{white} & \cite{13} &  &  &  &  &  &  &  &  &  &  & \cmark &  &  &  &  & \cmark &  & \cmark &  &  &  &  &  &  &  &  &  &  &  \\
\cellcolor{white} & \cite{15} &  &  &  & \cmark &  &  &  &  &  &  &  &  &  &  &  &  &  &  &  &  &  &  &  &  &  &  &  &  &  \\
\rowcolor{gray!15} \cellcolor{white} & \cite{16} &  &  &  &  &  &  &  &  &  &  &  &  &  &  &  &  &  &  &  &  &  &  & \cmark &  & \cmark &  &  &  &  \\
\cellcolor{white} & \cite{17} &  &  &  &  &  &  &  &  &  &  &  &  &  &  &  &  &  &  &  &  &  &  &  &  & \cmark & \cmark &  &  &  \\
\rowcolor{gray!15} \cellcolor{white} & \cite{18} &  &  &  &  &  &  &  &  &  &  &  &  &  &  & \cmark &  &  & \cmark &  &  & \cmark &  &  &  &  &  &  &  &  \\
\cellcolor{white} & \cite{20} &  &  &  &  &  &  &  &  &  &  &  &  &  &  & \cmark & \cmark &  & \cmark &  &  &  &  &  &  &  &  &  &  &  \\
\rowcolor{gray!15} \cellcolor{white} & \cite{21} &  &  &  &  &  &  &  &  &  &  & \cmark &  &  &  & \cmark & \cmark &  & \cmark &  &  &  &  &  &  &  &  &  &  &  \\
\cellcolor{white} & \cite{22} &  &  &  &  &  &  &  & \cmark &  &  & \cmark & \cmark & \cmark &  & \cmark &  & \cmark & \cmark & \cmark &  &  &  &  &  &  &  &  &  &  \\
\rowcolor{gray!15} \cellcolor{white} & \cite{24yildiz-etal-2025-benchmarking} &  &  &  &  &  &  &  &  &  &  &  &  &  & \cmark &  & \cmark &  &  &  &  &  &  &  &  &  &  &  &  &  \\
\cellcolor{white} & \cite{27} &  &  &  &  &  &  &  &  & \cmark &  & \cmark &  &  &  &  & \cmark &  &  &  &  &  &  &  &  &  &  &  &  &  \\
\rowcolor{gray!15} \cellcolor{white} & \cite{29} & \cmark & \cmark & \cmark &  & \cmark &  & \cmark &  &  &  &  &  &  &  &  &  & \cmark &  &  &  &  &  &  &  &  &  &  &  &  \\
\cellcolor{white} & \cite{30} &  &  &  &  &  &  &  & \cmark &  &  &  & \cmark &  &  &  &  & \cmark & \cmark &  &  &  &  &  &  &  &  &  &  &  \\
\rowcolor{gray!15} \cellcolor{white} & \cite{31} &  &  &  &  &  &  &  &  &  &  &  &  &  &  & \cmark &  &  &  &  & \cmark &  &  &  &  &  &  &  &  & \cmark \\
\cellcolor{white} & \cite{32} &  &  &  &  &  &  &  &  & \cmark &  &  &  &  &  & \cmark & \cmark &  &  &  & \cmark & \cmark &  &  &  &  &  &  &  &  \\
\rowcolor{gray!15} \cellcolor{white} & \cite{33} &  &  &  &  &  &  &  &  &  &  &  &  &  &  &  &  &  &  &  &  &  &  &  &  &  & \cmark &  &  &  \\
\cellcolor{white} & \cite{34} &  &  &  &  &  &  &  &  &  &  &  &  &  &  &  &  &  &  &  &  &  &  &  &  & \cmark & \cmark &  &  &  \\
\rowcolor{gray!15} \cellcolor{white} & \cite{35} &  &  &  &  &  &  &  &  &  &  & \cmark &  &  &  &  &  &  & \cmark &  &  &  &  &  &  &  &  &  &  &  \\
\cellcolor{white} & \cite{37app15169096} &  &  &  &  &  &  &  &  &  &  &  &  & \cmark &  & \cmark & \cmark & \cmark & \cmark &  &  &  &  &  &  & \cmark &  &  &  &  \\
\rowcolor{gray!15} \cellcolor{white} & \cite{38zhu2025cvebenchbenchmarkaiagents} &  &  &  &  &  &  &  & \cmark &  &  &  & \cmark &  & \cmark & \cmark &  &  &  &  &  &  &  &  &  &  &  &  &  &  \\
\cellcolor{white} & \cite{39} &  &  &  &  &  &  &  &  &  &  &  &  &  &  &  &  &  &  &  &  & \cmark &  & \cmark &  &  &  &  &  &  \\
\rowcolor{gray!15} \cellcolor{white} & \cite{41jiao2025deepvulhunter} &  &  &  &  &  &  &  &  &  &  &  &  &  &  &  & \cmark &  &  &  &  &  &  &  &  &  &  &  &  &  \\
\cellcolor{white} & \cite{42LOEVENICH2025111162} &  &  &  &  &  &  &  &  &  &  &  &  &  &  & \cmark & \cmark &  &  &  &  & \cmark &  & \cmark &  &  &  &  &  &  \\
\rowcolor{gray!15} \cellcolor{white} & \cite{45} &  &  &  &  &  &  &  &  &  &  &  & \cmark &  &  & \cmark &  &  &  &  &  &  & \cmark &  &  & \cmark &  &  &  &  \\
\cellcolor{white} & \cite{46abramovich2025enigma} &  &  &  &  & \cmark &  & \cmark &  &  & \cmark & \cmark &  &  & \cmark & \cmark &  & \cmark &  &  &  &  &  &  &  &  &  &  &  &  \\
\rowcolor{gray!15} \cellcolor{white} & \cite{47chowdhry2025evaluating} &  &  &  &  &  &  &  & \cmark &  &  &  &  &  &  &  &  &  &  &  &  &  &  & \cmark &  &  &  &  &  &  \\
\cellcolor{white} & \cite{48} &  &  &  &  &  &  &  & \cmark &  &  &  & \cmark &  &  &  &  &  &  &  &  &  &  & \cmark &  &  &  &  &  &  \\
\rowcolor{gray!15} \cellcolor{white} & \cite{49} &  &  &  &  &  &  &  &  & \cmark &  &  &  &  &  &  &  &  &  &  &  & \cmark &  & \cmark &  & \cmark &  &  &  &  \\
\cellcolor{white} & \cite{51} &  &  &  &  &  &  &  &  &  &  &  &  &  & \cmark &  &  &  &  &  &  &  &  &  &  &  &  &  &  &  \\
\rowcolor{gray!15} \cellcolor{white} & \cite{53} &  &  &  &  &  &  &  &  &  &  &  &  &  &  &  &  &  &  &  &  &  &  & \cmark &  &  &  &  &  &  \\
\cellcolor{white} & \cite{55} &  &  &  &  &  &  &  &  &  &  &  &  & \cmark &  &  &  &  & \cmark &  &  &  &  &  &  &  &  &  &  &  \\
\rowcolor{gray!15} \cellcolor{white} & \cite{56} &  &  &  &  &  &  &  & \cmark &  &  & \cmark & \cmark &  &  &  &  &  &  &  &  &  & \cmark &  &  &  &  &  &  & \cmark \\
\cellcolor{white} & \cite{58} & \cmark &  &  &  &  &  &  &  &  &  &  &  &  &  & \cmark & \cmark &  &  &  &  &  & \cmark &  &  &  &  &  &  &  \\
\rowcolor{gray!15} \cellcolor{white} & \cite{59} &  &  &  &  &  &  &  & \cmark &  &  &  &  &  &  &  &  & \cmark & \cmark & \cmark &  &  & \cmark &  &  &  &  &  &  &  \\
\cellcolor{white} & \cite{60} &  &  &  &  &  &  &  &  & \cmark &  &  &  &  &  &  &  &  &  &  &  &  &  &  &  & \cmark &  &  &  &  \\
\rowcolor{gray!15} \cellcolor{white} & \cite{61} &  &  &  &  &  &  &  &  &  &  &  &  &  &  &  &  &  &  &  &  &  &  &  &  & \cmark &  &  &  &  \\
\cellcolor{white} & \cite{62} &  &  &  &  &  &  &  &  &  &  &  &  &  &  &  & \cmark &  &  &  &  &  &  &  &  &  &  &  &  &  \\
\rowcolor{gray!15} \cellcolor{white} & \cite{65} &  &  &  &  & \cmark &  &  & \cmark &  & \cmark &  & \cmark &  &  &  &  & \cmark & \cmark & \cmark &  &  &  &  &  &  &  &  &  &  \\
\cellcolor{white} & \cite{67} & \cmark &  &  &  &  &  &  & \cmark &  &  &  &  &  &  &  &  &  &  &  & \cmark &  &  &  &  &  &  &  &  &  \\
\rowcolor{gray!15} \cellcolor{white} & \cite{68} &  &  &  &  &  &  &  &  &  &  &  &  &  &  & \cmark &  &  &  &  &  &  &  &  &  &  &  &  &  &  \\
\cellcolor{white} & \cite{69} & \cmark & \cmark &  &  &  & \cmark &  &  &  &  &  &  &  & \cmark &  &  &  &  &  &  &  &  &  &  &  &  &  &  &  \\
\rowcolor{gray!15} \cellcolor{white} & \cite{70} &  &  &  &  & \cmark &  &  & \cmark &  &  & \cmark &  &  &  & \cmark & \cmark & \cmark &  &  &  &  &  &  &  &  &  &  &  & \cmark \\
\cellcolor{white} & \cite{71} & \cmark &  &  &  &  &  & \cmark &  &  &  &  &  &  &  &  & \cmark &  &  &  &  &  &  &  &  & \cmark &  &  &  &  \\
\rowcolor{gray!15} \cellcolor{white} & \cite{72} &  &  &  &  & \cmark & \cmark &  &  & \cmark &  &  &  &  &  & \cmark & \cmark &  &  &  &  &  &  &  &  & \cmark &  &  &  &  \\
\cellcolor{white} & \cite{73} &  &  &  &  &  &  &  &  & \cmark &  &  &  &  &  & \cmark &  &  &  &  & \cmark &  & \cmark &  &  & \cmark &  &  &  &  \\
\rowcolor{gray!15} \cellcolor{white} & \cite{75} & \cmark &  &  &  &  & \cmark &  &  &  & \cmark &  &  &  & \cmark &  &  & \cmark &  &  &  &  &  &  &  &  &  &  &  &  \\
\cellcolor{white} & \cite{78} &  &  &  &  &  &  &  & \cmark &  &  &  & \cmark &  &  & \cmark & \cmark & \cmark & \cmark &  &  &  &  &  &  &  &  &  &  &  \\
\rowcolor{gray!15} \cellcolor{white} & \cite{80} &  &  &  &  &  &  &  &  &  &  &  &  &  &  & \cmark &  &  & \cmark &  &  &  & \cmark &  &  & \cmark &  &  &  &  \\
\cellcolor{white} & \cite{81Cao_Huang_Li_Huilin_He_Oo_Hooi_2025} &  &  &  &  &  &  &  &  &  &  &  &  &  &  & \cmark & \cmark &  &  &  &  &  &  &  &  &  &  &  &  &  \\
\rowcolor{gray!15} \cellcolor{white} & \cite{83} & \cmark &  &  & \cmark &  &  &  &  &  &  &  &  &  & \cmark & \cmark & \cmark &  &  &  &  &  &  &  &  &  &  &  &  &  \\
\cellcolor{white} & \cite{87} &  &  &  &  &  &  &  &  &  &  & \cmark &  &  &  & \cmark & \cmark &  & \cmark &  &  &  &  &  &  &  &  &  &  & \cmark \\
\rowcolor{gray!15} \cellcolor{white} & \cite{88} &  &  &  &  &  &  &  &  &  &  & \cmark &  &  & \cmark & \cmark &  &  &  &  &  &  &  &  &  &  &  &  &  &  \\
\cellcolor{white} & \cite{89guo2025repoauditautonomousllmagentrepositorylevel} & \cmark & \cmark &  &  &  & \cmark &  &  &  & \cmark &  &  &  & \cmark &  &  &  &  &  &  &  &  &  &  &  &  &  &  &  \\
\rowcolor{gray!15} \cellcolor{white} & \cite{90} &  &  &  &  &  &  &  &  &  &  &  &  &  &  & \cmark &  &  &  & \cmark &  &  &  &  & \cmark &  &  &  &  &  \\
\cellcolor{white} & \cite{93info16050365} &  &  &  &  &  &  &  &  &  &  &  &  &  &  &  &  &  &  &  & \cmark & \cmark &  &  &  &  &  &  &  & \cmark \\
\rowcolor{gray!15} \cellcolor{white} & \cite{94} &  &  &  &  &  &  &  &  & \cmark &  &  &  &  &  & \cmark &  &  &  &  & \cmark &  &  &  &  &  &  &  &  & \cmark \\
\cellcolor{white} & \cite{99} & \cmark &  &  &  &  &  &  &  &  &  &  &  &  &  & \cmark &  &  &  &  &  &  &  &  &  &  &  &  &  &  \\
\rowcolor{gray!15} \cellcolor{white} & \cite{100} & \cmark &  &  &  &  &  &  &  &  &  &  &  &  &  & \cmark &  &  &  &  &  &  &  &  &  &  &  &  &  &  \\
\hline
\cellcolor{white}2024 & \cite{1} &  &  &  &  &  &  &  &  & \cmark &  &  &  &  &  &  & \cmark &  &  &  &  &  &  &  &  &  &  &  &  &  \\
\rowcolor{gray!15} \cellcolor{white} & \cite{10debenedetti2024agentdojo} &  &  &  &  &  &  &  &  &  &  &  &  & \cmark &  &  &  &  &  &  &  &  &  &  &  &  &  &  &  &  \\
\cellcolor{white} & \cite{14chen2024agentpoison} &  &  &  &  &  &  &  &  &  &  &  &  &  &  &  & \cmark &  &  &  &  &  &  &  &  &  &  &  &  &  \\
\rowcolor{gray!15} \cellcolor{white} & \cite{28} &  &  &  &  &  &  &  &  &  &  &  &  & \cmark &  &  &  &  &  &  &  &  &  &  &  &  &  &  &  &  \\
\cellcolor{white} & \cite{43} &  &  &  &  &  &  &  &  &  &  &  &  &  &  &  & \cmark &  &  &  &  & \cmark &  & \cmark &  &  &  &  &  &  \\
\rowcolor{gray!15} \cellcolor{white} & \cite{50} &  &  &  &  &  &  &  &  &  &  &  &  &  &  &  &  &  & \cmark &  &  &  &  &  &  &  &  &  &  & \cmark \\
\cellcolor{white} & \cite{54} &  &  &  &  & \cmark &  &  &  &  &  &  &  &  &  &  &  & \cmark &  &  &  &  &  &  &  &  &  &  &  &  \\
\rowcolor{gray!15} \cellcolor{white} & \cite{63zemicheal2024llm} &  &  &  &  &  &  &  & \cmark &  &  &  & \cmark &  & \cmark & \cmark & \cmark &  &  &  &  &  & \cmark &  &  &  &  &  &  & \cmark \\
\cellcolor{white} & \cite{66} &  &  &  &  &  &  &  &  &  &  & \cmark &  &  &  &  &  &  & \cmark &  &  &  &  &  &  &  &  &  &  &  \\
\rowcolor{gray!15} \cellcolor{white} & \cite{74icaart24} &  &  &  &  &  &  &  &  &  &  &  &  &  &  &  &  &  &  &  &  &  &  & \cmark &  &  &  &  &  &  \\
\cellcolor{white} & \cite{76} &  &  &  &  &  &  &  &  &  &  &  &  &  &  & \cmark &  &  & \cmark &  &  &  &  &  &  &  &  &  &  &  \\
\rowcolor{gray!15} \cellcolor{white} & \cite{77} &  &  &  &  &  &  &  &  &  &  &  &  &  &  & \cmark &  &  & \cmark &  &  & \cmark &  &  &  &  &  &  &  &  \\
\cellcolor{white} & \cite{79} &  &  &  &  &  &  &  &  &  &  & \cmark &  &  & \cmark & \cmark &  & \cmark & \cmark &  &  &  &  &  &  &  &  &  & \cmark &  \\
\rowcolor{gray!15} \cellcolor{white} & \cite{82rigaki2024prompt} &  &  &  &  &  &  &  &  &  &  &  &  &  &  &  &  &  &  &  &  &  &  & \cmark &  &  &  &  &  &  \\
\cellcolor{white} & \cite{85} &  &  &  &  &  &  &  &  &  &  & \cmark &  &  &  &  &  &  & \cmark &  &  &  &  &  &  & \cmark &  &  &  &  \\
\rowcolor{gray!15} \cellcolor{white} & \cite{95} &  &  &  &  &  &  &  &  &  &  &  &  &  &  &  &  &  &  &  &  &  & \cmark & \cmark &  & \cmark &  &  &  &  \\
\cellcolor{white} & \cite{98} & \cmark &  &  &  &  &  &  &  &  &  &  &  &  &  & \cmark &  &  &  &  &  &  &  &  &  &  &  &  &  &  \\
\hline
\rowcolor{gray!15} \cellcolor{white}2023 & \cite{52} &  &  &  &  &  &  &  &  &  &  &  &  &  &  &  &  &  &  &  &  &  &  &  &  & \cmark & \cmark &  &  &  \\
\hline
    \end{tabular}%
\end{adjustbox}
  \label{tab:approach-agent-action-space}%
  \vspace{-18pt}
\end{table*}%

\noindent
{\bf Static analysis tools.}
\ul{Linters and pattern matchers} give agents lightweight static checks over
source code, configuration files, vulnerability templates, and suspicious code
idioms
\cite{4,23lbathaviator,84WANG2026103731,29,58,67,69,71,75,83,89guo2025repoauditautonomousllmagentrepositorylevel,99,100,98,9jie2025agent4vul}. These actions
usually serve as a first filter: they point the agent toward likely relevant
regions before deeper reasoning begins. \ul{Data-flow/taint analyzers}
go further by exposing how values move through a program
\cite{84WANG2026103731,101,29,69,
89guo2025repoauditautonomousllmagentrepositorylevel}. Guo et al.'s
\textsc{RepoAudit}~\cite{89guo2025repoauditautonomousllmagentrepositorylevel} uses demand-driven data-flow evidence and validation to decide whether a suspected bug is feasible along a program path.

\ul{Symbolic execution} is used to
reason about feasible paths and path conditions \cite{11,29}, while \ul{formal
verification tools} checks the generated or retrieved properties against the precise
semantic requirements \cite{11,23lbathaviator,97,83,15}. Liu et al.'s
\textsc{PropertyGPT}~\cite{83} generates
smart-contract properties, but the workflow does not stop at generation. It
uses compilation, static analysis, ranking, and a prover to determine whether
the properties are usable for formal verification.

\ul{Binary analysis frameworks} support reverse engineering and vulnerability
detection when source code is unavailable
\cite{29,46abramovich2025enigma,65,70,72,54}. Chen et al.'s
\textsc{ClearAgent}~\cite{29} shows how this action class changes the role of
the LLM. The agent interacts with binary-analysis interfaces, inspects
functions and control-flow structures, and attempts to validate candidate bugs,
so binary understanding is mediated through analysis actions rather than
unconstrained text generation.

\noindent
{\bf Dynamic analysis tools.}
\ul{Fuzzers} let agents generate or mutate inputs to expose crashes, protocol
violations, or hidden vulnerability conditions
\cite{69,72,75,89guo2025repoauditautonomousllmagentrepositorylevel}.
\ul{Debuggers} give agents a way to inspect intermediate execution states,
crashes, and binary behavior \cite{36ZOU2026104305,29,
46abramovich2025enigma,71}.

\ul{Execution sandboxes} provide controlled environments for running risky
commands, malware-like behaviors, exploit attempts, or benchmark tasks
\cite{4,25zhang2026bountybench,26RIGAKI2026129987,101,12,22,30,
38zhu2025cvebenchbenchmarkaiagents,47chowdhry2025evaluating,48,56,59,65,67,
70,78,63zemicheal2024llm}. In BountyBench
\cite{25zhang2026bountybench}, sandboxed environments are central to assessing
agents that must inspect code, exploit vulnerabilities, and produce patches in
realistic targets. \ul{Network traffic analysers} extend dynamic observation to
packets, flows, IDS evidence, and protocol behavior
\cite{19,27,32,49,60,72,73,94,1}.

\noindent
{\bf Code execution \& environment.}
\ul{Compilers / interpreters} enable agents to build, run, and check generated
code or specifications \cite{11,46abramovich2025enigma,65,75,
89guo2025repoauditautonomousllmagentrepositorylevel}. \ul{REPL / shell
execution} gives the agent direct command-level access to a target or analysis environment
\cite{4,25zhang2026bountybench,36ZOU2026104305,64happe2026llms,
84WANG2026103731,12,13,21,22,27,35,46abramovich2025enigma,56,70,87,88,66,79,
85}. \ul{Containerised environments} isolate vulnerable systems, reproduce
experiments, and standardize execution contexts
\cite{4,25zhang2026bountybench,22,30,38zhu2025cvebenchbenchmarkaiagents,45,
48,56,65,78,63zemicheal2024llm}. Shen et al.'s
\textsc{PentestAgent}~\cite{78} uses this kind of controlled setting to support
end-to-end penetration-testing workflows, where reconnaissance, vulnerability
analysis, and exploitation must be assessed against concrete targets.
\ul{Browser / web automation} appears in web exploitation, prompt-injection
assessment, and UI-driven security tasks
\cite{84WANG2026103731,22,37app15169096,55,10debenedetti2024agentdojo,2,28}.

\noindent
{\bf Information retrieval tools.}
\ul{Code search} lets agents locate functions, dependencies, vulnerable files,
proof targets, exploit-relevant snippets, or repair ingredients
\cite{11,25zhang2026bountybench,36ZOU2026104305,64happe2026llms,
91zhu-etal-2026-teams,8zhang2025agent,
24yildiz-etal-2025-benchmarking,38zhu2025cvebenchbenchmarkaiagents,46abramovich2025enigma,51,69,75,83,88,
89guo2025repoauditautonomousllmagentrepositorylevel,63zemicheal2024llm,79}.
Bouzenia et al.'s \textsc{RepairAgent}~\cite{88}, for example, searches the codebase for repair ingredients before constructing and validating candidate patches. \ul{Documentation retrieval} covers API references, tool manuals, exploit
writeups, vulnerability reports, cloud documentation, protocol standards, and
remediation guidance
\cite{4,23lbathaviator,36ZOU2026104305,84WANG2026103731,97,101,12,18,20,21,
22,31,32,37app15169096,38zhu2025cvebenchbenchmarkaiagents,
42LOEVENICH2025111162,45,46abramovich2025enigma,58,68,70,72,73,78,80,
81Cao_Huang_Li_Huilin_He_Oo_Hooi_2025,83,87,88,90,94,99,100,
63zemicheal2024llm,76,77,79,98}. \ul{Vector store / RAG retrieval} gives agents
a more targeted retrieval action over embedded corpora or knowledge bases
\cite{4,19,23lbathaviator,57,86,7,8zhang2025agent,13,20,21,
24yildiz-etal-2025-benchmarking,27,32,37app15169096,
41jiao2025deepvulhunter,42LOEVENICH2025111162,58,62,70,71,72,78,
81Cao_Huang_Li_Huilin_He_Oo_Hooi_2025,83,87,1,14chen2024agentpoison,43,
63zemicheal2024llm}. The difference is visible in \textsc{PropertyGPT}: Liu et
al.~\cite{83} retrieve similar human-written smart-contract properties from a
vector database and use them as references for generating new formal
properties. Dai et al.'s \textsc{RefPentester}~\cite{87} applies the same
general idea to penetration testing, retrieving stage-relevant knowledge before
choosing later actions.

\noindent
{\bf Security-specific platforms.}
\ul{CTF platforms} provide controlled exploitation, reverse-engineering, and
challenge-solving environments
\cite{36ZOU2026104305,40zhuo2026cyberzero,22,29,30,37app15169096,
46abramovich2025enigma,59,65,70,75,78,54,79}. \ul{Penetration testing
frameworks} embed agents into offensive-security workflows and toolchains
\cite{4,26RIGAKI2026129987,64happe2026llms,84WANG2026103731,12,13,18,20,21,
22,30,35,37app15169096,55,59,65,78,80,87,50,66,76,77,79,85}. \ul{Vulnerability
scanners} appear as actions or baselines that agents invoke, interpret, or
coordinate with \cite{4,101,22,59,65,90}. In these offensive settings, the
agent's action space resembles a security practitioner's toolbelt: scan,
enumerate, exploit, validate, and report.

\ul{SIEM / SOAR integration} places agents in security-operation workflows for
alert investigation, evidence synthesis, triage, and response support
\cite{31,32,67,73,93info16050365,94}. Sheikhi et al.'s
\textsc{Cognitive SOC}~\cite{32} is a representative defensive example: agents
enrich alerts, reason over evidence, and generate analyst-facing narratives.
\ul{MITRE ATT\&CK integration} gives agents a shared language for tactics,
techniques, and attack-chain reasoning
\cite{26RIGAKI2026129987,18,32,39,42LOEVENICH2025111162,49,93info16050365,43,
77}. Loevenich et al.~\cite{42LOEVENICH2025111162} connect CTI, CVEs, and
ATT\&CK-style tactics through cybersecurity knowledge graphs to support
autonomous cyber defense.

\ul{Cloud / IaC platform tools} support cloud defense, infrastructure policy,
IaC vulnerability analysis, and privilege-escalation workflows
\cite{92,45,56,58,59,73,80,63zemicheal2024llm,95}. \ul{Cyber range / network
simulation environments} provide controlled networks for assessing attack,
defense, and autonomous cyber operations
\cite{26RIGAKI2026129987,16,39,42LOEVENICH2025111162,
47chowdhry2025evaluating,48,49,53,43,74icaart24,82rigaki2024prompt,95}.
\ul{Asset search / OSINT platforms} connect agents to internet-exposed asset
discovery and vulnerability assessment \cite{90}. Together, these platform
actions show that security agents are often embedded in domain infrastructures,
not merely attached to generic programming tools.

\noindent
{\bf Inter-agent communication.}
\ul{Message passing between agents} lets specialized agents exchange
intermediate findings, divide labor, and coordinate multi-step workflows
\cite{5,97,3,12,16,17,37app15169096,45,49,60,61,71,72,73,80,85,95,6,34,52}. This is
common when no single agent owns the entire task, such as incident correlation,
multi-stage penetration testing, or collaborative vulnerability analysis.
\ul{Voting / aggregation protocols} combine multiple judgments to reduce
single-agent errors or resolve disagreement \cite{5,97,17,33}. \ul{Blockchain
consensus / audit trail} appears as a stronger accountability mechanism in one
multi-LLM network, where agent proposals and decisions are recorded and checked
through consensus \cite{5}. These communication actions matter because security
work often requires defensible coordination, not just individual prediction.

\noindent
{\bf Human interaction.}
\ul{Querying human expert} appears when the agent explicitly asks a human for
clarification, guidance, or execution support \cite{79}. Deng et al.'s
\textsc{PentestGPT}~\cite{79} reflects this interactive design: the human can
guide the penetration-testing loop and may execute or validate steps that the
agent proposes. \ul{Findings for human review} is broader, covering reports,
alerts, explanations, remediation recommendations, and analyst-facing summaries
\cite{19,36ZOU2026104305,96,3,31,56,70,87,93info16050365,94,50,
63zemicheal2024llm}.

\begin{wrapfigure}{r}{0.35\columnwidth}
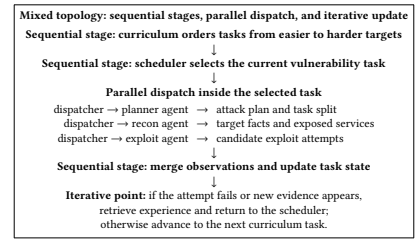

\centering
\vspace{-10pt}
\setlength{\fboxsep}{4pt}
\resizebox{0.35\textwidth}{!}{%
\fbox{\begin{varwidth}{\textwidth}
\centering
\textbf{Mixed topology: sequential stages, parallel dispatch, and iterative update}\\[4pt]
\begin{tabular}{c}
\textbf{Sequential stage: curriculum orders tasks from easier to harder targets}\\
$\downarrow$\\
\textbf{Sequential stage: scheduler selects the current vulnerability task}\\
$\downarrow$\\
\textbf{Parallel dispatch inside the selected task}\\[2pt]
\begin{tabular}{rcl}
dispatcher $\rightarrow$ planner agent & $\rightarrow$ & attack plan and task split\\
dispatcher $\rightarrow$ recon agent & $\rightarrow$ & target facts and exposed services\\
dispatcher $\rightarrow$ exploit agent & $\rightarrow$ & candidate exploit attempts
\end{tabular}\\[4pt]
$\downarrow$\\
\textbf{Sequential stage: merge observations and update task state}\\
$\downarrow$\\
\textbf{Iterative point:} if the attempt fails or new evidence appears,
\\retrieve experience and return to the scheduler; \\otherwise advance to the next curriculum task.
\end{tabular}
\end{varwidth}}}
\vspace{-10pt}
\caption{\textsc{CurriculumPT} workflow.}
\label{fig:workflow-mixed-topology}
\vspace{-8pt}
\end{wrapfigure}

\vspace{-15pt}
\subsubsection{Agent workflow \& orchestration}

Agent workflow and orchestration describe how an LLM agent organizes work over time, divides responsibilities among modules or agents, and decides where human users enter the loop. Table~\ref{tab:approach-agent-workflow-orchestration} summarizes the workflow topologies, coordination patterns, and forms of human involvement identified in the surveyed systems.
\begin{table*}[t]
  \centering
  \vspace{-10pt}
  \caption{Paper attribution for Approach: Agent workflow \& orchestration}
  \vspace{-10pt}
  \scriptsize
  \setlength{\tabcolsep}{1.5pt}
  \renewcommand{\arraystretch}{0.72}
\begin{adjustbox}{width=\textwidth,max totalheight=0.88\textheight,center}
    \begin{tabular}{|c|c|c|c|c|c|c|c|c|c|c|c|c|c|c|c|c|c|c|c|c|}
    \hline
\rowcolor{white} & & 
\multicolumn{6}{c|}{\makecell[c]{Workflow topology}} & \multicolumn{9}{c|}{\makecell[c]{Agent roles}} & \multicolumn{4}{c|}{\makecell[c]{HITL integration}} \\
\cline{3-8}\cline{9-17}\cline{18-21}
\rowcolor{white} 
\multicolumn{1}{|c|}{\multirow{-4}{*}[-0.2ex]{\makecell[c]{Year}}} & 
\multicolumn{1}{c|}{\multirow{-4}{*}[-0.2ex]{\makecell[c]{Paper}}} & 
\multicolumn{1}{c|}{\cellcolor{blue!60}\makecell[c]{Sequential\\pipeline \\ (90\%)}} & \multicolumn{1}{c|}{\cellcolor{blue!12}\makecell[c]{Parallel dispatch \\ (18\%)}} & \multicolumn{1}{c|}{\cellcolor{blue!52}\makecell[c]{Iterative loop \\ (78\%)}} & \multicolumn{1}{c|}{\cellcolor{blue!10}\makecell[c]{Hierarchical\\decomposition \\ (15\%)}} & \multicolumn{1}{c|}{\cellcolor{blue!5}\makecell[c]{Debate /\\discussion \\ (8\%)}} & \multicolumn{1}{c|}{\cellcolor{blue!5}\makecell[c]{Adversarial\\red / blue team \\ (6\%)}} & \multicolumn{1}{c|}{\cellcolor{blue!50}\makecell[c]{Planner \\ (74\%)}} & \multicolumn{1}{c|}{\cellcolor{blue!62}\makecell[c]{Analyst \\ (93\%)}} & \multicolumn{1}{c|}{\cellcolor{blue!61}\makecell[c]{Executor \\ (92\%)}} & \multicolumn{1}{c|}{\cellcolor{blue!5}\makecell[c]{Exploit\\developer \\ (7\%)}} & \multicolumn{1}{c|}{\cellcolor{blue!5}\makecell[c]{Patch\\generator \\ (6\%)}} & \multicolumn{1}{c|}{\cellcolor{blue!9}\makecell[c]{Verifier /\\tester \\ (14\%)}} & \multicolumn{1}{c|}{\cellcolor{blue!7}\makecell[c]{Reporter \\ (11\%)}} & \multicolumn{1}{c|}{\cellcolor{blue!5}\makecell[c]{Critic /\\judge \\ (6\%)}} & \multicolumn{1}{c|}{\cellcolor{blue!12}\makecell[c]{Orchestrator \\ (18\%)}} & \multicolumn{1}{c|}{\cellcolor{blue!63}\makecell[c]{Fully\\autonomous \\ (95\%)}} & \multicolumn{1}{c|}{\cellcolor{blue!5}\makecell[c]{Human-on-\\the-loop \\ (8\%)}} & \multicolumn{1}{c|}{\cellcolor{blue!5}\makecell[c]{Human-in-\\the-loop \\ (5\%)}} & \multicolumn{1}{c|}{\cellcolor{blue!5}\makecell[c]{Interactive /\\collaborative \\ (3\%)}} \\
    \hline

\cellcolor{white}2026 & \cite{4} & \cmark &  & \cmark &  &  &  & \cmark & \cmark & \cmark & \cmark &  &  &  &  &  & \cmark &  &  &  \\
\rowcolor{gray!15} \cellcolor{white} & \cite{5} & \cmark & \cmark & \cmark &  & \cmark &  & \cmark & \cmark & \cmark &  &  &  &  &  &  & \cmark &  &  &  \\
\cellcolor{white} & \cite{11} & \cmark &  & \cmark &  &  &  & \cmark & \cmark & \cmark &  &  & \cmark &  &  &  & \cmark &  &  &  \\
\rowcolor{gray!15} \cellcolor{white} & \cite{19} & \cmark &  & \cmark &  &  &  & \cmark & \cmark & \cmark &  &  &  &  &  &  & \cmark &  &  &  \\
\cellcolor{white} & \cite{23lbathaviator} & \cmark & \cmark & \cmark &  &  &  & \cmark & \cmark & \cmark &  &  & \cmark &  &  &  & \cmark &  &  &  \\
\rowcolor{gray!15} \cellcolor{white} & \cite{25zhang2026bountybench} & \cmark &  & \cmark &  &  &  & \cmark & \cmark & \cmark &  & \cmark &  &  &  &  & \cmark &  &  &  \\
\cellcolor{white} & \cite{26RIGAKI2026129987} & \cmark &  & \cmark &  &  & \cmark & \cmark & \cmark & \cmark &  &  &  &  &  &  & \cmark &  &  &  \\
\rowcolor{gray!15} \cellcolor{white} & \cite{36ZOU2026104305} & \cmark &  & \cmark & \cmark &  &  & \cmark & \cmark & \cmark &  &  &  &  &  & \cmark & \cmark &  & \cmark &  \\
\cellcolor{white} & \cite{40zhuo2026cyberzero} & \cmark &  & \cmark &  &  &  & \cmark & \cmark & \cmark &  &  &  &  &  &  & \cmark &  &  &  \\
\rowcolor{gray!15} \cellcolor{white} & \cite{57} &  &  & \cmark &  &  &  & \cmark & \cmark & \cmark &  &  &  &  &  &  & \cmark &  &  &  \\
\cellcolor{white} & \cite{64happe2026llms} & \cmark &  & \cmark &  &  &  &  & \cmark & \cmark & \cmark &  &  &  &  &  & \cmark &  &  &  \\
\rowcolor{gray!15} \cellcolor{white} & \cite{84WANG2026103731} & \cmark &  & \cmark & \cmark &  &  & \cmark & \cmark & \cmark &  &  &  &  &  & \cmark & \cmark &  &  &  \\
\cellcolor{white} & \cite{86} & \cmark &  &  &  &  &  & \cmark & \cmark & \cmark &  &  &  &  &  &  & \cmark &  &  &  \\
\rowcolor{gray!15} \cellcolor{white} & \cite{91zhu-etal-2026-teams} & \cmark &  &  & \cmark &  &  & \cmark & \cmark & \cmark & \cmark &  &  &  &  & \cmark & \cmark &  &  &  \\
\cellcolor{white} & \cite{92} & \cmark &  & \cmark &  &  &  & \cmark & \cmark & \cmark &  &  &  &  &  &  & \cmark &  &  &  \\
\rowcolor{gray!15} \cellcolor{white} & \cite{96} &  & \cmark &  &  & \cmark &  &  & \cmark & \cmark &  &  &  &  &  &  & \cmark &  &  &  \\
\cellcolor{white} & \cite{97} & \cmark & \cmark & \cmark & \cmark & \cmark &  & \cmark & \cmark & \cmark &  &  & \cmark &  &  & \cmark & \cmark &  &  &  \\
\rowcolor{gray!15} \cellcolor{white} & \cite{101} & \cmark &  & \cmark &  &  &  & \cmark & \cmark & \cmark &  &  & \cmark &  &  &  & \cmark &  &  &  \\
\hline
\cellcolor{white}2025 & \cite{2} & \cmark &  & \cmark &  &  &  & \cmark & \cmark & \cmark &  &  &  &  &  &  & \cmark &  &  &  \\
\rowcolor{gray!15} \cellcolor{white} & \cite{3} & \cmark & \cmark & \cmark &  &  &  &  & \cmark & \cmark &  &  &  & \cmark &  &  & \cmark &  &  &  \\
\cellcolor{white} & \cite{6} &  &  & \cmark &  &  &  &  & \cmark &  &  &  &  & \cmark & \cmark &  & \cmark &  &  &  \\
\rowcolor{gray!15} \cellcolor{white} & \cite{7} & \cmark &  & \cmark &  &  &  & \cmark & \cmark & \cmark &  &  &  &  &  &  & \cmark &  &  &  \\
\cellcolor{white} & \cite{8zhang2025agent} & \cmark &  &  &  &  & \cmark & \cmark & \cmark & \cmark &  &  &  &  &  &  & \cmark &  &  &  \\
\rowcolor{gray!15} \cellcolor{white} & \cite{9jie2025agent4vul} & \cmark &  &  &  &  &  &  & \cmark & \cmark &  &  &  &  &  &  & \cmark &  &  &  \\
\cellcolor{white} & \cite{12} & \cmark &  & \cmark &  &  & \cmark & \cmark & \cmark & \cmark &  &  &  &  &  & \cmark & \cmark & \cmark &  &  \\
\rowcolor{gray!15} \cellcolor{white} & \cite{13} & \cmark &  & \cmark &  &  &  & \cmark & \cmark & \cmark &  &  &  &  &  &  & \cmark &  &  &  \\
\cellcolor{white} & \cite{15} & \cmark &  &  &  &  &  &  & \cmark &  &  &  &  & \cmark &  &  & \cmark &  &  &  \\
\rowcolor{gray!15} \cellcolor{white} & \cite{16} &  &  & \cmark &  &  &  & \cmark & \cmark & \cmark &  &  &  &  &  &  & \cmark &  &  &  \\
\cellcolor{white} & \cite{17} & \cmark &  & \cmark & \cmark & \cmark &  & \cmark & \cmark & \cmark &  &  &  & \cmark &  & \cmark & \cmark &  &  &  \\
\rowcolor{gray!15} \cellcolor{white} & \cite{18} & \cmark &  & \cmark &  &  &  & \cmark & \cmark & \cmark &  &  &  &  &  &  & \cmark &  &  &  \\
\cellcolor{white} & \cite{20} & \cmark &  & \cmark &  &  &  & \cmark & \cmark & \cmark &  &  &  &  &  & \cmark & \cmark &  &  &  \\
\rowcolor{gray!15} \cellcolor{white} & \cite{21} & \cmark &  & \cmark &  &  &  & \cmark & \cmark & \cmark &  &  & \cmark & \cmark &  &  & \cmark &  &  &  \\
\cellcolor{white} & \cite{22} & \cmark &  & \cmark &  &  &  & \cmark & \cmark & \cmark &  &  &  &  &  &  & \cmark &  &  &  \\
\rowcolor{gray!15} \cellcolor{white} & \cite{24yildiz-etal-2025-benchmarking} & \cmark &  & \cmark &  &  &  & \cmark & \cmark & \cmark &  &  &  &  &  &  & \cmark &  &  &  \\
\cellcolor{white} & \cite{27} & \cmark &  & \cmark &  &  &  & \cmark & \cmark & \cmark &  &  &  &  &  &  & \cmark &  &  &  \\
\rowcolor{gray!15} \cellcolor{white} & \cite{29} & \cmark &  & \cmark &  &  &  & \cmark & \cmark & \cmark &  &  & \cmark &  &  &  & \cmark &  &  &  \\
\cellcolor{white} & \cite{30} & \cmark &  & \cmark &  &  & \cmark &  & \cmark &  &  &  &  &  &  &  & \cmark &  &  &  \\
\rowcolor{gray!15} \cellcolor{white} & \cite{31} & \cmark &  & \cmark & \cmark &  &  & \cmark & \cmark & \cmark &  &  &  &  &  & \cmark & \cmark &  &  &  \\
\cellcolor{white} & \cite{32} & \cmark &  &  &  &  &  &  & \cmark & \cmark &  &  &  & \cmark &  &  &  & \cmark &  & \cmark \\
\rowcolor{gray!15} \cellcolor{white} & \cite{33} &  & \cmark &  &  & \cmark &  &  & \cmark &  &  &  &  &  & \cmark &  & \cmark &  &  &  \\
\cellcolor{white} & \cite{34} & \cmark & \cmark & \cmark &  & \cmark &  &  & \cmark &  &  &  & \cmark &  & \cmark &  & \cmark &  &  &  \\
\rowcolor{gray!15} \cellcolor{white} & \cite{35} & \cmark &  &  & \cmark &  &  & \cmark & \cmark & \cmark &  &  &  &  &  & \cmark & \cmark &  &  &  \\
\cellcolor{white} & \cite{37app15169096} & \cmark & \cmark & \cmark & \cmark &  &  & \cmark & \cmark & \cmark & \cmark &  &  &  &  & \cmark & \cmark &  &  &  \\
\rowcolor{gray!15} \cellcolor{white} & \cite{38zhu2025cvebenchbenchmarkaiagents} & \cmark &  & \cmark &  &  &  & \cmark & \cmark & \cmark &  &  &  &  &  &  & \cmark &  &  &  \\
\cellcolor{white} & \cite{39} & \cmark &  & \cmark &  &  &  &  & \cmark & \cmark &  &  &  &  &  &  & \cmark &  &  &  \\
\rowcolor{gray!15} \cellcolor{white} & \cite{41jiao2025deepvulhunter} & \cmark &  & \cmark &  &  &  &  & \cmark &  &  &  &  &  &  &  & \cmark &  &  &  \\
\cellcolor{white} & \cite{42LOEVENICH2025111162} & \cmark &  & \cmark & \cmark &  &  & \cmark & \cmark & \cmark &  &  &  &  &  & \cmark & \cmark & \cmark &  &  \\
\rowcolor{gray!15} \cellcolor{white} & \cite{45} & \cmark &  &  &  &  &  & \cmark & \cmark & \cmark &  &  &  &  &  &  & \cmark &  &  &  \\
\cellcolor{white} & \cite{46abramovich2025enigma} & \cmark &  & \cmark &  &  &  & \cmark & \cmark & \cmark &  &  &  &  &  &  & \cmark &  &  &  \\
\rowcolor{gray!15} \cellcolor{white} & \cite{47chowdhry2025evaluating} & \cmark &  & \cmark &  &  &  & \cmark & \cmark & \cmark &  &  &  &  &  &  & \cmark &  &  &  \\
\cellcolor{white} & \cite{48} &  &  & \cmark &  &  &  & \cmark &  & \cmark &  &  &  &  &  &  & \cmark &  &  &  \\
\rowcolor{gray!15} \cellcolor{white} & \cite{49} & \cmark & \cmark &  &  & \cmark &  &  & \cmark & \cmark &  &  &  &  &  &  & \cmark &  &  &  \\
\cellcolor{white} & \cite{51} & \cmark &  & \cmark &  &  &  & \cmark & \cmark & \cmark &  &  &  & \cmark &  &  &  & \cmark &  &  \\
\rowcolor{gray!15} \cellcolor{white} & \cite{53} &  &  & \cmark &  &  &  &  & \cmark & \cmark &  &  &  &  &  &  & \cmark &  &  &  \\
\cellcolor{white} & \cite{55} & \cmark &  & \cmark &  &  &  & \cmark & \cmark & \cmark &  &  &  & \cmark &  &  & \cmark &  &  &  \\
\rowcolor{gray!15} \cellcolor{white} & \cite{56} & \cmark & \cmark & \cmark &  &  &  & \cmark & \cmark & \cmark &  & \cmark &  &  &  &  & \cmark &  &  &  \\
\cellcolor{white} & \cite{58} & \cmark &  & \cmark &  &  &  & \cmark & \cmark & \cmark &  & \cmark &  &  &  &  & \cmark &  &  &  \\
\rowcolor{gray!15} \cellcolor{white} & \cite{59} & \cmark &  &  &  &  &  & \cmark & \cmark & \cmark &  &  &  &  &  &  & \cmark &  &  &  \\
\cellcolor{white} & \cite{60} & \cmark & \cmark & \cmark &  &  &  & \cmark & \cmark & \cmark &  &  &  &  &  &  & \cmark &  &  &  \\
\rowcolor{gray!15} \cellcolor{white} & \cite{61} & \cmark & \cmark &  & \cmark &  &  & \cmark & \cmark & \cmark &  &  &  &  &  & \cmark & \cmark &  &  &  \\
\cellcolor{white} & \cite{62} & \cmark &  & \cmark &  &  &  &  & \cmark & \cmark &  &  &  &  &  &  & \cmark &  &  &  \\
\rowcolor{gray!15} \cellcolor{white} & \cite{65} & \cmark &  & \cmark &  &  &  & \cmark & \cmark & \cmark &  &  &  &  &  &  & \cmark &  &  &  \\
\cellcolor{white} & \cite{67} & \cmark &  & \cmark &  &  &  & \cmark & \cmark & \cmark &  &  &  &  &  &  & \cmark &  &  &  \\
\rowcolor{gray!15} \cellcolor{white} & \cite{68} & \cmark &  &  &  & \cmark &  &  & \cmark & \cmark &  &  &  & \cmark &  &  & \cmark &  &  &  \\
\cellcolor{white} & \cite{69} & \cmark &  & \cmark &  &  &  & \cmark & \cmark & \cmark &  &  &  &  &  &  & \cmark &  &  &  \\
\rowcolor{gray!15} \cellcolor{white} & \cite{70} & \cmark & \cmark & \cmark &  &  &  & \cmark & \cmark & \cmark &  &  &  &  &  &  & \cmark &  &  &  \\
\cellcolor{white} & \cite{71} & \cmark &  & \cmark & \cmark &  &  & \cmark & \cmark & \cmark &  &  &  &  &  & \cmark & \cmark &  &  & \cmark \\
\rowcolor{gray!15} \cellcolor{white} & \cite{72} & \cmark & \cmark & \cmark &  &  &  & \cmark &  & \cmark &  &  &  &  &  &  & \cmark &  &  &  \\
\cellcolor{white} & \cite{73} & \cmark & \cmark & \cmark & \cmark &  &  &  & \cmark & \cmark &  &  &  &  &  & \cmark & \cmark &  &  &  \\
\rowcolor{gray!15} \cellcolor{white} & \cite{75} & \cmark & \cmark & \cmark &  &  &  & \cmark & \cmark & \cmark & \cmark & \cmark & \cmark &  &  &  & \cmark &  &  &  \\
\cellcolor{white} & \cite{78} & \cmark &  & \cmark &  &  &  & \cmark & \cmark & \cmark & \cmark &  &  &  &  &  & \cmark &  &  &  \\
\rowcolor{gray!15} \cellcolor{white} & \cite{80} & \cmark &  & \cmark &  &  &  & \cmark & \cmark & \cmark &  &  &  &  &  &  & \cmark &  &  &  \\
\cellcolor{white} & \cite{81Cao_Huang_Li_Huilin_He_Oo_Hooi_2025} & \cmark &  & \cmark &  &  &  &  & \cmark & \cmark &  &  &  &  &  &  & \cmark &  &  &  \\
\rowcolor{gray!15} \cellcolor{white} & \cite{83} & \cmark &  & \cmark &  &  &  & \cmark & \cmark & \cmark &  &  & \cmark &  &  &  & \cmark &  &  &  \\
\cellcolor{white} & \cite{87} & \cmark &  & \cmark & \cmark &  &  & \cmark & \cmark & \cmark &  &  &  &  &  & \cmark &  &  & \cmark &  \\
\rowcolor{gray!15} \cellcolor{white} & \cite{88} & \cmark &  & \cmark &  &  &  & \cmark &  & \cmark &  & \cmark &  &  &  &  & \cmark &  &  &  \\
\cellcolor{white} & \cite{89guo2025repoauditautonomousllmagentrepositorylevel} & \cmark &  & \cmark &  &  &  & \cmark & \cmark & \cmark &  &  & \cmark &  &  &  & \cmark &  &  &  \\
\rowcolor{gray!15} \cellcolor{white} & \cite{90} & \cmark &  & \cmark &  &  &  & \cmark & \cmark & \cmark &  &  &  &  &  &  & \cmark &  &  &  \\
\cellcolor{white} & \cite{93info16050365} & \cmark &  & \cmark &  &  &  & \cmark & \cmark & \cmark &  &  &  &  &  &  & \cmark & \cmark &  &  \\
\rowcolor{gray!15} \cellcolor{white} & \cite{94} & \cmark &  &  &  &  &  & \cmark & \cmark & \cmark &  &  &  &  &  &  & \cmark &  &  &  \\
\cellcolor{white} & \cite{99} & \cmark &  &  &  &  &  &  & \cmark & \cmark &  & \cmark &  & \cmark & \cmark &  & \cmark &  &  &  \\
\rowcolor{gray!15} \cellcolor{white} & \cite{100} & \cmark &  & \cmark &  &  &  &  & \cmark & \cmark &  &  & \cmark &  &  &  & \cmark &  &  &  \\
\hline
\cellcolor{white}2024 & \cite{1} & \cmark &  &  &  &  &  &  & \cmark &  &  &  &  &  &  &  & \cmark &  &  &  \\
\rowcolor{gray!15} \cellcolor{white} & \cite{10debenedetti2024agentdojo} &  &  & \cmark &  &  & \cmark & \cmark & \cmark & \cmark &  &  &  &  &  &  & \cmark &  &  &  \\
\cellcolor{white} & \cite{14chen2024agentpoison} & \cmark &  &  &  &  &  & \cmark &  & \cmark &  &  &  &  &  &  & \cmark &  &  &  \\
\rowcolor{gray!15} \cellcolor{white} & \cite{28} &  &  & \cmark &  &  &  & \cmark & \cmark & \cmark &  &  &  &  &  &  & \cmark &  &  &  \\
\cellcolor{white} & \cite{43} & \cmark &  &  &  &  &  &  & \cmark & \cmark &  &  &  &  &  &  & \cmark & \cmark &  &  \\
\rowcolor{gray!15} \cellcolor{white} & \cite{50} & \cmark &  & \cmark &  &  &  &  & \cmark &  &  &  &  &  &  &  &  &  & \cmark &  \\
\cellcolor{white} & \cite{54} & \cmark &  & \cmark &  &  &  &  & \cmark & \cmark &  &  &  &  &  &  & \cmark &  & \cmark &  \\
\rowcolor{gray!15} \cellcolor{white} & \cite{63zemicheal2024llm} & \cmark &  & \cmark &  &  &  & \cmark & \cmark & \cmark &  &  &  & \cmark &  &  & \cmark &  &  &  \\
\cellcolor{white} & \cite{66} & \cmark &  &  &  &  &  & \cmark & \cmark & \cmark &  &  &  &  &  &  & \cmark &  &  &  \\
\rowcolor{gray!15} \cellcolor{white} & \cite{74icaart24} & \cmark &  &  &  &  &  & \cmark &  & \cmark &  &  &  &  &  &  & \cmark &  &  &  \\
\cellcolor{white} & \cite{76} & \cmark &  & \cmark &  &  &  & \cmark & \cmark & \cmark &  &  &  &  &  &  & \cmark &  &  &  \\
\rowcolor{gray!15} \cellcolor{white} & \cite{77} & \cmark &  & \cmark &  &  &  & \cmark & \cmark & \cmark & \cmark &  &  &  &  &  & \cmark &  &  &  \\
\cellcolor{white} & \cite{79} & \cmark &  & \cmark &  &  &  & \cmark & \cmark & \cmark &  &  &  &  &  &  &  & \cmark & \cmark & \cmark \\
\rowcolor{gray!15} \cellcolor{white} & \cite{82rigaki2024prompt} & \cmark &  & \cmark &  &  &  & \cmark &  & \cmark &  &  &  &  &  &  & \cmark &  &  &  \\
\cellcolor{white} & \cite{85} & \cmark &  & \cmark &  &  &  & \cmark &  & \cmark &  &  & \cmark &  &  & \cmark & \cmark &  &  &  \\
\rowcolor{gray!15} \cellcolor{white} & \cite{95} &  & \cmark &  & \cmark &  & \cmark &  & \cmark & \cmark &  &  &  &  &  & \cmark & \cmark & \cmark &  &  \\
\cellcolor{white} & \cite{98} & \cmark &  & \cmark & \cmark &  &  &  & \cmark & \cmark &  &  & \cmark &  & \cmark & \cmark & \cmark &  &  &  \\
\hline
\rowcolor{gray!15} \cellcolor{white}2023 & \cite{52} & \cmark & \cmark & \cmark &  &  &  & \cmark & \cmark & \cmark &  &  & \cmark &  & \cmark &  & \cmark &  &  &  \\
\hline
    \end{tabular}%
\end{adjustbox}
  \label{tab:approach-agent-workflow-orchestration}%
  \vspace{-18pt}
\end{table*}%

\noindent
{\bf Workflow topology.} Workflow topology captures the structural pattern of the agent process. A \ul{sequential pipeline} is the most basic form: the agent progresses through ordered stages such as collecting context, analyzing evidence, planning a response, executing commands, and reporting findings. For example, PentestGPT follows a staged penetration-testing workflow in which high-level reasoning, task generation, parsing of tool outputs, and next-step selection are connected in a serial process \cite{79}.

\ul{Parallel dispatch} is used when different agents, tools, or subtasks can run concurrently and then merge their results \cite{5,23lbathaviator,96,97,3,33,34,37app15169096,49,56,60,61,70,72,73,75,95,52}. Parallelism is useful when the target has multiple independent surfaces, such as several APIs, hosts, logs, or vulnerability hypotheses. \ul{Debate and discussion} are a related but more deliberative form of topology: agents exchange competing judgments before a final decision is made \cite{5,96,97,17,33,34,49,68}.

\ul{Iterative loops} are used when the agent must repeatedly act, observe, and revise its state. Such loops appear throughout the surveyed systems \cite{4,5,11,19,84WANG2026103731,92,97,101,6,13,16,17,18,20,21,22,24yildiz-etal-2025-benchmarking,27,46abramovich2025enigma,47chowdhry2025evaluating,48,51,58,60,67,69,71,81Cao_Huang_Li_Huilin_He_Oo_Hooi_2025,83,87,89guo2025repoauditautonomousllmagentrepositorylevel,90,77,79,82rigaki2024prompt,85,98,52}. \textsc{ClearAgent}~\cite{29}, for example,
iteratively explores binary code, updates its analysis context, and validates
candidate vulnerabilities with concrete inputs. \ul{Hierarchical decomposition} adds a manager, planner, controller, 
or chief
agent above specialized workers \cite{36ZOU2026104305,84WANG2026103731,91zhu-etal-2026-teams,97,17,31,35,37app15169096,42LOEVENICH2025111162,61,71,73,87,95,98}. For example, Zhu et al.'s HPTSA~\cite{91zhu-etal-2026-teams} illustrates the
pattern in zero-day exploitation: a planning agent analyzes the target and
launches specialized subagents for exploitation subtasks. \ul{Adversarial red/blue-team} topology appears when agents are explicitly assigned opposing attacker and defender roles \cite{26RIGAKI2026129987,8zhang2025agent,12,30,10debenedetti2024agentdojo,95}.

\begin{wraptable}{r}{0.4\textwidth}  
  \centering
  \vspace{-4pt}
  \caption{Paper attribution for Approach: Agent self-improvement \& learning}
  \vspace{-10pt}
  \scriptsize
  \setlength{\tabcolsep}{1.5pt}
  \renewcommand{\arraystretch}{0.70}
\begin{adjustbox}{width=\textwidth,max totalheight=0.88\textheight,center}
    \begin{tabular}{|c|c|c|c|c|c|c|c|}
    \hline
\rowcolor{white} & & 
\multicolumn{2}{c|}{\makecell[c]{In-context adaptation}} & \multicolumn{2}{c|}{\makecell[c]{Cross-session learning}} & \multicolumn{2}{c|}{\makecell[c]{RL-based improvement}} \\
\cline{3-4}\cline{5-6}\cline{7-8}
\rowcolor{white} 
\multicolumn{1}{|c|}{\multirow{-4}{*}[-0.2ex]{\makecell[c]{Year}}} & 
\multicolumn{1}{c|}{\multirow{-4}{*}[-0.2ex]{\makecell[c]{Paper}}} & 
\multicolumn{1}{c|}{\cellcolor{blue!21}\makecell[c]{Learning\\from feedback\\within session \\ (32\%)}} & \multicolumn{1}{c|}{\cellcolor{blue!5}\makecell[c]{Example\\-guided\\correction \\ (1\%)}} & \multicolumn{1}{c|}{\cellcolor{blue!5}\makecell[c]{Storing\\successful\\trajectories \\ (3\%)}} & \multicolumn{1}{c|}{\cellcolor{blue!5}\makecell[c]{Updating\\fine-tuning\\data \\ (2\%)}} & \multicolumn{1}{c|}{\cellcolor{blue!6}\makecell[c]{Reward\\signal\\design \\ (9\%)}} & \multicolumn{1}{c|}{\cellcolor{blue!6}\makecell[c]{Environment-\\based\\RL \\ (9\%)}} \\
    \hline

\cellcolor{white}2026 & \cite{4} & \cmark &  &  &  &  &  \\
\rowcolor{gray!15} \cellcolor{white} & \cite{5} & \cmark &  &  &  &  &  \\
\cellcolor{white} & \cite{11} &  &  &  &  &  &  \\
\rowcolor{gray!15} \cellcolor{white} & \cite{19} & \cmark &  & \cmark &  &  &  \\
\cellcolor{white} & \cite{23lbathaviator} &  &  &  &  &  &  \\
\rowcolor{gray!15} \cellcolor{white} & \cite{25zhang2026bountybench} &  &  &  &  &  &  \\
\cellcolor{white} & \cite{26RIGAKI2026129987} &  &  &  &  & \cmark &  \\
\rowcolor{gray!15} \cellcolor{white} & \cite{36ZOU2026104305} & \cmark &  &  &  &  &  \\
\cellcolor{white} & \cite{40zhuo2026cyberzero} &  &  & \cmark & \cmark &  &  \\
\rowcolor{gray!15} \cellcolor{white} & \cite{57} & \cmark &  &  &  &  &  \\
\cellcolor{white} & \cite{64happe2026llms} &  &  & \cmark &  &  &  \\
\rowcolor{gray!15} \cellcolor{white} & \cite{84WANG2026103731} & \cmark &  &  &  &  &  \\
\cellcolor{white} & \cite{86} &  &  &  &  &  &  \\
\rowcolor{gray!15} \cellcolor{white} & \cite{91zhu-etal-2026-teams} &  &  &  &  &  &  \\
\cellcolor{white} & \cite{92} & \cmark &  &  &  &  &  \\
\rowcolor{gray!15} \cellcolor{white} & \cite{96} &  &  &  &  &  &  \\
\cellcolor{white} & \cite{97} & \cmark &  &  &  &  &  \\
\rowcolor{gray!15} \cellcolor{white} & \cite{101} & \cmark &  &  &  &  &  \\
\hline
\cellcolor{white}2025 & \cite{2} & \cmark &  &  &  & \cmark & \cmark \\
\rowcolor{gray!15} \cellcolor{white} & \cite{3} &  &  &  &  &  &  \\
\cellcolor{white} & \cite{6} &  &  &  &  &  &  \\
\rowcolor{gray!15} \cellcolor{white} & \cite{7} &  &  &  &  &  &  \\
\cellcolor{white} & \cite{8zhang2025agent} &  &  &  &  &  &  \\
\rowcolor{gray!15} \cellcolor{white} & \cite{9jie2025agent4vul} &  &  &  &  &  &  \\
\cellcolor{white} & \cite{12} &  &  &  &  &  &  \\
\rowcolor{gray!15} \cellcolor{white} & \cite{13} &  &  &  &  &  &  \\
\cellcolor{white} & \cite{15} &  &  &  &  &  &  \\
\rowcolor{gray!15} \cellcolor{white} & \cite{16} & \cmark &  &  & \cmark &  &  \\
\cellcolor{white} & \cite{17} &  &  &  &  &  &  \\
\rowcolor{gray!15} \cellcolor{white} & \cite{18} &  &  &  &  &  &  \\
\cellcolor{white} & \cite{20} &  &  &  &  &  &  \\
\rowcolor{gray!15} \cellcolor{white} & \cite{21} & \cmark &  &  &  &  &  \\
\cellcolor{white} & \cite{22} &  &  &  &  &  &  \\
\rowcolor{gray!15} \cellcolor{white} & \cite{24yildiz-etal-2025-benchmarking} &  &  &  &  &  &  \\
\cellcolor{white} & \cite{27} &  &  &  &  &  &  \\
\rowcolor{gray!15} \cellcolor{white} & \cite{29} &  &  &  &  &  &  \\
\cellcolor{white} & \cite{30} &  &  &  &  &  &  \\
\rowcolor{gray!15} \cellcolor{white} & \cite{31} &  &  &  &  &  &  \\
\cellcolor{white} & \cite{32} &  &  &  &  &  &  \\
\rowcolor{gray!15} \cellcolor{white} & \cite{33} &  &  &  &  &  &  \\
\cellcolor{white} & \cite{34} &  &  &  &  &  &  \\
\rowcolor{gray!15} \cellcolor{white} & \cite{35} &  &  &  &  &  &  \\
\cellcolor{white} & \cite{37app15169096} & \cmark &  &  &  &  &  \\
\rowcolor{gray!15} \cellcolor{white} & \cite{38zhu2025cvebenchbenchmarkaiagents} &  &  &  &  &  &  \\
\cellcolor{white} & \cite{39} & \cmark &  &  &  & \cmark & \cmark \\
\rowcolor{gray!15} \cellcolor{white} & \cite{41jiao2025deepvulhunter} &  &  &  &  &  &  \\
\cellcolor{white} & \cite{42LOEVENICH2025111162} & \cmark &  &  &  & \cmark & \cmark \\
\rowcolor{gray!15} \cellcolor{white} & \cite{45} &  &  &  &  &  &  \\
\cellcolor{white} & \cite{46abramovich2025enigma} &  &  &  &  &  &  \\
\rowcolor{gray!15} \cellcolor{white} & \cite{47chowdhry2025evaluating} &  &  &  &  & \cmark & \cmark \\
\cellcolor{white} & \cite{48} &  &  &  &  & \cmark & \cmark \\
\rowcolor{gray!15} \cellcolor{white} & \cite{49} & \cmark &  &  &  &  &  \\
\cellcolor{white} & \cite{51} & \cmark & \cmark &  &  &  &  \\
\rowcolor{gray!15} \cellcolor{white} & \cite{53} &  &  &  &  &  &  \\
\cellcolor{white} & \cite{55} &  &  &  &  &  &  \\
\rowcolor{gray!15} \cellcolor{white} & \cite{56} & \cmark &  &  &  &  &  \\
\cellcolor{white} & \cite{58} & \cmark &  &  &  &  &  \\
\rowcolor{gray!15} \cellcolor{white} & \cite{59} &  &  &  &  &  &  \\
\cellcolor{white} & \cite{60} & \cmark &  &  &  &  &  \\
\rowcolor{gray!15} \cellcolor{white} & \cite{61} & \cmark &  &  &  &  &  \\
\cellcolor{white} & \cite{62} &  &  &  &  &  &  \\
\rowcolor{gray!15} \cellcolor{white} & \cite{65} &  &  &  &  &  &  \\
\cellcolor{white} & \cite{67} &  &  &  &  &  &  \\
\rowcolor{gray!15} \cellcolor{white} & \cite{68} &  &  &  &  &  &  \\
\cellcolor{white} & \cite{69} &  &  &  &  &  &  \\
\rowcolor{gray!15} \cellcolor{white} & \cite{70} &  &  &  &  &  &  \\
\cellcolor{white} & \cite{71} & \cmark &  &  &  &  &  \\
\rowcolor{gray!15} \cellcolor{white} & \cite{72} &  &  &  &  &  &  \\
\cellcolor{white} & \cite{73} & \cmark &  &  &  &  &  \\
\rowcolor{gray!15} \cellcolor{white} & \cite{75} &  &  &  &  &  &  \\
\cellcolor{white} & \cite{78} & \cmark &  &  &  &  &  \\
\rowcolor{gray!15} \cellcolor{white} & \cite{80} & \cmark &  &  &  &  &  \\
\cellcolor{white} & \cite{81Cao_Huang_Li_Huilin_He_Oo_Hooi_2025} &  &  &  &  &  &  \\
\rowcolor{gray!15} \cellcolor{white} & \cite{83} &  &  &  &  &  &  \\
\cellcolor{white} & \cite{87} & \cmark &  &  &  &  &  \\
\rowcolor{gray!15} \cellcolor{white} & \cite{88} &  &  &  &  &  &  \\
\cellcolor{white} & \cite{89guo2025repoauditautonomousllmagentrepositorylevel} &  &  &  &  &  &  \\
\rowcolor{gray!15} \cellcolor{white} & \cite{90} &  &  &  &  &  &  \\
\cellcolor{white} & \cite{93info16050365} & \cmark &  &  &  &  &  \\
\rowcolor{gray!15} \cellcolor{white} & \cite{94} &  &  &  &  &  &  \\
\cellcolor{white} & \cite{99} &  &  &  &  &  &  \\
\rowcolor{gray!15} \cellcolor{white} & \cite{100} &  &  &  &  &  &  \\
\hline
\cellcolor{white}2024 & \cite{1} &  &  &  &  &  &  \\
\rowcolor{gray!15} \cellcolor{white} & \cite{10debenedetti2024agentdojo} &  &  &  &  &  &  \\
\cellcolor{white} & \cite{14chen2024agentpoison} &  &  &  &  &  &  \\
\rowcolor{gray!15} \cellcolor{white} & \cite{28} & \cmark &  &  &  &  &  \\
\cellcolor{white} & \cite{43} & \cmark &  &  &  & \cmark & \cmark \\
\rowcolor{gray!15} \cellcolor{white} & \cite{50} &  &  &  &  &  &  \\
\cellcolor{white} & \cite{54} &  &  &  &  &  &  \\
\rowcolor{gray!15} \cellcolor{white} & \cite{63zemicheal2024llm} &  &  &  &  &  &  \\
\cellcolor{white} & \cite{66} &  &  &  &  &  &  \\
\rowcolor{gray!15} \cellcolor{white} & \cite{74icaart24} &  &  &  &  &  & \cmark \\
\cellcolor{white} & \cite{76} &  &  &  &  &  &  \\
\rowcolor{gray!15} \cellcolor{white} & \cite{77} &  &  &  &  &  &  \\
\cellcolor{white} & \cite{79} &  &  &  &  &  &  \\
\rowcolor{gray!15} \cellcolor{white} & \cite{82rigaki2024prompt} &  &  &  &  & \cmark & \cmark \\
\cellcolor{white} & \cite{85} & \cmark &  &  &  &  &  \\
\rowcolor{gray!15} \cellcolor{white} & \cite{95} & \cmark &  &  &  & \cmark & \cmark \\
\cellcolor{white} & \cite{98} & \cmark &  &  &  &  &  \\
\hline
\rowcolor{gray!15} \cellcolor{white}2023 & \cite{52} &  &  &  &  &  &  \\
\hline
    \end{tabular}%
\end{adjustbox}
  \label{tab:approach-agent-self-improvement-learning}%
  \vspace{-12pt}
\end{wraptable}
A mixed topology can combine several of these structures. As shown in Figure~\ref{fig:workflow-mixed-topology}, \textsc{CurriculumPT}
\cite{37app15169096} provides such an example: the overall curriculum proceeds
sequentially from easier to harder penetration-testing tasks; within a selected
task, work can be dispatched to specialized planning, reconnaissance, and
exploitation components; and the workflow iterates by storing successful
strategies and reusing experience in later tasks.

\noindent
{\bf Agent roles.}
Agent roles describe the functional responsibilities assigned inside the workflow. \ul{Planner}, \ul{analyst}, and \ul{executor} roles form the common backbone of agent workflows. A paper is tagged with a Planner role when it names or describes a component responsible for producing a plan, independent of which planning strategy that component uses; the two tags are related but answer different
questions, and a paper need not carry both.

More specialized roles encode security-domain labor. \ul{Exploit developer}
roles generate or refine attack steps
\cite{4,64happe2026llms,91zhu-etal-2026-teams,37app15169096,75,78,77}, whereas
\ul{patch generator} roles produce fixes or remediation artifacts
\cite{25zhang2026bountybench,56,58,75,88,99}. \ul{Verifier / tester} roles run
tests, validate exploit success, check generated properties, or reduce false
positives
\cite{11,23lbathaviator,97,101,21,29,34,75,83,
89guo2025repoauditautonomousllmagentrepositorylevel,100,85,98,52}. \ul{Reporter}
roles convert evidence into human-readable findings, incident narratives, or
remediation summaries \cite{3,6,15,17,21,32,51,55,68,99,63zemicheal2024llm}.
\ul{Critic / judge} roles provide a second 
opinion or quality filter
\cite{6,33,34,99,98,52}, and \ul{orchestrator} roles coordinate subagents,
maintain global state, or assign subtasks
\cite{36ZOU2026104305,84WANG2026103731,91zhu-etal-2026-teams,97,12,17,20,31,
35,37app15169096,42LOEVENICH2025111162,61,71,73,87,85,95,98}.

For example, in
\textsc{RepairAgent}~\cite{88}, the agent must search for repair ingredients,
propose a patch, execute validation, and revise after feedback. In
\textsc{PropertyGPT}~\cite{83}, generation is paired with verification: the
system generates formal properties for smart contracts, then checks whether they
compile, fit the target, and can be verified.

\noindent
{\bf HITL integration.}
\ul{Fully autonomous} workflows let the agent proceed without routine human intervention, which is common in benchmarked penetration testing, cyber-defense simulation, vulnerability scanning, repair, and analysis pipelines \cite{1,19,23lbathaviator,25zhang2026bountybench,26RIGAKI2026129987,36ZOU2026104305,40zhuo2026cyberzero,84WANG2026103731,8zhang2025agent,9jie2025agent4vul,12,13,15,16,24yildiz-etal-2025-benchmarking,27,29,30,31,33,34,35,37app15169096,38zhu2025cvebenchbenchmarkaiagents,39,41jiao2025deepvulhunter,42LOEVENICH2025111162,75,78,90,93info16050365,94,98,99,100,10debenedetti2024agentdojo,14chen2024agentpoison,28,63zemicheal2024llm,76,77,82rigaki2024prompt,95,52}. \ul{Human-on-the-loop} workflows keep the system mostly autonomous while allowing analysts to monitor, review, or approve high-level outcomes \cite{12,32,42LOEVENICH2025111162,51,93info16050365,43,79,95}. \ul{Human-in-the-loop} workflows require human input at particular decision points, such as executing commands, providing clarifications, or approving uncertain actions \cite{36ZOU2026104305,50,87,54,79}. \ul{Interactive or collaborative} workflows go further by treating the human as a continuing collaborator rather than only an approver \cite{32,71,79}. Deng et al.'s \textsc{PentestGPT}~\cite{79} combines these patterns. The system maintains a pentesting task tree and uses reasoning, generation, and parsing modules to propose next actions and interpret outputs. The human is on the loop when monitoring the agent's evolving task tree and strategic recommendations. The human is in the loop when executing commands in the target environment and feeding results back to the agent.


\vspace{-4pt}
\subsubsection{Agent self-improvement \& learning}

Agent self-improvement and learning describe whether a security agent changes its behavior from feedback, examples, stored experience, fine-tuning data, or reward-driven interaction. The identified learning and adaptation mechanisms are summarized in
Table~\ref{tab:approach-agent-self-improvement-learning}.

\noindent
{\bf In-context adaptation.}
\ul{Learning from feedback within session} refers to improvement that happens
inside the current task episode, without necessarily changing model weights
\cite{4,5,19,36ZOU2026104305,57,84WANG2026103731,92,97,101,2,16,21,
37app15169096,39,42LOEVENICH2025111162,49,51,56,58,60,61,71,73,78,80,87,
93info16050365,28,43,85,95,98}. The feedback may come from tool output, failed
commands, execution traces, analyst observations, retrieved evidence, peer-agent
critique, or changes in the target environment. In penetration testing, this
usually looks like an agent trying a command, reading the result, and adjusting
the next step. Ginige et al.'s \textsc{AutoPentester}~\cite{21}, for example,
uses the results of previous testing actions as context for later strategy
selection. Dai et al.'s \textsc{RefPentester}~\cite{87} makes the feedback loop
more explicit by reflecting on failed operations and using the result to guide
subsequent penetration-testing stages.

\ul{Example-guided correction} is a narrower form of in-context adaptation in
which the system uses examples, demonstrations, or corrected instances to steer
later outputs \cite{51}. Schuring et al.~\cite{51} use this idea for generating
software-security learning materials: traceability links and contextual examples
help the agent refine generated educational content so that it better aligns
with security requirements. Here, the correction signal is not a reward from an
environment, but an example of what a better output should resemble.

\noindent
{\bf Cross-session learning.}
\ul{Storing successful trajectories} preserves useful interaction traces beyond
a single run so that future tasks can reuse solved strategies, action sequences,
or reasoning patterns \cite{19,40zhuo2026cyberzero,64happe2026llms}. Zhuo et al.'s
\textsc{Cyber-Zero}~\cite{40zhuo2026cyberzero} is an example. Instead
of requiring access to the original runtime environments, it reconstructs
long-horizon cybersecurity trajectories from public CTF writeups and simulated
personas, then uses those trajectories as reusable experience for training
cybersecurity agents.

\ul{Updating fine-tuning data} carries improvement further by turning solved
trajectories, generated instructions, or adapted strategies into training data
for later model behavior \cite{40zhuo2026cyberzero,16}. In
\textsc{Cyber-Zero}, Zhuo et al.~\cite{40zhuo2026cyberzero} use synthesized
trajectories to train more capable open-weight cybersecurity agents. Qin et
al.~\cite{16} describe a self-evolving security framework for 6G
space-air-ground integrated networks, where updated instruction data helps the
security-strategy generator adapt to changing or unknown threats. In both cases,
improvement is not confined to the current prompt; the learning artifact is
carried into later deployments or training cycles.

\noindent
{\bf RL-based improvement.}
\ul{Reward signal design} defines what the agent should optimize, such as API
fault discovery, attack progress, service preservation, threat mitigation,
mission impact, or successful recovery
\cite{26RIGAKI2026129987,2,39,42LOEVENICH2025111162,
47chowdhry2025evaluating,48,43,82rigaki2024prompt,95}. Kim et al.'s
\textsc{AutoRestTest}~\cite{2} shows how reward design can be tied directly to
a software-security task: its API, dependency, parameter, and value agents
receive feedback from REST API responses and use multi-agent reinforcement
learning to coordinate exploration and improve fault detection.

\ul{Environment-based RL} uses a cyber range, simulator, network-security game,
or API-testing environment as the substrate for repeated interaction and policy
improvement \cite{2,39,42LOEVENICH2025111162,47chowdhry2025evaluating,48,43,
74icaart24,82rigaki2024prompt,95}. Rigaki et al.~\cite{82rigaki2024prompt}
place LLM agents in network-security environments and study their behavior under
repeated prompt-exploit-observe cycles. Loevenich et
al.~\cite{42LOEVENICH2025111162} assess autonomous cyber-defense agents in a
cyber-operations environment against attacker strategies, combining DRL,
augmented LLM components, and knowledge-graph support.


\begin{wraptable}{r}{0.6\textwidth}
  \centering
  \vspace{-40pt}
  \caption{Paper attribution for Application: Vulnerability detection \& analysis}
  \vspace{-10pt}
  \scriptsize
  \setlength{\tabcolsep}{1.5pt}
  \renewcommand{\arraystretch}{0.70}
\begin{adjustbox}{width=\textwidth,max totalheight=0.88\textheight,center}
    \begin{tabular}{|c|c|c|c|c|c|c|c|c|c|c|}
    \hline
\rowcolor{white} Year & Paper & \multicolumn{1}{c|}{\cellcolor{blue!21}\makecell[c]{Vulnerability\\detection \\ (32\%)}} & \multicolumn{1}{c|}{\cellcolor{blue!5}\makecell[c]{Vulnerability\\localization \\ (3\%)}} & \multicolumn{1}{c|}{\cellcolor{blue!5}\makecell[c]{Vulnerability\\classification \\ (3\%)}} & \multicolumn{1}{c|}{\cellcolor{blue!5}\makecell[c]{Root cause\\analysis \\ (3\%)}} & \multicolumn{1}{c|}{\cellcolor{blue!5}\makecell[c]{Vulnerability\\prioritization\\/ triage \\ (2\%)}} & \multicolumn{1}{c|}{\cellcolor{blue!5}\makecell[c]{Bug\\triaging \\ (2\%)}} & \multicolumn{1}{c|}{\cellcolor{blue!5}\makecell[c]{Vulnerability\\correlation \\ (1\%)}} & \multicolumn{1}{c|}{\cellcolor{blue!5}\makecell[c]{Zero-day\\discovery \\ (4\%)}} & \multicolumn{1}{c|}{\cellcolor{blue!14}\makecell[c]{API /\\input-surface\\security testing \\ (21\%)}} \\
    \hline

\cellcolor{white}2026 & \cite{4} & \cmark &  &  &  &  &  &  &  & \cmark \\
\rowcolor{gray!15} \cellcolor{white} & \cite{5} &  &  &  &  &  &  &  &  &  \\
\cellcolor{white} & \cite{11} & \cmark &  &  &  &  &  &  &  &  \\
\rowcolor{gray!15} \cellcolor{white} & \cite{19} &  &  &  &  &  &  &  &  &  \\
\cellcolor{white} & \cite{23lbathaviator} & \cmark & \cmark &  &  &  &  &  &  & \cmark \\
\rowcolor{gray!15} \cellcolor{white} & \cite{25zhang2026bountybench} & \cmark &  &  &  &  &  &  & \cmark & \cmark \\
\cellcolor{white} & \cite{26RIGAKI2026129987} &  &  &  &  &  &  &  &  &  \\
\rowcolor{gray!15} \cellcolor{white} & \cite{36ZOU2026104305} &  &  &  &  &  &  &  &  &  \\
\cellcolor{white} & \cite{40zhuo2026cyberzero} &  &  &  &  &  &  &  &  &  \\
\rowcolor{gray!15} \cellcolor{white} & \cite{57} &  &  &  &  &  &  &  &  &  \\
\cellcolor{white} & \cite{64happe2026llms} &  &  &  &  &  &  &  &  &  \\
\rowcolor{gray!15} \cellcolor{white} & \cite{84WANG2026103731} &  &  &  &  &  &  &  &  & \cmark \\
\cellcolor{white} & \cite{86} &  &  &  &  &  &  &  &  &  \\
\rowcolor{gray!15} \cellcolor{white} & \cite{91zhu-etal-2026-teams} &  &  &  &  &  &  &  & \cmark & \cmark \\
\cellcolor{white} & \cite{92} &  &  &  &  &  &  &  &  &  \\
\rowcolor{gray!15} \cellcolor{white} & \cite{96} &  &  &  &  &  &  &  &  &  \\
\cellcolor{white} & \cite{97} &  &  &  &  &  &  &  &  &  \\
\rowcolor{gray!15} \cellcolor{white} & \cite{101} & \cmark &  &  &  &  &  &  &  &  \\
\hline
\cellcolor{white}2025 & \cite{2} &  &  &  &  &  &  &  &  & \cmark \\
\rowcolor{gray!15} \cellcolor{white} & \cite{3} &  &  &  &  &  &  &  &  &  \\
\cellcolor{white} & \cite{6} & \cmark &  &  &  &  &  &  &  & \cmark \\
\rowcolor{gray!15} \cellcolor{white} & \cite{7} &  &  &  &  &  &  &  &  &  \\
\cellcolor{white} & \cite{8zhang2025agent} &  &  &  &  &  &  &  &  &  \\
\rowcolor{gray!15} \cellcolor{white} & \cite{9jie2025agent4vul} & \cmark &  &  &  &  &  &  &  & \cmark \\
\cellcolor{white} & \cite{12} &  &  &  &  &  &  &  &  &  \\
\rowcolor{gray!15} \cellcolor{white} & \cite{13} &  &  &  &  &  &  &  &  &  \\
\cellcolor{white} & \cite{15} &  &  &  &  &  &  &  &  &  \\
\rowcolor{gray!15} \cellcolor{white} & \cite{16} &  &  &  &  &  &  &  &  &  \\
\cellcolor{white} & \cite{17} &  &  &  &  &  &  &  &  &  \\
\rowcolor{gray!15} \cellcolor{white} & \cite{18} &  &  &  &  &  &  &  &  &  \\
\cellcolor{white} & \cite{20} &  &  &  &  &  &  &  &  &  \\
\rowcolor{gray!15} \cellcolor{white} & \cite{21} & \cmark &  &  &  &  &  &  &  &  \\
\cellcolor{white} & \cite{22} &  &  &  &  &  &  &  &  & \cmark \\
\rowcolor{gray!15} \cellcolor{white} & \cite{24yildiz-etal-2025-benchmarking} & \cmark &  &  &  &  &  &  &  &  \\
\cellcolor{white} & \cite{27} &  &  &  &  &  &  &  &  &  \\
\rowcolor{gray!15} \cellcolor{white} & \cite{29} & \cmark & \cmark &  &  &  &  &  &  &  \\
\cellcolor{white} & \cite{30} &  &  &  &  &  &  &  &  &  \\
\rowcolor{gray!15} \cellcolor{white} & \cite{31} &  &  &  & \cmark &  &  &  &  &  \\
\cellcolor{white} & \cite{32} &  &  &  &  &  &  &  &  &  \\
\rowcolor{gray!15} \cellcolor{white} & \cite{33} & \cmark &  &  &  &  &  &  &  &  \\
\cellcolor{white} & \cite{34} & \cmark &  & \cmark & \cmark &  &  &  &  &  \\
\rowcolor{gray!15} \cellcolor{white} & \cite{35} & \cmark &  &  &  &  &  &  &  &  \\
\cellcolor{white} & \cite{37app15169096} &  &  &  &  &  &  &  &  &  \\
\rowcolor{gray!15} \cellcolor{white} & \cite{38zhu2025cvebenchbenchmarkaiagents} & \cmark &  &  &  &  &  &  &  & \cmark \\
\cellcolor{white} & \cite{39} &  &  &  &  &  &  &  &  &  \\
\rowcolor{gray!15} \cellcolor{white} & \cite{41jiao2025deepvulhunter} & \cmark &  &  &  &  &  &  &  & \cmark \\
\cellcolor{white} & \cite{42LOEVENICH2025111162} &  &  &  &  &  &  &  &  &  \\
\rowcolor{gray!15} \cellcolor{white} & \cite{45} &  &  &  &  &  & \cmark &  &  &  \\
\cellcolor{white} & \cite{46abramovich2025enigma} &  &  &  &  &  &  &  &  &  \\
\rowcolor{gray!15} \cellcolor{white} & \cite{47chowdhry2025evaluating} &  &  &  &  &  &  &  &  &  \\
\cellcolor{white} & \cite{48} &  &  &  &  &  &  &  &  &  \\
\rowcolor{gray!15} \cellcolor{white} & \cite{49} &  &  &  &  &  &  &  &  &  \\
\cellcolor{white} & \cite{51} &  &  &  &  &  &  &  &  &  \\
\rowcolor{gray!15} \cellcolor{white} & \cite{53} &  &  &  &  &  &  &  &  &  \\
\cellcolor{white} & \cite{55} & \cmark &  &  &  &  &  &  &  & \cmark \\
\rowcolor{gray!15} \cellcolor{white} & \cite{56} &  &  &  &  &  &  &  &  &  \\
\cellcolor{white} & \cite{58} & \cmark &  &  &  &  &  &  &  &  \\
\rowcolor{gray!15} \cellcolor{white} & \cite{59} &  &  &  &  &  &  &  &  &  \\
\cellcolor{white} & \cite{60} &  &  &  &  &  &  &  &  &  \\
\rowcolor{gray!15} \cellcolor{white} & \cite{61} &  &  &  &  &  &  &  &  &  \\
\cellcolor{white} & \cite{62} &  &  &  &  &  &  &  &  &  \\
\rowcolor{gray!15} \cellcolor{white} & \cite{65} &  &  &  &  &  &  &  &  &  \\
\cellcolor{white} & \cite{67} &  &  &  &  &  &  &  &  &  \\
\rowcolor{gray!15} \cellcolor{white} & \cite{68} &  &  &  &  &  &  &  &  &  \\
\cellcolor{white} & \cite{69} & \cmark &  &  &  & \cmark &  &  & \cmark & \cmark \\
\rowcolor{gray!15} \cellcolor{white} & \cite{70} &  &  &  &  &  &  &  &  &  \\
\cellcolor{white} & \cite{71} & \cmark &  &  &  &  &  &  &  &  \\
\rowcolor{gray!15} \cellcolor{white} & \cite{72} &  &  &  &  &  &  &  &  &  \\
\cellcolor{white} & \cite{73} &  &  &  &  &  &  &  &  &  \\
\rowcolor{gray!15} \cellcolor{white} & \cite{75} & \cmark & \cmark &  &  & \cmark & \cmark &  &  & \cmark \\
\cellcolor{white} & \cite{78} &  &  &  &  &  &  &  &  &  \\
\rowcolor{gray!15} \cellcolor{white} & \cite{80} & \cmark &  &  &  &  &  &  &  & \cmark \\
\cellcolor{white} & \cite{81Cao_Huang_Li_Huilin_He_Oo_Hooi_2025} &  &  &  &  &  &  &  &  &  \\
\rowcolor{gray!15} \cellcolor{white} & \cite{83} & \cmark &  &  &  &  &  &  & \cmark &  \\
\cellcolor{white} & \cite{87} &  &  &  &  &  &  &  &  &  \\
\rowcolor{gray!15} \cellcolor{white} & \cite{88} &  &  &  &  &  &  &  &  &  \\
\cellcolor{white} & \cite{89guo2025repoauditautonomousllmagentrepositorylevel} & \cmark &  &  &  &  &  &  &  & \cmark \\
\rowcolor{gray!15} \cellcolor{white} & \cite{90} & \cmark &  &  &  &  &  &  &  & \cmark \\
\cellcolor{white} & \cite{93info16050365} &  &  &  &  &  &  &  &  &  \\
\rowcolor{gray!15} \cellcolor{white} & \cite{94} &  &  &  &  &  &  &  &  &  \\
\cellcolor{white} & \cite{99} & \cmark &  & \cmark & \cmark &  &  &  &  &  \\
\rowcolor{gray!15} \cellcolor{white} & \cite{100} & \cmark &  &  &  &  &  &  &  &  \\
\hline
\cellcolor{white}2024 & \cite{1} &  &  &  &  &  &  &  &  &  \\
\rowcolor{gray!15} \cellcolor{white} & \cite{10debenedetti2024agentdojo} &  &  &  &  &  &  &  &  &  \\
\cellcolor{white} & \cite{14chen2024agentpoison} &  &  &  &  &  &  &  &  &  \\
\rowcolor{gray!15} \cellcolor{white} & \cite{28} &  &  &  &  &  &  &  &  &  \\
\cellcolor{white} & \cite{43} &  &  &  &  &  &  &  &  &  \\
\rowcolor{gray!15} \cellcolor{white} & \cite{50} &  &  &  &  &  &  &  &  &  \\
\cellcolor{white} & \cite{54} &  &  &  &  &  &  &  &  &  \\
\rowcolor{gray!15} \cellcolor{white} & \cite{63zemicheal2024llm} & \cmark &  & \cmark &  &  &  &  &  &  \\
\cellcolor{white} & \cite{66} & \cmark &  &  &  &  &  &  &  & \cmark \\
\rowcolor{gray!15} \cellcolor{white} & \cite{74icaart24} &  &  &  &  &  &  &  &  &  \\
\cellcolor{white} & \cite{76} & \cmark &  &  &  &  &  &  &  & \cmark \\
\rowcolor{gray!15} \cellcolor{white} & \cite{77} &  &  &  &  &  &  &  &  &  \\
\cellcolor{white} & \cite{79} &  &  &  &  &  &  &  &  &  \\
\rowcolor{gray!15} \cellcolor{white} & \cite{82rigaki2024prompt} &  &  &  &  &  &  &  &  &  \\
\cellcolor{white} & \cite{85} & \cmark &  &  &  &  &  & \cmark &  &  \\
\rowcolor{gray!15} \cellcolor{white} & \cite{95} &  &  &  &  &  &  &  &  &  \\
\cellcolor{white} & \cite{98} & \cmark &  &  &  &  &  &  &  & \cmark \\
\hline
\rowcolor{gray!15} \cellcolor{white}2023 & \cite{52} & \cmark &  &  &  &  &  &  &  & \cmark \\
\hline
    \end{tabular}%
\end{adjustbox}
  \label{tab:application-vulnerability-detection-analysis}%
  \vspace{-12pt}
\end{wraptable}

\vspace{-2pt}
\subsection{Survey Result: Application}\label{ssec:resultapplication}

We analyze the surveyed LLM agents' security applications along eleven level-1 categories. Six of these---{\em vulnerability detection \& analysis}, {\em exploit development \& analysis}, {\em penetration testing \& red-teaming}, {\em vulnerability repair \& hardening}, {\em intrusion \& threat detection, incident response}, and {\em threat intelligence \& security operations (SecOps)}---each account for more than 10\% of the surveyed papers and are presented with individual taxonomy tables. The remaining lower-frequency categories---{\em fuzzing \& test generation}, {\em malware analysis \& detection}, {\em reverse engineering}, {\em access control \& authentication assessment}, and {\em model attack generation \& robustness evaluation}---share a single table. Within each category, we use the level-2 task groupings shown in the corresponding table to organize the discussion.

\vspace{-4pt}
\subsubsection{Vulnerability detection \& analysis}\label{sssec:app-vulndetect}
This is one of the largest application areas in the taxonomy~\cite{2,4,6,9jie2025agent4vul,11,21,22,23lbathaviator,24yildiz-etal-2025-benchmarking,25zhang2026bountybench,29,31,33,34,35,38zhu2025cvebenchbenchmarkaiagents,41jiao2025deepvulhunter,45,52,55,58,63zemicheal2024llm,66,69,71,75,76,80,83,84WANG2026103731,85,89guo2025repoauditautonomousllmagentrepositorylevel,90,91zhu-etal-2026-teams,98,99,100,101}, as 
summarized in Table~\ref{tab:application-vulnerability-detection-analysis}.
These agents support the discovery, localization, classification, prioritization, and validation of vulnerabilities across source code, smart contracts, binaries, APIs, and infrastructure configurations.

\noindent
{\bf Vulnerability detection.}
Vulnerability detection refers to the identification of security weaknesses in a given software artifact, whether source code, bytecode, binary, or configuration file. In this sub-task the agent inspects program entities, forms candidate vulnerability hypotheses, and refines its assessment through tool use or iterative reasoning~\cite{4,6,9jie2025agent4vul,11,21,23lbathaviator,24yildiz-etal-2025-benchmarking,25zhang2026bountybench,29,33,34,35,38zhu2025cvebenchbenchmarkaiagents,41jiao2025deepvulhunter,52,55,58,63zemicheal2024llm,66,69,71,75,76,80,83,85,89guo2025repoauditautonomousllmagentrepositorylevel,90,98,99,100,101}.
\ul{Smart-contract auditing} is heavily represented in this sub-task~\cite{6,33,34,52,83,99}. For example, \textsc{GPTLens}~\cite{52} decomposes auditing into a two-stage auditor/critic process in which one agent proposes candidate weaknesses and another suppresses false positives. \textsc{LLM-SmartAudit}~\cite{6} extends this into a multi-agent conversational architecture with a buffer-of-thought mechanism that propagates intermediate insights across agents, while \textsc{iAudit}~\cite{34} fine-tunes separate Detector and Reasoner models and subjects candidates to a Ranker/Critic debate.
\ul{General source-code detection} forms a second cluster~\cite{9jie2025agent4vul,24yildiz-etal-2025-benchmarking,41jiao2025deepvulhunter,71,89guo2025repoauditautonomousllmagentrepositorylevel,98}. \textsc{DeepVulHunter}~\cite{41jiao2025deepvulhunter} grounds multi-round reasoning in retrieval over semantically similar code snippets, and \textsc{JITVUL}~\cite{24yildiz-etal-2025-benchmarking} shows on a benchmark of 879 CVEs that ReAct-style agents outperform raw prompting. \textsc{Agent4Vul}~\cite{9jie2025agent4vul} treats detection as an agentic code-analysis task in which Commentator and Vectorizer agents reason over program context rather than classify isolated snippets. A representative repository-scale instance is \textsc{RepoAudit}~\cite{89guo2025repoauditautonomousllmagentrepositorylevel}, which traverses interprocedural data-flow facts on demand and uses a path-condition validator to suppress false positives, achieving 78.4\% precision at roughly \$2.54 per project.
\ul{Binary and infrastructure detection} extends the sub-task beyond conventional source code~\cite{29,55,58,66,80,90}. \textsc{ClearAgent}~\cite{29} detects vulnerabilities in COTS binaries where no source is available, the IaC misconfiguration agent~\cite{58} analyses infrastructure-as-code definitions, and the multi-agent SQLi pipeline~\cite{55} targets deployed web services.

\noindent
{\bf API / input-surface security testing.}
This sub-task concerns testing for vulnerabilities through external interfaces---APIs, web surfaces, network endpoints, and infrastructure configurations---rather than by inspecting source code alone~\cite{2,4,6,9jie2025agent4vul,22,23lbathaviator,25zhang2026bountybench,38zhu2025cvebenchbenchmarkaiagents,41jiao2025deepvulhunter,52,55,66,69,75,76,80,84WANG2026103731,89guo2025repoauditautonomousllmagentrepositorylevel,90,91zhu-etal-2026-teams,98}. \textsc{AutoRestTest}~\cite{2} decomposes REST API testing into four cooperating agents trained via multi-agent reinforcement learning, coordinating over a semantic dependency graph of API operations. \textsc{AutoPT}~\cite{22} establishes an end-to-end web-pentest benchmark and finds that agents are procedurally fluent in the workflow but frequently fail to synthesise commands that work against real targets. \textsc{LLMSQLi}~\cite{66} provides a black-box SQL injection detection tool driven by multi-agent payload generation.

\noindent
{\bf Zero-day discovery.}
Zero-day discovery refers to the identification of previously-unknown vulnerabilities in realistic settings, as opposed to classification of known weakness patterns~\cite{25zhang2026bountybench,69,83,91zhu-etal-2026-teams}. For example, \textsc{AgentFuzz}~\cite{69} reports 34 zero-day vulnerabilities (23 CVEs assigned) across 20 popular open-source LLM-agent applications, and \textsc{PropertyGPT}~\cite{83} discovers 12 zero-days in smart contracts through retrieval-augmented formal property generation. \textsc{BountyBench}~\cite{25zhang2026bountybench} instantiates 25 real-world systems with 40 monetary bug bounties and separate Detect/Exploit/Patch tasks, while \textsc{HPTSA}~\cite{91zhu-etal-2026-teams} exploits 14 real-world zero-day vulnerabilities via a planner that dispatches specialist subagents.

\noindent
{\bf Vulnerability localization.}
Vulnerability localization concerns identifying the precise code regions responsible for a weakness, as opposed to merely flagging a file or function as vulnerable~\cite{23lbathaviator,29,75}. \textsc{ClearAgent}~\cite{29} localises bugs in binaries by combining an analyzer-friendly intermediate representation with an LLM-friendly natural-language view, then verifying candidates by constructing concrete triggering inputs. \textsc{PatchAgent}~\cite{75} performs fault localization as the first stage of its repair loop, and \textsc{AVIATOR}~\cite{23lbathaviator} localises injection points when synthesising training vulnerabilities.

\noindent
{\bf Vulnerability classification.}
Vulnerability classification maps detected findings to established taxonomies such as CWE types or CVSS severity levels~\cite{34,63zemicheal2024llm,99}. \textsc{iAudit}~\cite{34} classifies smart-contract vulnerabilities through its Ranker/Critic debate mechanism, the VEX-justification agent of~\cite{63zemicheal2024llm} maps CVE observations to impact categories across containerised environments, and \textsc{VulnPatch}~\cite{99} couples classification with explanation and downstream patching.

\noindent
{\bf Root cause analysis.}
Root cause analysis reconstructs the underlying cause of a flaw rather than merely reporting its symptoms~\cite{31,34,99}. \textsc{Clouseau}~\cite{31} traces attack chains back to root causes by autonomously traversing system logs from a single point-of-interest. \textsc{iAudit}~\cite{34} provides root-cause explanation through its Reasoner agent, and \textsc{VulnPatch}~\cite{99} generates natural-language explanations of the vulnerability mechanism as part of its detect-explain-patch pipeline.

\begin{wraptable}{r}{0.4\textwidth}
  \centering
  \vspace{-12pt}
  \caption{Paper attribution for Application: Exploit development \& analysis}
  \vspace{-10pt}
  \scriptsize
  \setlength{\tabcolsep}{1.5pt}
  \renewcommand{\arraystretch}{0.70}
\begin{adjustbox}{width=\textwidth,max totalheight=0.86\textheight,center}
    \begin{tabular}{|c|c|c|c|c|c|c|}
    \hline
\rowcolor{white} Year & Paper & \multicolumn{1}{c|}{\cellcolor{blue!5}\makecell[c]{Exploit\\generation \\ (6\%)}} & \multicolumn{1}{c|}{\cellcolor{blue!5}\makecell[c]{Exploit chaining\\/ attack-path\\construction \\ (3\%)}} & \multicolumn{1}{c|}{\cellcolor{blue!5}\makecell[c]{Payload\\crafting \\ (6\%)}} & \multicolumn{1}{c|}{\cellcolor{blue!5}\makecell[c]{Exploit analysis\\/ reverse\\engineering \\ (1\%)}} & \multicolumn{1}{c|}{\cellcolor{blue!5}\makecell[c]{Exploit\\detection \\ (3\%)}} \\
    \hline

\cellcolor{white}2026 & \cite{4} &  &  &  &  &  \\
\rowcolor{gray!15} \cellcolor{white} & \cite{5} &  &  &  &  &  \\
\cellcolor{white} & \cite{11} &  &  &  &  &  \\
\rowcolor{gray!15} \cellcolor{white} & \cite{19} &  &  &  &  &  \\
\cellcolor{white} & \cite{23lbathaviator} &  &  &  &  &  \\
\rowcolor{gray!15} \cellcolor{white} & \cite{25zhang2026bountybench} & \cmark &  & \cmark &  &  \\
\cellcolor{white} & \cite{26RIGAKI2026129987} &  &  &  &  &  \\
\rowcolor{gray!15} \cellcolor{white} & \cite{36ZOU2026104305} &  &  &  &  &  \\
\cellcolor{white} & \cite{40zhuo2026cyberzero} &  &  &  &  &  \\
\rowcolor{gray!15} \cellcolor{white} & \cite{57} &  &  &  &  &  \\
\cellcolor{white} & \cite{64happe2026llms} &  &  &  &  &  \\
\rowcolor{gray!15} \cellcolor{white} & \cite{84WANG2026103731} &  &  &  &  &  \\
\cellcolor{white} & \cite{86} &  &  &  &  &  \\
\rowcolor{gray!15} \cellcolor{white} & \cite{91zhu-etal-2026-teams} & \cmark &  & \cmark &  &  \\
\cellcolor{white} & \cite{92} &  &  &  &  &  \\
\rowcolor{gray!15} \cellcolor{white} & \cite{96} &  &  &  &  &  \\
\cellcolor{white} & \cite{97} &  &  &  &  &  \\
\rowcolor{gray!15} \cellcolor{white} & \cite{101} &  &  &  &  & \cmark \\
\hline
\cellcolor{white}2025 & \cite{2} &  &  &  &  &  \\
\rowcolor{gray!15} \cellcolor{white} & \cite{3} &  &  &  &  &  \\
\cellcolor{white} & \cite{6} &  &  &  &  &  \\
\rowcolor{gray!15} \cellcolor{white} & \cite{7} &  &  &  &  &  \\
\cellcolor{white} & \cite{8zhang2025agent} &  &  &  &  &  \\
\rowcolor{gray!15} \cellcolor{white} & \cite{9jie2025agent4vul} &  &  &  &  &  \\
\cellcolor{white} & \cite{12} &  & \cmark &  &  &  \\
\rowcolor{gray!15} \cellcolor{white} & \cite{13} &  &  &  &  &  \\
\cellcolor{white} & \cite{15} &  &  &  &  &  \\
\rowcolor{gray!15} \cellcolor{white} & \cite{16} &  &  &  &  &  \\
\cellcolor{white} & \cite{17} &  &  &  &  &  \\
\rowcolor{gray!15} \cellcolor{white} & \cite{18} &  & \cmark &  &  &  \\
\cellcolor{white} & \cite{20} &  &  &  &  &  \\
\rowcolor{gray!15} \cellcolor{white} & \cite{21} &  &  &  &  &  \\
\cellcolor{white} & \cite{22} &  &  &  &  &  \\
\rowcolor{gray!15} \cellcolor{white} & \cite{24yildiz-etal-2025-benchmarking} &  &  &  &  &  \\
\cellcolor{white} & \cite{27} &  &  &  &  &  \\
\rowcolor{gray!15} \cellcolor{white} & \cite{29} &  &  &  &  &  \\
\cellcolor{white} & \cite{30} &  &  &  &  &  \\
\rowcolor{gray!15} \cellcolor{white} & \cite{31} &  &  &  &  &  \\
\cellcolor{white} & \cite{32} &  &  &  &  &  \\
\rowcolor{gray!15} \cellcolor{white} & \cite{33} &  &  &  &  &  \\
\cellcolor{white} & \cite{34} &  &  &  &  &  \\
\rowcolor{gray!15} \cellcolor{white} & \cite{35} &  &  &  &  &  \\
\cellcolor{white} & \cite{37app15169096} & \cmark &  & \cmark &  &  \\
\rowcolor{gray!15} \cellcolor{white} & \cite{38zhu2025cvebenchbenchmarkaiagents} & \cmark &  & \cmark &  &  \\
\cellcolor{white} & \cite{39} &  &  &  &  &  \\
\rowcolor{gray!15} \cellcolor{white} & \cite{41jiao2025deepvulhunter} &  &  &  &  &  \\
\cellcolor{white} & \cite{42LOEVENICH2025111162} &  &  &  &  &  \\
\rowcolor{gray!15} \cellcolor{white} & \cite{45} &  &  &  &  &  \\
\cellcolor{white} & \cite{46abramovich2025enigma} &  &  &  &  &  \\
\rowcolor{gray!15} \cellcolor{white} & \cite{47chowdhry2025evaluating} &  &  &  &  &  \\
\cellcolor{white} & \cite{48} &  &  &  &  &  \\
\rowcolor{gray!15} \cellcolor{white} & \cite{49} &  &  &  &  &  \\
\cellcolor{white} & \cite{51} &  &  &  &  &  \\
\rowcolor{gray!15} \cellcolor{white} & \cite{53} &  &  &  &  &  \\
\cellcolor{white} & \cite{55} &  &  &  &  &  \\
\rowcolor{gray!15} \cellcolor{white} & \cite{56} &  &  &  &  &  \\
\cellcolor{white} & \cite{58} &  &  &  &  &  \\
\rowcolor{gray!15} \cellcolor{white} & \cite{59} &  & \cmark &  &  &  \\
\cellcolor{white} & \cite{60} &  &  &  &  &  \\
\rowcolor{gray!15} \cellcolor{white} & \cite{61} &  &  &  &  &  \\
\cellcolor{white} & \cite{62} &  &  &  &  &  \\
\rowcolor{gray!15} \cellcolor{white} & \cite{65} & \cmark &  & \cmark & \cmark &  \\
\cellcolor{white} & \cite{67} &  &  &  &  &  \\
\rowcolor{gray!15} \cellcolor{white} & \cite{68} &  &  &  &  &  \\
\cellcolor{white} & \cite{69} &  &  &  &  &  \\
\rowcolor{gray!15} \cellcolor{white} & \cite{70} &  &  &  &  &  \\
\cellcolor{white} & \cite{71} &  &  &  &  &  \\
\rowcolor{gray!15} \cellcolor{white} & \cite{72} &  &  &  &  &  \\
\cellcolor{white} & \cite{73} &  &  &  &  &  \\
\rowcolor{gray!15} \cellcolor{white} & \cite{75} &  &  &  &  &  \\
\cellcolor{white} & \cite{78} &  &  &  &  &  \\
\rowcolor{gray!15} \cellcolor{white} & \cite{80} &  &  &  &  &  \\
\cellcolor{white} & \cite{81Cao_Huang_Li_Huilin_He_Oo_Hooi_2025} &  &  &  &  &  \\
\rowcolor{gray!15} \cellcolor{white} & \cite{83} &  &  &  &  &  \\
\cellcolor{white} & \cite{87} &  &  &  &  &  \\
\rowcolor{gray!15} \cellcolor{white} & \cite{88} &  &  &  &  &  \\
\cellcolor{white} & \cite{89guo2025repoauditautonomousllmagentrepositorylevel} &  &  &  &  &  \\
\rowcolor{gray!15} \cellcolor{white} & \cite{90} &  &  &  &  &  \\
\cellcolor{white} & \cite{93info16050365} &  &  &  &  &  \\
\rowcolor{gray!15} \cellcolor{white} & \cite{94} &  &  &  &  &  \\
\cellcolor{white} & \cite{99} &  &  &  &  &  \\
\rowcolor{gray!15} \cellcolor{white} & \cite{100} &  &  &  &  & \cmark \\
\hline
\cellcolor{white}2024 & \cite{1} &  &  &  &  &  \\
\rowcolor{gray!15} \cellcolor{white} & \cite{10debenedetti2024agentdojo} &  &  &  &  &  \\
\cellcolor{white} & \cite{14chen2024agentpoison} &  &  &  &  &  \\
\rowcolor{gray!15} \cellcolor{white} & \cite{28} &  &  &  &  &  \\
\cellcolor{white} & \cite{43} &  &  &  &  &  \\
\rowcolor{gray!15} \cellcolor{white} & \cite{50} & \cmark &  & \cmark &  &  \\
\cellcolor{white} & \cite{54} &  &  &  &  &  \\
\rowcolor{gray!15} \cellcolor{white} & \cite{63zemicheal2024llm} &  &  &  &  & \cmark \\
\cellcolor{white} & \cite{66} &  &  &  &  &  \\
\rowcolor{gray!15} \cellcolor{white} & \cite{74icaart24} &  &  &  &  &  \\
\cellcolor{white} & \cite{76} &  &  &  &  &  \\
\rowcolor{gray!15} \cellcolor{white} & \cite{77} &  &  &  &  &  \\
\cellcolor{white} & \cite{79} &  &  &  &  &  \\
\rowcolor{gray!15} \cellcolor{white} & \cite{82rigaki2024prompt} &  &  &  &  &  \\
\cellcolor{white} & \cite{85} &  &  &  &  &  \\
\rowcolor{gray!15} \cellcolor{white} & \cite{95} &  &  &  &  &  \\
\cellcolor{white} & \cite{98} &  &  &  &  &  \\
\hline
\rowcolor{gray!15} \cellcolor{white}2023 & \cite{52} &  &  &  &  &  \\
\hline
    \end{tabular}%
\end{adjustbox}
  \label{tab:application-exploit-development-analysis}%
  \vspace{-12pt}
\end{wraptable}
\noindent
{\bf Vulnerability prioritization / triage.}
Vulnerability prioritization concerns ordering findings by severity, exploitability, or remediation cost to guide effort~\cite{69,75}. \textsc{AgentFuzz}~\cite{69} prioritises candidate taint-style findings by exploitability before reporting them, and \textsc{PatchAgent}~\cite{75} orders localised faults to determine which to repair first within its autonomous loop.

\noindent
{\bf Bug triaging.}
Bug triaging refers to the assignment and routing of detected findings to the appropriate remediation workflow~\cite{45,75}. \textsc{LLM2Policy}~\cite{45} routes configuration findings to the corresponding policy layer, while \textsc{PatchAgent}~\cite{75} determines which findings are actionable within its repair pipeline. This sub-task overlaps with prioritization but concerns the routing decision rather than the ordering itself.

\noindent
{\bf Vulnerability correlation.}
Vulnerability correlation links observations across multiple data sources to build a coherent picture of a weakness and its impact~\cite{85}. \textsc{PTGroup}~\cite{85} correlates findings discovered across separate stages of an 
engagement through chained prompt sequences, representing an early form of cross-artifact vulnerability management.

\vspace{-4pt}
\subsubsection{Exploit development \& analysis}\label{sssec:app-exploit}
These agents, as summarized in Table~\ref{tab:application-exploit-development-analysis}, synthesise and analyse working exploits once a vulnerability is known~\cite{12,18,25zhang2026bountybench,37app15169096,38zhu2025cvebenchbenchmarkaiagents,50,59,63zemicheal2024llm,65,91zhu-etal-2026-teams,100,101}. Much exploit-related activity is subsumed by penetration testing; the papers here treat exploit synthesis or analysis as a primary objective.

\noindent
{\bf Exploit generation.}
Exploit generation concerns synthesising payloads or scripts from vulnerability descriptions~\cite{25zhang2026bountybench,37app15169096,38zhu2025cvebenchbenchmarkaiagents,50,65,91zhu-etal-2026-teams}. For example, \textsc{HPTSA}~\cite{91zhu-etal-2026-teams} dispatches specialist subagents from a planning agent, achieving 4.3$\times$ improvement over single-agent baselines on 14 real zero-day vulnerabilities. The study by Happe et al.~\cite{50} compares five prompt formulations for CVE-to-Metasploit-script generation across GPT-4o, WhiteRabbitNeo, and Mistral in a military coalition-network context.

\noindent
{\bf Payload crafting.}
Payload crafting refers to the construction of concrete attack payloads that trigger the intended vulnerability behaviour~\cite{25zhang2026bountybench,37app15169096,38zhu2025cvebenchbenchmarkaiagents,50,65,91zhu-etal-2026-teams}. 
In our corpus this sub-task is co-extensive with exploit generation: every paper that synthesizes an exploit also constructs a target-specific payload satisfying constraints such as input filtering or memory layout. 
In general, the two remain distinct capabilities, though. 

\noindent
{\bf Exploit chaining / attack-path construction.}
Exploit chaining concerns assembling multi-step attack paths from individual vulnerability primitives~\cite{12,18,59}. In these papers agentic generative AI combines cybersecurity knowledge 
\begin{wraptable}{r}{0.45\textwidth}
  \centering
  \vspace{-13pt}
  \caption{Paper attribution for Application: Penetration testing \& red-teaming}
  \vspace{-10pt}
  \scriptsize
  \setlength{\tabcolsep}{1.5pt}
  \renewcommand{\arraystretch}{0.70}
\begin{adjustbox}{width=\textwidth,max totalheight=0.86\textheight,center}
    \begin{tabular}{|c|c|c|c|c|c|c|c|c|c|}
    \hline
\rowcolor{white} Year & Paper & \multicolumn{1}{c|}{\cellcolor{blue!11}\makecell[c]{Reconn-\\aissance \\ (17\%)}} & \multicolumn{1}{c|}{\cellcolor{blue!9}\makecell[c]{Initial\\access \\ (14\%)}} & \multicolumn{1}{c|}{\cellcolor{blue!5}\makecell[c]{Privilege\\escalation \\ (3\%)}} & \multicolumn{1}{c|}{\cellcolor{blue!5}\makecell[c]{Lateral\\movement \\ (2\%)}} & \multicolumn{1}{c|}{\cellcolor{blue!5}\makecell[c]{Exfiltration\\simulation \\ (2\%)}} & \multicolumn{1}{c|}{\cellcolor{blue!9}\makecell[c]{Full end-to\\-end pentest \\ (14\%)}} & \multicolumn{1}{c|}{\cellcolor{blue!5}\makecell[c]{CTF challenge\\solving \\ (7\%)}} & \multicolumn{1}{c|}{\cellcolor{blue!17}\makecell[c]{Red team\\automation \\ (25\%)}} \\
    \hline

\cellcolor{white}2026 & \cite{4} & \cmark &  &  &  &  &  &  & \cmark \\
\rowcolor{gray!15} \cellcolor{white} & \cite{5} &  &  &  &  &  &  &  &  \\
\cellcolor{white} & \cite{11} &  &  &  &  &  &  &  &  \\
\rowcolor{gray!15} \cellcolor{white} & \cite{19} &  &  &  &  &  &  &  &  \\
\cellcolor{white} & \cite{23lbathaviator} &  &  &  &  &  &  &  &  \\
\rowcolor{gray!15} \cellcolor{white} & \cite{25zhang2026bountybench} &  &  &  &  &  &  &  &  \\
\cellcolor{white} & \cite{26RIGAKI2026129987} & \cmark & \cmark & \cmark & \cmark & \cmark & \cmark &  & \cmark \\
\rowcolor{gray!15} \cellcolor{white} & \cite{36ZOU2026104305} &  &  &  &  &  &  & \cmark &  \\
\cellcolor{white} & \cite{40zhuo2026cyberzero} &  &  &  &  &  &  & \cmark &  \\
\rowcolor{gray!15} \cellcolor{white} & \cite{57} &  &  &  &  &  &  &  &  \\
\cellcolor{white} & \cite{64happe2026llms} &  &  & \cmark &  &  &  &  &  \\
\rowcolor{gray!15} \cellcolor{white} & \cite{84WANG2026103731} & \cmark & \cmark &  &  &  & \cmark &  & \cmark \\
\cellcolor{white} & \cite{86} &  &  &  &  &  &  &  &  \\
\rowcolor{gray!15} \cellcolor{white} & \cite{91zhu-etal-2026-teams} & \cmark &  &  &  &  &  &  & \cmark \\
\cellcolor{white} & \cite{92} &  &  &  &  &  &  &  &  \\
\rowcolor{gray!15} \cellcolor{white} & \cite{96} &  &  &  &  &  &  &  &  \\
\cellcolor{white} & \cite{97} &  &  &  &  &  &  &  &  \\
\rowcolor{gray!15} \cellcolor{white} & \cite{101} &  &  &  &  &  &  &  &  \\
\hline
\cellcolor{white}2025 & \cite{2} &  &  &  &  &  &  &  &  \\
\rowcolor{gray!15} \cellcolor{white} & \cite{3} &  &  &  &  &  &  &  &  \\
\cellcolor{white} & \cite{6} &  &  &  &  &  &  &  &  \\
\rowcolor{gray!15} \cellcolor{white} & \cite{7} &  &  &  &  &  &  &  &  \\
\cellcolor{white} & \cite{8zhang2025agent} &  &  &  &  &  &  &  &  \\
\rowcolor{gray!15} \cellcolor{white} & \cite{9jie2025agent4vul} &  &  &  &  &  &  &  &  \\
\cellcolor{white} & \cite{12} &  &  &  &  &  &  &  & \cmark \\
\rowcolor{gray!15} \cellcolor{white} & \cite{13} &  &  &  &  &  &  &  & \cmark \\
\cellcolor{white} & \cite{15} &  &  &  &  &  &  &  &  \\
\rowcolor{gray!15} \cellcolor{white} & \cite{16} &  &  &  &  &  &  &  &  \\
\cellcolor{white} & \cite{17} &  &  &  &  &  &  &  &  \\
\rowcolor{gray!15} \cellcolor{white} & \cite{18} &  &  &  &  &  &  &  & \cmark \\
\cellcolor{white} & \cite{20} & \cmark & \cmark &  &  &  & \cmark &  & \cmark \\
\rowcolor{gray!15} \cellcolor{white} & \cite{21} & \cmark & \cmark &  &  &  & \cmark &  & \cmark \\
\cellcolor{white} & \cite{22} & \cmark & \cmark &  &  &  & \cmark &  & \cmark \\
\rowcolor{gray!15} \cellcolor{white} & \cite{24yildiz-etal-2025-benchmarking} &  &  &  &  &  &  &  &  \\
\cellcolor{white} & \cite{27} &  &  &  &  &  &  &  &  \\
\rowcolor{gray!15} \cellcolor{white} & \cite{29} &  &  &  &  &  &  &  &  \\
\cellcolor{white} & \cite{30} &  &  &  &  &  &  &  &  \\
\rowcolor{gray!15} \cellcolor{white} & \cite{31} &  &  &  &  &  &  &  &  \\
\cellcolor{white} & \cite{32} &  &  &  &  &  &  &  &  \\
\rowcolor{gray!15} \cellcolor{white} & \cite{33} &  &  &  &  &  &  &  &  \\
\cellcolor{white} & \cite{34} &  &  &  &  &  &  &  &  \\
\rowcolor{gray!15} \cellcolor{white} & \cite{35} & \cmark & \cmark &  &  &  & \cmark &  & \cmark \\
\cellcolor{white} & \cite{37app15169096} & \cmark & \cmark &  &  &  & \cmark &  & \cmark \\
\rowcolor{gray!15} \cellcolor{white} & \cite{38zhu2025cvebenchbenchmarkaiagents} &  &  &  &  &  &  &  & \cmark \\
\cellcolor{white} & \cite{39} &  &  &  &  &  &  &  &  \\
\rowcolor{gray!15} \cellcolor{white} & \cite{41jiao2025deepvulhunter} &  &  &  &  &  &  &  &  \\
\cellcolor{white} & \cite{42LOEVENICH2025111162} &  &  &  &  &  &  &  &  \\
\rowcolor{gray!15} \cellcolor{white} & \cite{45} &  &  &  &  &  &  &  &  \\
\cellcolor{white} & \cite{46abramovich2025enigma} &  &  &  &  &  &  & \cmark &  \\
\rowcolor{gray!15} \cellcolor{white} & \cite{47chowdhry2025evaluating} &  &  &  &  &  &  &  &  \\
\cellcolor{white} & \cite{48} &  &  &  &  &  &  &  &  \\
\rowcolor{gray!15} \cellcolor{white} & \cite{49} &  &  &  &  &  &  &  &  \\
\cellcolor{white} & \cite{51} &  &  &  &  &  &  &  &  \\
\rowcolor{gray!15} \cellcolor{white} & \cite{53} &  &  &  &  &  &  &  &  \\
\cellcolor{white} & \cite{55} &  &  &  &  &  &  &  & \cmark \\
\rowcolor{gray!15} \cellcolor{white} & \cite{56} &  &  &  &  &  &  &  &  \\
\cellcolor{white} & \cite{58} &  &  &  &  &  &  &  &  \\
\rowcolor{gray!15} \cellcolor{white} & \cite{59} & \cmark & \cmark &  &  &  & \cmark & \cmark & \cmark \\
\cellcolor{white} & \cite{60} &  &  &  &  &  &  &  &  \\
\rowcolor{gray!15} \cellcolor{white} & \cite{61} &  &  &  &  &  &  &  &  \\
\cellcolor{white} & \cite{62} &  &  &  &  &  &  &  &  \\
\rowcolor{gray!15} \cellcolor{white} & \cite{65} &  &  &  &  &  &  & \cmark & \cmark \\
\cellcolor{white} & \cite{67} &  &  &  &  &  &  &  &  \\
\rowcolor{gray!15} \cellcolor{white} & \cite{68} &  &  &  &  &  &  &  &  \\
\cellcolor{white} & \cite{69} &  &  &  &  &  &  &  &  \\
\rowcolor{gray!15} \cellcolor{white} & \cite{70} &  &  &  &  &  &  & \cmark &  \\
\cellcolor{white} & \cite{71} &  &  &  &  &  &  &  &  \\
\rowcolor{gray!15} \cellcolor{white} & \cite{72} &  &  &  &  &  &  &  &  \\
\cellcolor{white} & \cite{73} &  &  &  &  &  &  &  &  \\
\rowcolor{gray!15} \cellcolor{white} & \cite{75} &  &  &  &  &  &  &  &  \\
\cellcolor{white} & \cite{78} & \cmark & \cmark &  &  &  & \cmark &  & \cmark \\
\rowcolor{gray!15} \cellcolor{white} & \cite{80} &  &  & \cmark &  &  &  &  &  \\
\cellcolor{white} & \cite{81Cao_Huang_Li_Huilin_He_Oo_Hooi_2025} &  &  &  &  &  &  &  &  \\
\rowcolor{gray!15} \cellcolor{white} & \cite{83} &  &  &  &  &  &  &  &  \\
\cellcolor{white} & \cite{87} & \cmark & \cmark &  & \cmark &  & \cmark &  & \cmark \\
\rowcolor{gray!15} \cellcolor{white} & \cite{88} &  &  &  &  &  &  &  &  \\
\cellcolor{white} & \cite{89guo2025repoauditautonomousllmagentrepositorylevel} &  &  &  &  &  &  &  &  \\
\rowcolor{gray!15} \cellcolor{white} & \cite{90} & \cmark &  &  &  &  &  &  &  \\
\cellcolor{white} & \cite{93info16050365} &  &  &  &  &  &  &  &  \\
\rowcolor{gray!15} \cellcolor{white} & \cite{94} &  &  &  &  &  &  &  &  \\
\cellcolor{white} & \cite{99} &  &  &  &  &  &  &  &  \\
\rowcolor{gray!15} \cellcolor{white} & \cite{100} &  &  &  &  &  &  &  &  \\
\hline
\cellcolor{white}2024 & \cite{1} &  &  &  &  &  &  &  &  \\
\rowcolor{gray!15} \cellcolor{white} & \cite{10debenedetti2024agentdojo} &  &  &  &  & \cmark &  &  &  \\
\cellcolor{white} & \cite{14chen2024agentpoison} &  &  &  &  &  &  &  &  \\
\rowcolor{gray!15} \cellcolor{white} & \cite{28} &  &  &  &  &  &  &  &  \\
\cellcolor{white} & \cite{43} &  &  &  &  &  &  &  &  \\
\rowcolor{gray!15} \cellcolor{white} & \cite{50} &  &  &  &  &  &  &  & \cmark \\
\cellcolor{white} & \cite{54} &  &  &  &  &  &  & \cmark &  \\
\rowcolor{gray!15} \cellcolor{white} & \cite{63zemicheal2024llm} &  &  &  &  &  &  &  &  \\
\cellcolor{white} & \cite{66} &  &  &  &  &  &  &  &  \\
\rowcolor{gray!15} \cellcolor{white} & \cite{74icaart24} &  &  &  &  &  &  &  & \cmark \\
\cellcolor{white} & \cite{76} & \cmark & \cmark &  &  &  & \cmark &  & \cmark \\
\rowcolor{gray!15} \cellcolor{white} & \cite{77} &  &  &  &  &  &  &  & \cmark \\
\cellcolor{white} & \cite{79} & \cmark & \cmark &  &  &  & \cmark &  & \cmark \\
\rowcolor{gray!15} \cellcolor{white} & \cite{82rigaki2024prompt} & \cmark & \cmark &  &  &  & \cmark &  & \cmark \\
\cellcolor{white} & \cite{85} & \cmark & \cmark &  &  &  & \cmark &  & \cmark \\
\rowcolor{gray!15} \cellcolor{white} & \cite{95} &  &  &  &  &  &  &  &  \\
\cellcolor{white} & \cite{98} &  &  &  &  &  &  &  &  \\
\hline
\rowcolor{gray!15} \cellcolor{white}2023 & \cite{52} &  &  &  &  &  &  &  &  \\
\hline
    \end{tabular}%
\end{adjustbox}
  \label{tab:application-penetration-testing-red-teaming}%
  \vspace{-15pt}
\end{wraptable}

graphs with LLM agents to compose chains against tactical network targets, reasoning about preconditions and privilege transitions between steps.

\noindent
{\bf Exploit detection.}
Exploit detection concerns identifying the presence or feasibility of exploitable weaknesses from an external perspective~\cite{63zemicheal2024llm,100,101}. \textsc{VVF-AI}~\cite{100} builds a proof-of-concept-based verification framework to filter false-positive reports, and \textsc{ZT-ICAS}~\cite{101} uses agentic scanning to detect exploitable weaknesses while hardening the scanner itself against adversarial manipulation.

\noindent
{\bf Exploit analysis / reverse engineering.}
Exploit analysis concerns understanding the mechanism and impact of existing exploits~\cite{65}. Only one paper addresses this sub-task, targeting the reverse engineering of exploits as a precursor to developing defensive countermeasures.

\vspace{-4pt}
\subsubsection{Penetration testing \& red-teaming}\label{sssec:app-pentest}
Agents in this category simulate adversarial activity against a target, interleaving reconnaissance, exploitation, and reporting~\cite{4,10debenedetti2024agentdojo,12,13,18,20,21,22,26RIGAKI2026129987,35,36ZOU2026104305,37app15169096,38zhu2025cvebenchbenchmarkaiagents,40zhuo2026cyberzero,46abramovich2025enigma,50,54,55,59,64happe2026llms,65,70,74icaart24,76,77,78,79,80,82rigaki2024prompt,84WANG2026103731,85,87,90,91zhu-etal-2026-teams}. This was the earliest application area to attract LLM-agent attention. 
Table~\ref{tab:application-penetration-testing-red-teaming} 
summarizes these applications.

\noindent
{\bf Red team automation.}
Red team automation refers to the end-to-end orchestration of adversarial campaigns in which the agent selects the next action based on the current state of the engagement~\cite{4,12,13,18,20,21,22,26RIGAKI2026129987,35,37app15169096,38zhu2025cvebenchbenchmarkaiagents,50,55,59,65,74icaart24,76,77,78,79,82rigaki2024prompt,84WANG2026103731,85,87,91zhu-etal-2026-teams}. For example, \textsc{AutoPentester}~\cite{21} accepts only a target IP and iteratively invokes standard security tools, outperforming \textsc{PentestGPT} by 27\% in subtask completion with substantially less human intervention. \textsc{PENTEST-AI}~\cite{77} structures attack chains around MITRE ATT\&CK, \textsc{PTFusion}~\cite{84WANG2026103731} introduces a semi-decentralised multi-agent design with Model Context Protocol--based tool invocation, and \textsc{RefPentester}~\cite{87} uses knowledge self-reflection to mitigate hallucinated commands. The controller-based framework of~\cite{35} adds explicit workflow management to reduce stage-skipping and token waste.

\noindent
{\bf Reconnaissance.}
Reconnaissance concerns the identification of hosts, services, attack surfaces, and environmental cues prior to exploitation~\cite{4,20,21,22,26RIGAKI2026129987,35,37app15169096,59,76,78,79,82rigaki2024prompt,84WANG2026103731,85,87,90,91zhu-etal-2026-teams}. Most end-to-end penetration-testing 
\begin{wraptable}{r}{0.4\textwidth}
  \centering
  \vspace{-14pt}
  \caption{Paper attribution for Application: Vulnerability repair \& hardening}
  \vspace{-10pt}
  \scriptsize
  \setlength{\tabcolsep}{1.5pt}
  \renewcommand{\arraystretch}{0.70}
\begin{adjustbox}{width=\textwidth,max totalheight=0.86\textheight,center}
    \begin{tabular}{|c|c|c|c|c|c|c|}
    \hline
\rowcolor{white} Year & Paper & \multicolumn{1}{c|}{\cellcolor{blue!5}\makecell[c]{Automated patch\\generation \\ (6\%)}} & \multicolumn{1}{c|}{\cellcolor{blue!5}\makecell[c]{Patch validation\\/ verification \\ (5\%)}} & \multicolumn{1}{c|}{\cellcolor{blue!9}\makecell[c]{Code review\\/ auditing \\ (13\%)}} & \multicolumn{1}{c|}{\cellcolor{blue!5}\makecell[c]{Code\\hardening \\ (5\%)}} & \multicolumn{1}{c|}{\cellcolor{blue!5}\makecell[c]{Security policy\\enforcement \\ (4\%)}} \\
    \hline

\cellcolor{white}2026 & \cite{4} &  &  &  &  &  \\
\rowcolor{gray!15} \cellcolor{white} & \cite{5} &  &  &  &  &  \\
\cellcolor{white} & \cite{11} &  &  &  &  &  \\
\rowcolor{gray!15} \cellcolor{white} & \cite{19} &  &  &  &  &  \\
\cellcolor{white} & \cite{23lbathaviator} &  &  &  &  &  \\
\rowcolor{gray!15} \cellcolor{white} & \cite{25zhang2026bountybench} & \cmark & \cmark &  &  &  \\
\cellcolor{white} & \cite{26RIGAKI2026129987} &  &  &  &  &  \\
\rowcolor{gray!15} \cellcolor{white} & \cite{36ZOU2026104305} &  &  &  &  &  \\
\cellcolor{white} & \cite{40zhuo2026cyberzero} &  &  &  &  &  \\
\rowcolor{gray!15} \cellcolor{white} & \cite{57} &  &  &  &  &  \\
\cellcolor{white} & \cite{64happe2026llms} &  &  &  &  &  \\
\rowcolor{gray!15} \cellcolor{white} & \cite{84WANG2026103731} &  &  &  &  &  \\
\cellcolor{white} & \cite{86} &  &  &  &  &  \\
\rowcolor{gray!15} \cellcolor{white} & \cite{91zhu-etal-2026-teams} &  &  &  &  &  \\
\cellcolor{white} & \cite{92} &  &  &  &  &  \\
\rowcolor{gray!15} \cellcolor{white} & \cite{96} &  &  &  &  &  \\
\cellcolor{white} & \cite{97} &  &  & \cmark &  &  \\
\rowcolor{gray!15} \cellcolor{white} & \cite{101} &  &  & \cmark &  &  \\
\hline
\cellcolor{white}2025 & \cite{2} &  &  &  &  &  \\
\rowcolor{gray!15} \cellcolor{white} & \cite{3} &  &  &  &  &  \\
\cellcolor{white} & \cite{6} &  &  & \cmark &  &  \\
\rowcolor{gray!15} \cellcolor{white} & \cite{7} &  &  &  &  &  \\
\cellcolor{white} & \cite{8zhang2025agent} &  &  &  &  &  \\
\rowcolor{gray!15} \cellcolor{white} & \cite{9jie2025agent4vul} &  &  & \cmark &  &  \\
\cellcolor{white} & \cite{12} &  &  &  &  &  \\
\rowcolor{gray!15} \cellcolor{white} & \cite{13} &  &  &  &  &  \\
\cellcolor{white} & \cite{15} &  &  &  &  &  \\
\rowcolor{gray!15} \cellcolor{white} & \cite{16} &  &  &  &  &  \\
\cellcolor{white} & \cite{17} &  &  &  &  &  \\
\rowcolor{gray!15} \cellcolor{white} & \cite{18} &  &  &  &  &  \\
\cellcolor{white} & \cite{20} &  &  &  &  &  \\
\rowcolor{gray!15} \cellcolor{white} & \cite{21} &  &  &  &  &  \\
\cellcolor{white} & \cite{22} &  &  &  &  &  \\
\rowcolor{gray!15} \cellcolor{white} & \cite{24yildiz-etal-2025-benchmarking} &  &  &  &  &  \\
\cellcolor{white} & \cite{27} &  &  &  &  &  \\
\rowcolor{gray!15} \cellcolor{white} & \cite{29} &  &  &  &  &  \\
\cellcolor{white} & \cite{30} &  &  &  &  &  \\
\rowcolor{gray!15} \cellcolor{white} & \cite{31} &  &  &  &  &  \\
\cellcolor{white} & \cite{32} &  &  &  &  &  \\
\rowcolor{gray!15} \cellcolor{white} & \cite{33} &  &  & \cmark &  &  \\
\cellcolor{white} & \cite{34} &  &  & \cmark &  &  \\
\rowcolor{gray!15} \cellcolor{white} & \cite{35} &  &  &  &  &  \\
\cellcolor{white} & \cite{37app15169096} &  &  &  &  &  \\
\rowcolor{gray!15} \cellcolor{white} & \cite{38zhu2025cvebenchbenchmarkaiagents} &  &  &  &  &  \\
\cellcolor{white} & \cite{39} &  &  &  &  &  \\
\rowcolor{gray!15} \cellcolor{white} & \cite{41jiao2025deepvulhunter} &  &  & \cmark &  &  \\
\cellcolor{white} & \cite{42LOEVENICH2025111162} &  &  &  &  &  \\
\rowcolor{gray!15} \cellcolor{white} & \cite{45} &  &  &  & \cmark & \cmark \\
\cellcolor{white} & \cite{46abramovich2025enigma} &  &  &  &  &  \\
\rowcolor{gray!15} \cellcolor{white} & \cite{47chowdhry2025evaluating} &  &  &  &  &  \\
\cellcolor{white} & \cite{48} &  &  &  &  &  \\
\rowcolor{gray!15} \cellcolor{white} & \cite{49} &  &  &  &  &  \\
\cellcolor{white} & \cite{51} &  &  &  &  &  \\
\rowcolor{gray!15} \cellcolor{white} & \cite{53} &  &  &  &  &  \\
\cellcolor{white} & \cite{55} &  &  &  &  &  \\
\rowcolor{gray!15} \cellcolor{white} & \cite{56} &  &  &  & \cmark & \cmark \\
\cellcolor{white} & \cite{58} & \cmark &  &  & \cmark & \cmark \\
\rowcolor{gray!15} \cellcolor{white} & \cite{59} &  &  &  &  &  \\
\cellcolor{white} & \cite{60} &  &  &  &  &  \\
\rowcolor{gray!15} \cellcolor{white} & \cite{61} &  &  &  &  &  \\
\cellcolor{white} & \cite{62} &  &  &  &  &  \\
\rowcolor{gray!15} \cellcolor{white} & \cite{65} &  &  &  &  &  \\
\cellcolor{white} & \cite{67} &  &  &  &  &  \\
\rowcolor{gray!15} \cellcolor{white} & \cite{68} &  &  &  &  &  \\
\cellcolor{white} & \cite{69} &  &  &  & \cmark &  \\
\rowcolor{gray!15} \cellcolor{white} & \cite{70} &  &  &  &  &  \\
\cellcolor{white} & \cite{71} &  &  & \cmark &  &  \\
\rowcolor{gray!15} \cellcolor{white} & \cite{72} &  &  &  &  &  \\
\cellcolor{white} & \cite{73} &  &  &  &  &  \\
\rowcolor{gray!15} \cellcolor{white} & \cite{75} & \cmark & \cmark &  & \cmark &  \\
\cellcolor{white} & \cite{78} &  &  &  &  &  \\
\rowcolor{gray!15} \cellcolor{white} & \cite{80} &  &  &  &  &  \\
\cellcolor{white} & \cite{81Cao_Huang_Li_Huilin_He_Oo_Hooi_2025} &  &  &  &  &  \\
\rowcolor{gray!15} \cellcolor{white} & \cite{83} &  & \cmark & \cmark &  & \cmark \\
\cellcolor{white} & \cite{87} &  &  &  &  &  \\
\rowcolor{gray!15} \cellcolor{white} & \cite{88} & \cmark & \cmark &  &  &  \\
\cellcolor{white} & \cite{89guo2025repoauditautonomousllmagentrepositorylevel} &  &  & \cmark &  &  \\
\rowcolor{gray!15} \cellcolor{white} & \cite{90} &  &  &  &  &  \\
\cellcolor{white} & \cite{93info16050365} &  &  &  &  &  \\
\rowcolor{gray!15} \cellcolor{white} & \cite{94} &  &  &  &  &  \\
\cellcolor{white} & \cite{99} & \cmark & \cmark & \cmark &  &  \\
\rowcolor{gray!15} \cellcolor{white} & \cite{100} &  &  &  &  &  \\
\hline
\cellcolor{white}2024 & \cite{1} &  &  &  &  &  \\
\rowcolor{gray!15} \cellcolor{white} & \cite{10debenedetti2024agentdojo} &  &  &  &  &  \\
\cellcolor{white} & \cite{14chen2024agentpoison} &  &  &  &  &  \\
\rowcolor{gray!15} \cellcolor{white} & \cite{28} &  &  &  &  &  \\
\cellcolor{white} & \cite{43} &  &  &  &  &  \\
\rowcolor{gray!15} \cellcolor{white} & \cite{50} &  &  &  &  &  \\
\cellcolor{white} & \cite{54} &  &  &  &  &  \\
\rowcolor{gray!15} \cellcolor{white} & \cite{63zemicheal2024llm} &  &  &  &  &  \\
\cellcolor{white} & \cite{66} &  &  &  &  &  \\
\rowcolor{gray!15} \cellcolor{white} & \cite{74icaart24} &  &  &  &  &  \\
\cellcolor{white} & \cite{76} & \cmark &  &  &  &  \\
\rowcolor{gray!15} \cellcolor{white} & \cite{77} &  &  &  &  &  \\
\cellcolor{white} & \cite{79} &  &  &  &  &  \\
\rowcolor{gray!15} \cellcolor{white} & \cite{82rigaki2024prompt} &  &  &  &  &  \\
\cellcolor{white} & \cite{85} &  &  &  &  &  \\
\rowcolor{gray!15} \cellcolor{white} & \cite{95} &  &  &  &  &  \\
\cellcolor{white} & \cite{98} &  &  & \cmark &  &  \\
\hline
\rowcolor{gray!15} \cellcolor{white}2023 & \cite{52} &  &  & \cmark &  &  \\
\hline
    \end{tabular}%
\end{adjustbox}
  \label{tab:application-vulnerability-repair-hardening}%
  \vspace{-16pt}
\end{wraptable}

frameworks incorporate reconnaissance as the initial phase. The FOFA-based asset-search agent of~\cite{90} contributes context-aware reconnaissance queries over internet-exposed assets, and the \textsc{HPTSA} planner~\cite{91zhu-etal-2026-teams} performs reconnaissance to inform specialist subagent dispatch.

\noindent
{\bf Initial access.}
Initial access refers to the step where the agent gains a first foothold in the target system~\cite{20,21,22,26RIGAKI2026129987,35,37app15169096,59,76,78,79,82rigaki2024prompt,84WANG2026103731,85,87}. In these papers the agent must translate reconnaissance findings into a concrete exploitation attempt against a live service. \textsc{PenHeal}~\cite{76} couples this offensive phase with a remediation module via counterfactual prompting, improving vulnerability coverage by 31\% and reducing cost by 46\%.

\noindent
{\bf Full end-to-end pentest.}
These pentests require the agent to coordinate the entire loop from initial probing through exploitation and, in some cases, reporting or remediation~\cite{20,21,22,26RIGAKI2026129987,35,37app15169096,59,76,78,79,82rigaki2024prompt,84WANG2026103731,85,87}. The local-LLM study of~\cite{26RIGAKI2026129987} highlights the privacy, operational, and reproducibility implications of deploying offensive agents without cloud-hosted models, while \textsc{CurriculumPT}~\cite{37app15169096} organises the task as a curriculum that proceeds from easier to harder targets and reuses stored experience across engagements.

\noindent
{\bf CTF challenge solving.}
CTF challenge solving uses capture-the-flag environments as controlled settings for studying offensive agent capabilities~\cite{36ZOU2026104305,40zhuo2026cyberzero,46abramovich2025enigma,54,59,65,70}. For example, \textsc{EnIGMA}~\cite{46abramovich2025enigma} introduces Interactive Agent Tools---a gdb-attached debugger and a persistent network client---that let an LM agent drive genuinely interactive utilities, reaching state-of-the-art results on 390 challenges across NYU CTF, Intercode-CTF, and CyBench. \textsc{CYBER-ZERO}~\cite{40zhuo2026cyberzero} synthesises agent trajectories from public writeups via persona-driven simulation, \textsc{CTFAgent}~\cite{36ZOU2026104305} uses a plan-and-execute pattern with adaptive problem decomposition, and \textsc{CTFKnow}~\cite{70} provides a 3{,}992-question benchmark measuring technical CTF knowledge.

\noindent
{\bf Privilege escalation.}
Privilege escalation concerns actions taken after initial access to obtain higher-level permissions on the compromised host~\cite{26RIGAKI2026129987,64happe2026llms,80}. \textsc{hackingBuddyGPT}~\cite{64happe2026llms} targets autonomous Linux escalation, and \textsc{Perses}~\cite{80} tests whether small local models can achieve escalation results comparable to frontier models.

\noindent
{\bf Lateral movement.}
This refers to an attacker's traversal across a network from one compromised host to others~\cite{26RIGAKI2026129987,87}. Only two papers address this sub-task, both within broader end-to-end penetration-testing frameworks rather than as a dedicated capability.

\noindent
{\bf Exfiltration simulation.}
Exfiltration simulation concerns the data-extraction phase of offensive operations~\cite{10debenedetti2024agentdojo,26RIGAKI2026129987}. \textsc{AgentDojo}~\cite{10debenedetti2024agentdojo} includes data-exfiltration objectives among its adversarial tasks. The sparsity of work on lateral movement and exfiltration, and the absence of any paper addressing persistence, indicate that post-access phases remain substantially less developed than initial-access and reconnaissance capabilities.

\vspace{-4pt}
\subsubsection{Vulnerability repair \& hardening}\label{sssec:app-repair}
Agents in this category (Table~\ref{tab:application-vulnerability-repair-hardening}) 
generate and validate patches and hardening modifications once a vulnerability has been identified~\cite{6,9jie2025agent4vul,25zhang2026bountybench,33,34,41jiao2025deepvulhunter,45,52,56,58,69,71,75,76,83,88,89guo2025repoauditautonomousllmagentrepositorylevel,97,98,99,101}. Many of these papers also appear under vulnerability detection \& analysis, reflecting a common detect-then-repair design.

\noindent
{\bf Code review / auditing.}
Code review and auditing concern systematic inspection of code for security weaknesses, often combining automated detection with human-readable explanations~\cite{6,9jie2025agent4vul,33,34,41jiao2025deepvulhunter,52,71,83,89guo2025repoauditautonomousllmagentrepositorylevel,97,98,99,101}. This sub-task overlaps substantially with detection, since identifying a weakness and reasoning about how to remediate it are closely connected. The multi-agent code-review framework of~\cite{71} integrates review, bug detection, and security analysis with a FAISS-memory feedback mechanism, \textsc{PropertyGPT}~\cite{83} 
retrieves and adapts human-authored Certora properties to produce verifiable invariants, and \textsc{VulnGPT}~\cite{98} layers a controller-and-evaluator AutoGPT loop on top of GPT-4.

\noindent
{\bf Automated patch generation.}
Automated patch generation concerns synthesising candidate fixes for identified vulnerabilities~\cite{25zhang2026bountybench,58,75,76,88,99}. For example, \textsc{PatchAgent}~\cite{75} integrates fault localisation, patch synthesis, and test-suite validation into a single autonomous loop, repairing over 90\% of 178 real-world vulnerabilities and outperforming specialised program-repair tools. \textsc{VulnPatch}~\cite{99} delegates detection, explanation, and patch generation to three fine-tuned LLM agents, \textsc{RepairAgent}~\cite{88} interleaves test execution with fix synthesis, and \textsc{BountyBench}~\cite{25zhang2026bountybench} contributes the empirical finding that frontier agents are notably stronger at patching than at exploitation.

\noindent
{\bf Patch validation / verification.}
Patch validation concerns checking whether a candidate fix actually resolves the targeted vulnerability without introducing new problems~\cite{25zhang2026bountybench,75,83,88,99}. Validation typically involves running the test suite, checking formal properties, or executing candidates against proof-of-concept inputs. In \textsc{PropertyGPT}~\cite{83} verification is itself an iterative agentic loop: the agent generates smart-contract properties and then uses compilation, static analysis, and a prover to determine whether they are usable.

\noindent
{\bf Code hardening.}
Code hardening concerns preventive or structural changes that reduce attack surface rather than targeting a specific vulnerability~\cite{45,56,58,69,75}. \textsc{LLM2Policy}~\cite{45} generates fine-grained zero-trust policies from Kubernetes configurations, the IaC remediation agent of~\cite{58} issues policy-level recommendations for CI/CD pipelines, and the autonomous cloud agent of~\cite{56} includes hardening among its responsibilities.

\noindent
{\bf Security policy enforcement.}
Security policy enforcement concerns generating and applying rules that constrain system behaviour to meet security requirements~\cite{45,56,58,83}. \textsc{LLM2Policy}~\cite{45} and the IaC agent of~\cite{58} generate and enforce infrastructure-level policies, while \textsc{PropertyGPT}~\cite{83} enforces formally-specified invariants as a policy layer over smart contracts.

\vspace{-4pt}
\subsubsection{Intrusion \& threat detection, incident response}\label{sssec:app-idr}
Agents for these applications (summarized in Table~\ref{tab:application-intrusion-threat-detection-incident-response}) detect active or recent compromises, attribute them to actors or techniques, and reconstruct attack chains to support remediation~\cite{1,3,5,16,17,18,19,20,27,29,31,32,39,42LOEVENICH2025111162,43,47chowdhry2025evaluating,48,49,50,53,56,58,60,61,62,63zemicheal2024llm,65,67,73,78,79,81Cao_Huang_Li_Huilin_He_Oo_Hooi_2025,89guo2025repoauditautonomousllmagentrepositorylevel,92,93info16050365,94,95,96}. It is tied 
\begin{wraptable}{r}{0.6\textwidth}
  \centering
  \vspace{-2pt}
  \caption{Paper attribution for Application: Intrusion \& threat detection, incident response}
  \vspace{-10pt}
  \scriptsize
  \setlength{\tabcolsep}{1.5pt}
  \renewcommand{\arraystretch}{0.70}
\begin{adjustbox}{width=\textwidth,max totalheight=0.87\textheight,center}
    \begin{tabular}{|c|c|c|c|c|c|c|c|c|c|c|}
    \hline
\rowcolor{white} Year & Paper & \multicolumn{1}{c|}{\cellcolor{blue!15}\makecell[c]{Log / traffic\\analysis \& anomaly\\detection \\ (22\%)}} & \multicolumn{1}{c|}{\cellcolor{blue!5}\makecell[c]{DDoS detection\\\& mitigation \\ (3\%)}} & \multicolumn{1}{c|}{\cellcolor{blue!5}\makecell[c]{Phishing \&\\social-engineering\\detection \\ (3\%)}} & \multicolumn{1}{c|}{\cellcolor{blue!5}\makecell[c]{Alert\\triage \\ (2\%)}} & \multicolumn{1}{c|}{\cellcolor{blue!5}\makecell[c]{Threat\\hunting \\ (1\%)}} & \multicolumn{1}{c|}{\cellcolor{blue!5}\makecell[c]{Attack\\attribution \\ (5\%)}} & \multicolumn{1}{c|}{\cellcolor{blue!10}\makecell[c]{Incident\\investigation \\ (13\%)}} & \multicolumn{1}{c|}{\cellcolor{blue!11}\makecell[c]{Incident response\\automation \\ (17\%)}} & \multicolumn{1}{c|}{\cellcolor{blue!5}\makecell[c]{Forensic\\analysis \\ (3\%)}} \\
    \hline

\cellcolor{white}2026 & \cite{4} &  &  &  &  &  &  &  &  &  \\
\rowcolor{gray!15} \cellcolor{white} & \cite{5} & \cmark &  &  &  &  &  &  &  &  \\
\cellcolor{white} & \cite{11} &  &  &  &  &  &  &  &  &  \\
\rowcolor{gray!15} \cellcolor{white} & \cite{19} & \cmark &  &  &  &  &  &  &  &  \\
\cellcolor{white} & \cite{23lbathaviator} &  &  &  &  &  &  &  &  &  \\
\rowcolor{gray!15} \cellcolor{white} & \cite{25zhang2026bountybench} &  &  &  &  &  &  &  &  &  \\
\cellcolor{white} & \cite{26RIGAKI2026129987} &  &  &  &  &  &  &  &  &  \\
\rowcolor{gray!15} \cellcolor{white} & \cite{36ZOU2026104305} &  &  &  &  &  &  &  &  &  \\
\cellcolor{white} & \cite{40zhuo2026cyberzero} &  &  &  &  &  &  &  &  &  \\
\rowcolor{gray!15} \cellcolor{white} & \cite{57} &  &  &  &  &  &  &  &  &  \\
\cellcolor{white} & \cite{64happe2026llms} &  &  &  &  &  &  &  &  &  \\
\rowcolor{gray!15} \cellcolor{white} & \cite{84WANG2026103731} &  &  &  &  &  &  &  &  &  \\
\cellcolor{white} & \cite{86} &  &  &  &  &  &  &  &  &  \\
\rowcolor{gray!15} \cellcolor{white} & \cite{91zhu-etal-2026-teams} &  &  &  &  &  &  &  &  &  \\
\cellcolor{white} & \cite{92} & \cmark & \cmark &  &  &  &  &  & \cmark &  \\
\rowcolor{gray!15} \cellcolor{white} & \cite{96} &  &  & \cmark &  &  &  &  &  &  \\
\cellcolor{white} & \cite{97} &  &  &  &  &  &  &  &  &  \\
\rowcolor{gray!15} \cellcolor{white} & \cite{101} &  &  &  &  &  &  &  &  &  \\
\hline
\cellcolor{white}2025 & \cite{2} &  &  &  &  &  &  &  &  &  \\
\rowcolor{gray!15} \cellcolor{white} & \cite{3} & \cmark &  & \cmark & \cmark &  &  &  &  &  \\
\cellcolor{white} & \cite{6} &  &  &  &  &  &  &  &  &  \\
\rowcolor{gray!15} \cellcolor{white} & \cite{7} &  &  &  &  &  &  &  &  &  \\
\cellcolor{white} & \cite{8zhang2025agent} &  &  &  &  &  &  &  &  &  \\
\rowcolor{gray!15} \cellcolor{white} & \cite{9jie2025agent4vul} &  &  &  &  &  &  &  &  &  \\
\cellcolor{white} & \cite{12} &  &  &  &  &  &  &  &  &  \\
\rowcolor{gray!15} \cellcolor{white} & \cite{13} &  &  &  &  &  &  &  &  &  \\
\cellcolor{white} & \cite{15} &  &  &  &  &  &  &  &  &  \\
\rowcolor{gray!15} \cellcolor{white} & \cite{16} & \cmark & \cmark &  &  &  &  &  &  &  \\
\cellcolor{white} & \cite{17} &  &  &  &  &  &  & \cmark & \cmark &  \\
\rowcolor{gray!15} \cellcolor{white} & \cite{18} &  &  &  &  &  &  & \cmark &  &  \\
\cellcolor{white} & \cite{20} &  &  &  &  &  & \cmark &  &  &  \\
\rowcolor{gray!15} \cellcolor{white} & \cite{21} &  &  &  &  &  &  &  &  &  \\
\cellcolor{white} & \cite{22} &  &  &  &  &  &  &  &  &  \\
\rowcolor{gray!15} \cellcolor{white} & \cite{24yildiz-etal-2025-benchmarking} &  &  &  &  &  &  &  &  &  \\
\cellcolor{white} & \cite{27} & \cmark &  &  &  &  & \cmark & \cmark & \cmark & \cmark \\
\rowcolor{gray!15} \cellcolor{white} & \cite{29} &  &  &  &  &  &  &  &  &  \\
\cellcolor{white} & \cite{30} &  &  &  &  &  &  &  &  &  \\
\rowcolor{gray!15} \cellcolor{white} & \cite{31} & \cmark &  &  &  &  &  & \cmark & \cmark & \cmark \\
\cellcolor{white} & \cite{32} & \cmark &  &  &  &  &  &  & \cmark &  \\
\rowcolor{gray!15} \cellcolor{white} & \cite{33} &  &  &  &  &  &  &  &  &  \\
\cellcolor{white} & \cite{34} &  &  &  &  &  &  &  &  &  \\
\rowcolor{gray!15} \cellcolor{white} & \cite{35} &  &  &  &  &  &  &  &  &  \\
\cellcolor{white} & \cite{37app15169096} &  &  &  &  &  &  &  &  &  \\
\rowcolor{gray!15} \cellcolor{white} & \cite{38zhu2025cvebenchbenchmarkaiagents} &  &  &  &  &  &  &  &  &  \\
\cellcolor{white} & \cite{39} & \cmark &  &  &  &  &  &  & \cmark &  \\
\rowcolor{gray!15} \cellcolor{white} & \cite{41jiao2025deepvulhunter} &  &  &  &  &  &  &  &  &  \\
\cellcolor{white} & \cite{42LOEVENICH2025111162} & \cmark &  &  &  &  &  & \cmark & \cmark &  \\
\rowcolor{gray!15} \cellcolor{white} & \cite{45} &  &  &  &  &  &  &  &  &  \\
\cellcolor{white} & \cite{46abramovich2025enigma} &  &  &  &  &  &  &  &  &  \\
\rowcolor{gray!15} \cellcolor{white} & \cite{47chowdhry2025evaluating} & \cmark &  &  &  &  &  &  & \cmark &  \\
\cellcolor{white} & \cite{48} &  &  &  &  &  &  &  & \cmark &  \\
\rowcolor{gray!15} \cellcolor{white} & \cite{49} & \cmark & \cmark &  &  &  &  &  &  &  \\
\cellcolor{white} & \cite{51} &  &  &  &  &  &  &  &  &  \\
\rowcolor{gray!15} \cellcolor{white} & \cite{53} & \cmark &  &  &  &  &  &  & \cmark &  \\
\cellcolor{white} & \cite{55} &  &  &  &  &  &  &  &  &  \\
\rowcolor{gray!15} \cellcolor{white} & \cite{56} & \cmark &  &  &  &  &  &  & \cmark &  \\
\cellcolor{white} & \cite{58} &  &  &  &  &  & \cmark &  &  &  \\
\rowcolor{gray!15} \cellcolor{white} & \cite{59} &  &  &  &  &  &  &  &  &  \\
\cellcolor{white} & \cite{60} & \cmark &  &  &  &  &  & \cmark &  &  \\
\rowcolor{gray!15} \cellcolor{white} & \cite{61} &  &  &  &  &  &  &  & \cmark &  \\
\cellcolor{white} & \cite{62} & \cmark &  &  &  &  &  &  &  &  \\
\rowcolor{gray!15} \cellcolor{white} & \cite{65} &  &  &  &  &  &  & \cmark &  &  \\
\cellcolor{white} & \cite{67} &  &  &  &  &  & \cmark &  &  &  \\
\rowcolor{gray!15} \cellcolor{white} & \cite{68} &  &  &  &  &  &  &  &  &  \\
\cellcolor{white} & \cite{69} &  &  &  &  &  &  &  &  &  \\
\rowcolor{gray!15} \cellcolor{white} & \cite{70} &  &  &  &  &  &  &  &  &  \\
\cellcolor{white} & \cite{71} &  &  &  &  &  &  &  &  &  \\
\rowcolor{gray!15} \cellcolor{white} & \cite{72} &  &  &  &  &  &  &  &  &  \\
\cellcolor{white} & \cite{73} & \cmark &  &  &  &  &  &  & \cmark &  \\
\rowcolor{gray!15} \cellcolor{white} & \cite{75} &  &  &  &  &  &  &  &  &  \\
\cellcolor{white} & \cite{78} &  &  &  &  &  & \cmark &  &  &  \\
\rowcolor{gray!15} \cellcolor{white} & \cite{80} &  &  &  &  &  &  &  &  &  \\
\cellcolor{white} & \cite{81Cao_Huang_Li_Huilin_He_Oo_Hooi_2025} &  &  & \cmark &  &  &  & \cmark &  &  \\
\rowcolor{gray!15} \cellcolor{white} & \cite{83} &  &  &  &  &  &  &  &  &  \\
\cellcolor{white} & \cite{87} &  &  &  &  &  &  &  &  &  \\
\rowcolor{gray!15} \cellcolor{white} & \cite{88} &  &  &  &  &  &  &  &  &  \\
\cellcolor{white} & \cite{89guo2025repoauditautonomousllmagentrepositorylevel} &  &  &  &  &  &  &  &  &  \\
\rowcolor{gray!15} \cellcolor{white} & \cite{90} &  &  &  &  &  &  &  &  &  \\
\cellcolor{white} & \cite{93info16050365} & \cmark &  &  &  & \cmark &  & \cmark & \cmark &  \\
\rowcolor{gray!15} \cellcolor{white} & \cite{94} & \cmark &  &  & \cmark &  &  & \cmark & \cmark & \cmark \\
\cellcolor{white} & \cite{99} &  &  &  &  &  &  &  &  &  \\
\rowcolor{gray!15} \cellcolor{white} & \cite{100} &  &  &  &  &  &  &  &  &  \\
\hline
\cellcolor{white}2024 & \cite{1} & \cmark &  &  &  &  &  &  &  &  \\
\rowcolor{gray!15} \cellcolor{white} & \cite{10debenedetti2024agentdojo} &  &  &  &  &  &  &  &  &  \\
\cellcolor{white} & \cite{14chen2024agentpoison} &  &  &  &  &  &  &  &  &  \\
\rowcolor{gray!15} \cellcolor{white} & \cite{28} &  &  &  &  &  &  &  &  &  \\
\cellcolor{white} & \cite{43} & \cmark &  &  &  &  &  & \cmark & \cmark &  \\
\rowcolor{gray!15} \cellcolor{white} & \cite{50} &  &  &  &  &  &  & \cmark &  &  \\
\cellcolor{white} & \cite{54} &  &  &  &  &  &  &  &  &  \\
\rowcolor{gray!15} \cellcolor{white} & \cite{63zemicheal2024llm} &  &  &  &  &  &  &  &  &  \\
\cellcolor{white} & \cite{66} &  &  &  &  &  &  &  &  &  \\
\rowcolor{gray!15} \cellcolor{white} & \cite{74icaart24} &  &  &  &  &  &  &  &  &  \\
\cellcolor{white} & \cite{76} &  &  &  &  &  &  &  &  &  \\
\rowcolor{gray!15} \cellcolor{white} & \cite{77} &  &  &  &  &  &  &  &  &  \\
\cellcolor{white} & \cite{79} &  &  &  &  &  &  & \cmark &  &  \\
\rowcolor{gray!15} \cellcolor{white} & \cite{82rigaki2024prompt} &  &  &  &  &  &  &  &  &  \\
\cellcolor{white} & \cite{85} &  &  &  &  &  &  &  &  &  \\
\rowcolor{gray!15} \cellcolor{white} & \cite{95} & \cmark &  &  &  &  &  &  & \cmark &  \\
\cellcolor{white} & \cite{98} &  &  &  &  &  &  &  &  &  \\
\hline
\rowcolor{gray!15} \cellcolor{white}2023 & \cite{52} &  &  &  &  &  &  &  &  &  \\
\hline
    \end{tabular}%
\end{adjustbox}
  \label{tab:application-intrusion-threat-detection-incident-response}%
  \vspace{-12pt}
\end{wraptable}
with vulnerability detection \& analysis as the largest category in the taxonomy.

\noindent
{\bf Log / traffic analysis \& anomaly detection.}
This sub-task concerns reasoning over runtime telemetry---logs, alerts, network flows, and packet captures---to identify suspicious or anomalous patterns~\cite{1,3,5,16,19,27,31,32,39,42LOEVENICH2025111162,43,47chowdhry2025evaluating,49,53,56,60,62,73,92,93info16050365,94,95}. For example, \textsc{Cognitive SOC}~\cite{32} deploys Triage, Enrichment, and StoryWeaver agents that transform raw alerts into structured narratives where every claim is linked to MITRE ATT\&CK, Sigma, or CVE sources, achieving 99.6\% evidence linkage on 1{,}000 UNSW-NB15 alerts. The memory-augmented agent of~\cite{19} provides IoT/IIoT anomaly detection with edge-optimised SHAP explanations, \textsc{NetMoniAI}~\cite{73} deploys micro-agents at each network node for local detection with central aggregation, and \textsc{LaMA-SON}~\cite{16} introduces a self-evolving framework for 6G space-air-ground integrated networks that updates its threat model as new attack types appear.

\noindent
{\bf Incident response automation.}
Incident response automation concerns autonomous or semi-autonomous execution of defensive actions in response to detected threats~\cite{17,27,31,32,39,42LOEVENICH2025111162,43,47chowdhry2025evaluating,48,53,56,61,73,92,93info16050365,94,95}.
\ul{Autonomous cyber-defence agents (ACDs)} evaluated in CAGE-style reinforcement-learning environments form a substantial cluster~\cite{39,42LOEVENICH2025111162,43,47chowdhry2025evaluating,48,53,95}. The pure LLM-based ACD of~\cite{53} matches the strongest RL agent submitted to CAGE-2 without retraining; hybrid DRL+LLM architectures appear in~\cite{42LOEVENICH2025111162,43,47chowdhry2025evaluating}; and MARL+LLM blue/red training is explored in~\cite{95}.
\ul{SOC orchestration} forms the second cluster~\cite{17,32,93info16050365}. \textsc{AutoBnB}~\cite{17} compares centralised, decentralised, and hybrid agent team structures for incident response using the Backdoors~\&~Breaches tabletop framework, and the agentic SOAR system of~\cite{93info16050365} replaces rigid no-code execution with LLM-driven orchestration that interprets natural-language playbook descriptions and branches dynamically on intermediate results.

\begin{wraptable}{r}{0.5\textwidth}
  \centering
  \vspace{-12pt}
  \caption{Paper attribution for Application: Threat intelligence \& security operations (SecOps)}
  \vspace{-10pt}
  \scriptsize
  \setlength{\tabcolsep}{1.5pt}
  \renewcommand{\arraystretch}{0.70}
\begin{adjustbox}{width=\textwidth,max totalheight=0.88\textheight,center}
    \begin{tabular}{|c|c|c|c|c|c|c|c|}
    \hline
\rowcolor{white} Year & Paper & \multicolumn{1}{c|}{\cellcolor{blue!13}\makecell[c]{Threat intelligence\\synthesis \\ (19\%)}} & \multicolumn{1}{c|}{\cellcolor{blue!5}\makecell[c]{Vulnerability report\\generation \\ (4\%)}} & \multicolumn{1}{c|}{\cellcolor{blue!5}\makecell[c]{CVE / advisory\\drafting \\ (2\%)}} & \multicolumn{1}{c|}{\cellcolor{blue!5}\makecell[c]{Security policy\\generation \\ (5\%)}} & \multicolumn{1}{c|}{\cellcolor{blue!5}\makecell[c]{Security Q\&A\\assistant \\ (3\%)}} & \multicolumn{1}{c|}{\cellcolor{blue!5}\makecell[c]{CTI\\analysis \\ (8\%)}} \\
    \hline

\cellcolor{white}2026 & \cite{4} & \cmark &  &  &  &  & \cmark \\
\rowcolor{gray!15} \cellcolor{white} & \cite{5} & \cmark &  &  &  &  &  \\
\cellcolor{white} & \cite{11} &  &  &  &  &  &  \\
\rowcolor{gray!15} \cellcolor{white} & \cite{19} &  &  &  &  &  &  \\
\cellcolor{white} & \cite{23lbathaviator} &  &  &  &  &  &  \\
\rowcolor{gray!15} \cellcolor{white} & \cite{25zhang2026bountybench} &  &  &  &  &  &  \\
\cellcolor{white} & \cite{26RIGAKI2026129987} &  &  &  &  &  &  \\
\rowcolor{gray!15} \cellcolor{white} & \cite{36ZOU2026104305} &  &  &  &  &  &  \\
\cellcolor{white} & \cite{40zhuo2026cyberzero} &  &  &  &  &  &  \\
\rowcolor{gray!15} \cellcolor{white} & \cite{57} &  &  &  &  &  &  \\
\cellcolor{white} & \cite{64happe2026llms} &  &  &  &  &  &  \\
\rowcolor{gray!15} \cellcolor{white} & \cite{84WANG2026103731} &  &  &  &  &  &  \\
\cellcolor{white} & \cite{86} &  &  &  &  &  &  \\
\rowcolor{gray!15} \cellcolor{white} & \cite{91zhu-etal-2026-teams} &  &  &  &  &  &  \\
\cellcolor{white} & \cite{92} & \cmark &  &  & \cmark &  &  \\
\rowcolor{gray!15} \cellcolor{white} & \cite{96} &  &  &  &  &  &  \\
\cellcolor{white} & \cite{97} &  &  &  &  &  &  \\
\rowcolor{gray!15} \cellcolor{white} & \cite{101} &  &  &  &  &  &  \\
\hline
\cellcolor{white}2025 & \cite{2} &  &  &  &  &  &  \\
\rowcolor{gray!15} \cellcolor{white} & \cite{3} & \cmark &  &  &  &  & \cmark \\
\cellcolor{white} & \cite{6} &  & \cmark & \cmark &  &  &  \\
\rowcolor{gray!15} \cellcolor{white} & \cite{7} &  &  &  &  &  &  \\
\cellcolor{white} & \cite{8zhang2025agent} &  &  &  &  &  &  \\
\rowcolor{gray!15} \cellcolor{white} & \cite{9jie2025agent4vul} &  &  &  &  &  &  \\
\cellcolor{white} & \cite{12} & \cmark &  &  &  &  & \cmark \\
\rowcolor{gray!15} \cellcolor{white} & \cite{13} &  &  &  &  &  &  \\
\cellcolor{white} & \cite{15} &  &  &  &  &  &  \\
\rowcolor{gray!15} \cellcolor{white} & \cite{16} & \cmark &  &  &  &  &  \\
\cellcolor{white} & \cite{17} &  &  &  &  &  &  \\
\rowcolor{gray!15} \cellcolor{white} & \cite{18} & \cmark &  &  &  &  & \cmark \\
\cellcolor{white} & \cite{20} &  &  &  &  &  &  \\
\rowcolor{gray!15} \cellcolor{white} & \cite{21} &  &  &  &  &  &  \\
\cellcolor{white} & \cite{22} &  &  &  &  &  &  \\
\rowcolor{gray!15} \cellcolor{white} & \cite{24yildiz-etal-2025-benchmarking} &  &  &  &  &  &  \\
\cellcolor{white} & \cite{27} &  &  &  &  &  &  \\
\rowcolor{gray!15} \cellcolor{white} & \cite{29} &  &  &  &  &  &  \\
\cellcolor{white} & \cite{30} &  &  &  &  &  &  \\
\rowcolor{gray!15} \cellcolor{white} & \cite{31} &  &  &  &  &  &  \\
\cellcolor{white} & \cite{32} & \cmark &  &  &  &  & \cmark \\
\rowcolor{gray!15} \cellcolor{white} & \cite{33} &  &  &  &  &  &  \\
\cellcolor{white} & \cite{34} &  &  &  &  &  &  \\
\rowcolor{gray!15} \cellcolor{white} & \cite{35} &  &  &  &  &  &  \\
\cellcolor{white} & \cite{37app15169096} &  &  &  &  &  &  \\
\rowcolor{gray!15} \cellcolor{white} & \cite{38zhu2025cvebenchbenchmarkaiagents} &  &  &  &  &  &  \\
\cellcolor{white} & \cite{39} & \cmark &  &  &  &  &  \\
\rowcolor{gray!15} \cellcolor{white} & \cite{41jiao2025deepvulhunter} &  &  &  &  &  &  \\
\cellcolor{white} & \cite{42LOEVENICH2025111162} & \cmark &  &  &  & \cmark & \cmark \\
\rowcolor{gray!15} \cellcolor{white} & \cite{45} &  &  &  & \cmark &  &  \\
\cellcolor{white} & \cite{46abramovich2025enigma} &  &  &  &  &  &  \\
\rowcolor{gray!15} \cellcolor{white} & \cite{47chowdhry2025evaluating} & \cmark &  &  &  &  &  \\
\cellcolor{white} & \cite{48} &  &  &  &  &  &  \\
\rowcolor{gray!15} \cellcolor{white} & \cite{49} & \cmark &  &  &  &  &  \\
\cellcolor{white} & \cite{51} &  &  &  &  & \cmark &  \\
\rowcolor{gray!15} \cellcolor{white} & \cite{53} & \cmark &  &  &  &  &  \\
\cellcolor{white} & \cite{55} &  & \cmark &  &  &  &  \\
\rowcolor{gray!15} \cellcolor{white} & \cite{56} &  &  &  & \cmark &  &  \\
\cellcolor{white} & \cite{58} &  &  &  & \cmark &  &  \\
\rowcolor{gray!15} \cellcolor{white} & \cite{59} &  &  &  &  &  &  \\
\cellcolor{white} & \cite{60} &  &  &  &  &  &  \\
\rowcolor{gray!15} \cellcolor{white} & \cite{61} &  &  &  &  &  &  \\
\cellcolor{white} & \cite{62} &  &  &  &  &  &  \\
\rowcolor{gray!15} \cellcolor{white} & \cite{65} &  &  &  &  &  &  \\
\cellcolor{white} & \cite{67} & \cmark &  &  &  &  & \cmark \\
\rowcolor{gray!15} \cellcolor{white} & \cite{68} & \cmark &  &  &  &  & \cmark \\
\cellcolor{white} & \cite{69} &  &  &  &  &  &  \\
\rowcolor{gray!15} \cellcolor{white} & \cite{70} &  &  &  &  &  &  \\
\cellcolor{white} & \cite{71} &  &  &  &  &  &  \\
\rowcolor{gray!15} \cellcolor{white} & \cite{72} &  &  &  &  &  &  \\
\cellcolor{white} & \cite{73} &  &  &  &  &  &  \\
\rowcolor{gray!15} \cellcolor{white} & \cite{75} &  &  &  &  &  &  \\
\cellcolor{white} & \cite{78} &  &  &  &  &  &  \\
\rowcolor{gray!15} \cellcolor{white} & \cite{80} &  &  &  &  &  &  \\
\cellcolor{white} & \cite{81Cao_Huang_Li_Huilin_He_Oo_Hooi_2025} &  &  &  &  &  &  \\
\rowcolor{gray!15} \cellcolor{white} & \cite{83} &  &  &  &  &  &  \\
\cellcolor{white} & \cite{87} &  &  &  & \cmark &  &  \\
\rowcolor{gray!15} \cellcolor{white} & \cite{88} &  &  &  &  &  &  \\
\cellcolor{white} & \cite{89guo2025repoauditautonomousllmagentrepositorylevel} &  &  &  &  &  &  \\
\rowcolor{gray!15} \cellcolor{white} & \cite{90} & \cmark &  &  &  &  &  \\
\cellcolor{white} & \cite{93info16050365} &  &  &  &  &  &  \\
\rowcolor{gray!15} \cellcolor{white} & \cite{94} &  &  &  &  &  &  \\
\cellcolor{white} & \cite{99} &  & \cmark &  &  &  &  \\
\rowcolor{gray!15} \cellcolor{white} & \cite{100} &  &  &  &  &  &  \\
\hline
\cellcolor{white}2024 & \cite{1} &  &  &  &  &  &  \\
\rowcolor{gray!15} \cellcolor{white} & \cite{10debenedetti2024agentdojo} &  &  &  &  &  &  \\
\cellcolor{white} & \cite{14chen2024agentpoison} &  &  &  &  &  &  \\
\rowcolor{gray!15} \cellcolor{white} & \cite{28} &  &  &  &  &  &  \\
\cellcolor{white} & \cite{43} & \cmark &  &  &  & \cmark &  \\
\rowcolor{gray!15} \cellcolor{white} & \cite{50} &  &  &  &  &  &  \\
\cellcolor{white} & \cite{54} &  &  &  &  &  &  \\
\rowcolor{gray!15} \cellcolor{white} & \cite{63zemicheal2024llm} & \cmark & \cmark & \cmark &  &  &  \\
\cellcolor{white} & \cite{66} &  &  &  &  &  &  \\
\rowcolor{gray!15} \cellcolor{white} & \cite{74icaart24} &  &  &  &  &  &  \\
\cellcolor{white} & \cite{76} &  &  &  &  &  &  \\
\rowcolor{gray!15} \cellcolor{white} & \cite{77} &  &  &  &  &  &  \\
\cellcolor{white} & \cite{79} &  &  &  &  &  &  \\
\rowcolor{gray!15} \cellcolor{white} & \cite{82rigaki2024prompt} &  &  &  &  &  &  \\
\cellcolor{white} & \cite{85} &  &  &  &  &  &  \\
\rowcolor{gray!15} \cellcolor{white} & \cite{95} & \cmark &  &  &  &  &  \\
\cellcolor{white} & \cite{98} &  &  &  &  &  &  \\
\hline
\rowcolor{gray!15} \cellcolor{white}2023 & \cite{52} &  &  &  &  &  &  \\
\hline
    \end{tabular}%
\end{adjustbox}
  \label{tab:application-threat-intelligence-security-operations-secops}%
  \vspace{-12pt}
\end{wraptable}

\noindent
{\bf Incident investigation.}
This task concerns the reconstruction of what happened during or after a security event, including timeline recovery, evidence assembly, and causal chain analysis~\cite{17,18,27,31,42LOEVENICH2025111162,43,50,60,63zemicheal2024llm,65,79,81Cao_Huang_Li_Huilin_He_Oo_Hooi_2025,89guo2025repoauditautonomousllmagentrepositorylevel,93info16050365,94}. \textsc{Clouseau}~\cite{31} starts from a single point-of-interest and reconstructs attack chains without prior training or predefined heuristics, achieving an average F1 of 99.78\% across 21 scenarios including DARPA OpTC and ATLAS APT cases. \textsc{CFA-Bench}~\cite{27} provides a controlled benchmark for evaluating investigation capability, and the workflow of~\cite{94} 
automates first-stage triage using structured queries over Suricata logs.

\noindent
{\bf Attack attribution.}
This task concerns linking observed attack evidence to a responsible actor, campaign, or technique~\cite{20,27,29,58,63zemicheal2024llm,67,78}. \textsc{MAD-Agent}~\cite{67} produces family-attribution reports via MCP-exposed threat-intelligence tools, and \textsc{CFA-Bench}~\cite{27} benchmarks agent performance on attribution across controlled scenarios.

\noindent
{\bf Forensic analysis.}
Forensic analysis concerns the systematic collection, preservation, and interpretation of digital evidence for post-incident use~\cite{27,31,94}. \textsc{CFA-Bench}~\cite{27} provides a controlled forensic benchmark, \textsc{Clouseau}~\cite{31} performs autonomous forensic investigation from a single seed event, and the alert-investigation workflow of~\cite{94} assembles structured evidence for analyst review.

\noindent
{\bf DDoS detection \& mitigation.}
This sub-task concerns identifying and responding to volumetric or protocol-level denial-of-service attacks~\cite{16,49,92}. \textsc{LaMA-SON}~\cite{16} addresses DDoS within its self-evolving 6G defence framework, 
the zero-trust SAGIN framework of~\cite{49} integrates detection into distributed network defence, and the proactive cloud defence of~\cite{92} combines anomaly detection with automated mitigation.

\noindent
{\bf Phishing \& social-engineering detection.}
This sub-task concerns detection of attacks that manipulate users through deceptive communications or counterfeit interfaces~\cite{3,81Cao_Huang_Li_Huilin_He_Oo_Hooi_2025,96}. \textsc{PhishAgent}~\cite{81Cao_Huang_Li_Huilin_He_Oo_Hooi_2025} combines online brand records with offline cached fingerprints and a multimodal LLM to improve recall on novel brand-impersonation attacks, and \textsc{TrustNet}~\cite{96} uses a hybrid Random Forest and LLM-feature-characterisation pipeline for scam-website detection, reporting 96.94\% accuracy.

\noindent
{\bf Alert triage.}
This concerns the initial classification and prioritization of security alerts before deeper investigation~\cite{3,94}. The cross-domain multi-agent of~\cite{3} correlates email-verification, log-analysis, and IP-scanning agents through a contextual-recommendation layer, achieving 93.6\% detection accuracy, and the workflow of~\cite{94} automates first-stage triage over Suricata logs.

\noindent
{\bf Threat hunting.}
This refers to proactive, hypothesis-driven searching for threats that evade automated detection~\cite{93info16050365}. The agentic SOAR system of~\cite{93info16050365} supports proactive hunting by autonomously orchestrating investigative playbooks over collected telemetry, representing one of the few genuinely proactive defensive capabilities in the surveyed corpus.

\vspace{-4pt}
\subsubsection{Threat intelligence \& security operations (SecOps)}\label{sssec:app-secops}
Agents serving SecOps automate human-in-the-loop security workflows, including threat-intelligence ingestion, reporting, policy generation, and analyst support~\cite{3,4,5,6,12,16,18,32,39,42LOEVENICH2025111162,43,45,47chowdhry2025evaluating,49,51,53,55,56,58,63zemicheal2024llm,67,68,87,90,92,95,99}, as summarized in Table~\ref{tab:application-threat-intelligence-security-operations-secops}.

\noindent
{\bf Threat intelligence synthesis.}
Threat intelligence synthesis concerns the collection, correlation, and structuring of security information from heterogeneous sources such as advisories, dark-web material, system telemetry, and vulnerability reports~\cite{3,4,5,12,16,18,32,39,42LOEVENICH2025111162,43,47chowdhry2025evaluating,49,53,63zemicheal2024llm,67,68,90,92,95}. The cross-domain multi-agent of~\cite{3} correlates email-verification, log-analysis, and IP-scanning agents through a contextual-recommendation system, and the trustworthy multi-LLM network of~\cite{5} 
\noindent addresses collaborative reasoning under adversarial-host risks using blockchain-based audit trails and Byzantine Fault Tolerance consensus.

\noindent
{\bf CTI analysis.}
CTI analysis concerns extracting indicators of compromise, threat-actor tactics, and attack-chain structures from textual and semi-structured intelligence sources~\cite{3,4,12,18,32,42LOEVENICH2025111162,67,68}. \textsc{MAD-CTI}~\cite{68} scrapes and classifies dark-web sources---hacker forums, breach announcements, advisory posts---for structured extraction, the CTI annotation system of~\cite{18} builds attack chains via cybersecurity knowledge graphs, and Loevenich et al.~\cite{42LOEVENICH2025111162} construct knowledge graphs from network logs, CTI reports, and vulnerability frameworks to support autonomous cyber defence.

\noindent
{\bf Security policy generation.}
Security policy generation concerns translating natural-language security objectives or system configurations into enforceable rules~\cite{45,56,58,87,92}. \textsc{LLM2Policy}~\cite{45} generates zero-trust policies from Kubernetes configurations, the IaC agent of~\cite{58} issues policy-level recommendations, the autonomous cloud agent of~\cite{56} includes policy generation among its duties, and the proactive cloud defence of~\cite{92} translates objectives into firewall rules.

\noindent
{\bf Vulnerability report generation.}
Vulnerability report generation concerns producing structured, human-readable reports from detection and analysis outputs~\cite{6,55,63zemicheal2024llm,99}. The VEX-justification agent of~\cite{63zemicheal2024llm} automates CVE impact analysis in containerised environments via plan-and-execute LLM agents operating over SBOM, source, and documentation, representing one of the earliest end-to-end agentic vulnerability-management deployments.

\noindent
{\bf Security Q\&A assistant.}
Security Q\&A assistants answer analyst or learner queries by drawing on knowledge graphs, documentation, prior incident records, or security curricula~\cite{42LOEVENICH2025111162,43,51}. The system of~\cite{51} uses trace links and a multi-agent design to convert real vulnerability scan results into contextualised security learning materials, functioning as an interactive educational assistant.

\noindent
{\bf CVE / advisory drafting.}
CVE and advisory drafting concerns producing formal advisories from detection and classification outputs~\cite{6,63zemicheal2024llm}. These agents complement report generation with an output format suitable for distribution through CVE or VEX channels.

\begin{wraptable}{r}{0.45\textwidth}
  \centering
  \vspace{-2pt}
  \caption{Paper attribution for Application: other domains}
  \vspace{-10pt}
  \scriptsize
  \setlength{\tabcolsep}{1.5pt}
  \renewcommand{\arraystretch}{0.70}
\begin{adjustbox}{width=\textwidth,max totalheight=0.88\textheight,center}
    \begin{tabular}{|c|c|c|c|c|c|c|}
    \hline
\rowcolor{white} Year & Paper & \multicolumn{1}{c|}{\cellcolor{blue!5}\makecell[c]{Fuzzing\\\& test\\generation \\ (5\%)}} & \multicolumn{1}{c|}{\cellcolor{blue!5}\makecell[c]{Malware\\analysis\\\& detection \\ (2\%)}} & \multicolumn{1}{c|}{\cellcolor{blue!5}\makecell[c]{Reverse\\engineering \\ (5\%)}} & \multicolumn{1}{c|}{\cellcolor{blue!5}\makecell[c]{Access control\\\& authentication\\assessment \\ (1\%)}} & \multicolumn{1}{c|}{\cellcolor{blue!7}\makecell[c]{Model attack\\generation \&\\robustness evaluation \\ (10\%)}} \\
    \hline

\rowcolor{gray!15} \cellcolor{white}2026 & \cite{4} &  &  &  &  &  \\
\cellcolor{white} & \cite{5} &  &  &  &  &  \\
\rowcolor{gray!15} \cellcolor{white} & \cite{11} & \cmark &  &  &  &  \\
\cellcolor{white} & \cite{19} &  &  &  &  &  \\
\rowcolor{gray!15} \cellcolor{white} & \cite{23lbathaviator} &  &  &  &  &  \\
\cellcolor{white} & \cite{25zhang2026bountybench} &  &  &  &  &  \\
\rowcolor{gray!15} \cellcolor{white} & \cite{26RIGAKI2026129987} &  &  &  &  &  \\
\cellcolor{white} & \cite{36ZOU2026104305} &  &  &  &  &  \\
\rowcolor{gray!15} \cellcolor{white} & \cite{40zhuo2026cyberzero} &  &  &  &  &  \\
\cellcolor{white} & \cite{57} &  &  &  &  & \cmark \\
\rowcolor{gray!15} \cellcolor{white} & \cite{64happe2026llms} &  &  &  &  &  \\
\cellcolor{white} & \cite{84WANG2026103731} &  &  &  &  &  \\
\rowcolor{gray!15} \cellcolor{white} & \cite{86} &  &  &  &  & \cmark \\
\cellcolor{white} & \cite{91zhu-etal-2026-teams} &  &  &  &  &  \\
\rowcolor{gray!15} \cellcolor{white} & \cite{92} &  &  &  &  &  \\
\cellcolor{white} & \cite{96} &  &  &  &  &  \\
\rowcolor{gray!15} \cellcolor{white} & \cite{97} & \cmark &  & \cmark &  &  \\
\cellcolor{white} & \cite{101} &  &  &  &  & \cmark \\
\hline
\rowcolor{gray!15} \cellcolor{white}2025 & \cite{2} & \cmark &  &  &  &  \\
\cellcolor{white} & \cite{3} &  &  &  &  &  \\
\rowcolor{gray!15} \cellcolor{white} & \cite{6} &  &  &  &  &  \\
\cellcolor{white} & \cite{7} &  & \cmark &  &  & \cmark \\
\rowcolor{gray!15} \cellcolor{white} & \cite{8zhang2025agent} &  &  &  &  & \cmark \\
\cellcolor{white} & \cite{9jie2025agent4vul} &  &  &  &  &  \\
\rowcolor{gray!15} \cellcolor{white} & \cite{12} &  &  &  &  &  \\
\cellcolor{white} & \cite{13} &  &  &  &  &  \\
\rowcolor{gray!15} \cellcolor{white} & \cite{15} &  &  & \cmark &  &  \\
\cellcolor{white} & \cite{16} &  &  &  &  &  \\
\rowcolor{gray!15} \cellcolor{white} & \cite{17} &  &  &  &  &  \\
\cellcolor{white} & \cite{18} &  &  &  &  &  \\
\rowcolor{gray!15} \cellcolor{white} & \cite{20} &  &  &  &  &  \\
\cellcolor{white} & \cite{21} &  &  &  &  &  \\
\rowcolor{gray!15} \cellcolor{white} & \cite{22} &  &  &  &  &  \\
\cellcolor{white} & \cite{24yildiz-etal-2025-benchmarking} &  &  &  &  &  \\
\rowcolor{gray!15} \cellcolor{white} & \cite{27} &  &  &  &  &  \\
\cellcolor{white} & \cite{29} &  &  & \cmark &  &  \\
\rowcolor{gray!15} \cellcolor{white} & \cite{30} &  &  &  &  & \cmark \\
\cellcolor{white} & \cite{31} &  &  &  &  &  \\
\rowcolor{gray!15} \cellcolor{white} & \cite{32} &  &  &  &  &  \\
\cellcolor{white} & \cite{33} &  &  &  &  &  \\
\rowcolor{gray!15} \cellcolor{white} & \cite{34} &  &  &  &  &  \\
\cellcolor{white} & \cite{35} &  &  &  &  &  \\
\rowcolor{gray!15} \cellcolor{white} & \cite{37app15169096} &  &  &  &  &  \\
\cellcolor{white} & \cite{38zhu2025cvebenchbenchmarkaiagents} &  &  &  &  &  \\
\rowcolor{gray!15} \cellcolor{white} & \cite{39} &  &  &  &  &  \\
\cellcolor{white} & \cite{41jiao2025deepvulhunter} &  &  &  &  &  \\
\rowcolor{gray!15} \cellcolor{white} & \cite{42LOEVENICH2025111162} &  &  &  &  &  \\
\cellcolor{white} & \cite{45} &  &  &  & \cmark &  \\
\rowcolor{gray!15} \cellcolor{white} & \cite{46abramovich2025enigma} &  &  & \cmark &  &  \\
\cellcolor{white} & \cite{47chowdhry2025evaluating} &  &  &  &  &  \\
\rowcolor{gray!15} \cellcolor{white} & \cite{48} &  &  &  &  &  \\
\cellcolor{white} & \cite{49} &  &  &  &  &  \\
\rowcolor{gray!15} \cellcolor{white} & \cite{51} &  &  &  &  &  \\
\cellcolor{white} & \cite{53} &  &  &  &  &  \\
\rowcolor{gray!15} \cellcolor{white} & \cite{55} &  &  &  &  &  \\
\cellcolor{white} & \cite{56} &  &  &  &  &  \\
\rowcolor{gray!15} \cellcolor{white} & \cite{58} &  &  &  &  &  \\
\cellcolor{white} & \cite{59} &  &  &  &  &  \\
\rowcolor{gray!15} \cellcolor{white} & \cite{60} &  &  &  &  &  \\
\cellcolor{white} & \cite{61} &  &  &  &  &  \\
\rowcolor{gray!15} \cellcolor{white} & \cite{62} &  &  &  &  &  \\
\cellcolor{white} & \cite{65} &  &  & \cmark &  &  \\
\rowcolor{gray!15} \cellcolor{white} & \cite{67} &  & \cmark &  &  &  \\
\cellcolor{white} & \cite{68} &  &  &  &  &  \\
\rowcolor{gray!15} \cellcolor{white} & \cite{69} & \cmark &  &  &  & \cmark \\
\cellcolor{white} & \cite{70} &  &  &  &  &  \\
\rowcolor{gray!15} \cellcolor{white} & \cite{71} &  &  &  &  &  \\
\cellcolor{white} & \cite{72} & \cmark &  &  &  &  \\
\rowcolor{gray!15} \cellcolor{white} & \cite{73} &  &  &  &  &  \\
\cellcolor{white} & \cite{75} &  &  &  &  &  \\
\rowcolor{gray!15} \cellcolor{white} & \cite{78} &  &  &  &  &  \\
\cellcolor{white} & \cite{80} &  &  &  &  &  \\
\rowcolor{gray!15} \cellcolor{white} & \cite{81Cao_Huang_Li_Huilin_He_Oo_Hooi_2025} &  &  &  &  &  \\
\cellcolor{white} & \cite{83} &  &  &  &  &  \\
\rowcolor{gray!15} \cellcolor{white} & \cite{87} &  &  &  &  &  \\
\cellcolor{white} & \cite{88} &  &  &  &  &  \\
\rowcolor{gray!15} \cellcolor{white} & \cite{89guo2025repoauditautonomousllmagentrepositorylevel} &  &  &  &  &  \\
\cellcolor{white} & \cite{90} &  &  &  &  &  \\
\rowcolor{gray!15} \cellcolor{white} & \cite{93info16050365} &  &  &  &  &  \\
\cellcolor{white} & \cite{94} &  &  &  &  &  \\
\rowcolor{gray!15} \cellcolor{white} & \cite{99} &  &  &  &  &  \\
\cellcolor{white} & \cite{100} &  &  &  &  &  \\
\hline
\rowcolor{gray!15} \cellcolor{white}2024 & \cite{1} &  &  &  &  &  \\
\cellcolor{white} & \cite{10debenedetti2024agentdojo} &  &  &  &  & \cmark \\
\rowcolor{gray!15} \cellcolor{white} & \cite{14chen2024agentpoison} &  &  &  &  & \cmark \\
\cellcolor{white} & \cite{28} &  &  &  &  & \cmark \\
\rowcolor{gray!15} \cellcolor{white} & \cite{43} &  &  &  &  &  \\
\cellcolor{white} & \cite{50} &  &  &  &  &  \\
\rowcolor{gray!15} \cellcolor{white} & \cite{54} &  &  &  &  &  \\
\cellcolor{white} & \cite{63zemicheal2024llm} &  &  &  &  &  \\
\rowcolor{gray!15} \cellcolor{white} & \cite{66} &  &  &  &  &  \\
\cellcolor{white} & \cite{74icaart24} &  &  &  &  &  \\
\rowcolor{gray!15} \cellcolor{white} & \cite{76} &  &  &  &  &  \\
\cellcolor{white} & \cite{77} &  &  &  &  &  \\
\rowcolor{gray!15} \cellcolor{white} & \cite{79} &  &  &  &  &  \\
\cellcolor{white} & \cite{82rigaki2024prompt} &  &  &  &  &  \\
\rowcolor{gray!15} \cellcolor{white} & \cite{85} &  &  &  &  &  \\
\cellcolor{white} & \cite{95} &  &  &  &  &  \\
\rowcolor{gray!15} \cellcolor{white} & \cite{98} &  &  &  &  &  \\
\hline
\cellcolor{white}2023 & \cite{52} &  &  &  &  &  \\
\hline
    \end{tabular}%
\end{adjustbox}
  \label{tab:application-lower-frequency-domains}%
  \vspace{-12pt}
\end{wraptable} 
\vspace{-4pt}
\subsubsection{Other (Lower-frequency) security domains}\label{sssec:app-lowfreq}
The remaining application areas~\cite{2,7,8zhang2025agent,10debenedetti2024agentdojo,11,14chen2024agentpoison,15,28,29,30,45,46abramovich2025enigma,57,65,67,69,72,86,97,101} each account for a small share of the corpus and are consolidated into a single taxonomy table (Table~\ref{tab:application-lower-frequency-domains}). 
They nonetheless indicate the breadth of emerging agentic security applications beyond the dominant vulnerability-analysis and offensive-security workflows.

\noindent
{\bf Model attack generation \& robustness evaluation.}
This category concerns generating adversarial inputs against learned models and evaluating their robustness~\cite{7,8zhang2025agent,10debenedetti2024agentdojo,14chen2024agentpoison,28,30,57,69,86,101}.
\ul{Jailbreak generation} and \ul{prompt injection generation / detection} are addressed by the same four papers~\cite{8zhang2025agent,10debenedetti2024agentdojo,86,101}. \textsc{AgentDojo}~\cite{10debenedetti2024agentdojo} provides a dynamic environment for evaluating prompt-injection attacks and defences, \textsc{Agent Security Bench}~\cite{8zhang2025agent} formalises attacks and defences across 10 scenarios and 10 agents, and \textsc{RedAgent}~\cite{86} is an autonomous context-aware red-teamer that jailbreaks deployed LLM applications within five queries.
\ul{Agent safety \& guardrail evaluation} is the most represented sub-task here~\cite{8zhang2025agent,10debenedetti2024agentdojo,14chen2024agentpoison,30,69,86,101}. \textsc{ZT-ICAS}~\cite{101} proposes zero-trust defence with input neutralisation and capability-bounded tool invocation, and the deception framework of~\cite{30} exploits LLM biases and tokenisation quirks to cloak assets and lure attacking agents.
\ul{Adversarial example generation against ML/LLM components} covers attacks on non-agent learned models~\cite{7,8zhang2025agent,14chen2024agentpoison,28,57}. \textsc{AgentPoison}~\cite{14chen2024agentpoison} demonstrates that poisoning a retrieval knowledge base compromises downstream behaviour, \textsc{CheatAgent}~\cite{28} constructs black-box adversarial perturbations in user--recommender dialogues, and \textsc{AgentSA}~\cite{57} deploys LLM user-agents with coherent fake personas for training-time shilling attacks.

\noindent
{\bf Fuzzing \& test generation.}
This category concerns generating test inputs that drive programs toward unexpected or security-relevant states~\cite{2,11,45,69,72,97}. \ul{Test case generation for security properties} is the dominant sub-task~\cite{2,11,69,72,97}: \textsc{ConcoLLMic}~\cite{11} uses an LLM agent as a symboliser and natural-language constraint reasoner for concolic execution, achieving 115--233\% higher coverage than KLEE across twelve real-world subjects, and \textsc{MultiFuzz}~\cite{72} couples dense retrieval over protocol RFCs with multi-agent decomposition for network-protocol fuzzing. \ul{Smart seed mutation} appears in~\cite{69,72}, \ul{symbolic execution guidance} in \textsc{ConcoLLMic}~\cite{11}, and \ul{crash triage \& deduplication} in~\cite{45}. The absence of work on fuzzing harness generation, seed input generation, and coverage-guided orchestration indicates that agents are currently applied to the reasoning-heavy parts of fuzzing rather than its infrastructure.

\noindent
{\bf Reverse engineering.}
This category concerns recovering structure and semantics from compiled artifacts or undocumented protocols~\cite{15,29,46abramovich2025enigma,65,97}. \ul{Disassembly \& decompilation} is addressed by three papers~\cite{29,46abramovich2025enigma,65}: \textsc{ClearAgent}~\cite{29} exposes both an analyzer-friendly intermediate representation and an LLM-friendly natural-language view of a binary, letting the agent form vulnerability hypotheses linguistically and verify them by constructing concrete triggering inputs, and \textsc{EnIGMA}~\cite{46abramovich2025enigma} supports binary reverse engineering through Interactive Agent Tools in CTF settings. \ul{Protocol reverse engineering} appears in~\cite{15,97}, where \textsc{FlowFSM}~\cite{15} uses prompt chaining to extract finite-state machines from RFC documents with reduced hallucinated transitions. \ul{Variable \& type recovery} is addressed only by \textsc{ClearAgent}~\cite{29}.

\noindent
{\bf Malware analysis \& detection.}
This category concerns the analysis and classification of malicious software~\cite{7,67}. \ul{Malware detection} is addressed by both papers~\cite{7,67}, while \ul{malware classification / family attribution} and \ul{malware behavior analysis} appear only in \textsc{MAD-Agent}~\cite{67}, which orchestrates static analysis, dynamic sandbox analysis, and threat intelligence through MCP-exposed tools. \ul{Anti-malware evasion analysis} is addressed by \textsc{LAMLAD}~\cite{7}, which uses a dual-agent manipulator/analyzer loop to perturb malicious Android APKs while preserving functionality, achieving a 97\% attack success rate in an average of three queries. The absence of work on obfuscation detection and malware generation marks this as the least developed area in the corpus.

\noindent
{\bf Access control \& authentication assessment.}
This category concerns evaluating whether access-control and authentication mechanisms are correctly specified and enforced~\cite{45}. \textsc{LLM2Policy}~\cite{45} is the sole representative, contributing to both \ul{firewall / ACL policy analysis} and \ul{access control policy generation / enforcement} by extracting microservice entity information from Kubernetes manifests and generating fine-grained zero-trust access policies. The 
absence of any work on authentication or authorization testing indicates that this domain remains largely untouched by agentic approaches.

\subsection{Survey Result: Assessment}\label{ssec:resultassessment}

\subsubsection{Metric}
An assessment metric is an observable measure used to judge an agent's outcome, trajectory, safety, or comparative advantage. We organize the reported measures into four attributes: task effectiveness, agentic process/trajectory, safety/reliability, and comparative/baseline evidence.

The tables use non-exclusive coding: a paper receives every attribute directly supported by its assessment, so percentages within a table need not sum to 100\%. We audited the full text of every paper with a blank row and retained a blank only when the paper reports no measure in that table's scope; the blank therefore means \emph{unspecified for that attribute}, not that the paper lacks an assessment altogether. After this audit, six task-effectiveness rows remain blank: four safety-focused studies~\cite{86,101,8zhang2025agent,14chen2024agentpoison}, Cognitive SOC's process/report assessment~\cite{32}, and an attack-chain demonstration without a formal outcome metric~\cite{12}. Twenty-nine process rows remain blank because those papers report outcomes without a trajectory measure, 70 safety rows remain blank because the papers do not assess the agent's boundedness or adversarial reliability, and four comparative rows remain blank for the specific reasons given below.

\noindent
{\bf Task Effectiveness.}
Task effectiveness is the outcome-level attribute that measures whether an agent accomplishes its intended security task, such as finding a vulnerability, completing an exploit, producing a valid patch, classifying an alert, or preserving service availability. 
As shown in Table~\ref{tab:assessment-task-effectiveness-metrics}, the surveyed papers use ten task-metric families. Unlike process-oriented metrics, these measures judge the result rather than how the agent reasoned or used tools.

\ul{Accuracy and classification}.
Accuracy-oriented metrics are used when the agent's output can be judged against a discrete label, verdict, class, or action. This family includes top-$k$ accuracy and CWE classification for vulnerability 
\columnratio{0.28}
\globalcounter{table}
\begin{paracol}{2}
tasks, intrusion or log-detection accuracy, malware detection and family classification, web/ phishing/scam classification accuracy, property or exploitability verification accuracy, response-action accuracy, asset-search accuracy, and investigation verdict accuracy. Such metrics appear in papers on vulnerability detection, malware analysis, phishing/scam detection, incident response, cloud operations, asset discovery, command generation, forensics, CTF knowledge, and CTI extraction~\cite{96,7,13,16,17,27,31,41jiao2025deepvulhunter,45,49,56,60,67,68,69,70,81Cao_Huang_Li_Huilin_He_Oo_Hooi_2025,90,93info16050365,94,100,52,63zemicheal2024llm}. For example, TrustNet~\cite{96} assesses a hybrid machine-learning and multi-agent LLM detector for scam websites with the standard classification suite of accuracy, precision, recall, and F1, reporting that the LLM-agent team improves over the Random Forest baseline on the labeled scam-detection task. In a different setting, MAD-Agent~\cite{67} treats malware analysis as a combination of malware detection, behavior classification, family attribution, and report generation, so task effectiveness is assessed not only by whether the sample is detected but also by whether the agent assigns the right behavioral interpretation.

\switchcolumn
\begin{table}[t]
  \centering
  \vspace{2pt}
  \caption{Paper attribution for Assessment: Task effectiveness metric families}
  \vspace{-10pt}
  \scriptsize
  \setlength{\tabcolsep}{3.2pt}
  \renewcommand{\arraystretch}{0.70}
\begin{adjustbox}{max totalsize={\textwidth}{0.86\textheight},center}
    \begin{tabular}{|c|l|c|c|c|c|c|c|c|c|c|c|}
    \hline
\rowcolor{white} \makecell[c]{Year} & \makecell[c]{Paper} & \multicolumn{1}{c|}{\cellcolor{blue!15}\makecell[c]{Accuracy/\\classification \\ (23\%)}} & \multicolumn{1}{c|}{\cellcolor{blue!21}\makecell[c]{Precision/\\recall/F1 \\ (34\%)}} & \multicolumn{1}{c|}{\cellcolor{blue!9}\makecell[c]{Error\\rates \\ (12\%)}} & \multicolumn{1}{c|}{\cellcolor{blue!33}\makecell[c]{Success/\\completion \\ (51\%)}} & \multicolumn{1}{c|}{\cellcolor{blue!5}\makecell[c]{Findings/\\yield \\ (7\%)}} & \multicolumn{1}{c|}{\cellcolor{blue!5}\makecell[c]{Correctness/\\validity \\ (8\%)}} & \multicolumn{1}{c|}{\cellcolor{blue!7}\makecell[c]{Coverage \\ (10\%)}} & \multicolumn{1}{c|}{\cellcolor{blue!16}\makecell[c]{Time/\\responsiveness \\ (25\%)}} & \multicolumn{1}{c|}{\cellcolor{blue!5}\makecell[c]{Availability/\\reward \\ (7\%)}} & \multicolumn{1}{c|}{\cellcolor{blue!5}\makecell[c]{Qualitative/\\user assess. \\ (5\%)}} \\
    \hline
\cellcolor{white}2026 & \cite{4} &  & \cmark &  &  &  &  &  &  &  &  \\
\rowcolor{gray!15} \cellcolor{white} & \cite{5} &  &  &  & \cmark &  &  &  & \cmark &  &  \\
\cellcolor{white} & \cite{11} &  &  &  &  &  &  & \cmark & \cmark &  &  \\
\rowcolor{gray!15} \cellcolor{white} & \cite{19} &  & \cmark & \cmark &  &  &  &  & \cmark &  &  \\
\cellcolor{white} & \cite{23lbathaviator} &  & \cmark &  &  &  &  &  &  &  &  \\
\rowcolor{gray!15} \cellcolor{white} & \cite{25zhang2026bountybench} &  &  &  & \cmark &  & \cmark &  &  &  &  \\
\cellcolor{white} & \cite{26RIGAKI2026129987} &  &  &  & \cmark &  &  &  &  &  &  \\
\rowcolor{gray!15} \cellcolor{white} & \cite{36ZOU2026104305} &  &  &  & \cmark & \cmark &  &  &  &  &  \\
\cellcolor{white} & \cite{40zhuo2026cyberzero} &  &  &  &  & \cmark &  &  &  &  &  \\
\rowcolor{gray!15} \cellcolor{white} & \cite{57} &  &  & \cmark & \cmark &  &  &  &  &  &  \\
\cellcolor{white} & \cite{64happe2026llms} &  &  &  & \cmark &  &  &  & \cmark &  &  \\
\rowcolor{gray!15} \cellcolor{white} & \cite{84WANG2026103731} &  &  &  & \cmark &  &  &  &  &  &  \\
\cellcolor{white} & \cite{86} &  &  &  &  &  &  &  &  &  &  \\
\rowcolor{gray!15} \cellcolor{white} & \cite{91zhu-etal-2026-teams} &  &  &  & \cmark &  &  &  &  &  &  \\
\cellcolor{white} & \cite{92} &  &  &  & \cmark &  &  &  & \cmark &  &  \\
\rowcolor{gray!15} \cellcolor{white} & \cite{96} & \cmark & \cmark &  &  &  &  &  &  &  &  \\
\cellcolor{white} & \cite{97} &  &  &  &  &  & \cmark & \cmark &  &  &  \\
\rowcolor{gray!15} \cellcolor{white} & \cite{101} &  &  &  &  &  &  &  &  &  &  \\
\hline
\cellcolor{white}2025 & \cite{2} &  &  &  &  &  &  & \cmark &  &  &  \\
\rowcolor{gray!15} \cellcolor{white} & \cite{3} &  & \cmark & \cmark &  &  &  &  & \cmark &  &  \\
\cellcolor{white} & \cite{6} &  & \cmark &  &  &  &  &  & \cmark &  &  \\
\rowcolor{gray!15} \cellcolor{white} & \cite{7} & \cmark &  &  &  &  &  &  &  &  &  \\
\cellcolor{white} & \cite{8zhang2025agent} &  &  &  &  &  &  &  &  &  &  \\
\rowcolor{gray!15} \cellcolor{white} & \cite{9jie2025agent4vul} &  & \cmark &  &  &  &  &  &  &  &  \\
\cellcolor{white} & \cite{12} &  &  &  &  &  &  &  &  &  &  \\
\rowcolor{gray!15} \cellcolor{white} & \cite{13} & \cmark &  &  & \cmark &  &  &  &  &  &  \\
\cellcolor{white} & \cite{15} &  & \cmark &  &  &  &  &  &  &  &  \\
\rowcolor{gray!15} \cellcolor{white} & \cite{16} & \cmark &  &  & \cmark &  &  &  &  &  &  \\
\cellcolor{white} & \cite{17} & \cmark &  &  &  &  &  &  &  &  &  \\
\rowcolor{gray!15} \cellcolor{white} & \cite{18} &  & \cmark &  &  &  & \cmark &  &  &  &  \\
\cellcolor{white} & \cite{20} &  &  &  & \cmark &  &  &  &  &  &  \\
\rowcolor{gray!15} \cellcolor{white} & \cite{21} &  &  &  & \cmark &  &  &  &  &  &  \\
\cellcolor{white} & \cite{22} &  &  &  & \cmark &  &  &  &  &  &  \\
\rowcolor{gray!15} \cellcolor{white} & \cite{24yildiz-etal-2025-benchmarking} &  & \cmark &  &  &  &  &  &  &  &  \\
\cellcolor{white} & \cite{27} & \cmark &  &  &  &  &  &  &  &  &  \\
\rowcolor{gray!15} \cellcolor{white} & \cite{29} &  &  &  & \cmark &  &  &  &  &  &  \\
\cellcolor{white} & \cite{30} &  &  &  & \cmark &  &  &  &  &  &  \\
\rowcolor{gray!15} \cellcolor{white} & \cite{31} & \cmark &  &  &  &  &  &  &  &  &  \\
\cellcolor{white} & \cite{32} &  &  &  &  &  &  &  &  &  &  \\
\rowcolor{gray!15} \cellcolor{white} & \cite{33} &  & \cmark &  &  &  &  &  & \cmark &  &  \\
\cellcolor{white} & \cite{34} &  & \cmark &  &  &  &  &  &  &  &  \\
\rowcolor{gray!15} \cellcolor{white} & \cite{35} &  &  &  & \cmark &  &  &  &  &  &  \\
\cellcolor{white} & \cite{37app15169096} &  &  &  & \cmark &  &  &  & \cmark &  &  \\
\rowcolor{gray!15} \cellcolor{white} & \cite{38zhu2025cvebenchbenchmarkaiagents} &  &  &  & \cmark &  &  &  &  &  &  \\
\cellcolor{white} & \cite{39} &  &  &  & \cmark &  &  &  &  & \cmark &  \\
\rowcolor{gray!15} \cellcolor{white} & \cite{41jiao2025deepvulhunter} & \cmark & \cmark &  &  &  &  &  &  &  &  \\
\cellcolor{white} & \cite{42LOEVENICH2025111162} &  & \cmark &  & \cmark &  &  &  &  & \cmark &  \\
\rowcolor{gray!15} \cellcolor{white} & \cite{45} & \cmark &  & \cmark & \cmark &  & \cmark &  & \cmark &  &  \\
\cellcolor{white} & \cite{46abramovich2025enigma} &  &  &  & \cmark & \cmark &  & \cmark &  &  &  \\
\rowcolor{gray!15} \cellcolor{white} & \cite{47chowdhry2025evaluating} &  & \cmark &  & \cmark &  &  &  & \cmark & \cmark &  \\
\cellcolor{white} & \cite{48} &  &  &  & \cmark &  &  &  &  & \cmark &  \\
\rowcolor{gray!15} \cellcolor{white} & \cite{49} & \cmark &  &  & \cmark &  &  &  & \cmark &  &  \\
\cellcolor{white} & \cite{51} &  &  &  &  &  &  &  &  &  & \cmark \\
\rowcolor{gray!15} \cellcolor{white} & \cite{53} &  &  &  & \cmark &  &  &  &  & \cmark &  \\
\cellcolor{white} & \cite{55} &  &  &  & \cmark &  &  &  &  &  &  \\
\rowcolor{gray!15} \cellcolor{white} & \cite{56} & \cmark &  &  & \cmark &  &  &  & \cmark &  &  \\
\cellcolor{white} & \cite{58} &  & \cmark &  & \cmark &  &  &  & \cmark &  &  \\
\rowcolor{gray!15} \cellcolor{white} & \cite{59} &  &  &  & \cmark &  &  & \cmark &  &  &  \\
\cellcolor{white} & \cite{60} & \cmark & \cmark &  &  &  &  &  &  &  &  \\
\rowcolor{gray!15} \cellcolor{white} & \cite{61} &  &  &  & \cmark &  &  &  &  &  &  \\
\cellcolor{white} & \cite{62} &  & \cmark &  &  &  &  &  &  &  &  \\
\rowcolor{gray!15} \cellcolor{white} & \cite{65} &  &  &  & \cmark & \cmark &  & \cmark &  &  &  \\
\cellcolor{white} & \cite{67} & \cmark & \cmark &  &  &  &  &  & \cmark &  & \cmark \\
\rowcolor{gray!15} \cellcolor{white} & \cite{68} & \cmark & \cmark &  &  &  &  &  &  &  &  \\
\cellcolor{white} & \cite{69} & \cmark & \cmark &  &  &  &  & \cmark & \cmark &  &  \\
\rowcolor{gray!15} \cellcolor{white} & \cite{70} & \cmark &  &  &  & \cmark &  & \cmark &  &  &  \\
\cellcolor{white} & \cite{71} &  & \cmark &  &  &  &  &  & \cmark &  &  \\
\rowcolor{gray!15} \cellcolor{white} & \cite{72} &  &  &  &  &  & \cmark & \cmark &  &  &  \\
\cellcolor{white} & \cite{73} &  &  &  &  &  &  &  & \cmark &  &  \\
\rowcolor{gray!15} \cellcolor{white} & \cite{75} &  &  & \cmark & \cmark &  & \cmark &  & \cmark &  &  \\
\cellcolor{white} & \cite{78} &  &  &  & \cmark &  &  &  & \cmark &  &  \\
\rowcolor{gray!15} \cellcolor{white} & \cite{80} &  &  &  & \cmark &  &  &  & \cmark &  &  \\
\cellcolor{white} & \cite{81Cao_Huang_Li_Huilin_He_Oo_Hooi_2025} & \cmark &  & \cmark &  &  &  &  & \cmark &  &  \\
\rowcolor{gray!15} \cellcolor{white} & \cite{83} &  & \cmark &  &  & \cmark &  &  &  &  &  \\
\cellcolor{white} & \cite{87} &  &  &  & \cmark &  &  &  &  &  & \cmark \\
\rowcolor{gray!15} \cellcolor{white} & \cite{88} &  &  &  & \cmark &  & \cmark &  &  &  &  \\
\cellcolor{white} & \cite{89guo2025repoauditautonomousllmagentrepositorylevel} &  & \cmark &  &  &  &  &  &  &  &  \\
\rowcolor{gray!15} \cellcolor{white} & \cite{90} & \cmark & \cmark &  &  &  &  &  & \cmark &  &  \\
\cellcolor{white} & \cite{93info16050365} & \cmark &  &  &  &  &  &  & \cmark &  &  \\
\rowcolor{gray!15} \cellcolor{white} & \cite{94} & \cmark & \cmark & \cmark &  &  &  &  &  &  &  \\
\cellcolor{white} & \cite{99} &  & \cmark &  & \cmark &  & \cmark &  &  &  &  \\
\rowcolor{gray!15} \cellcolor{white} & \cite{100} & \cmark & \cmark & \cmark &  &  &  &  &  &  &  \\
\hline
\cellcolor{white}2024 & \cite{1} &  & \cmark & \cmark &  &  &  &  &  &  &  \\
\rowcolor{gray!15} \cellcolor{white} & \cite{10debenedetti2024agentdojo} &  &  &  & \cmark &  &  &  &  &  &  \\
\cellcolor{white} & \cite{14chen2024agentpoison} &  &  &  &  &  &  &  &  &  &  \\
\rowcolor{gray!15} \cellcolor{white} & \cite{28} &  &  &  & \cmark &  &  &  &  &  &  \\
\cellcolor{white} & \cite{43} &  &  &  & \cmark &  &  &  &  & \cmark &  \\
\rowcolor{gray!15} \cellcolor{white} & \cite{50} &  &  &  & \cmark &  &  &  &  &  &  \\
\cellcolor{white} & \cite{54} &  &  &  & \cmark & \cmark &  & \cmark & \cmark &  &  \\
\rowcolor{gray!15} \cellcolor{white} & \cite{63zemicheal2024llm} & \cmark & \cmark &  &  &  &  &  &  &  &  \\
\cellcolor{white} & \cite{66} &  & \cmark &  &  &  &  &  &  &  &  \\
\rowcolor{gray!15} \cellcolor{white} & \cite{74icaart24} &  &  &  & \cmark &  &  &  &  &  &  \\
\cellcolor{white} & \cite{76} &  & \cmark &  & \cmark &  &  &  &  &  &  \\
\rowcolor{gray!15} \cellcolor{white} & \cite{77} &  &  &  & \cmark &  &  &  &  &  & \cmark \\
\cellcolor{white} & \cite{79} &  &  &  & \cmark &  &  &  &  &  & \cmark \\
\rowcolor{gray!15} \cellcolor{white} & \cite{82rigaki2024prompt} &  &  &  & \cmark &  &  &  &  &  &  \\
\cellcolor{white} & \cite{85} &  &  &  & \cmark &  &  &  &  &  &  \\
\rowcolor{gray!15} \cellcolor{white} & \cite{95} &  &  & \cmark & \cmark &  &  &  &  & \cmark &  \\
\cellcolor{white} & \cite{98} &  & \cmark & \cmark &  &  &  &  &  &  &  \\
\hline
\rowcolor{gray!15} \cellcolor{white}2023 & \cite{52} & \cmark & \cmark & \cmark &  &  &  &  &  &  &  \\
\hline
    \end{tabular}
\end{adjustbox}
  \label{tab:assessment-task-effectiveness-metrics}
  \vspace{-12pt}
\end{table}

\end{paracol}

\ul{Precision, recall, and F1}.
Precision, recall, and F1 are the dominant metrics for detection-style tasks, especially vulnerability detection and classification, intrusion detection, malware or phishing detection, property generation, repository auditing, state-machine extraction, and CTI analysis~\cite{4,19,23lbathaviator,96,3,6,9jie2025agent4vul,15,18,24yildiz-etal-2025-benchmarking,33,34,41jiao2025deepvulhunter,42LOEVENICH2025111162,47chowdhry2025evaluating,58,60,62,67,68,69,71,83,89guo2025repoauditautonomousllmagentrepositorylevel,90,94,99,100,1,52,63zemicheal2024llm,66,76,98}. These metrics are useful when false alarms and missed findings have different operational costs: precision reflects whether reported findings are trustworthy, while recall or detection rate reflects whether important security issues are being missed. Agent4Vul~\cite{9jie2025agent4vul}, for instance, assesses a multi-agent smart-contract vulnerability detector on more than 40,000 contracts and reports F1 improvements over 19 baselines across four vulnerability types. PropertyGPT~\cite{83} illustrates the same family in formal-verification settings: it assesses generated smart-contract properties using precision and recall against ground-truth properties, and further reports how many known CVEs and attack incidents are detected from the generated specifications.

\ul{Error rates}.
Error-rate metrics provide the complementary view of task effectiveness by measuring how often the agent is wrong. Papers in this group report false-positive rate, false-negative rate, RMSE/MAE, or related error rates in vulnerability detection, intrusion detection, recommender-system attacks, access-policy generation, and web-security classification~\cite{19,57,3,45,75,81Cao_Huang_Li_Huilin_He_Oo_Hooi_2025,94,100,1,52,95,98}. These metrics matter because some agentic systems are intended for analyst-facing use: a high false-positive rate can overload analysts, while a high false-negative rate can make an autonomous detector unsafe to rely on. For example, vulnerability-detection papers that report false positives make clear whether an agent is merely increasing the number of warnings or actually improving the quality of surfaced findings; web and intrusion-detection papers similarly use FP/FN rates to expose the tradeoff between aggressive detection and operational noise.

\ul{Success and completion}.
Success and completion metrics are the largest task-effectiveness family. They are used when a security task has an executable or benchmark-defined endpoint, such as exploit or attack success, CTF or penetration-testing completion, patch success, test passing, remediation completion, or successful policy enforcement~\cite{5,25zhang2026bountybench,26RIGAKI2026129987,36ZOU2026104305,57,64happe2026llms,84WANG2026103731,91zhu-etal-2026-teams,92,13,16,20,21,22,29,30,35,37app15169096,38zhu2025cvebenchbenchmarkaiagents,39,42LOEVENICH2025111162,45,46abramovich2025enigma,47chowdhry2025evaluating,48,49,53,55,56,58,59,61,65,75,78,80,87,88,99,10debenedetti2024agentdojo,28,43,50,54,74icaart24,76,77,79,82rigaki2024prompt,85,95}. In offensive settings, success typically means that the agent compromises the target, captures the flag, reaches the vulnerable state, executes a proof-of-concept exploit, or achieves a defined adversarial objective. CVE-Bench~\cite{38zhu2025cvebenchbenchmarkaiagents} is a representative example: it assesses agents against real-world critical-severity CVEs in sandboxed web applications and scores whether the agent can complete exploitation against a reference target. BountyBench~\cite{25zhang2026bountybench} broadens the idea by assessing detect, exploit, and patch tasks over real bug-bounty issues, so completion is measured separately for recognizing the bug, weaponizing it, and repairing it. In repair settings, success is instead tied to whether the generated patch applies, passes tests, and fixes the vulnerability; RepairAgent~\cite{88} follows this pattern by validating autonomous patches on Defects4J bugs and reporting repaired-bug counts as the primary task outcome.

\ul{Findings and yield}.
Finding-yield metrics count the security artifacts produced by an agent: valid bugs, confirmed vulnerabilities, captured flags, CVEs, exploit chains, or bounty-relevant findings. This family appears in work on penetration testing, exploit generation, formal verification, vulnerability discovery, and security education or challenge environments~\cite{36ZOU2026104305,40zhuo2026cyberzero,46abramovich2025enigma,65,70,83,54}. These metrics are most informative when the task is exploratory and the target set is not naturally reduced to a single label. For example, PropertyGPT~\cite{83} reports not only precision and recall for generated formal properties, but also the number of known CVEs, historical attack incidents, and zero-day bugs surfaced by the agent-generated properties. This makes the assessment more security-relevant than a pure classification score, because it connects property generation to concrete vulnerability discovery.

\ul{Correctness and validity}.
Correctness and validity metrics ask whether the produced artifact is semantically acceptable, executable, or security-relevant. This family includes exploitability-verification accuracy, attack-chain correctness, policy exact match, patch correctness, test-pass or regression-preservation metrics, valid-test ratios, and other validity checks for generated outputs~\cite{25zhang2026bountybench,97,18,45,72,75,88,99}. These metrics are especially important for agentic systems that generate executable artifacts, because a plausible-looking exploit, test case, policy, or patch can still be invalid. MultiFuzz~\cite{72}, for instance, assesses generated protocol-fuzzing inputs in terms of their ability to exercise valid protocol behavior and reach meaningful states rather than merely producing syntactically varied messages. RepairAgent~\cite{88} similarly uses patch validation through compilation and tests, making the metric about whether the agent's edit actually repairs the program rather than whether the proposed fix appears reasonable in natural language.

\ul{Coverage}.
Coverage metrics measure how much of the target space the agent explores. The surveyed papers use code coverage, branch coverage, protocol-state or transition coverage, challenge-category coverage, human-expert comparison, and other exploration-oriented measures~\cite{11,97,2,46abramovich2025enigma,59,65,69,70,72,54}. This family is common in fuzzing, testing, exploit exploration, and benchmark design, where a single success outcome can hide shallow exploration. MultiFuzz~\cite{72} provides a clear example: it assesses whether a retrieval-augmented multi-agent fuzzer covers more branches and protocol states than baselines such as NSFuzz, AFLNet, and ChatAFL. In penetration-testing and challenge settings, category coverage serves a related role by showing whether the agent can handle diverse vulnerability classes rather than overfitting to one familiar exploit pattern.

\ul{Time and responsiveness}.
Time-oriented task metrics measure whether the agent completes a security task quickly enough for operational use. Papers in this family report response time, latency, mean time to respond or remediate, report-generation time, exploit time, and asset-discovery speed~\cite{5,11,19,64happe2026llms,92,3,6,33,37app15169096,45,47chowdhry2025evaluating,49,56,58,67,69,71,73,75,78,80,81Cao_Huang_Li_Huilin_He_Oo_Hooi_2025,90,93info16050365,54}. These metrics are particularly relevant for incident response, SOC automation, cloud operations, malware analysis, network monitoring, vulnerability analysis, and penetration testing, where a technically correct response may still be ineffective if it arrives too late. For example, the cloud-automation agent in~\cite{56} assesses mean time to respond and remediation latency alongside accuracy and policy-adherence metrics, reflecting the operational requirement that an autonomous cloud-security agent detect an anomalous event, select an action, and apply mitigation promptly. Similarly, MAD-Agent~\cite{67} reports the end-to-end time needed for tool use, reasoning, and report generation, while NetMoniAI~\cite{73} measures detection latency under degraded links and multi-node simulations.

\ul{Availability and reward}.
These metrics appear when security effectiveness is measured by the state of an environment rather than by a single prediction or generated artifact. They include cumulative reward, mission or service availability, 
and defender-performance scores in cyber-defense simulations and network-security games~\cite{39,42LOEVENICH2025111162,47chowdhry2025evaluating,48,53,43,95}. These metrics are useful for defensive agents because the correct action is often not simply to block every suspicious event; the agent must also preserve normal service. For example, the CAGE-2 cyber-defense study~\cite{53} assesses LLM-based defenders by balancing attacker disruption against network functionality, which makes task effectiveness a joint measure of security control and operational continuity. Other simulation-based defender studies similarly use reward signals to capture whether the agent's sequence of actions improves the environment state over time.

\ul{Qualitative and user assessment}.
Some papers assess task effectiveness with expert judgments, user ratings, report-quality assessments, or learning-gain measures~\cite{51,67,87,77,79}. These metrics are used when the output is difficult to reduce to a single executable success condition, such as a security report, educational artifact, analyst-facing explanation, or interactive penetration-testing support. For example, MAD-Agent~\cite{67} includes report generation quality as part of malware-analysis effectiveness, because a malware analyst needs a coherent account of observed behavior and threat indicators rather than only a binary malicious/benign label. PentestGPT~\cite{79} is another example: because the system is designed for interactive penetration-testing assistance, user-facing usefulness and human assessment complement task-completion metrics by capturing whether the agent's reasoning and recommendations are practically helpful.

\noindent
{\bf Agentic process / trajectory.}
Agentic process and trajectory is the attribute that assesses the path by which an agent reaches an outcome, including its steps, tool calls, hallucinations, resource use, iterations, intermediate artifacts, explanations, and economic value. 
These metrics are particularly important because many security tasks are long-horizon, tool-mediated, and partially observable. A final success or failure score does not reveal whether the agent used 
tools correctly, fabricated evidence, wasted iterations, produced a weak intermediate artifact, or generated an auditable explanation. Table~\ref{tab:assessment-agentic-process-trajectory} groups these process metrics into eight families.

\columnratio{0.5}
\begin{paracol}{2}
\ul{Steps to completion}.
Step-based metrics measure how many actions, subtasks, commands, or reasoning steps an agent needs before reaching a task endpoint. They appear in penetration testing, software testing, cyber-defense, vulnerability repair, fuzzing, and agent-security benchmarks~\cite{64happe2026llms,84WANG2026103731,86,92,8zhang2025agent,21,22,24yildiz-etal-2025-benchmarking,27,35,37app15169096,39,46abramovich2025enigma,55,61,70,88,10debenedetti2024agentdojo,74icaart24,82rigaki2024prompt,85}. These metrics are useful because two agents with the same final success rate may differ substantially in operational efficiency and risk exposure: a shorter trajectory may reduce cost and attack surface, while an excessively short trajectory may indicate shallow exploration. AgentDojo~\cite{10debenedetti2024agentdojo}, for example, assesses agents in stateful tool-calling tasks under adversarial data, where progress through the environment matters because prompt injection can alter not only the answer but also the sequence of tool calls. The privilege-escalation study of~\cite{64happe2026llms} similarly records command counts and contrasts the trajectories of autonomous agents with professional penetration testers.

\ul{Tool-call accuracy}.
Tool-call accuracy measures whether the agent invokes the right tool, target, or environment operation for the current task state. This metric is currently rare in the surveyed literature, with CVE-Bench~\cite{38zhu2025cvebenchbenchmarkaiagents} explicitly foregrounding the need to assess agents that interact with sandboxed web applications and execute attacks against concrete targets. The rarity is notable: many security agents are advertised as tool-using systems, but most assessments score only the final result. Tool-call accuracy gives a more diagnostic signal by distinguishing reasoning failure from execution failure, such as choosing an irrelevant scanner, issuing a malformed command, or interacting with the wrong component of a web application.

\ul{Hallucination rate}.
Hallucination metrics measure unsupported claims, fabricated evidence, invalid vulnerability explanations, or inconsistent security reasoning in the agent trajectory. They are used in vulnerability analysis, red teaming, SOC narrative generation, fuzzing, repository 
\switchcolumn
\begin{table}[t]
  \centering
  \vspace{2pt}
  \caption{Paper attribution for Assessment: Agentic process / trajectory metrics}
  \vspace{-10pt}
  \scriptsize
  \setlength{\tabcolsep}{1.5pt}
  \renewcommand{\arraystretch}{0.70}
\begin{adjustbox}{max totalsize={\textwidth}{0.86\textheight},center}
    \begin{tabular}{|c|c|c|c|c|c|c|c|c|c|}
    \hline
\rowcolor{white} Year & Paper & \multicolumn{1}{c|}{\cellcolor{blue!14}\makecell[c]{Steps to\\compl-\\etion \\ (21\%)}} & \multicolumn{1}{c|}{\cellcolor{blue!5}\makecell[c]{Tool call\\accuracy \\ (1\%)}} & \multicolumn{1}{c|}{\cellcolor{blue!7}\makecell[c]{Halluc-\\ination\\rate \\ (11\%)}} & \multicolumn{1}{c|}{\cellcolor{blue!20}\makecell[c]{Cost /\\token\\efficiency \\ (32\%)}} & \multicolumn{1}{c|}{\cellcolor{blue!7}\makecell[c]{Iteration\\count \\ (10\%)}} & \multicolumn{1}{c|}{\cellcolor{blue!5}\makecell[c]{Artifact\\correctness \\ (3\%)}} & \multicolumn{1}{c|}{\cellcolor{blue!7}\makecell[c]{Report /\\explanation \\ (11\%)}} & \multicolumn{1}{c|}{\cellcolor{blue!5}\makecell[c]{Economic\\ /bounty\\ value \\ (2\%)}} \\
    \hline

\cellcolor{white}2026 & \cite{4} &  &  & \cmark &  &  &  &  &  \\
\rowcolor{gray!15} \cellcolor{white} & \cite{5} &  &  & \cmark & \cmark &  &  &  &  \\
\cellcolor{white} & \cite{11} &  &  &  & \cmark &  &  &  &  \\
\rowcolor{gray!15} \cellcolor{white} & \cite{19} &  &  &  & \cmark &  &  &  &  \\
\cellcolor{white} & \cite{23lbathaviator} &  &  &  &  &  & \cmark &  &  \\
\rowcolor{gray!15} \cellcolor{white} & \cite{25zhang2026bountybench} &  &  &  &  &  &  &  & \cmark \\
\cellcolor{white} & \cite{26RIGAKI2026129987} &  &  & \cmark &  &  &  &  &  \\
\rowcolor{gray!15} \cellcolor{white} & \cite{36ZOU2026104305} &  &  &  & \cmark & \cmark &  &  &  \\
\cellcolor{white} & \cite{40zhuo2026cyberzero} &  &  &  & \cmark &  &  &  &  \\
\rowcolor{gray!15} \cellcolor{white} & \cite{57} &  &  &  &  &  &  &  &  \\
\cellcolor{white} & \cite{64happe2026llms} & \cmark &  &  & \cmark &  &  &  &  \\
\rowcolor{gray!15} \cellcolor{white} & \cite{84WANG2026103731} & \cmark &  &  &  &  &  &  &  \\
\cellcolor{white} & \cite{86} & \cmark &  &  &  &  &  &  &  \\
\rowcolor{gray!15} \cellcolor{white} & \cite{91zhu-etal-2026-teams} &  &  &  & \cmark &  &  &  &  \\
\cellcolor{white} & \cite{92} & \cmark &  &  &  &  &  &  &  \\
\rowcolor{gray!15} \cellcolor{white} & \cite{96} &  &  &  &  &  &  &  &  \\
\cellcolor{white} & \cite{97} &  &  &  &  &  &  &  &  \\
\rowcolor{gray!15} \cellcolor{white} & \cite{101} &  &  &  & \cmark &  &  &  &  \\
\hline
\cellcolor{white}2025 & \cite{2} &  &  &  &  &  &  &  &  \\
\rowcolor{gray!15} \cellcolor{white} & \cite{3} &  &  &  &  &  &  &  &  \\
\cellcolor{white} & \cite{6} &  &  &  & \cmark & \cmark &  &  &  \\
\rowcolor{gray!15} \cellcolor{white} & \cite{7} &  &  &  &  &  &  &  &  \\
\cellcolor{white} & \cite{8zhang2025agent} & \cmark &  &  &  &  &  &  &  \\
\rowcolor{gray!15} \cellcolor{white} & \cite{9jie2025agent4vul} &  &  &  &  &  &  &  &  \\
\cellcolor{white} & \cite{12} &  &  &  &  &  &  &  &  \\
\rowcolor{gray!15} \cellcolor{white} & \cite{13} &  &  & \cmark &  &  &  &  &  \\
\cellcolor{white} & \cite{15} &  &  &  &  &  & \cmark &  &  \\
\rowcolor{gray!15} \cellcolor{white} & \cite{16} &  &  &  &  &  &  &  &  \\
\cellcolor{white} & \cite{17} &  &  &  &  &  &  & \cmark &  \\
\rowcolor{gray!15} \cellcolor{white} & \cite{18} &  &  &  &  &  & \cmark &  &  \\
\cellcolor{white} & \cite{20} &  &  &  & \cmark &  &  &  &  \\
\rowcolor{gray!15} \cellcolor{white} & \cite{21} & \cmark &  &  &  & \cmark &  &  &  \\
\cellcolor{white} & \cite{22} & \cmark &  &  &  & \cmark &  &  &  \\
\rowcolor{gray!15} \cellcolor{white} & \cite{24yildiz-etal-2025-benchmarking} & \cmark &  &  &  &  &  &  &  \\
\cellcolor{white} & \cite{27} & \cmark &  &  &  & \cmark &  &  &  \\
\rowcolor{gray!15} \cellcolor{white} & \cite{29} &  &  &  &  &  &  &  &  \\
\cellcolor{white} & \cite{30} &  &  &  &  &  &  &  &  \\
\rowcolor{gray!15} \cellcolor{white} & \cite{31} &  &  &  &  &  &  & \cmark &  \\
\cellcolor{white} & \cite{32} &  &  & \cmark & \cmark &  &  & \cmark &  \\
\rowcolor{gray!15} \cellcolor{white} & \cite{33} &  &  &  & \cmark &  &  &  &  \\
\cellcolor{white} & \cite{34} &  &  &  &  &  &  &  &  \\
\rowcolor{gray!15} \cellcolor{white} & \cite{35} & \cmark &  &  &  &  &  &  &  \\
\cellcolor{white} & \cite{37app15169096} & \cmark &  &  & \cmark &  &  &  &  \\
\rowcolor{gray!15} \cellcolor{white} & \cite{38zhu2025cvebenchbenchmarkaiagents} &  & \cmark &  &  &  &  &  &  \\
\cellcolor{white} & \cite{39} & \cmark &  &  &  & \cmark &  &  &  \\
\rowcolor{gray!15} \cellcolor{white} & \cite{41jiao2025deepvulhunter} &  &  & \cmark &  &  &  &  &  \\
\cellcolor{white} & \cite{42LOEVENICH2025111162} &  &  &  &  &  &  & \cmark &  \\

\rowcolor{gray!15} \cellcolor{white} & \cite{45} &  &  &  & \cmark &  &  &  &  \\
\cellcolor{white} & \cite{46abramovich2025enigma} & \cmark &  &  & \cmark &  &  &  &  \\
\rowcolor{gray!15} \cellcolor{white} & \cite{47chowdhry2025evaluating} &  &  &  &  &  &  & \cmark &  \\
\cellcolor{white} & \cite{48} &  &  &  &  &  &  & \cmark &  \\
\rowcolor{gray!15} \cellcolor{white} & \cite{49} &  &  &  & \cmark &  &  &  &  \\
\cellcolor{white} & \cite{51} &  &  &  &  &  &  & \cmark &  \\
\rowcolor{gray!15} \cellcolor{white} & \cite{53} &  &  &  &  &  &  &  &  \\
\cellcolor{white} & \cite{55} & \cmark &  &  &  &  &  &  &  \\
\rowcolor{gray!15} \cellcolor{white} & \cite{56} &  &  &  & \cmark &  &  &  &  \\
\cellcolor{white} & \cite{58} &  &  &  & \cmark &  &  &  &  \\
\rowcolor{gray!15} \cellcolor{white} & \cite{59} &  &  &  &  &  &  &  &  \\
\cellcolor{white} & \cite{60} &  &  &  & \cmark &  &  &  &  \\
\rowcolor{gray!15} \cellcolor{white} & \cite{61} & \cmark &  &  &  &  &  &  &  \\
\cellcolor{white} & \cite{62} &  &  &  &  &  &  &  &  \\
\rowcolor{gray!15} \cellcolor{white} & \cite{65} &  &  &  &  &  &  &  &  \\
\cellcolor{white} & \cite{67} &  &  &  & \cmark &  &  &  &  \\
\rowcolor{gray!15} \cellcolor{white} & \cite{68} &  &  &  &  &  &  & \cmark &  \\
\cellcolor{white} & \cite{69} &  &  &  & \cmark &  &  &  &  \\
\rowcolor{gray!15} \cellcolor{white} & \cite{70} & \cmark &  & \cmark &  & \cmark &  &  &  \\
\cellcolor{white} & \cite{71} &  &  &  & \cmark &  &  &  &  \\
\rowcolor{gray!15} \cellcolor{white} & \cite{72} &  &  & \cmark &  &  &  &  &  \\
\cellcolor{white} & \cite{73} &  &  &  & \cmark &  &  &  &  \\
\rowcolor{gray!15} \cellcolor{white} & \cite{75} &  &  &  & \cmark &  &  &  &  \\
\cellcolor{white} & \cite{78} &  &  &  & \cmark &  &  &  &  \\
\rowcolor{gray!15} \cellcolor{white} & \cite{80} &  &  &  & \cmark & \cmark &  &  &  \\
\cellcolor{white} & \cite{81Cao_Huang_Li_Huilin_He_Oo_Hooi_2025} &  &  &  & \cmark &  &  &  &  \\
\rowcolor{gray!15} \cellcolor{white} & \cite{83} &  &  &  &  &  &  &  & \cmark \\
\cellcolor{white} & \cite{87} &  &  & \cmark &  &  &  &  &  \\
\rowcolor{gray!15} \cellcolor{white} & \cite{88} & \cmark &  &  & \cmark & \cmark &  &  &  \\
\cellcolor{white} & \cite{89guo2025repoauditautonomousllmagentrepositorylevel} &  &  & \cmark & \cmark &  &  &  &  \\
\rowcolor{gray!15} \cellcolor{white} & \cite{90} &  &  &  &  &  &  &  &  \\
\cellcolor{white} & \cite{93info16050365} &  &  &  &  &  &  & \cmark &  \\
\rowcolor{gray!15} \cellcolor{white} & \cite{94} &  &  &  &  &  &  &  &  \\
\cellcolor{white} & \cite{99} &  &  & \cmark &  &  &  & \cmark &  \\
\rowcolor{gray!15} \cellcolor{white} & \cite{100} &  &  &  &  &  &  &  &  \\
\hline
\cellcolor{white}2024 & \cite{1} &  &  &  &  &  &  &  &  \\
\rowcolor{gray!15} \cellcolor{white} & \cite{10debenedetti2024agentdojo} & \cmark &  &  &  &  &  &  &  \\
\cellcolor{white} & \cite{14chen2024agentpoison} &  &  &  &  &  &  &  &  \\
\rowcolor{gray!15} \cellcolor{white} & \cite{28} &  &  &  &  &  &  &  &  \\
\cellcolor{white} & \cite{43} &  &  &  &  &  &  &  &  \\
\rowcolor{gray!15} \cellcolor{white} & \cite{50} &  &  &  &  &  &  &  &  \\
\cellcolor{white} & \cite{54} &  &  &  & \cmark &  &  &  &  \\
\rowcolor{gray!15} \cellcolor{white} & \cite{63zemicheal2024llm} &  &  &  &  &  &  & \cmark &  \\
\cellcolor{white} & \cite{66} &  &  &  &  &  &  &  &  \\
\rowcolor{gray!15} \cellcolor{white} & \cite{74icaart24} & \cmark &  &  &  &  &  &  &  \\
\cellcolor{white} & \cite{76} &  &  &  & \cmark &  &  &  &  \\
\rowcolor{gray!15} \cellcolor{white} & \cite{77} &  &  &  &  &  &  &  &  \\
\cellcolor{white} & \cite{79} &  &  &  & \cmark &  &  &  &  \\
\rowcolor{gray!15} \cellcolor{white} & \cite{82rigaki2024prompt} & \cmark &  &  &  &  &  &  &  \\
\cellcolor{white} & \cite{85} & \cmark &  &  &  & \cmark &  &  &  \\
\rowcolor{gray!15} \cellcolor{white} & \cite{95} &  &  &  &  &  &  &  &  \\
\cellcolor{white} & \cite{98} &  &  &  &  &  &  &  &  \\
\hline
\rowcolor{gray!15} \cellcolor{white}2023 & \cite{52} &  &  &  &  &  &  &  &  \\
\hline

    \end{tabular}%
\end{adjustbox}
  \label{tab:assessment-agentic-process-trajectory}%
  \vspace{-12pt}
\end{table}%

\end{paracol}
\noindent
auditing, and repair~\cite{4,5,26RIGAKI2026129987,13,32,41jiao2025deepvulhunter,70,72,87,89guo2025repoauditautonomousllmagentrepositorylevel,99}. This family is central for security agents because a hallucinated finding can become a false alarm, an unsafe patch, or an incorrect incident narrative. Cognitive SOC~\cite{32} is a clear example: it assesses evidence-backed security narratives with evidence linkage, citation coverage, and hallucination rate, reporting a 0.1\% hallucination rate on 1,000 alerts while producing citation-backed narratives. RepoAudit~\cite{89guo2025repoauditautonomousllmagentrepositorylevel} attacks the same problem in repository-level auditing by adding a validator that checks data-flow facts and path-condition satisfiability before bug reports are accepted, thereby treating hallucination mitigation as part of the agent trajectory.

\ul{Cost and token efficiency}.
Cost and token-efficiency metrics measure the resource burden of running an agent, including token usage, wall-clock time, monetary cost, and assessment-time reductions. Papers in this family study trajectory synthesis, CTF solving, SOC automation, cloud operations, edge deployment, malware analysis, repository auditing, program repair, fuzzing, and penetration testing~\cite{5,11,19,36ZOU2026104305,40zhuo2026cyberzero,64happe2026llms,91zhu-etal-2026-teams,101,6,20,32,33,37app15169096,45,46abramovich2025enigma,49,56,58,60,67,69,71,73,75,78,80,81Cao_Huang_Li_Huilin_He_Oo_Hooi_2025,88,89guo2025repoauditautonomousllmagentrepositorylevel,54,76,79}. These metrics matter because many security workflows operate over large codebases, long logs, interactive targets, or repeated benchmark trials. CTFAgent~\cite{36ZOU2026104305}, for example, reports interaction rounds, token volumes, and monetary cost per solved challenge, while the privilege-escalation study of~\cite{64happe2026llms} reports token and dollar cost per exploited machine. RepoAudit~\cite{89guo2025repoauditautonomousllmagentrepositorylevel} gives a complementary repository-scale example, reporting an average of 0.44 hours and \$2.54 per project while auditing real-world codebases.

\ul{Iteration count}.
Iteration-count metrics capture how many revise-act-observe cycles, conversational turns, rounds, or repeated attempts the agent needs. They appear in multi-step vulnerability workflows, cyber-defense simulation, exploit or CTF tasks, program repair, and agent orchestration studies~\cite{36ZOU2026104305,6,21,22,27,39,70,80,88,85}. Iteration counts are useful when the agent's main advantage is adaptivity: the assessment should show whether repeated interaction actually helps or whether the agent is looping unproductively. RepairAgent~\cite{88} is representative because the agent must propose a patch, validate it, and use test feedback to revise the candidate; the number of iterations is therefore tied to both efficiency and the quality of feedback use. In offensive or challenge environments, the same metric distinguishes agents that converge on a viable exploit path from agents that repeatedly explore irrelevant actions.

\ul{Artifact correctness}.
Artifact-correctness metrics assess intermediate or final artifacts produced by the trajectory, such as generated tests, exploits, prompts, proofs, or analysis artifacts~\cite{23lbathaviator,15,18}. This family differs from task success because it asks whether a generated artifact is itself well-formed, executable, or semantically faithful, even before measuring the downstream security outcome. For agentic systems, this is an important diagnostic layer: an exploit attempt can fail because the target is not vulnerable, or because the generated artifact is invalid; a test can fail because it reveals a bug, or because it is malformed. Artifact-correctness metrics help separate these cases.

\ul{Report and explanation}.
Report and explanation metrics assess whether the agent's output is traceable, complete, understandable, and useful for security practitioners. These metrics appear in papers on SOC automation, incident response, vulnerability reporting, security education, cloud operations, cyber-defense analysis, and auditing~\cite{17,31,32,42LOEVENICH2025111162,47chowdhry2025evaluating,48,51,68,93info16050365,99,63zemicheal2024llm}. Cognitive SOC~\cite{32} is again illustrative because it treats evidence-backed narrative generation as the core output: the assessment measures evidence linkage and citation coverage, not only classification. In SOAR-style systems~\cite{93info16050365}, explanation quality is similarly tied to operational usefulness, because analysts need to understand why an investigation verdict or response recommendation is justified before acting on it.

\ul{Economic and bounty value}.
Economic metrics map security outcomes to monetary value, which can make benchmark progress more interpretable for real-world vulnerability workflows. They are reported by BountyBench~\cite{25zhang2026bountybench} and PropertyGPT~\cite{83}. BountyBench assesses agents over real-world bug-bounty systems and reports not only detect, exploit, and patch success rates, but also dollar-valued outcomes derived from bounty awards; PropertyGPT reports that a previously unknown bug found through its generated properties received an \$8,256 bounty. This metric is useful because two vulnerabilities with the same binary success label may differ substantially in practical impact; dollar value provides a coarse but concrete proxy for the economic importance of agent-discovered or agent-repaired vulnerabilities.

\columnratio{0.5}
\begin{paracol}{2}
\noindent
{\bf Safety \& reliability.}
Safety and reliability is the attribute that measures whether the agent remains bounded and dependable under adversarial inputs, repeated execution, prompt injection, poisoned memory, or constrained operational settings. 
This group treats the agent itself as part of the security boundary. As Table~\ref{tab:assessment-safety-reliability} shows, seven metric families cover containment, robustness, consistency, attack success, defense effectiveness, utility tradeoffs, and overhead.

\ul{Scope containment}.
Scope-containment metrics ask if the agent stays within authorized goals, tools, data access, and execution boundaries. They appear in work on jailbreak red teaming, zero-trust vulnerability scanning, executable exploit benchmarks, prompt-injection-aware agents, and constrained security workflows~\cite{86,101,8zhang2025agent,38zhu2025cvebenchbenchmarkaiagents,49}. These metrics are important because security agents often interact with code, logs, metadata, web applications, or external tools that may contain adversarial instructions. ZT-ICAS~\cite{101} is a direct example: it designs agentic vulnerability scanning around semantic input neutralization, ephemeral context sharding, and capability-bounded tool invocation, then assesses whether those constraints preserve analysis integrity when code comments or dependency metadata are adversarially manipulated. CVE-Bench~\cite{38zhu2025cvebenchbenchmarkaiagents} complements this defensive view by recording whether an exploiting agent takes actions outside the authorized target scope.

\ul{Prompt-injection robustness}.
Prompt-injection robustness measures whether the agent resists malicious instructions embedded in data, tool outputs, retrieved documents, code comments, or environment state. It is assessed in prompt-injection benchmarks, context-aware red teaming, agentic scanners, and other tool-using security systems~\cite{86,101,8zhang2025agent,69,81Cao_Huang_Li_Huilin_He_Oo_Hooi_2025,10debenedetti2024agentdojo}. AgentDojo~\cite{10debenedetti2024agentdojo} is a useful benchmark example because it assesses agents that execute tools over untrusted data with 97 realistic tasks and 629 security test cases. Its design makes robustness observable through both utility checks and attacker-goal checks over 
\switchcolumn
\begin{table}[t]
  \centering
  \vspace{4pt}
  \caption{Paper attribution for Assessment: Safety \& reliability metrics}
  \vspace{-12pt}
  \scriptsize
  \setlength{\tabcolsep}{1.5pt}
  \renewcommand{\arraystretch}{0.70}
\begin{adjustbox}{max totalsize={\textwidth}{0.88\textheight},center}
    \begin{tabular}{|c|c|c|c|c|c|c|c|c|}
    \hline
\rowcolor{white} Year & Paper & \multicolumn{1}{c|}{\cellcolor{blue!5}\makecell[c]{Scope\\containment \\ (5\%)}} & \multicolumn{1}{c|}{\cellcolor{blue!5}\makecell[c]{Prompt\\injection\\robustness \\ (6\%)}} & \multicolumn{1}{c|}{\cellcolor{blue!12}\makecell[c]{Consistency \\ (18\%)}} & \multicolumn{1}{c|}{\cellcolor{blue!7}\makecell[c]{Attack\\success rate (ASR) \\ (10\%)}} & \multicolumn{1}{c|}{\cellcolor{blue!5}\makecell[c]{Defense\\success \\ (8\%)}} & \multicolumn{1}{c|}{\cellcolor{blue!5}\makecell[c]{Utility-\\security\\tradeoff \\ (4\%)}} & \multicolumn{1}{c|}{\cellcolor{blue!5}\makecell[c]{Overhead \\ (3\%)}} \\
    \hline

\cellcolor{white}2026 & \cite{4} &  &  &  &  &  &  &  \\
\rowcolor{gray!15} \cellcolor{white} & \cite{5} &  &  &  &  &  &  &  \\
\cellcolor{white} & \cite{11} &  &  &  &  &  &  &  \\
\rowcolor{gray!15} \cellcolor{white} & \cite{19} &  &  &  &  &  &  &  \\
\cellcolor{white} & \cite{23lbathaviator} &  &  & \cmark &  &  &  &  \\
\rowcolor{gray!15} \cellcolor{white} & \cite{25zhang2026bountybench} &  &  &  &  &  &  &  \\
\cellcolor{white} & \cite{26RIGAKI2026129987} &  &  &  &  &  &  &  \\
\rowcolor{gray!15} \cellcolor{white} & \cite{36ZOU2026104305} &  &  &  &  &  &  &  \\
\cellcolor{white} & \cite{40zhuo2026cyberzero} &  &  & \cmark &  &  &  &  \\
\rowcolor{gray!15} \cellcolor{white} & \cite{57} &  &  &  &  &  &  &  \\
\cellcolor{white} & \cite{64happe2026llms} &  &  &  &  &  &  &  \\
\rowcolor{gray!15} \cellcolor{white} & \cite{84WANG2026103731} &  &  & \cmark &  &  &  &  \\
\cellcolor{white} & \cite{86} & \cmark & \cmark &  & \cmark &  &  &  \\
\rowcolor{gray!15} \cellcolor{white} & \cite{91zhu-etal-2026-teams} &  &  &  &  &  &  &  \\
\cellcolor{white} & \cite{92} &  &  &  &  &  &  &  \\
\rowcolor{gray!15} \cellcolor{white} & \cite{96} &  &  &  &  &  &  &  \\
\cellcolor{white} & \cite{97} &  &  & \cmark &  &  &  &  \\
\rowcolor{gray!15} \cellcolor{white} & \cite{101} & \cmark & \cmark &  & \cmark & \cmark & \cmark & \cmark \\
\hline
\cellcolor{white}2025 & \cite{2} &  &  &  &  &  &  &  \\
\rowcolor{gray!15} \cellcolor{white} & \cite{3} &  &  &  &  &  &  &  \\
\cellcolor{white} & \cite{6} &  &  & \cmark &  &  &  &  \\
\rowcolor{gray!15} \cellcolor{white} & \cite{7} &  &  &  & \cmark & \cmark &  &  \\
\cellcolor{white} & \cite{8zhang2025agent} & \cmark & \cmark &  & \cmark & \cmark & \cmark &  \\
\rowcolor{gray!15} \cellcolor{white} & \cite{9jie2025agent4vul} &  &  &  &  &  &  &  \\
\cellcolor{white} & \cite{12} &  &  &  &  &  &  &  \\
\rowcolor{gray!15} \cellcolor{white} & \cite{13} &  &  &  &  &  &  &  \\
\cellcolor{white} & \cite{15} &  &  &  &  &  &  &  \\
\rowcolor{gray!15} \cellcolor{white} & \cite{16} &  &  &  &  &  &  &  \\
\cellcolor{white} & \cite{17} &  &  &  &  &  &  &  \\
\rowcolor{gray!15} \cellcolor{white} & \cite{18} &  &  &  &  &  &  &  \\
\cellcolor{white} & \cite{20} &  &  &  &  &  &  &  \\
\rowcolor{gray!15} \cellcolor{white} & \cite{21} &  &  &  &  &  &  &  \\
\cellcolor{white} & \cite{22} &  &  &  &  &  &  &  \\
\rowcolor{gray!15} \cellcolor{white} & \cite{24yildiz-etal-2025-benchmarking} &  &  &  &  &  &  &  \\
\cellcolor{white} & \cite{27} &  &  & \cmark &  &  &  &  \\
\rowcolor{gray!15} \cellcolor{white} & \cite{29} &  &  &  &  &  &  &  \\
\cellcolor{white} & \cite{30} &  &  &  & \cmark & \cmark &  & \cmark \\
\rowcolor{gray!15} \cellcolor{white} & \cite{31} &  &  &  &  &  &  &  \\
\cellcolor{white} & \cite{32} &  &  &  &  &  &  &  \\
\rowcolor{gray!15} \cellcolor{white} & \cite{33} &  &  & \cmark &  &  &  &  \\
\cellcolor{white} & \cite{34} &  &  & \cmark &  &  &  &  \\
\rowcolor{gray!15} \cellcolor{white} & \cite{35} &  &  &  &  &  &  &  \\
\cellcolor{white} & \cite{37app15169096} &  &  &  &  &  &  &  \\
\rowcolor{gray!15} \cellcolor{white} & \cite{38zhu2025cvebenchbenchmarkaiagents} & \cmark &  &  &  &  &  &  \\
\cellcolor{white} & \cite{39} &  &  &  &  &  &  &  \\
\rowcolor{gray!15} \cellcolor{white} & \cite{41jiao2025deepvulhunter} &  &  &  &  &  &  &  \\
\cellcolor{white} & \cite{42LOEVENICH2025111162} &  &  &  &  &  &  &  \\

\rowcolor{gray!15} \cellcolor{white} & \cite{45} &  &  &  &  &  &  &  \\
\cellcolor{white} & \cite{46abramovich2025enigma} &  &  &  &  &  &  &  \\
\rowcolor{gray!15} \cellcolor{white} & \cite{47chowdhry2025evaluating} &  &  & \cmark &  &  &  &  \\
\cellcolor{white} & \cite{48} &  &  &  &  &  &  &  \\
\rowcolor{gray!15} \cellcolor{white} & \cite{49} & \cmark &  &  &  & \cmark &  &  \\
\cellcolor{white} & \cite{51} &  &  &  &  &  &  &  \\
\rowcolor{gray!15} \cellcolor{white} & \cite{53} &  &  & \cmark &  &  &  &  \\
\cellcolor{white} & \cite{55} &  &  &  &  &  &  &  \\
\rowcolor{gray!15} \cellcolor{white} & \cite{56} &  &  & \cmark &  &  &  &  \\
\cellcolor{white} & \cite{58} &  &  & \cmark &  &  &  &  \\
\rowcolor{gray!15} \cellcolor{white} & \cite{59} &  &  &  &  &  &  &  \\
\cellcolor{white} & \cite{60} &  &  &  &  &  &  &  \\
\rowcolor{gray!15} \cellcolor{white} & \cite{61} &  &  &  &  &  &  &  \\
\cellcolor{white} & \cite{62} &  &  &  &  &  &  &  \\
\rowcolor{gray!15} \cellcolor{white} & \cite{65} &  &  &  &  &  &  &  \\
\cellcolor{white} & \cite{67} &  &  &  &  &  &  &  \\
\rowcolor{gray!15} \cellcolor{white} & \cite{68} &  &  &  &  &  &  &  \\
\cellcolor{white} & \cite{69} &  & \cmark & \cmark & \cmark &  &  &  \\
\rowcolor{gray!15} \cellcolor{white} & \cite{70} &  &  &  &  &  &  &  \\
\cellcolor{white} & \cite{71} &  &  &  &  &  &  &  \\
\rowcolor{gray!15} \cellcolor{white} & \cite{72} &  &  &  &  &  &  &  \\
\cellcolor{white} & \cite{73} &  &  & \cmark &  &  &  &  \\
\rowcolor{gray!15} \cellcolor{white} & \cite{75} &  &  &  &  &  &  &  \\
\cellcolor{white} & \cite{78} &  &  &  &  &  &  &  \\
\rowcolor{gray!15} \cellcolor{white} & \cite{80} &  &  & \cmark &  &  &  &  \\
\cellcolor{white} & \cite{81Cao_Huang_Li_Huilin_He_Oo_Hooi_2025} &  & \cmark &  &  &  &  &  \\
\rowcolor{gray!15} \cellcolor{white} & \cite{83} &  &  &  &  &  &  &  \\
\cellcolor{white} & \cite{87} &  &  &  &  &  &  &  \\
\rowcolor{gray!15} \cellcolor{white} & \cite{88} &  &  &  &  &  &  &  \\
\cellcolor{white} & \cite{89guo2025repoauditautonomousllmagentrepositorylevel} &  &  &  &  &  &  &  \\
\rowcolor{gray!15} \cellcolor{white} & \cite{90} &  &  &  &  &  &  &  \\
\cellcolor{white} & \cite{93info16050365} &  &  &  &  &  &  &  \\
\rowcolor{gray!15} \cellcolor{white} & \cite{94} &  &  & \cmark &  &  &  &  \\
\cellcolor{white} & \cite{99} &  &  &  &  &  &  &  \\
\rowcolor{gray!15} \cellcolor{white} & \cite{100} &  &  &  &  &  &  &  \\
\hline
\cellcolor{white}2024 & \cite{1} &  &  &  &  &  &  &  \\
\rowcolor{gray!15} \cellcolor{white} & \cite{10debenedetti2024agentdojo} &  & \cmark &  & \cmark & \cmark & \cmark &  \\
\cellcolor{white} & \cite{14chen2024agentpoison} &  &  &  & \cmark & \cmark & \cmark &  \\
\rowcolor{gray!15} \cellcolor{white} & \cite{28} &  &  &  & \cmark &  &  &  \\
\cellcolor{white} & \cite{43} &  &  &  &  &  &  &  \\
\rowcolor{gray!15} \cellcolor{white} & \cite{50} &  &  &  &  &  &  &  \\
\cellcolor{white} & \cite{54} &  &  &  &  &  &  &  \\
\rowcolor{gray!15} \cellcolor{white} & \cite{63zemicheal2024llm} &  &  & \cmark &  &  &  &  \\
\cellcolor{white} & \cite{66} &  &  &  &  &  &  &  \\
\rowcolor{gray!15} \cellcolor{white} & \cite{74icaart24} &  &  &  &  &  &  &  \\
\cellcolor{white} & \cite{76} &  &  &  &  &  &  &  \\
\rowcolor{gray!15} \cellcolor{white} & \cite{77} &  &  &  &  &  &  &  \\
\cellcolor{white} & \cite{79} &  &  &  &  &  &  &  \\
\rowcolor{gray!15} \cellcolor{white} & \cite{82rigaki2024prompt} &  &  &  &  &  &  &  \\
\cellcolor{white} & \cite{85} &  &  &  &  &  &  &  \\
\rowcolor{gray!15} \cellcolor{white} & \cite{95} &  &  &  & \cmark & \cmark &  & \cmark \\
\cellcolor{white} & \cite{98} &  &  & \cmark &  &  &  &  \\
\hline
\rowcolor{gray!15} \cellcolor{white}2023 & \cite{52} &  &  &  &  &  &  &  \\

\hline
    \end{tabular}%
\end{adjustbox}
  \label{tab:assessment-safety-reliability}%
  \vspace{-16pt}
\end{table}%

\end{paracol}
\noindent
the environment state. ZT-ICAS~\cite{101} addresses the same 
risk in code-security settings, where adversarial comments and poisoned metadata can manipulate the scanner's reasoning.

\ul{Consistency}.
Consistency metrics assess whether repeated runs, alternative prompts, inference budgets, or model configurations produce stable security conclusions. Papers in this family apply repeated-run or variance analysis to vulnerability detection, cyber defense, cloud automation, remediation, network monitoring, penetration testing, and prompt-injection-robust workflows~\cite{23lbathaviator,40zhuo2026cyberzero,84WANG2026103731,97,6,27,33,34,47chowdhry2025evaluating,53,56,58,69,73,80,94,63zemicheal2024llm,98}. Consistency is a reliability concern because security users need to know whether a finding is robust or merely an artifact of stochastic generation. VulnGPT~\cite{98}, for example, repeats identical vulnerability queries, documents non-deterministic findings, and assesses whether a self-reflection step changes those conclusions. PTFusion~\cite{84WANG2026103731} instead uses a reasoning-similarity score across runs, while NetMoniAI~\cite{73} reports stable sub-five-second latency over repeated tests.

\ul{Attack success rate}.
Attack success rate (ASR) is used when the assessment is adversarial: the metric measures how often an attack against the agent, model, recommender, or agent-mediated system succeeds. This family includes jailbreaks, prompt injection, memory/RAG poisoning, adversarial training data, recommender manipulation, and agent red-teaming studies~\cite{86,101,7,8zhang2025agent,30,69,10debenedetti2024agentdojo,14chen2024agentpoison,28,95}. RedAgent~\cite{86} assesses context-aware jailbreak generation and reports that context-specific attacks substantially improve ASR against customized LLM applications. AgentPoison~\cite{14chen2024agentpoison} studies a different attack surface: it poisons long-term memory or RAG knowledge bases so triggered queries retrieve malicious demonstrations, reporting high retrieval and end-to-end attack success with very low poisoning rates. Together these papers show why ASR is essential for agentic security assessment: attacks can target the prompt, context, memory, retrieval layer, training data, recommendation dialogue, or tool-mediated action path.

\ul{Defense success}.
Defense-success metrics measure whether a mitigation reduces adversarial compromise while preserving useful behavior. They appear in papers that assess defenses against jailbreaks, prompt injection, memory poisoning, poisoned samples or training data, and unsafe agent actions~\cite{101,7,8zhang2025agent,30,49,10debenedetti2024agentdojo,14chen2024agentpoison,95}. AgentDojo~\cite{10debenedetti2024agentdojo} provides a concrete example by comparing attacks and defenses over stateful tool-calling tasks; in its initial environment, deploying existing prompt-injection defenses reduces attack success but does not eliminate the broader utility-security tension. AgentPoison~\cite{14chen2024agentpoison} assesses its attack under perplexity filtering and query rephrasing, while LLM-SA~\cite{49} measures how accurately its analysis remains under poisoning and the ACD-training study~\cite{95} defines poisoned-data and backdoor-trigger detection rates for defensive training pipelines.

\ul{Utility-security tradeoff}.
Utility-security tradeoff metrics assess whether robustness gains come at the cost of task completion or user utility. This family is currently small but important~\cite{101,8zhang2025agent,10debenedetti2024agentdojo,14chen2024agentpoison}. AgentDojo~\cite{10debenedetti2024agentdojo} explicitly motivates this tradeoff by assessing both user-task utility and attacker-goal success over the same environment state. AgentPoison~\cite{14chen2024agentpoison} measures benign accuracy alongside attack success, and ZT-ICAS~\cite{101} uses a utility-resilience test that penalizes over-defensive behavior. These are stronger assessment patterns than reporting defense success alone: an overly restrictive defense can appear secure simply because the agent stops doing useful work.

\ul{Overhead}.
Overhead metrics measure the added cost of safety mechanisms, such as latency, compute, token usage, or reduced throughput~\cite{101,30,95}. These metrics are necessary because safety constraints are unlikely to be adopted in operational security workflows if they impose prohibitive delay or cost. ZT-ICAS~\cite{101}, for example, reports that its zero-trust constraints reduce analysis compromise with manageable performance overhead, while the ACD-training study~\cite{95} explicitly includes the compute and memory burden of robustness controls. This kind of measurement is important because defended agents must still scale to realistic codebases, networks, and security pipelines.

\noindent
{\bf Comparative \& baseline.}
Comparative and baseline is the attribute that assesses an agent relative to traditional tools, prior learned methods, a non-agent LLM, humans, other agent systems, or its own ablated variants. 
These comparisons 
make raw task scores interpretable across benchmarks, vulnerability classes, and target environments. Table~\ref{tab:assessment-comparative-baseline} groups them into six families. Four blank rows remain for papers that provide no matching comparator---for example, a bench-
\columnratio{0.72}
\begin{paracol}{2}
\noindent
mark that compares attacks and defenses rather than agent approaches~\cite{8zhang2025agent}, two assessments without a conventional baseline~\cite{67,73}, and a conceptual design study~\cite{95}.

\ul{Comparison with non-AI tools}.
The most common comparison is against traditional security tools, program-analysis tools, fuzzers, scanners, policy generators, runbooks, or other non-agent baselines~\cite{4,11,23lbathaviator,64happe2026llms,84WANG2026103731,97,2,3,6,9jie2025agent4vul,12,13,15,18,20,32,33,45,55,56,58,59,61,65,69,72,81Cao_Huang_Li_Huilin_He_Oo_Hooi_2025,87,89guo2025repoauditautonomousllmagentrepositorylevel,90,93info16050365,98,100,1,50,66}. These comparisons clarify whether agentic behavior improves over established automation rather than merely producing plausible natural-language output. AutoRestTest~\cite{2} is a strong example: it compares its multi-agent REST API tester against leading black-box REST testing tools, including RESTler, EvoMaster, MoRest, and ARAT-RL, and reports improvements in coverage and fault detection. ClearAgent~\cite{29} is instead counted under agent-system comparisons because its reported baselines are CRAKEN and EnIGMA, whereas VulnGPT~\cite{98} directly compares its scorecard with conventional SAST tools.

\ul{Comparison with prior ML/DL methods}.
Some papers compare agentic systems with earlier machine-learning, deep-learning, reinforcement-learning, or learned attack approaches~\cite{5,19,26RIGAKI2026129987,40zhuo2026cyberzero,57,92,96,6,7,9jie2025agent4vul,14chen2024agentpoison,16,31,37app15169096,39,42LOEVENICH2025111162,47chowdhry2025evaluating,49,53,61,62,68,75,88,43,74icaart24,82rigaki2024prompt}. These comparisons are useful when the task was already studied before LLM agents, such as intrusion detection, cyber-defense simulation, vulnerability repair, CTF solving, or adversarial attack generation. AgentPoison~\cite{14chen2024agentpoison}, for example, compares with four learned trigger-optimization and poisoning attacks, while PatchAgent~\cite{75} compares an end-to-end repair agent with prior automated program repair methods.

\ul{Comparison with plain LLMs}.
Plain-LLM baselines hold the underlying model family close to the agent under test but remove agentic scaffolding such as iterative planning, memory, tool feedback, role decomposition, or critic stages~\cite{6,9jie2025agent4vul,24yildiz-etal-2025-benchmarking,41jiao2025deepvulhunter,59,60,99,76,98}. They provide the most direct evidence that a gain comes from the agent design rather than from access to a stronger language model. For example, the penetration-testing assessment in~\cite{59} compares its interactive architecture with a single-shot GPT-4 plan, while VulnGPT~\cite{98} contrasts one-shot ChatGPT with the same models augmented by controller-driven self-reflection.

\ul{Comparison with humans}.
Human baselines are used when the task is traditionally performed by analysts, penetration testers, educators, CTF participants, bug-bounty participants, or human system designers~\cite{25zhang2026bountybench,36ZOU2026104305,64happe2026llms,51,59,70,80,87,54,63zemicheal2024llm,77,79}. These comparisons are valuable but difficult to standardize 
because expertise varies and human solutions often include tacit strategy not captured in benchmark artifacts.
\switchcolumn
\begin{table}[t]
  \centering
  \vspace{4pt}
  \caption{Paper attribution for Assessment: Comparative \& baseline metrics}
  \vspace{-12pt}
  \scriptsize
  \setlength{\tabcolsep}{1.5pt}
  \renewcommand{\arraystretch}{0.70}
\begin{adjustbox}{max totalsize={\textwidth}{0.86\textheight},center}
    \begin{tabular}{|c|c|c|c|c|c|c|c|}
    \hline
\rowcolor{white} Year & Paper & \multicolumn{1}{c|}{\cellcolor{blue!23}\makecell[c]{vs.\\non-AI\\tools \\ (36\%)}} & \multicolumn{1}{c|}{\cellcolor{blue!17}\makecell[c]{vs.\\prior\\ML/DL \\ (27\%)}} & \multicolumn{1}{c|}{\cellcolor{blue!5}\makecell[c]{vs.\\plain\\LLMs \\ (9\%)}} & \multicolumn{1}{c|}{\cellcolor{blue!9}\makecell[c]{vs.\\humans \\ (12\%)}} & \multicolumn{1}{c|}{\cellcolor{blue!19}\makecell[c]{vs.\\agent\\systems \\ (28\%)}} & \multicolumn{1}{c|}{\cellcolor{blue!17}\makecell[c]{Ablation\\studies \\ (26\%)}} \\
    \hline

\cellcolor{white}2026 & \cite{4} & \cmark &  &  &  & \cmark & \cmark \\
\rowcolor{gray!15} \cellcolor{white} & \cite{5} &  & \cmark &  &  &  &  \\
\cellcolor{white} & \cite{11} & \cmark &  &  &  &  &  \\
\rowcolor{gray!15} \cellcolor{white} & \cite{19} &  & \cmark &  &  &  & \cmark \\
\cellcolor{white} & \cite{23lbathaviator} & \cmark &  &  &  &  &  \\
\rowcolor{gray!15} \cellcolor{white} & \cite{25zhang2026bountybench} &  &  &  & \cmark & \cmark &  \\
\cellcolor{white} & \cite{26RIGAKI2026129987} &  & \cmark &  &  &  &  \\
\rowcolor{gray!15} \cellcolor{white} & \cite{36ZOU2026104305} &  &  &  & \cmark &  &  \\
\cellcolor{white} & \cite{40zhuo2026cyberzero} &  & \cmark &  &  &  &  \\
\rowcolor{gray!15} \cellcolor{white} & \cite{57} &  & \cmark &  &  &  &  \\
\cellcolor{white} & \cite{64happe2026llms} & \cmark &  &  & \cmark &  &  \\
\rowcolor{gray!15} \cellcolor{white} & \cite{84WANG2026103731} & \cmark &  &  &  &  &  \\
\cellcolor{white} & \cite{86} &  &  &  &  & \cmark & \cmark \\
\rowcolor{gray!15} \cellcolor{white} & \cite{91zhu-etal-2026-teams} &  &  &  &  & \cmark &  \\
\cellcolor{white} & \cite{92} &  & \cmark &  &  &  &  \\
\rowcolor{gray!15} \cellcolor{white} & \cite{96} &  & \cmark &  &  &  &  \\
\cellcolor{white} & \cite{97} & \cmark &  &  &  &  &  \\
\rowcolor{gray!15} \cellcolor{white} & \cite{101} &  &  &  &  & \cmark &  \\
\hline
\cellcolor{white}2025 & \cite{2} & \cmark &  &  &  &  & \cmark \\
\rowcolor{gray!15} \cellcolor{white} & \cite{3} & \cmark &  &  &  & \cmark &  \\
\cellcolor{white} & \cite{6} & \cmark & \cmark & \cmark &  & \cmark & \cmark \\
\rowcolor{gray!15} \cellcolor{white} & \cite{7} &  & \cmark &  &  &  &  \\
\cellcolor{white} & \cite{8zhang2025agent} &  &  &  &  &  &  \\
\rowcolor{gray!15} \cellcolor{white} & \cite{9jie2025agent4vul} & \cmark & \cmark & \cmark &  &  & \cmark \\
\cellcolor{white} & \cite{12} & \cmark &  &  &  &  &  \\
\rowcolor{gray!15} \cellcolor{white} & \cite{13} & \cmark &  &  &  &  &  \\
\cellcolor{white} & \cite{15} & \cmark &  &  &  &  &  \\
\rowcolor{gray!15} \cellcolor{white} & \cite{16} &  & \cmark &  &  &  &  \\
\cellcolor{white} & \cite{17} &  &  &  &  &  & \cmark \\
\rowcolor{gray!15} \cellcolor{white} & \cite{18} & \cmark &  &  &  &  &  \\
\cellcolor{white} & \cite{20} & \cmark &  &  &  & \cmark &  \\
\rowcolor{gray!15} \cellcolor{white} & \cite{21} &  &  &  &  & \cmark &  \\
\cellcolor{white} & \cite{22} &  &  &  &  & \cmark &  \\
\rowcolor{gray!15} \cellcolor{white} & \cite{24yildiz-etal-2025-benchmarking} &  &  & \cmark &  &  &  \\
\cellcolor{white} & \cite{27} &  &  &  &  & \cmark &  \\
\rowcolor{gray!15} \cellcolor{white} & \cite{29} &  &  &  &  & \cmark &  \\
\cellcolor{white} & \cite{30} &  &  &  &  & \cmark &  \\
\rowcolor{gray!15} \cellcolor{white} & \cite{31} &  & \cmark &  &  &  &  \\
\cellcolor{white} & \cite{32} & \cmark &  &  &  & \cmark &  \\
\rowcolor{gray!15} \cellcolor{white} & \cite{33} & \cmark &  &  &  &  &  \\
\cellcolor{white} & \cite{34} &  &  &  &  &  & \cmark \\
\rowcolor{gray!15} \cellcolor{white} & \cite{35} &  &  &  &  & \cmark &  \\
\cellcolor{white} & \cite{37app15169096} &  & \cmark &  &  &  &  \\
\rowcolor{gray!15} \cellcolor{white} & \cite{38zhu2025cvebenchbenchmarkaiagents} &  &  &  &  & \cmark &  \\
\cellcolor{white} & \cite{39} &  & \cmark &  &  &  &  \\
\rowcolor{gray!15} \cellcolor{white} & \cite{41jiao2025deepvulhunter} &  &  & \cmark &  &  & \cmark \\
\cellcolor{white} & \cite{42LOEVENICH2025111162} &  & \cmark &  &  &  &  \\

\rowcolor{gray!15} \cellcolor{white} & \cite{45} & \cmark &  &  &  &  &  \\
\cellcolor{white} & \cite{46abramovich2025enigma} &  &  &  &  & \cmark &  \\
\rowcolor{gray!15} \cellcolor{white} & \cite{47chowdhry2025evaluating} &  & \cmark &  &  &  &  \\
\cellcolor{white} & \cite{48} &  &  &  &  &  & \cmark \\
\rowcolor{gray!15} \cellcolor{white} & \cite{49} &  & \cmark &  &  & \cmark &  \\
\cellcolor{white} & \cite{51} &  &  &  & \cmark &  &  \\
\rowcolor{gray!15} \cellcolor{white} & \cite{53} &  & \cmark &  &  &  &  \\
\cellcolor{white} & \cite{55} & \cmark &  &  &  &  &  \\
\rowcolor{gray!15} \cellcolor{white} & \cite{56} & \cmark &  &  &  &  &  \\
\cellcolor{white} & \cite{58} & \cmark &  &  &  &  &  \\
\rowcolor{gray!15} \cellcolor{white} & \cite{59} & \cmark &  & \cmark & \cmark &  &  \\
\cellcolor{white} & \cite{60} &  &  & \cmark &  &  &  \\
\rowcolor{gray!15} \cellcolor{white} & \cite{61} & \cmark & \cmark &  &  & \cmark &  \\
\cellcolor{white} & \cite{62} &  & \cmark &  &  &  &  \\
\rowcolor{gray!15} \cellcolor{white} & \cite{65} & \cmark &  &  &  &  &  \\
\cellcolor{white} & \cite{67} &  &  &  &  &  &  \\
\rowcolor{gray!15} \cellcolor{white} & \cite{68} &  & \cmark &  &  & \cmark & \cmark \\
\cellcolor{white} & \cite{69} & \cmark &  &  &  &  & \cmark \\
\rowcolor{gray!15} \cellcolor{white} & \cite{70} &  &  &  & \cmark &  &  \\
\cellcolor{white} & \cite{71} &  &  &  &  & \cmark & \cmark \\
\rowcolor{gray!15} \cellcolor{white} & \cite{72} & \cmark &  &  &  &  &  \\
\cellcolor{white} & \cite{73} &  &  &  &  &  &  \\
\rowcolor{gray!15} \cellcolor{white} & \cite{75} &  & \cmark &  &  &  & \cmark \\
\cellcolor{white} & \cite{78} &  &  &  &  & \cmark & \cmark \\
\rowcolor{gray!15} \cellcolor{white} & \cite{80} &  &  &  & \cmark & \cmark & \cmark \\
\cellcolor{white} & \cite{81Cao_Huang_Li_Huilin_He_Oo_Hooi_2025} & \cmark &  &  &  &  &  \\
\rowcolor{gray!15} \cellcolor{white} & \cite{83} &  &  &  &  &  & \cmark \\
\cellcolor{white} & \cite{87} & \cmark &  &  & \cmark &  &  \\
\rowcolor{gray!15} \cellcolor{white} & \cite{88} &  & \cmark &  &  &  &  \\
\cellcolor{white} & \cite{89guo2025repoauditautonomousllmagentrepositorylevel} & \cmark &  &  &  & \cmark & \cmark \\
\rowcolor{gray!15} \cellcolor{white} & \cite{90} & \cmark &  &  &  &  &  \\
\cellcolor{white} & \cite{93info16050365} & \cmark &  &  &  &  &  \\
\rowcolor{gray!15} \cellcolor{white} & \cite{94} &  &  &  &  & \cmark &  \\
\cellcolor{white} & \cite{99} &  &  & \cmark &  &  & \cmark \\
\rowcolor{gray!15} \cellcolor{white} & \cite{100} & \cmark &  &  &  &  &  \\
\hline
\cellcolor{white}2024 & \cite{1} & \cmark &  &  &  &  &  \\
\rowcolor{gray!15} \cellcolor{white} & \cite{10debenedetti2024agentdojo} &  &  &  &  &  & \cmark \\
\cellcolor{white} & \cite{14chen2024agentpoison} &  & \cmark &  &  &  & \cmark \\
\rowcolor{gray!15} \cellcolor{white} & \cite{28} &  &  &  &  &  & \cmark \\
\cellcolor{white} & \cite{43} &  & \cmark &  &  &  &  \\
\rowcolor{gray!15} \cellcolor{white} & \cite{50} & \cmark &  &  &  &  &  \\
\cellcolor{white} & \cite{54} &  &  &  & \cmark &  &  \\
\rowcolor{gray!15} \cellcolor{white} & \cite{63zemicheal2024llm} &  &  &  & \cmark &  &  \\
\cellcolor{white} & \cite{66} & \cmark &  &  &  &  &  \\
\rowcolor{gray!15} \cellcolor{white} & \cite{74icaart24} &  & \cmark &  &  &  &  \\
\cellcolor{white} & \cite{76} &  &  & \cmark &  & \cmark & \cmark \\
\rowcolor{gray!15} \cellcolor{white} & \cite{77} &  &  &  & \cmark &  &  \\
\cellcolor{white} & \cite{79} &  &  &  & \cmark & \cmark &  \\
\rowcolor{gray!15} \cellcolor{white} & \cite{82rigaki2024prompt} &  & \cmark &  &  &  &  \\
\cellcolor{white} & \cite{85} &  &  &  &  & \cmark & \cmark \\
\rowcolor{gray!15} \cellcolor{white} & \cite{95} &  &  &  &  &  &  \\
\cellcolor{white} & \cite{98} & \cmark &  & \cmark &  &  & \cmark \\
\hline
\rowcolor{gray!15} \cellcolor{white}2023 & \cite{52} &  &  &  &  &  & \cmark \\
\hline

    \end{tabular}%
\end{adjustbox}
  \label{tab:assessment-comparative-baseline}%
  \vspace{-18pt}
\end{table}%

\end{paracol}
\noindent
The privilege-escalation study of~\cite{64happe2026llms} compares success, command behavior, latency, and cost with professional penetration testers, while CTFAgent~\cite{36ZOU2026104305} and CTFKnow~\cite{70} compare to competition teams or human CTF participants.

\columnratio{0.60}
\begin{paracol}{2}

\ul{Comparison with other agent systems}.
Many assessments compare one agent design with other agent frameworks, model-backed agents, or scaffold variants~\cite{4,25zhang2026bountybench,86,91zhu-etal-2026-teams,101,3,6,20,21,22,27,29,30,32,35,38zhu2025cvebenchbenchmarkaiagents,46abramovich2025enigma,49,61,68,71,78,80,89guo2025repoauditautonomousllmagentrepositorylevel,94,76,79,85}. This comparison family is becoming more important as memory, decomposition, multi-agent orchestration, retrieval, and tool-use policies can affect outcomes as much as the underlying LLM. ClearAgent~\cite{29}, for example, compares on the same NYU CTF subset with CRAKEN and EnIGMA, while BountyBench~\cite{25zhang2026bountybench} assesses 10 agents across detect, exploit, and patch tasks under shared systems and task definitions.

\ul{Ablation studies}.
Ablation studies isolate which components of an agent contribute to performance. They appear across REST API testing, vulnerability detection, agent security, repair, formal verification, fuzzing, and penetration testing~\cite{4,19,86,2,6,9jie2025agent4vul,14chen2024agentpoison,17,34,41jiao2025deepvulhunter,48,68,69,71,75,78,80,83,89guo2025repoauditautonomousllmagentrepositorylevel,98,99,10debenedetti2024agentdojo,28,76,85,52}. AutoRestTest~\cite{2} removes its semantic dependency graph, LLM value generation, and reinforcement-learning component; AgentPoison~\cite{14chen2024agentpoison} disables each loss term; and VulnGPT~\cite{98} contrasts one-shot analysis with its self-reflection stage. Such ablations are important because the agent label can hide many independent design choices.

\subsubsection{Dataset}
An assessment dataset is the collection of subjects, environments, tasks, or observations on which an agent is tested. We characterize datasets by their origin, the security task they exercise, and cross-cutting properties such as realism, label quality, scale, modality, coverage, and accessibility. Dataset origin and security task are integral attributes, so every paper has at least one supported mapping in both tables. Within the characterisation table, however, a blank cell 
means that the paper does not provide enough evidence for that particular axis; we treat the information as \emph{unspecified} rather than infer a label, scale, modality, or access condition.

\noindent
{\bf Origin}.
Dataset origin is the attribute that records where assessed subjects, environments, alerts, programs, or tasks come from: production systems, competitions, threat feeds, curated public artifacts, or controlled/generated sources. 
Origin determines whether an assessment measures operationally realistic systems, reusable benchmark tasks, adversarially controlled settings, or synthetic artifacts whose behavior is easier to validate. Table~\ref{tab:assessment-dataset-origin} separates real-world from synthetic or controlled sources.

\switchcolumn
\begin{table}[t]
  \centering
  \vspace{4pt}
  \caption{Paper attribution for Assessment: Dataset origin}
  \vspace{-12pt}
  \scriptsize
  \setlength{\tabcolsep}{1.5pt}
  \renewcommand{\arraystretch}{0.70}
\begin{adjustbox}{max totalsize={\textwidth}{0.86\textheight},center}
    \begin{tabular}{|c|c|c|c|c|c|c|c|c|}
    \hline
\rowcolor{white} & & 
\multicolumn{4}{c|}{\makecell[c]{Real-world}} & \multicolumn{3}{c|}{\makecell[c]{Synthetic / controlled}} \\
\cline{3-6}\cline{7-9}
\rowcolor{white} 
\multicolumn{1}{|c|}{\multirow{-4}{*}[-0.2ex]{\makecell[c]{Year}}} & 
\multicolumn{1}{c|}{\multirow{-4}{*}[-0.2ex]{\makecell[c]{Paper}}} & 
\multicolumn{1}{c|}{\cellcolor{blue!17}\makecell[c]{Production\\systems \\ (28\%)}} & \multicolumn{1}{c|}{\cellcolor{blue!7}\makecell[c]{Compet-\\itions \\ (11\%)}} & \multicolumn{1}{c|}{\cellcolor{blue!11}\makecell[c]{Threat\\feeds \\ (17\%)}} & \multicolumn{1}{c|}{\cellcolor{blue!15}\makecell[c]{Curated\\public\\artifacts \\ (23\%)}} & \multicolumn{1}{c|}{\cellcolor{blue!33}\makecell[c]{Purpose\\-built\\programs \\ (50\%)}} & \multicolumn{1}{c|}{\cellcolor{blue!5}\makecell[c]{Mutated\\corpora \\ (5\%)}} & \multicolumn{1}{c|}{\cellcolor{blue!7}\makecell[c]{LLM-\\generated \\ (10\%)}} \\
    \hline

\cellcolor{white}2026 & \cite{4} &  &  & \cmark &  & \cmark &  &  \\
\rowcolor{gray!15} \cellcolor{white} & \cite{5} &  &  &  &  & \cmark &  &  \\
\cellcolor{white} & \cite{11} &  &  &  & \cmark &  &  &  \\
\rowcolor{gray!15} \cellcolor{white} & \cite{19} &  &  &  & \cmark &  &  &  \\
\cellcolor{white} & \cite{23lbathaviator} &  &  &  &  &  &  & \cmark \\
\rowcolor{gray!15} \cellcolor{white} & \cite{25zhang2026bountybench} & \cmark &  &  &  &  &  &  \\
\cellcolor{white} & \cite{26RIGAKI2026129987} &  &  &  &  & \cmark &  &  \\
\rowcolor{gray!15} \cellcolor{white} & \cite{36ZOU2026104305} &  & \cmark &  &  &  &  &  \\
\cellcolor{white} & \cite{40zhuo2026cyberzero} &  & \cmark &  &  & \cmark &  &  \\
\rowcolor{gray!15} \cellcolor{white} & \cite{57} &  &  &  &  & \cmark &  &  \\
\cellcolor{white} & \cite{64happe2026llms} &  &  &  &  & \cmark &  & \cmark \\
\rowcolor{gray!15} \cellcolor{white} & \cite{84WANG2026103731} &  &  &  &  & \cmark &  &  \\
\cellcolor{white} & \cite{86} &  &  &  & \cmark &  &  & \cmark \\
\rowcolor{gray!15} \cellcolor{white} & \cite{91zhu-etal-2026-teams} & \cmark &  &  &  &  &  &  \\
\cellcolor{white} & \cite{92} &  &  &  &  & \cmark &  &  \\
\rowcolor{gray!15} \cellcolor{white} & \cite{96} &  &  &  &  & \cmark &  &  \\
\cellcolor{white} & \cite{97} &  &  &  & \cmark &  &  & \cmark \\
\rowcolor{gray!15} \cellcolor{white} & \cite{101} &  &  &  & \cmark &  &  &  \\
\hline
\cellcolor{white}2025 & \cite{2} & \cmark &  &  & \cmark &  &  & \cmark \\
\rowcolor{gray!15} \cellcolor{white} & \cite{3} &  &  &  &  & \cmark &  & \cmark \\
\cellcolor{white} & \cite{6} & \cmark &  & \cmark &  &  &  &  \\
\rowcolor{gray!15} \cellcolor{white} & \cite{7} &  &  &  & \cmark &  & \cmark &  \\
\cellcolor{white} & \cite{8zhang2025agent} &  &  &  &  & \cmark &  &  \\
\rowcolor{gray!15} \cellcolor{white} & \cite{9jie2025agent4vul} & \cmark &  &  &  &  &  &  \\
\cellcolor{white} & \cite{12} &  &  & \cmark &  & \cmark &  &  \\
\rowcolor{gray!15} \cellcolor{white} & \cite{13} &  &  & \cmark &  &  &  &  \\
\cellcolor{white} & \cite{15} &  &  &  & \cmark &  &  & \cmark \\
\rowcolor{gray!15} \cellcolor{white} & \cite{16} &  &  &  &  & \cmark &  &  \\
\cellcolor{white} & \cite{17} &  &  &  &  & \cmark &  &  \\
\rowcolor{gray!15} \cellcolor{white} & \cite{18} &  &  & \cmark &  & \cmark &  &  \\
\cellcolor{white} & \cite{20} &  &  &  &  & \cmark &  &  \\
\rowcolor{gray!15} \cellcolor{white} & \cite{21} &  &  &  &  & \cmark &  &  \\
\cellcolor{white} & \cite{22} & \cmark &  &  &  & \cmark &  &  \\
\rowcolor{gray!15} \cellcolor{white} & \cite{24yildiz-etal-2025-benchmarking} & \cmark &  & \cmark & \cmark &  &  &  \\
\cellcolor{white} & \cite{27} &  &  & \cmark &  & \cmark &  &  \\
\rowcolor{gray!15} \cellcolor{white} & \cite{29} &  & \cmark &  & \cmark &  &  &  \\
\cellcolor{white} & \cite{30} &  & \cmark &  &  &  &  &  \\
\rowcolor{gray!15} \cellcolor{white} & \cite{31} &  &  &  &  & \cmark &  &  \\
\cellcolor{white} & \cite{32} &  &  & \cmark &  &  &  &  \\
\rowcolor{gray!15} \cellcolor{white} & \cite{33} &  &  &  & \cmark &  & \cmark &  \\
\cellcolor{white} & \cite{34} & \cmark &  &  &  &  &  & \cmark \\
\rowcolor{gray!15} \cellcolor{white} & \cite{35} & \cmark &  &  &  &  &  &  \\
\cellcolor{white} & \cite{37app15169096} & \cmark &  & \cmark &  &  &  &  \\
\rowcolor{gray!15} \cellcolor{white} & \cite{38zhu2025cvebenchbenchmarkaiagents} & \cmark &  & \cmark & \cmark &  &  &  \\
\cellcolor{white} & \cite{39} &  &  &  &  & \cmark &  &  \\
\rowcolor{gray!15} \cellcolor{white} & \cite{41jiao2025deepvulhunter} &  &  &  & \cmark &  &  &  \\
\cellcolor{white} & \cite{42LOEVENICH2025111162} &  &  & \cmark &  & \cmark &  &  \\

\rowcolor{gray!15} \cellcolor{white} & \cite{45} &  &  &  & \cmark & \cmark &  &  \\
\cellcolor{white} & \cite{46abramovich2025enigma} &  & \cmark &  &  & \cmark &  &  \\
\rowcolor{gray!15} \cellcolor{white} & \cite{47chowdhry2025evaluating} &  &  &  &  & \cmark &  &  \\
\cellcolor{white} & \cite{48} &  &  &  &  & \cmark &  &  \\
\rowcolor{gray!15} \cellcolor{white} & \cite{49} &  &  &  &  & \cmark &  &  \\
\cellcolor{white} & \cite{51} & \cmark &  &  &  &  &  &  \\
\rowcolor{gray!15} \cellcolor{white} & \cite{53} &  &  &  &  & \cmark &  &  \\
\cellcolor{white} & \cite{55} &  &  &  &  & \cmark &  &  \\
\rowcolor{gray!15} \cellcolor{white} & \cite{56} & \cmark &  &  &  & \cmark &  &  \\
\cellcolor{white} & \cite{58} &  &  &  &  & \cmark &  &  \\
\rowcolor{gray!15} \cellcolor{white} & \cite{59} &  & \cmark &  &  & \cmark &  & \cmark \\
\cellcolor{white} & \cite{60} & \cmark &  &  &  &  &  &  \\
\rowcolor{gray!15} \cellcolor{white} & \cite{61} & \cmark &  &  &  &  &  &  \\
\cellcolor{white} & \cite{62} & \cmark &  &  &  &  &  &  \\
\rowcolor{gray!15} \cellcolor{white} & \cite{65} &  & \cmark &  &  & \cmark &  &  \\
\cellcolor{white} & \cite{67} &  &  &  & \cmark &  &  &  \\
\rowcolor{gray!15} \cellcolor{white} & \cite{68} &  &  & \cmark &  &  &  &  \\
\cellcolor{white} & \cite{69} & \cmark &  & \cmark &  &  & \cmark &  \\
\rowcolor{gray!15} \cellcolor{white} & \cite{70} &  & \cmark &  & \cmark &  &  & \cmark \\
\cellcolor{white} & \cite{71} &  &  &  &  & \cmark &  &  \\
\rowcolor{gray!15} \cellcolor{white} & \cite{72} &  &  &  &  &  & \cmark &  \\
\cellcolor{white} & \cite{73} & \cmark &  &  &  & \cmark &  &  \\
\rowcolor{gray!15} \cellcolor{white} & \cite{75} & \cmark &  &  & \cmark &  &  &  \\
\cellcolor{white} & \cite{78} & \cmark & \cmark &  &  &  &  &  \\
\rowcolor{gray!15} \cellcolor{white} & \cite{80} &  &  &  &  & \cmark &  &  \\
\cellcolor{white} & \cite{81Cao_Huang_Li_Huilin_He_Oo_Hooi_2025} & \cmark &  &  &  &  &  &  \\
\rowcolor{gray!15} \cellcolor{white} & \cite{83} &  &  &  & \cmark &  &  &  \\
\cellcolor{white} & \cite{87} & \cmark &  &  &  &  &  &  \\
\rowcolor{gray!15} \cellcolor{white} & \cite{88} &  &  &  & \cmark &  &  &  \\
\cellcolor{white} & \cite{89guo2025repoauditautonomousllmagentrepositorylevel} &  &  &  & \cmark &  &  &  \\
\rowcolor{gray!15} \cellcolor{white} & \cite{90} &  &  & \cmark &  &  &  &  \\
\cellcolor{white} & \cite{93info16050365} &  &  &  &  & \cmark &  &  \\
\rowcolor{gray!15} \cellcolor{white} & \cite{94} & \cmark &  &  &  &  &  &  \\
\cellcolor{white} & \cite{99} &  &  &  & \cmark &  &  &  \\
\rowcolor{gray!15} \cellcolor{white} & \cite{100} &  &  & \cmark &  & \cmark &  &  \\
\hline
\cellcolor{white}2024 & \cite{1} &  &  & \cmark &  &  &  &  \\
\rowcolor{gray!15} \cellcolor{white} & \cite{10debenedetti2024agentdojo} &  &  &  &  & \cmark &  &  \\
\cellcolor{white} & \cite{14chen2024agentpoison} &  &  &  & \cmark &  & \cmark &  \\
\rowcolor{gray!15} \cellcolor{white} & \cite{28} & \cmark &  &  &  &  &  &  \\
\cellcolor{white} & \cite{43} &  &  &  &  & \cmark &  &  \\
\rowcolor{gray!15} \cellcolor{white} & \cite{50} &  &  &  &  & \cmark &  &  \\
\cellcolor{white} & \cite{54} &  & \cmark &  &  &  &  &  \\
\rowcolor{gray!15} \cellcolor{white} & \cite{63zemicheal2024llm} & \cmark &  & \cmark &  & \cmark &  &  \\
\cellcolor{white} & \cite{66} & \cmark &  &  &  &  &  &  \\
\rowcolor{gray!15} \cellcolor{white} & \cite{74icaart24} &  &  &  &  & \cmark &  &  \\
\cellcolor{white} & \cite{76} &  &  &  &  & \cmark &  &  \\
\rowcolor{gray!15} \cellcolor{white} & \cite{77} &  &  &  &  & \cmark &  &  \\
\cellcolor{white} & \cite{79} & \cmark & \cmark &  &  & \cmark &  &  \\
\rowcolor{gray!15} \cellcolor{white} & \cite{82rigaki2024prompt} &  &  &  &  & \cmark &  &  \\
\cellcolor{white} & \cite{85} &  &  &  &  & \cmark &  &  \\
\rowcolor{gray!15} \cellcolor{white} & \cite{95} &  &  &  &  & \cmark &  &  \\
\cellcolor{white} & \cite{98} &  &  &  & \cmark & \cmark &  &  \\
\hline
\rowcolor{gray!15} \cellcolor{white}2023 & \cite{52} & \cmark &  &  &  &  &  &  \\

\hline
    \end{tabular}%
\end{adjustbox}
  \label{tab:assessment-dataset-origin}%
  \vspace{-18pt}
\end{table}%

\end{paracol}

\ul{Real-world sources}.
Production systems are used when the assessment is built from deployed services, realistic repositories, cloud environments, bug-bounty systems, web applications, penetration-testing targets, or other systems with operational structure~\cite{25zhang2026bountybench,91zhu-etal-2026-teams,2,6,9jie2025agent4vul,22,24yildiz-etal-2025-benchmarking,34,35,37app15169096,38zhu2025cvebenchbenchmarkaiagents,51,56,60,61,62,69,73,75,78,81Cao_Huang_Li_Huilin_He_Oo_Hooi_2025,87,94,28,63zemicheal2024llm,66,79,52}. These datasets are valuable because agent behavior depends heavily on real system structure: repository layout, service state, execution dependencies, and operational constraints can all affect the agent's trajectory. BountyBench~\cite{25zhang2026bountybench} is representative because it assesses agents on 25 real-world systems with bug-bounty tasks that cover detection, exploitation, and patching. CVE-Bench~\cite{38zhu2025cvebenchbenchmarkaiagents} similarly grounds exploitation assessment in vulnerable real-world web applications reproduced inside sandboxes.

Competition-derived datasets are common in offensive-security work, especially CTF and cyber-range settings~\cite{29,36ZOU2026104305,40zhuo2026cyberzero,30,46abramovich2025enigma,59,65,70,78,54,79}. These datasets offer well-defined success conditions, reusable tasks, and rich community writeups, but they may also reward puzzle-solving behavior that differs from real production intrusion. ClearAgent~\cite{29}, for example, is assessed on binary challenges from NYU CTF Bench, while Cyber-Zero~\cite{40zhuo2026cyberzero} leverages public CTF writeups to synthesize long-horizon agent trajectories when the original runtime environments are unavailable.

Threat feeds and security databases provide another real-world source, including CVE/NVD data, vulnerability advisories and proof-of-concept feeds, attack frameworks, threat-intelligence resources, and incident-related feeds~\cite{4,6,12,13,18,24yildiz-etal-2025-benchmarking,27,32,37app15169096,38zhu2025cvebenchbenchmarkaiagents,42LOEVENICH2025111162,68,69,90,100,1,63zemicheal2024llm}. These sources are useful for vulnerability classification, exploitability assessment, SOC enrichment, and asset-discovery tasks because they connect agent outputs to recognized security knowledge. For example, Cognitive SOC~\cite{32} enriches alerts with threat-intelligence sources such as MITRE ATT\&CK, Sigma rules, and CVE databases, making dataset origin part of the evidence chain for generated incident narratives.

Curated public artifacts include open-source repositories, malware corpora, standards, benchmark applications, human-written properties, and other reusable public collections~\cite{11,19,86,97,101,2,7,15,24yildiz-etal-2025-benchmarking,29,33,38zhu2025cvebenchbenchmarkaiagents,41jiao2025deepvulhunter,45,67,70,75,83,88,89guo2025repoauditautonomousllmagentrepositorylevel,98,99,14chen2024agentpoison}. This broader label avoids treating public malware, RFCs, CTF writeups, prompts, or smart-contract properties as software repositories. These datasets balance realism and repeatability by retaining real artifacts while allowing researchers to package, label, and rerun assessments.

\ul{Synthetic and controlled sources}.
Purpose-built programs and environments are used when researchers need tight control over vulnerabilities, attack paths, services, state transitions, or agent-security tasks~\cite{4,5,26RIGAKI2026129987,40zhuo2026cyberzero,57,64happe2026llms,84WANG2026103731,92,96,3,8zhang2025agent,12,16,17,18,20,21,22,27,31,39,42LOEVENICH2025111162,45,46abramovich2025enigma,47chowdhry2025evaluating,48,49,53,55,56,58,59,65,71,73,80,93info16050365,100,10debenedetti2024agentdojo,43,50,63zemicheal2024llm,74icaart24,76,77,79,82rigaki2024prompt,85,95,98}. This is common in cyber-defense simulation, penetration-testing benchmarks, adversarial-agent assessment, and controlled vulnerability tasks. The advantage is measurability: the assessor can know the intended flag, exploit condition, network state, or successful defense outcome. The limitation is that the resulting task may underrepresent the messiness of production systems.

Mutated corpora and generated variants are less common but important for stress-testing coverage and robustness. Mutated corpora appear when studies adversarially alter applications, contracts, inputs, or retrieved content~\cite{7,33,69,72,14chen2024agentpoison}. LLM-generated datasets are used when a task requires scalable prompts, examples, protocol models, negative samples, test inputs, or challenge-like artifacts~\cite{23lbathaviator,64happe2026llms,86,97,2,3,15,34,59,70}. These are multi-label origins: e.g., CTFKnow~\cite{70} generates questions from public competition writeups and then filters and manually verifies them.

\noindent
{\bf Dataset by security task.}
Dataset by security task is the attribute that records the security capability an assessment exercises, such as auditing, penetration testing, smart-contract analysis, repair, defense, fuzzing, agent security, malware analysis, or SOC operations.
\begin{table*}[t]
  \centering
  \vspace{-10pt}
  \caption{Paper attribution for Assessment: Dataset by security task}
  \vspace{-10pt}
  \scriptsize
  \setlength{\tabcolsep}{1.5pt}
  \renewcommand{\arraystretch}{0.70}
\begin{adjustbox}{max totalsize={\textwidth}{0.88\textheight},center}
    \begin{tabular}{|c|c|c|c|c|c|c|c|c|c|c|c|c|c|c|c|c|c|c|c|c|c|c|}
    \hline
\rowcolor{white} \multirow{2}[1]{*}{Year} & \multirow{2}[1]{*}{Paper}
  & \multicolumn{5}{c|}{\makecell[c]{Vuln. detection \&\\classification}}
  & \multicolumn{6}{c|}{\makecell[c]{Pentest \& CTF}}
  & \multicolumn{1}{c|}{\cellcolor{blue!7}}
  & \multicolumn{1}{c|}{\cellcolor{blue!5}}
  & \multicolumn{1}{c|}{\cellcolor{blue!5}}
  & \multicolumn{1}{c|}{\cellcolor{blue!11}}
  & \multicolumn{1}{c|}{\cellcolor{blue!5}}
  & \multicolumn{1}{c|}{\cellcolor{blue!7}}
  & \multicolumn{1}{c|}{\cellcolor{blue!5}}
  & \multicolumn{1}{c|}{\cellcolor{blue!5}}
  & \multicolumn{1}{c|}{\cellcolor{blue!7}}
  & \multicolumn{1}{c|}{\cellcolor{blue!5}} \\
\cline{3-7}\cline{8-13}
\rowcolor{white} &
  & \multicolumn{1}{c|}{\cellcolor{blue!5}\makecell[c]{BigVul \\ (2\%)}}
  & \multicolumn{1}{c|}{\cellcolor{blue!5}\makecell[c]{ReVeal \\ (1\%)}}
  & \multicolumn{1}{c|}{\cellcolor{blue!5}\makecell[c]{PrimeVul \\ (1\%)}}
  & \multicolumn{1}{c|}{\cellcolor{blue!9}\makecell[c]{NVD /\\CVE \\database \\ (12\%)}}
  & \multicolumn{1}{c|}{\cellcolor{blue!5}\makecell[c]{Other vuln.\\ /audit\\corpora \\ (6\%)}}
  & \multicolumn{1}{c|}{\cellcolor{blue!5}\makecell[c]{InterCode\\-CTF \\ (3\%)}}
  & \multicolumn{1}{c|}{\cellcolor{blue!5}\makecell[c]{PicoCTF \\ (2\%)}}
  & \multicolumn{1}{c|}{\cellcolor{blue!5}\makecell[c]{HackTheBox \\ (5\%)}}
  & \multicolumn{1}{c|}{\cellcolor{blue!5}\makecell[c]{VulnHub \\ (6\%)}}
  & \multicolumn{1}{c|}{\cellcolor{blue!5}\makecell[c]{AutoAttacker\\benchmark \\ (6\%)}}
  & \multicolumn{1}{c|}{\cellcolor{blue!9}\makecell[c]{Other\\pentest /\\CTF \\ (13\%)}}
  & \multicolumn{1}{c|}{\cellcolor{blue!7}\multirow{-2}{*}[8pt]{\makecell[c]{Smart\\contracts \\ (8\%)}}}
  & \multicolumn{1}{c|}{\cellcolor{blue!5}\multirow{-2}{*}[8pt]{\makecell[c]{Web\\/phishing\\/scam \\ (2\%)}}}
  & \multicolumn{1}{c|}{\cellcolor{blue!5}\multirow{-2}{*}[8pt]{\makecell[c]{Repair\\/ bug-fix \\ (4\%)}}}
  & \multicolumn{1}{c|}{\cellcolor{blue!11}\multirow{-2}{*}[8pt]{\makecell[c]{Cyber\\defense /\\simulation \\ (16\%)}}}
  & \multicolumn{1}{c|}{\cellcolor{blue!5}\multirow{-2}{*}[8pt]{\makecell[c]{Security\\testing /\\fuzzing \\ (5\%)}}}
  & \multicolumn{1}{c|}{\cellcolor{blue!7}\multirow{-2}{*}[8pt]{\makecell[c]{AI /\\agent\\security \\ (9\%)}}}
  & \multicolumn{1}{c|}{\cellcolor{blue!5}\multirow{-2}{*}[8pt]{\makecell[c]{Asset\\discovery /\\OSINT \\ (1\%)}}}
  & \multicolumn{1}{c|}{\cellcolor{blue!5}\multirow{-2}{*}[8pt]{\makecell[c]{Malware\\/ Android\\apps \\ (2\%)}}}
  & \multicolumn{1}{c|}{\cellcolor{blue!7}\multirow{-2}{*}[8pt]{\makecell[c]{Intrusion /\\SOC /\\ SOAR \\ (12\%)}}}
  & \multicolumn{1}{c|}{\cellcolor{blue!5}\multirow{-2}{*}[8pt]{\makecell[c]{Other\\security\\task \\ (2\%)}}} \\
    \hline

\cellcolor{white}2026 & \cite{4} &  &  &  & \cmark &  &  &  &  &  &  &  &  &  &  &  &  &  &  &  &  &  \\
\rowcolor{gray!15} \cellcolor{white} & \cite{5} &  &  &  &  &  &  &  &  &  &  &  &  &  &  & \cmark &  &  &  &  &  &  \\
\cellcolor{white} & \cite{11} &  &  &  &  &  &  &  &  &  &  &  &  &  &  &  & \cmark &  &  &  &  &  \\
\rowcolor{gray!15} \cellcolor{white} & \cite{19} &  &  &  &  &  &  &  &  &  &  &  &  &  &  &  &  &  &  &  & \cmark &  \\
\cellcolor{white} & \cite{23lbathaviator} & \cmark & \cmark & \cmark & \cmark &  &  &  &  &  &  &  &  &  &  &  &  &  &  &  &  &  \\
\rowcolor{gray!15} \cellcolor{white} & \cite{25zhang2026bountybench} &  &  &  &  & \cmark &  &  &  &  &  & \cmark &  &  & \cmark &  &  &  &  &  &  &  \\
\cellcolor{white} & \cite{26RIGAKI2026129987} &  &  &  &  &  &  &  &  &  &  &  &  &  &  & \cmark &  &  &  &  &  &  \\
\rowcolor{gray!15} \cellcolor{white} & \cite{36ZOU2026104305} &  &  &  &  &  &  & \cmark &  &  &  &  &  &  &  &  &  &  &  &  &  &  \\
\cellcolor{white} & \cite{40zhuo2026cyberzero} &  &  &  &  &  & \cmark &  &  &  &  &  &  &  &  &  &  &  &  &  &  &  \\
\rowcolor{gray!15} \cellcolor{white} & \cite{57} &  &  &  &  &  &  &  &  &  &  &  &  &  &  &  &  & \cmark &  &  &  &  \\
\cellcolor{white} & \cite{64happe2026llms} &  &  &  &  &  &  &  &  &  &  & \cmark &  &  &  &  &  &  &  &  &  &  \\
\rowcolor{gray!15} \cellcolor{white} & \cite{84WANG2026103731} &  &  &  &  &  &  &  &  & \cmark &  &  &  &  &  &  &  &  &  &  &  &  \\
\cellcolor{white} & \cite{86} &  &  &  &  &  &  &  &  &  &  &  &  &  &  &  &  & \cmark &  &  &  &  \\
\rowcolor{gray!15} \cellcolor{white} & \cite{91zhu-etal-2026-teams} &  &  &  & \cmark &  &  &  &  &  &  & \cmark &  &  &  &  &  &  &  &  &  &  \\
\cellcolor{white} & \cite{92} &  &  &  &  &  &  &  &  &  &  &  &  &  &  & \cmark &  &  &  &  &  &  \\
\rowcolor{gray!15} \cellcolor{white} & \cite{96} &  &  &  &  &  &  &  &  &  &  &  &  & \cmark &  &  &  &  &  &  &  &  \\
\cellcolor{white} & \cite{97} &  &  &  &  &  &  &  &  &  &  &  &  &  &  &  & \cmark &  &  &  &  &  \\
\rowcolor{gray!15} \cellcolor{white} & \cite{101} &  &  &  &  &  &  &  &  &  &  &  &  &  &  &  &  & \cmark &  &  &  &  \\
\hline
\cellcolor{white}2025 & \cite{2} &  &  &  &  &  &  &  &  &  &  &  &  &  &  &  & \cmark &  &  &  &  &  \\
\rowcolor{gray!15} \cellcolor{white} & \cite{3} &  &  &  & \cmark &  &  &  &  &  &  &  &  &  &  &  &  &  &  &  &  &  \\
\cellcolor{white} & \cite{6} &  &  &  & \cmark &  &  &  &  &  &  &  & \cmark &  &  &  &  &  &  &  &  &  \\
\rowcolor{gray!15} \cellcolor{white} & \cite{7} &  &  &  &  &  &  &  &  &  &  &  &  &  &  &  &  &  &  & \cmark &  &  \\
\cellcolor{white} & \cite{8zhang2025agent} &  &  &  &  &  &  &  &  &  &  &  &  &  &  &  &  & \cmark &  &  &  &  \\
\rowcolor{gray!15} \cellcolor{white} & \cite{9jie2025agent4vul} &  &  &  &  &  &  &  &  &  &  &  & \cmark &  &  &  &  &  &  &  &  &  \\
\cellcolor{white} & \cite{12} &  &  &  &  &  &  &  &  &  &  & \cmark &  &  &  &  &  &  &  &  &  &  \\
\rowcolor{gray!15} \cellcolor{white} & \cite{13} &  &  &  & \cmark &  &  &  &  &  &  &  &  &  &  &  &  &  &  &  &  &  \\
\cellcolor{white} & \cite{15} &  &  &  &  &  &  &  &  &  &  &  &  &  &  &  & \cmark &  &  &  &  &  \\
\rowcolor{gray!15} \cellcolor{white} & \cite{16} &  &  &  &  &  &  &  &  &  &  &  &  &  &  & \cmark &  &  &  &  &  &  \\
\cellcolor{white} & \cite{17} &  &  &  &  &  &  &  &  &  &  &  &  &  &  &  &  &  &  &  & \cmark &  \\
\rowcolor{gray!15} \cellcolor{white} & \cite{18} &  &  &  & \cmark &  &  &  &  &  &  &  &  &  &  &  &  &  &  &  &  &  \\
\cellcolor{white} & \cite{20} &  &  &  &  &  &  &  &  & \cmark &  &  &  &  &  &  &  &  &  &  &  &  \\
\rowcolor{gray!15} \cellcolor{white} & \cite{21} &  &  &  &  &  &  &  & \cmark &  &  &  &  &  &  &  &  &  &  &  &  &  \\
\cellcolor{white} & \cite{22} &  &  &  &  &  &  &  & \cmark & \cmark &  &  &  &  &  &  &  &  &  &  &  &  \\
\rowcolor{gray!15} \cellcolor{white} & \cite{24yildiz-etal-2025-benchmarking} &  &  &  & \cmark &  &  &  &  &  &  &  & \cmark &  &  &  &  &  &  &  &  &  \\
\cellcolor{white} & \cite{27} &  &  &  & \cmark &  &  &  &  &  &  &  &  &  &  &  &  &  &  &  &  &  \\
\rowcolor{gray!15} \cellcolor{white} & \cite{29} &  &  &  &  &  &  &  &  &  &  & \cmark &  &  &  &  &  &  &  &  &  &  \\
\cellcolor{white} & \cite{30} &  &  &  &  &  &  &  &  &  &  &  &  &  &  &  &  & \cmark &  &  &  &  \\
\rowcolor{gray!15} \cellcolor{white} & \cite{31} &  &  &  &  &  &  &  &  &  &  &  &  &  &  &  &  &  &  &  & \cmark &  \\
\cellcolor{white} & \cite{32} &  &  &  & \cmark &  &  &  &  &  &  &  &  &  &  &  &  &  &  &  & \cmark &  \\
\rowcolor{gray!15} \cellcolor{white} & \cite{33} &  &  &  &  &  &  &  &  &  &  &  & \cmark &  &  &  &  &  &  &  &  &  \\
\cellcolor{white} & \cite{34} &  &  &  &  &  &  &  &  &  &  &  & \cmark &  &  &  &  &  &  &  &  &  \\
\rowcolor{gray!15} \cellcolor{white} & \cite{35} &  &  &  &  &  &  &  &  & \cmark &  & \cmark &  &  &  &  &  &  &  &  &  &  \\
\cellcolor{white} & \cite{37app15169096} &  &  &  &  &  &  &  &  &  & \cmark &  &  &  &  &  &  &  &  &  &  &  \\
\rowcolor{gray!15} \cellcolor{white} & \cite{38zhu2025cvebenchbenchmarkaiagents} &  &  &  & \cmark &  &  &  &  &  &  &  &  &  &  &  &  &  &  &  &  &  \\
\cellcolor{white} & \cite{39} &  &  &  &  &  &  &  &  &  &  &  &  &  &  & \cmark &  &  &  &  &  &  \\
\rowcolor{gray!15} \cellcolor{white} & \cite{41jiao2025deepvulhunter} & \cmark &  &  &  &  &  &  &  &  &  &  &  &  &  &  &  &  &  &  &  &  \\
\cellcolor{white} & \cite{42LOEVENICH2025111162} &  &  &  &  &  &  &  &  &  &  &  &  &  &  & \cmark &  &  &  &  &  &  \\
\rowcolor{gray!15} \cellcolor{white} & \cite{45} &  &  &  &  &  &  &  &  &  &  &  &  &  &  &  &  &  &  &  &  & \cmark \\
\cellcolor{white} & \cite{46abramovich2025enigma} &  &  &  &  &  & \cmark &  &  &  &  &  &  &  &  &  &  &  &  &  &  &  \\
\rowcolor{gray!15} \cellcolor{white} & \cite{47chowdhry2025evaluating} &  &  &  &  &  &  &  &  &  &  &  &  &  &  & \cmark &  &  &  &  &  &  \\
\cellcolor{white} & \cite{48} &  &  &  &  &  &  &  &  &  &  &  &  &  &  & \cmark &  &  &  &  &  &  \\
\rowcolor{gray!15} \cellcolor{white} & \cite{49} &  &  &  &  &  &  &  &  &  &  &  &  &  &  & \cmark &  &  &  &  &  &  \\
\cellcolor{white} & \cite{51} &  &  &  &  &  &  &  &  &  &  &  &  &  &  &  &  &  &  &  &  & \cmark \\
\rowcolor{gray!15} \cellcolor{white} & \cite{53} &  &  &  &  &  &  &  &  &  &  &  &  &  &  & \cmark &  &  &  &  &  &  \\
\cellcolor{white} & \cite{55} &  &  &  &  &  &  &  &  &  & \cmark &  &  &  &  &  &  &  &  &  &  &  \\
\rowcolor{gray!15} \cellcolor{white} & \cite{56} &  &  &  &  &  &  &  &  &  &  &  &  &  &  &  &  &  &  &  & \cmark &  \\
\cellcolor{white} & \cite{58} &  &  &  &  & \cmark &  &  &  &  &  &  &  &  & \cmark &  &  &  &  &  &  &  \\
\rowcolor{gray!15} \cellcolor{white} & \cite{59} &  &  &  &  &  &  &  &  &  &  & \cmark &  &  &  &  &  &  &  &  &  &  \\
\cellcolor{white} & \cite{60} &  &  &  &  &  &  &  &  &  &  &  &  &  &  &  &  &  &  &  & \cmark &  \\
\rowcolor{gray!15} \cellcolor{white} & \cite{61} &  &  &  &  &  &  &  &  &  &  &  &  &  &  & \cmark &  &  &  &  & \cmark &  \\
\cellcolor{white} & \cite{62} &  &  &  &  &  &  &  &  &  &  &  &  &  &  & \cmark &  &  &  &  &  &  \\
\rowcolor{gray!15} \cellcolor{white} & \cite{65} &  &  &  &  &  &  &  &  &  &  & \cmark &  &  &  &  &  &  &  &  &  &  \\
\cellcolor{white} & \cite{67} &  &  &  &  &  &  &  &  &  &  &  &  &  &  &  &  &  &  & \cmark &  &  \\
\rowcolor{gray!15} \cellcolor{white} & \cite{68} &  &  &  &  &  &  &  &  &  &  &  &  &  &  &  &  &  &  &  & \cmark &  \\
\cellcolor{white} & \cite{69} &  &  &  &  &  &  &  &  &  &  &  &  &  &  &  &  & \cmark &  &  &  &  \\
\rowcolor{gray!15} \cellcolor{white} & \cite{70} &  &  &  &  &  & \cmark & \cmark &  &  &  &  &  &  &  &  &  &  &  &  &  &  \\
\cellcolor{white} & \cite{71} &  &  &  &  & \cmark &  &  &  &  &  &  &  &  &  &  &  &  &  &  &  &  \\
\rowcolor{gray!15} \cellcolor{white} & \cite{72} &  &  &  &  &  &  &  &  &  &  &  &  &  &  &  & \cmark &  &  &  &  &  \\
\cellcolor{white} & \cite{73} &  &  &  &  &  &  &  &  &  &  &  &  &  &  &  &  &  &  &  & \cmark &  \\
\rowcolor{gray!15} \cellcolor{white} & \cite{75} &  &  &  &  &  &  &  &  &  &  &  &  &  & \cmark &  &  &  &  &  &  &  \\
\cellcolor{white} & \cite{78} &  &  &  &  &  &  &  & \cmark &  & \cmark &  &  &  &  &  &  &  &  &  &  &  \\
\rowcolor{gray!15} \cellcolor{white} & \cite{80} &  &  &  &  &  &  &  &  &  & \cmark &  &  &  &  &  &  &  &  &  &  &  \\
\cellcolor{white} & \cite{81Cao_Huang_Li_Huilin_He_Oo_Hooi_2025} &  &  &  &  &  &  &  &  &  &  &  &  & \cmark &  &  &  &  &  &  &  &  \\
\rowcolor{gray!15} \cellcolor{white} & \cite{83} &  &  &  &  &  &  &  &  &  &  &  & \cmark &  &  &  &  &  &  &  &  &  \\
\cellcolor{white} & \cite{87} &  &  &  &  &  &  &  & \cmark &  & \cmark &  &  &  &  &  &  &  &  &  &  &  \\
\rowcolor{gray!15} \cellcolor{white} & \cite{88} &  &  &  &  &  &  &  &  &  &  &  &  &  & \cmark &  &  &  &  &  &  &  \\
\cellcolor{white} & \cite{89guo2025repoauditautonomousllmagentrepositorylevel} &  &  &  &  & \cmark &  &  &  &  &  &  &  &  &  &  &  &  &  &  &  &  \\
\rowcolor{gray!15} \cellcolor{white} & \cite{90} &  &  &  &  &  &  &  &  &  &  &  &  &  &  &  &  &  & \cmark &  &  &  \\
\cellcolor{white} & \cite{93info16050365} &  &  &  &  &  &  &  &  &  &  &  &  &  &  &  &  &  &  &  & \cmark &  \\
\rowcolor{gray!15} \cellcolor{white} & \cite{94} &  &  &  &  &  &  &  &  &  &  &  &  &  &  &  &  &  &  &  & \cmark &  \\
\cellcolor{white} & \cite{99} &  &  &  &  &  &  &  &  &  &  &  & \cmark &  &  &  &  &  &  &  &  &  \\
\rowcolor{gray!15} \cellcolor{white} & \cite{100} &  &  &  &  & \cmark &  &  &  &  &  &  &  &  &  &  &  &  &  &  &  &  \\
\hline
\cellcolor{white}2024 & \cite{1} &  &  &  &  &  &  &  &  &  &  &  &  &  &  &  &  &  &  &  & \cmark &  \\
\rowcolor{gray!15} \cellcolor{white} & \cite{10debenedetti2024agentdojo} &  &  &  &  &  &  &  &  &  &  &  &  &  &  &  &  & \cmark &  &  &  &  \\
\cellcolor{white} & \cite{14chen2024agentpoison} &  &  &  &  &  &  &  &  &  &  &  &  &  &  &  &  & \cmark &  &  &  &  \\
\rowcolor{gray!15} \cellcolor{white} & \cite{28} &  &  &  &  &  &  &  &  &  &  &  &  &  &  &  &  & \cmark &  &  &  &  \\
\cellcolor{white} & \cite{43} &  &  &  &  &  &  &  &  &  &  &  &  &  &  & \cmark &  &  &  &  &  &  \\
\rowcolor{gray!15} \cellcolor{white} & \cite{50} &  &  &  &  &  &  &  &  &  &  & \cmark &  &  &  &  &  &  &  &  &  &  \\
\cellcolor{white} & \cite{54} &  &  &  &  &  &  &  &  &  &  & \cmark &  &  &  &  &  &  &  &  &  &  \\
\rowcolor{gray!15} \cellcolor{white} & \cite{63zemicheal2024llm} &  &  &  & \cmark &  &  &  &  &  &  &  &  &  &  &  &  &  &  &  &  &  \\
\cellcolor{white} & \cite{66} &  &  &  &  &  &  &  &  &  &  & \cmark &  &  &  &  &  &  &  &  &  &  \\
\rowcolor{gray!15} \cellcolor{white} & \cite{74icaart24} &  &  &  &  &  &  &  &  &  &  &  &  &  &  & \cmark &  &  &  &  &  &  \\
\cellcolor{white} & \cite{76} &  &  &  &  &  &  &  &  &  & \cmark &  &  &  &  &  &  &  &  &  &  &  \\
\rowcolor{gray!15} \cellcolor{white} & \cite{77} &  &  &  &  &  &  &  &  &  &  & \cmark &  &  &  &  &  &  &  &  &  &  \\
\cellcolor{white} & \cite{79} &  &  &  &  &  &  &  & \cmark & \cmark &  &  &  &  &  &  &  &  &  &  &  &  \\
\rowcolor{gray!15} \cellcolor{white} & \cite{82rigaki2024prompt} &  &  &  &  &  &  &  &  &  &  &  &  &  &  & \cmark &  &  &  &  &  &  \\
\cellcolor{white} & \cite{85} &  &  &  &  &  &  &  &  & \cmark &  & \cmark &  &  &  &  &  &  &  &  &  &  \\
\rowcolor{gray!15} \cellcolor{white} & \cite{95} &  &  &  &  &  &  &  &  &  &  &  &  &  &  & \cmark &  &  &  &  &  &  \\
\cellcolor{white} & \cite{98} &  &  &  &  & \cmark &  &  &  &  &  &  &  &  &  &  &  &  &  &  &  &  \\
\hline
\rowcolor{gray!15} \cellcolor{white}2023 & \cite{52} &  &  &  &  &  &  &  &  &  &  &  & \cmark &  &  &  &  &  &  &  &  &  \\
\hline

    \end{tabular}%
\end{adjustbox}
  \label{tab:assessment-dataset-by-security-task}%
  \vspace{-12pt}
\end{table*}%

This view distinguishes assessments that use similarly realistic artifacts for different capabilities. Table~\ref{tab:assessment-dataset-by-security-task} uses named benchmark families where possible and broader other-task columns when a paper uses a task-relevant corpus outside those named families.

\ul{Vulnerability detection \& classification}.
Vulnerability-detection datasets include named corpora such as BigVul, ReVeal, and PrimeVul~\cite{23lbathaviator,41jiao2025deepvulhunter}, CVE/NVD-derived data~\cite{4,23lbathaviator,91zhu-etal-2026-teams,3,6,13,18,24yildiz-etal-2025-benchmarking,27,32,38zhu2025cvebenchbenchmarkaiagents,63zemicheal2024llm}, and other vulnerability or audit corpora~\cite{25zhang2026bountybench,58,71,89guo2025repoauditautonomousllmagentrepositorylevel,100,98}. The broader other-audit column prevents custom Infrastructure-as-Code, repository, PoC, and benchmark collections from being left unmapped merely because they do not use one of the three named corpora.

\ul{Penetration testing \& CTF}.
Penetration-testing and CTF datasets cover InterCode-CTF~\cite{40zhuo2026cyberzero,46abramovich2025enigma,70}, PicoCTF~\cite{36ZOU2026104305,70}, HackTheBox~\cite{21,22,78,87,79}, VulnHub~\cite{84WANG2026103731,20,22,35,79,85}, AutoAttacker-style benchmarks~\cite{37app15169096,55,78,80,87,76}, and other custom pentest or CTF suites~\cite{25zhang2026bountybench,64happe2026llms,91zhu-etal-2026-teams,12,29,35,59,65,50,54,66,77,85}. These datasets are well suited to long-horizon assessment because the agent must gather evidence, choose tools, execute actions, and adapt to observations. A paper may legitimately receive both a named-benchmark and an other-pentest mark when it is assessed across multiple suites.

\ul{Smart contracts}.
Smart-contract datasets form a distinct application domain spanning contract vulnerability detection, auditing, and property verification~\cite{6,9jie2025agent4vul,24yildiz-etal-2025-benchmarking,33,34,83,99,52}. This attribute was added because these eight papers previously had no integral task-domain column even though their datasets and assessment targets were explicit.

\ul{Web, phishing, and scam}.
Web, phishing, and scam datasets assess whether agents can identify malicious sites, web threats, or scam behavior~\cite{96,81Cao_Huang_Li_Huilin_He_Oo_Hooi_2025}. TrustNet~\cite{96}, for example, assesses a hybrid ML and LLM-agent detector for scam websites using labeled examples and standard classification metrics. This category differs from CVE exploitation because the task is often judgment or classification over web-facing evidence, not interaction with a vulnerable application.

\ul{Repair and bug-fix}.
Repair datasets assess whether agents can modify code correctly, usually with tests, proof-of-concept inputs, or known bug labels~\cite{25zhang2026bountybench,58,75,88}. PatchAgent~\cite{75} uses real-world vulnerability cases with proof-of-concept tests and functional test suites, while BountyBench~\cite{25zhang2026bountybench} couples repair with detection and exploitation tasks; its multiple task marks therefore reflect a genuinely multi-task benchmark.

\ul{Cyber defense and simulation}.
Cyber-defense datasets and environments support defender agents, network-security games, attack-defense simulations, and autonomous response assessment~\cite{5,26RIGAKI2026129987,92,16,39,42LOEVENICH2025111162,47chowdhry2025evaluating,48,49,53,61,62,43,74icaart24,82rigaki2024prompt,95}. These datasets often define stateful environments where the agent must preserve service availability while detecting or mitigating an attacker. The CAGE-2 study~\cite{53} is representative because it assesses LLM-based cyber defenders in a simulated network environment where reward reflects both security and operational continuity.

\ul{Security testing and fuzzing}.
Security-testing datasets cover REST API and program-level testing as well as RFC-derived protocol models and fuzzing targets~\cite{11,97,2,15,72}. MultiFuzz~\cite{72} uses RFC-derived knowledge for protocol exploration, whereas AutoRestTest~\cite{2} assesses graph- and agent-guided REST API input generation. The broadened label records the shared testing objective without incorrectly describing every assessment as protocol fuzzing.

\ul{AI and agent security}.
AI- and agent-security datasets assess prompt injection, jailbreaks, memory poisoning, unsafe tool use, taint-style agent vulnerabilities, and attacks on other learned systems~\cite{57,86,101,8zhang2025agent,30,69,10debenedetti2024agentdojo,14chen2024agentpoison,28}. AgentDojo~\cite{10debenedetti2024agentdojo} provides realistic tool-calling tasks with prompt-injection tests, while AgentFuzz~\cite{69} targets vulnerabilities in LLM applications and~\cite{57,28} assess agent-generated attacks on recommender systems.

\ul{Asset discovery and OSINT}.
Asset-discovery and OSINT datasets assess whether an agent can search, correlate, or prioritize exposed assets and vulnerability-relevant information~\cite{90}. This category is small in the current taxonomy, but it is practically important as many security workflows begin with asset visibility rather than code or exploit generation. Dataset design in this category needs to capture both retrieval accuracy and the operational speed of discovery.

\ul{Malware and Android apps}.
Malware and Android-app datasets support agentic analysis of suspicious software, app behavior, family attribution, or malware reports~\cite{7,67}. MAD-Agent~\cite{67} illustrates this task family by combining static and dynamic analysis tools with threat-intelligence access, then assessing detection, behavioral classification, family attribution, and report quality. Compared with vulnerability datasets, malware datasets often require interpreting multiple evidence streams rather than only matching a code location to a vulnerability label.

\ul{Intrusion, SOC, and SOAR}.
This broader family covers intrusion or traffic detection as well as alert triage, enrichment, reporting, response recommendation, and automated investigation~\cite{19,17,31,32,56,60,61,68,73,93info16050365,94,1}. Cognitive SOC~\cite{32} assesses evidence linkage, citation coverage, hallucination, and time-to-recommendation on 1,000 alerts. The broader label prevents intrusion datasets from being excluded merely because they are not packaged as SOC alerts.

\ul{Other security tasks}.
The residual task category is used for two explicit but otherwise uncovered assessment targets: cloud-native access-control policy generation~\cite{45} and context-aware software-security education~\cite{51}. It is not an unknown or unspecified label; it records a defined task that falls outside the listed families.

{\bf Dataset characterisation axes.}
Dataset characterisation axes are cross-cutting attributes of the assessment data: realism, label quality, scale, artifact type, vulnerability-type coverage, and accessibility.
\begin{table*}[t]
  \centering
  \vspace{-10pt}
  \caption{Paper attribution for Assessment: Dataset characterisation axes}
  \vspace{-10pt}
  \scriptsize
  \setlength{\tabcolsep}{1.5pt}
  \renewcommand{\arraystretch}{0.70}
\begin{adjustbox}{max totalsize={\textwidth}{0.88\textheight},center}
    \begin{tabular}{|c|c|c|c|c|c|c|c|c|c|c|c|c|c|c|c|c|c|c|c|}
    \hline
\rowcolor{white} \multirow{2}[1]{*}{Year} & \multirow{2}[1]{*}{Paper} & \multicolumn{3}{c|}{\makecell[c]{Realism}} & \multicolumn{4}{c|}{\makecell[c]{Label quality}} & \multicolumn{3}{c|}{\makecell[c]{Scale}} & \multicolumn{3}{c|}{\makecell[c]{Language / artifact}} & \multicolumn{1}{c|}{\makecell[c]{Vulnerability\\type coverage}} & \multicolumn{4}{c|}{\makecell[c]{Accessibility}} \\
\cline{3-5}\cline{6-9}\cline{10-12}\cline{13-15}\cline{16-16}\cline{17-20}
\rowcolor{white} & & \multicolumn{1}{c|}{\cellcolor{blue!5}\makecell[c]{Fully\\synthetic \\ (9\%)}} & \multicolumn{1}{c|}{\cellcolor{blue!31}\makecell[c]{Semi-\\synthetic \\ (47\%)}} & \multicolumn{1}{c|}{\cellcolor{blue!23}\makecell[c]{Fully\\real-world \\ (36\%)}} & \multicolumn{1}{c|}{\cellcolor{blue!5}\makecell[c]{Auto-\\labelled \\ (4\%)}} & \multicolumn{1}{c|}{\cellcolor{blue!7}\makecell[c]{Competition\\verified \\ (11\%)}} & \multicolumn{1}{c|}{\cellcolor{blue!5}\makecell[c]{Noisy /\\weakly\\labelled \\ (2\%)}} & \multicolumn{1}{c|}{\cellcolor{blue!11}\makecell[c]{Expert /\\reference\\verified \\ (16\%)}} & \multicolumn{1}{c|}{\cellcolor{blue!5}\makecell[c]{Small \\ (8\%)}} & \multicolumn{1}{c|}{\cellcolor{blue!5}\makecell[c]{Medium \\ (8\%)}} & \multicolumn{1}{c|}{\cellcolor{blue!5}\makecell[c]{Large \\ (2\%)}} & \multicolumn{1}{c|}{\cellcolor{blue!7}\makecell[c]{Single\\language \\ (10\%)}} & \multicolumn{1}{c|}{\cellcolor{blue!5}\makecell[c]{Binary /\\compiled \\ (7\%)}} & \multicolumn{1}{c|}{\cellcolor{blue!32}\makecell[c]{Network /\\log \\ (49\%)}} & \multicolumn{1}{c|}{\cellcolor{blue!15}\makecell[c]{Multi-type \\coverage \\ (23\%)}} & \multicolumn{1}{c|}{\cellcolor{blue!33}\makecell[c]{Fully\\public \\ (50\%)}} & \multicolumn{1}{c|}{\cellcolor{blue!5}\makecell[c]{Public\\w/\\registration \\ (3\%)}} & \multicolumn{1}{c|}{\cellcolor{blue!5}\makecell[c]{Restricted \\ (4\%)}} & \multicolumn{1}{c|}{\cellcolor{blue!5}\makecell[c]{Proprietary /\\internal \\ (1\%)}} \\
    \hline

\cellcolor{white}2026 & \cite{4} &  & \cmark &  &  &  &  &  &  &  &  &  &  & \cmark &  &  &  &  &  \\
\rowcolor{gray!15} \cellcolor{white} & \cite{5} & \cmark &  &  &  &  &  &  &  &  &  &  &  & \cmark &  &  &  &  &  \\
\cellcolor{white} & \cite{11} &  &  & \cmark &  &  &  &  &  &  &  & \cmark &  &  &  & \cmark &  &  &  \\
\rowcolor{gray!15} \cellcolor{white} & \cite{19} &  &  & \cmark &  &  &  &  &  &  &  &  &  & \cmark &  & \cmark &  &  &  \\
\cellcolor{white} & \cite{23lbathaviator} &  & \cmark &  &  &  & \cmark & \cmark &  &  & \cmark &  &  &  &  & \cmark &  &  &  \\
\rowcolor{gray!15} \cellcolor{white} & \cite{25zhang2026bountybench} &  &  & \cmark &  &  &  &  &  &  &  & \cmark &  &  & \cmark & \cmark &  &  &  \\
\cellcolor{white} & \cite{26RIGAKI2026129987} &  & \cmark &  &  &  &  &  &  &  &  &  &  & \cmark &  &  &  &  &  \\
\rowcolor{gray!15} \cellcolor{white} & \cite{36ZOU2026104305} &  &  & \cmark &  & \cmark &  &  &  &  &  &  & \cmark & \cmark & \cmark & \cmark &  &  &  \\
\cellcolor{white} & \cite{40zhuo2026cyberzero} &  & \cmark &  &  & \cmark &  &  & \cmark &  &  &  &  &  & \cmark & \cmark &  & \cmark &  \\
\rowcolor{gray!15} \cellcolor{white} & \cite{57} & \cmark &  &  &  &  &  &  &  &  &  &  &  &  &  & \cmark &  &  &  \\
\cellcolor{white} & \cite{64happe2026llms} &  & \cmark &  &  &  &  &  &  &  &  &  &  &  & \cmark & \cmark &  &  &  \\
\rowcolor{gray!15} \cellcolor{white} & \cite{84WANG2026103731} &  & \cmark &  &  &  &  &  &  &  &  &  &  & \cmark &  & \cmark &  &  &  \\
\cellcolor{white} & \cite{86} &  &  & \cmark &  &  &  & \cmark &  &  &  &  &  &  &  &  &  &  &  \\
\rowcolor{gray!15} \cellcolor{white} & \cite{91zhu-etal-2026-teams} &  &  & \cmark &  &  &  &  &  &  &  &  &  &  &  & \cmark &  &  &  \\
\cellcolor{white} & \cite{92} &  & \cmark &  &  &  &  &  &  &  &  &  &  & \cmark &  &  &  &  &  \\
\rowcolor{gray!15} \cellcolor{white} & \cite{96} &  & \cmark &  &  &  &  &  &  & \cmark &  &  &  & \cmark &  & \cmark &  &  &  \\
\cellcolor{white} & \cite{97} &  &  &  &  &  &  &  &  &  &  &  &  & \cmark &  & \cmark &  &  &  \\
\rowcolor{gray!15} \cellcolor{white} & \cite{101} &  &  & \cmark &  &  &  &  &  &  &  &  &  &  &  &  &  &  &  \\
\hline
\cellcolor{white}2025 & \cite{2} &  & \cmark &  &  &  &  &  &  &  &  &  &  &  &  & \cmark &  &  &  \\
\rowcolor{gray!15} \cellcolor{white} & \cite{3} &  & \cmark &  &  &  &  &  &  &  &  &  &  & \cmark &  &  &  & \cmark &  \\
\cellcolor{white} & \cite{6} &  &  & \cmark &  &  &  &  &  & \cmark &  &  &  &  &  & \cmark &  &  &  \\
\rowcolor{gray!15} \cellcolor{white} & \cite{7} &  &  & \cmark &  &  &  &  &  &  &  &  & \cmark &  &  &  &  &  &  \\
\cellcolor{white} & \cite{8zhang2025agent} &  & \cmark &  &  &  &  &  &  & \cmark &  &  &  &  & \cmark & \cmark &  &  &  \\
\rowcolor{gray!15} \cellcolor{white} & \cite{9jie2025agent4vul} &  &  & \cmark &  &  &  &  &  &  & \cmark & \cmark & \cmark &  &  &  &  &  &  \\
\cellcolor{white} & \cite{12} &  & \cmark &  &  &  &  &  &  &  &  &  &  & \cmark &  &  &  &  &  \\
\rowcolor{gray!15} \cellcolor{white} & \cite{13} &  &  & \cmark &  &  &  &  &  &  &  &  &  &  &  &  &  &  &  \\
\cellcolor{white} & \cite{15} &  &  &  &  &  &  & \cmark &  &  &  &  &  & \cmark &  & \cmark &  &  &  \\
\rowcolor{gray!15} \cellcolor{white} & \cite{16} &  & \cmark &  &  &  &  &  &  &  &  &  &  & \cmark &  &  &  &  &  \\
\cellcolor{white} & \cite{17} & \cmark &  &  &  &  &  &  &  &  &  &  &  &  &  &  &  &  &  \\
\rowcolor{gray!15} \cellcolor{white} & \cite{18} &  & \cmark &  &  &  &  &  &  &  &  &  &  & \cmark &  &  &  &  &  \\
\cellcolor{white} & \cite{20} &  & \cmark &  &  &  &  &  &  &  &  &  &  & \cmark &  & \cmark &  &  &  \\
\rowcolor{gray!15} \cellcolor{white} & \cite{21} &  & \cmark &  &  &  &  &  &  &  &  &  &  &  &  &  &  &  &  \\
\cellcolor{white} & \cite{22} &  & \cmark &  &  &  &  &  &  &  &  &  &  &  & \cmark & \cmark & \cmark &  &  \\
\rowcolor{gray!15} \cellcolor{white} & \cite{24yildiz-etal-2025-benchmarking} &  &  & \cmark &  &  &  &  &  &  &  &  &  &  &  &  &  &  &  \\
\cellcolor{white} & \cite{27} &  & \cmark &  &  &  &  &  &  &  &  &  &  & \cmark &  &  &  &  &  \\
\rowcolor{gray!15} \cellcolor{white} & \cite{29} &  & \cmark &  &  & \cmark &  &  & \cmark &  &  &  & \cmark &  & \cmark & \cmark &  &  &  \\
\cellcolor{white} & \cite{30} &  & \cmark &  &  & \cmark &  &  &  &  &  &  &  & \cmark & \cmark & \cmark &  &  &  \\
\rowcolor{gray!15} \cellcolor{white} & \cite{31} &  & \cmark &  &  &  &  & \cmark &  &  &  &  &  & \cmark &  & \cmark &  & \cmark & \cmark \\
\cellcolor{white} & \cite{32} &  & \cmark &  &  &  &  &  &  & \cmark &  &  &  & \cmark &  &  &  &  &  \\
\rowcolor{gray!15} \cellcolor{white} & \cite{33} & \cmark &  &  &  &  &  & \cmark &  &  &  &  &  &  &  & \cmark &  &  &  \\
\cellcolor{white} & \cite{34} &  &  &  & \cmark &  &  &  &  &  &  & \cmark &  &  &  & \cmark &  &  &  \\
\rowcolor{gray!15} \cellcolor{white} & \cite{35} &  &  & \cmark &  &  &  &  &  &  &  &  &  &  & \cmark &  &  &  &  \\
\cellcolor{white} & \cite{37app15169096} &  &  & \cmark &  &  &  &  &  &  &  &  &  & \cmark & \cmark &  &  &  &  \\
\rowcolor{gray!15} \cellcolor{white} & \cite{38zhu2025cvebenchbenchmarkaiagents} &  &  & \cmark &  & \cmark &  &  &  &  &  &  &  &  & \cmark &  &  &  &  \\
\cellcolor{white} & \cite{39} &  & \cmark &  &  &  &  &  &  &  &  &  &  & \cmark &  & \cmark &  &  &  \\
\rowcolor{gray!15} \cellcolor{white} & \cite{41jiao2025deepvulhunter} &  &  & \cmark &  &  &  &  &  &  &  &  &  &  &  &  &  &  &  \\
\cellcolor{white} & \cite{42LOEVENICH2025111162} &  & \cmark &  &  &  &  &  & \cmark &  &  &  &  & \cmark &  & \cmark &  &  &  \\

\rowcolor{gray!15} \cellcolor{white} & \cite{45} &  & \cmark &  &  &  &  & \cmark & \cmark &  &  &  &  & \cmark &  & \cmark &  &  &  \\
\cellcolor{white} & \cite{46abramovich2025enigma} &  & \cmark &  &  & \cmark &  &  & \cmark &  &  &  & \cmark & \cmark & \cmark & \cmark &  &  &  \\
\rowcolor{gray!15} \cellcolor{white} & \cite{47chowdhry2025evaluating} & \cmark &  &  &  &  &  &  & \cmark &  &  &  &  & \cmark &  &  &  &  &  \\
\cellcolor{white} & \cite{48} & \cmark &  &  &  &  &  &  &  &  &  &  &  & \cmark &  &  &  &  &  \\
\rowcolor{gray!15} \cellcolor{white} & \cite{49} &  & \cmark &  &  &  &  &  &  &  &  &  &  & \cmark &  & \cmark &  &  &  \\
\cellcolor{white} & \cite{51} &  & \cmark &  &  &  &  &  &  &  &  &  &  &  &  &  &  &  &  \\
\rowcolor{gray!15} \cellcolor{white} & \cite{53} &  & \cmark &  &  &  &  &  &  &  &  &  &  & \cmark &  &  &  &  &  \\
\cellcolor{white} & \cite{55} & \cmark &  &  &  &  &  &  &  &  &  &  &  &  &  &  &  &  &  \\
\rowcolor{gray!15} \cellcolor{white} & \cite{56} &  & \cmark &  &  &  &  &  &  &  &  &  &  &  &  &  &  &  &  \\
\cellcolor{white} & \cite{58} &  &  &  &  &  &  &  &  &  &  & \cmark &  &  &  &  &  &  &  \\
\rowcolor{gray!15} \cellcolor{white} & \cite{59} &  & \cmark &  &  & \cmark & \cmark &  &  &  &  &  &  &  & \cmark &  &  &  &  \\
\cellcolor{white} & \cite{60} &  &  & \cmark &  &  &  &  &  &  &  &  &  & \cmark & \cmark &  &  &  &  \\
\rowcolor{gray!15} \cellcolor{white} & \cite{61} &  &  & \cmark &  &  &  &  &  &  &  &  &  &  &  &  &  &  &  \\
\cellcolor{white} & \cite{62} &  &  & \cmark &  &  &  &  &  &  &  &  &  & \cmark &  &  &  &  &  \\
\rowcolor{gray!15} \cellcolor{white} & \cite{65} &  & \cmark &  &  & \cmark &  &  & \cmark &  &  &  & \cmark & \cmark & \cmark & \cmark &  &  &  \\
\cellcolor{white} & \cite{67} &  &  & \cmark &  &  &  & \cmark & \cmark &  &  &  & \cmark &  &  & \cmark &  &  &  \\
\rowcolor{gray!15} \cellcolor{white} & \cite{68} &  &  & \cmark &  &  &  &  &  &  &  &  &  &  &  &  &  &  &  \\
\cellcolor{white} & \cite{69} &  & \cmark &  &  &  &  & \cmark &  &  &  &  &  & \cmark &  & \cmark &  &  &  \\
\rowcolor{gray!15} \cellcolor{white} & \cite{70} &  & \cmark &  &  & \cmark &  & \cmark &  & \cmark &  &  &  & \cmark & \cmark & \cmark &  & \cmark &  \\
\cellcolor{white} & \cite{71} &  &  &  &  &  &  &  &  &  &  & \cmark &  &  &  &  &  &  &  \\
\rowcolor{gray!15} \cellcolor{white} & \cite{72} &  & \cmark &  &  &  &  &  &  &  &  &  &  & \cmark &  &  &  &  &  \\
\cellcolor{white} & \cite{73} &  & \cmark &  &  &  &  &  &  &  &  &  &  & \cmark &  & \cmark &  &  &  \\
\rowcolor{gray!15} \cellcolor{white} & \cite{75} &  &  & \cmark & \cmark &  &  &  &  &  &  & \cmark &  &  & \cmark & \cmark &  &  &  \\
\cellcolor{white} & \cite{78} &  &  & \cmark &  & \cmark &  &  &  &  &  &  &  &  & \cmark & \cmark & \cmark &  &  \\
\rowcolor{gray!15} \cellcolor{white} & \cite{80} &  & \cmark &  &  &  &  &  &  &  &  &  &  &  &  &  &  &  &  \\
\cellcolor{white} & \cite{81Cao_Huang_Li_Huilin_He_Oo_Hooi_2025} &  &  & \cmark &  &  &  & \cmark &  &  &  &  &  &  &  & \cmark &  &  &  \\
\rowcolor{gray!15} \cellcolor{white} & \cite{83} &  &  & \cmark & \cmark &  &  & \cmark &  &  &  & \cmark &  &  &  & \cmark &  &  &  \\
\cellcolor{white} & \cite{87} &  &  & \cmark &  &  &  &  &  &  &  &  &  & \cmark &  &  &  &  &  \\
\rowcolor{gray!15} \cellcolor{white} & \cite{88} &  &  & \cmark &  &  &  &  &  &  &  &  &  &  &  & \cmark &  &  &  \\
\cellcolor{white} & \cite{89guo2025repoauditautonomousllmagentrepositorylevel} &  &  & \cmark &  &  &  &  &  &  &  &  &  &  &  & \cmark &  &  &  \\
\rowcolor{gray!15} \cellcolor{white} & \cite{90} &  &  & \cmark &  &  &  &  &  &  &  &  &  & \cmark &  &  &  &  &  \\
\cellcolor{white} & \cite{93info16050365} &  & \cmark &  &  &  &  &  &  &  &  &  &  & \cmark &  &  &  &  &  \\
\rowcolor{gray!15} \cellcolor{white} & \cite{94} &  &  & \cmark &  &  &  &  &  &  &  &  &  & \cmark &  & \cmark &  &  &  \\
\cellcolor{white} & \cite{99} &  &  & \cmark &  &  &  & \cmark &  &  &  & \cmark &  &  &  & \cmark &  &  &  \\
\rowcolor{gray!15} \cellcolor{white} & \cite{100} &  & \cmark &  &  &  &  & \cmark &  & \cmark &  &  &  & \cmark & \cmark &  &  &  &  \\
\hline
\cellcolor{white}2024 & \cite{1} &  & \cmark &  &  &  &  &  &  &  &  &  &  & \cmark &  & \cmark &  &  &  \\
\rowcolor{gray!15} \cellcolor{white} & \cite{10debenedetti2024agentdojo} &  & \cmark &  & \cmark &  &  &  &  & \cmark &  &  &  &  &  & \cmark &  &  &  \\
\cellcolor{white} & \cite{14chen2024agentpoison} &  &  & \cmark &  &  &  &  &  &  &  &  &  &  &  & \cmark &  &  &  \\
\rowcolor{gray!15} \cellcolor{white} & \cite{28} &  &  & \cmark &  &  &  &  &  &  &  &  &  &  &  &  &  &  &  \\
\cellcolor{white} & \cite{43} &  & \cmark &  &  &  &  &  &  &  &  &  &  & \cmark &  &  &  &  &  \\
\rowcolor{gray!15} \cellcolor{white} & \cite{50} &  &  &  &  &  &  & \cmark &  &  &  &  &  &  &  & \cmark &  &  &  \\
\cellcolor{white} & \cite{54} &  &  & \cmark &  & \cmark &  &  &  &  &  &  &  & \cmark & \cmark & \cmark &  &  &  \\
\rowcolor{gray!15} \cellcolor{white} & \cite{63zemicheal2024llm} &  & \cmark &  &  &  &  & \cmark &  &  &  &  &  &  &  & \cmark &  &  &  \\
\cellcolor{white} & \cite{66} &  &  & \cmark &  &  &  &  &  &  &  &  &  &  &  &  &  &  &  \\
\rowcolor{gray!15} \cellcolor{white} & \cite{74icaart24} &  &  &  &  &  &  &  &  &  &  &  &  & \cmark &  &  &  &  &  \\
\cellcolor{white} & \cite{76} &  &  &  &  &  &  &  &  &  &  &  &  & \cmark & \cmark &  &  &  &  \\
\rowcolor{gray!15} \cellcolor{white} & \cite{77} &  & \cmark &  &  &  &  &  &  &  &  &  &  &  &  &  &  &  &  \\
\cellcolor{white} & \cite{79} &  & \cmark &  &  &  &  &  &  &  &  &  &  & \cmark & \cmark & \cmark & \cmark &  &  \\
\rowcolor{gray!15} \cellcolor{white} & \cite{82rigaki2024prompt} &  & \cmark &  &  &  &  &  &  &  &  &  &  & \cmark & \cmark &  &  &  &  \\
\cellcolor{white} & \cite{85} &  & \cmark &  &  &  &  &  &  &  &  &  &  & \cmark &  &  &  &  &  \\
\rowcolor{gray!15} \cellcolor{white} & \cite{95} & \cmark &  &  &  &  &  &  &  &  &  &  &  & \cmark &  &  &  &  &  \\
\cellcolor{white} & \cite{98} & \cmark &  &  &  &  &  & \cmark &  & \cmark &  & \cmark &  &  &  & \cmark &  &  &  \\
\hline
\rowcolor{gray!15} \cellcolor{white}2023 & \cite{52} &  &  & \cmark &  &  &  &  &  &  &  &  &  &  &  & \cmark &  &  &  \\

\hline
    \end{tabular}%
\end{adjustbox}
  \label{tab:assessment-dataset-characterisation-axes}%
  \vspace{-12pt}
\end{table*}%

These axes explain why results may not be directly comparable even for the same application. Table~\ref{tab:assessment-dataset-characterisation-axes} applies every supported item to each paper; multiple checkmarks are therefore expected when an assessment combines modalities or access conditions. Realism and scale are coded at the level of the assessed collection, while unsupported or unreported axis values remain blank and are interpreted as unspecified.

\ul{Realism}.
The surveyed datasets span fully synthetic, semi-synthetic, and fully real-world settings. Fully synthetic datasets are used when the authors construct all or most of the assessed artifacts or environment states~\cite{5,57,17,33,47chowdhry2025evaluating,48,55,95,98}. Semi-synthetic datasets combine real components with controlled modifications, generated tasks, simulated environments, or benchmark wrappers~\cite{4,23lbathaviator,26RIGAKI2026129987,29,40zhuo2026cyberzero,64happe2026llms,84WANG2026103731,92,96,2,3,8zhang2025agent,12,16,18,20,21,22,27,30,31,32,39,42LOEVENICH2025111162,45,46abramovich2025enigma,49,51,53,56,59,65,69,70,72,73,80,93info16050365,1,10debenedetti2024agentdojo,43,63zemicheal2024llm,77,79,82rigaki2024prompt,85,100}. Fully real-world datasets use real vulnerabilities, applications, repositories, malware, alerts, or production-style targets~\cite{11,19,25zhang2026bountybench,36ZOU2026104305,86,91zhu-etal-2026-teams,101,6,7,9jie2025agent4vul,13,24yildiz-etal-2025-benchmarking,35,37app15169096,38zhu2025cvebenchbenchmarkaiagents,41jiao2025deepvulhunter,60,61,62,67,68,75,78,81Cao_Huang_Li_Huilin_He_Oo_Hooi_2025,83,87,88,89guo2025repoauditautonomousllmagentrepositorylevel,90,94,99,14chen2024agentpoison,28,54,66,52}. The tradeoff is consistent across the survey: real-world datasets improve ecological validity, while synthetic and semi-synthetic datasets improve control, repeatability, and safety.

\ul{Label quality}.
Label quality determines whether the assessment target is trustworthy. Some datasets rely on automated labels, such as generated labels, oracle checks, or test-derived labels~\cite{34,75,83,10debenedetti2024agentdojo}. Competition-verified labels appear in CTF and cyber-range datasets where a flag, official solution, or challenge infrastructure validates success~\cite{29,36ZOU2026104305,40zhuo2026cyberzero,30,38zhu2025cvebenchbenchmarkaiagents,46abramovich2025enigma,59,65,70,78,54}. Expert/reference-verified labels are grounded in official behavior, human-written references, manual annotation, or review by domain or security experts~\cite{15,23lbathaviator,31,33,45,50,63zemicheal2024llm,67,69,70,81Cao_Huang_Li_Huilin_He_Oo_Hooi_2025,83,86,98,99,100}. Noisy or weak labels appear when large vulnerability corpora or generated benchmark artifacts may contain imperfect ground truth~\cite{23lbathaviator,59}. When no evidence supports one of these label sources, the label-quality value is left unspecified rather than guessed.

\ul{Scale}.
Dataset scale varies widely. Small datasets are common in expensive, environment-heavy assessments such as CTF, cyber-defense, malware analysis, policy generation, and exploit benchmarks~\cite{29,40zhuo2026cyberzero,42LOEVENICH2025111162,45,46abramovich2025enigma,47chowdhry2025evaluating,65,67}. Medium-scale datasets appear in phishing, vulnerability, SOC, prompt-injection, and CTF-knowledge assessment~\cite{96,6,8zhang2025agent,32,70,10debenedetti2024agentdojo,98,100}. Large datasets are relatively rare in the verified taxonomy and appear mainly in large vulnerability-detection corpora and smart-contract analysis~\cite{23lbathaviator,9jie2025agent4vul}. This imbalance reflects a core assessment challenge: agentic security tasks are often costly to execute because they require live environments, tool interaction, sandboxing, or human validation, so scale is often traded for realism and measurement depth.

\ul{Language and artifact type}.
Some datasets are single-language code datasets, which are common in vulnerability detection, smart-contract analysis, repair, and formal-verification tasks~\cite{11,25zhang2026bountybench,9jie2025agent4vul,34,58,71,75,83,99,98}. Others involve binary or compiled artifacts, including malware, reverse engineering, CTF binaries, and exploit-development tasks~\cite{36ZOU2026104305,7,9jie2025agent4vul,29,46abramovich2025enigma,65,67}. The largest artifact family is network/log data, which appears in cyber-defense, intrusion detection, protocol fuzzing, penetration testing, SOC alerting, and web-security tasks~\cite{4,5,19,26RIGAKI2026129987,36ZOU2026104305,84WANG2026103731,92,96,97,3,12,15,16,18,20,27,30,31,32,37app15169096,39,42LOEVENICH2025111162,45,46abramovich2025enigma,47chowdhry2025evaluating,48,49,53,60,62,65,69,70,72,73,87,90,93info16050365,94,1,43,54,74icaart24,76,79,82rigaki2024prompt,85,95,100}. Artifact type strongly shapes the agent's required capabilities: code datasets emphasize semantic program reasoning, binary datasets require lower-level analysis, and network/log datasets emphasize temporal correlation and environment state.

\ul{Vulnerability type coverage}.
Some datasets explicitly cover multiple vulnerability classes, attack types, or benchmark categories~\cite{25zhang2026bountybench,36ZOU2026104305,40zhuo2026cyberzero,64happe2026llms,8zhang2025agent,22,29,30,35,37app15169096,38zhu2025cvebenchbenchmarkaiagents,46abramovich2025enigma,59,60,65,70,75,78,54,76,79,82rigaki2024prompt,100}. Coverage matters because many agents perform well on familiar vulnerability patterns but fail when the target requires a different exploitation strategy, input modality, or repair mechanism. BountyBench~\cite{25zhang2026bountybench} is illustrative because it spans real bug-bounty vulnerabilities across OWASP risk categories and assesses detect, exploit, and patch tasks. PentestGPT~\cite{79} similarly uses targets that cover OWASP top-10 vulnerabilities and multiple CWE items, making subtask performance more informative than a single aggregate score.

\ul{Accessibility}.
Accessibility determines whether assessment results can be reproduced. Fully public datasets are the dominant accessibility category~\cite{11,19,23lbathaviator,25zhang2026bountybench,36ZOU2026104305,40zhuo2026cyberzero,57,64happe2026llms,84WANG2026103731,91zhu-etal-2026-teams,96,97,2,6,8zhang2025agent,15,20,22,29,30,31,33,34,39,42LOEVENICH2025111162,45,46abramovich2025enigma,49,65,67,69,70,73,75,78,81Cao_Huang_Li_Huilin_He_Oo_Hooi_2025,83,88,89guo2025repoauditautonomousllmagentrepositorylevel,94,98,99,1,10debenedetti2024agentdojo,14chen2024agentpoison,50,54,63zemicheal2024llm,79,52}. A smaller set is public but requires registration or platform access, often because targets live on hosted penetration-testing platforms~\cite{22,78,79}. Restricted datasets appear when environments, challenge assets, or collected data cannot be fully released~\cite{40zhuo2026cyberzero,3,31,70}, and proprietary or internal data appears rarely~\cite{31}. If a paper does not disclose access conditions, accessibility is recorded as unspecified; public source artifacts are not taken to imply that the authors' complete assessment dataset is public.

\section{Analysis and Discussion of Survey Results} \label{sec:sra}
%

\subsection{Agentic Approach to Software/Systems Security}\label{subsec:approach-gaps-future}
The survey results on Approach ($\S$\ref{ssec:resultapproach}) show that LLM-based security agents are systems in which an LLM plans, acts, observes, and revises. This is a significant shift from treating the LLM as a standalone text generator. At the same time, this domain remains unevenly developed. Architecture and workflow design are relatively mature, while explicit uncertainty handling, safety controls, human governance, and long-term learning are still limited.

\subsubsection{Limitations and Challenges of Existing Works}\label{sssec:approach-limitations}
Existing approaches are highly modular, but their modularity is often operational rather than principled. Table~\ref{tab:approach-agent-architecture} shows that 91\% of the surveyed papers use a modular pipeline, and 88\% place the backbone LLM in a planner+actor role. This indicates that most systems now separate planning, execution, tool use, observation, or reporting in some form. However, fewer papers include explicit critic/verifier/judge roles (15\%), guardrail or policy-enforcement layers (13\%), fine-tuned LLM components (7\%), mixture-of-agents designs (4\%), or all-in-one role definitions (4\%), and a further 3\% do not describe their internal organization in enough detail to classify at all. Systems such as \textsc{RepoAudit}~\cite{89guo2025repoauditautonomousllmagentrepositorylevel} and \textsc{ZT-ICAS}~\cite{101} show that validators and guardrails can be architectural components rather than external checks, but they remain the exception. Thus, many agents are modular enough to operate but not yet sufficiently modular to make their own failure modes auditable. Verification, policy enforcement, and role boundaries are still treated as optional enhancements rather than core architectural requirements.

Memory and perception are broad but shallow. Table~\ref{tab:approach-agent-memory} shows that in-context working memory (60\%) and long-term/persistent memory (57\%) are common, and all systems rely on parametric memory in the broad sense of model-internal knowledge. Security-specific memory is also visible, including tool documentation (49\%), execution traces and fuzzing logs (33\%), vulnerability knowledge bases (25\%), and code repositories or patch histories (23\%). However, these stores are consumed far more often than they are curated: read operations appear in 61\% of papers and update/reflection in 42\%, but explicit write operations in only 7\%, and knowledge-graph memory in 10\%. \textsc{Cognitive SOC}~\cite{32}, which converts alert evidence into a durable incident narrative, is among the few that treat memory as something the agent produces rather than only consumes. This asymmetry suggests that many papers use memory as context storage or retrieval support, but fewer define memory as a governed stateful subsystem with clear read/write/update semantics, provenance, forgetting, conflict handling, or contamination defenses.

Perception follows a similar pattern. Table~\ref{tab:approach-agent-perception} shows universal direct LLM ingestion (100\%), frequent tool-mediated observation (68\%), and substantial use of environment observations (66\%), structured data (60\%), system logs (55\%), and natural language (42\%). However, retrieval-augmented perception appears in 30\%, summarisation/compression in 35\%, embedding-based semantic processing in 10\%, and compiled binaries or decompiled outputs in only 5\%. Security agents therefore consume many kinds of inputs, but their perception pipelines often depend on ad hoc prompt ingestion and tool output parsing. This is risky in adversarial settings where logs, web pages, code comments, repository files, and tool outputs may contain misleading or malicious instructions---precisely the threat model that AgentDojo~\cite{10debenedetti2024agentdojo} and \textsc{AgentPoison}~\cite{14chen2024agentpoison} demonstrate against the ingestion and retrieval paths, respectively.

Reasoning and planning are dominated by reactive control rather than robust decision-making. Table~\ref{tab:approach-agent-reasoning-planning} shows that dynamic/reactive planning appears in 80\% of papers, while least-to-most/task decomposition appears in 46\% and hypothesis-driven reasoning in 29\%. These patterns are well matched to security tasks where observations arrive incrementally. However, explicit mechanisms for decision-making under uncertainty are much less common: confidence-based branching appears in 23\%, majority voting in 7\%, and human-in-the-loop checkpoints in only 4\%. This is a central limitation because security agents often act under uncertainty. A related reporting gap compounds it: 18\% of the surveyed papers do not identify their reasoning paradigm at all, and 16\% of the papers do not identify their planning paradigm, which makes it difficult to tell whether a result follows from the agent scaffold or from the backbone model.

The action space reveals a strong dependence on retrieval and practitioner tools, but uneven integration with deeper security analysis. Table~\ref{tab:approach-agent-action-space} shows that documentation retrieval (38\%), vector/RAG retrieval (29\%), penetration-testing frameworks (25\%), REPL/shell execution (19\%), code search (17\%), execution sandboxes (17\%), CTF platforms (14\%), and linters/pattern matchers (15\%) are among the most common actions. These actions make agents practically useful because they connect LLM reasoning to external evidence and executable environments. Still, more specialized tools are less consistently represented, including symbolic execution (2\%), formal verification (5\%), taint analysis (5\%), binary analysis, cloud/IaC tooling, SIEM/SOAR integration, and cyber-range infrastructure. Where they are used, as in \textsc{PropertyGPT}~\cite{83}, a sound external analysis rather than the model itself decides whether a candidate finding survives. As a result, many agents resemble tool-using assistants rather than deeply integrated security-analysis systems.

Workflow results show that autonomy has outpaced governance. Table~\ref{tab:approach-agent-workflow-orchestration} shows that fully autonomous workflows appear in 95\% of papers, with analyst, executor, and planner roles appearing in 93\%, 92\%, and 74\%, respectively. Sequential pipelines (90\%) and iterative loops (78\%) are also dominant. By contrast, human-on-the-loop workflows appear in 8\%, human-in-the-loop workflows in 5\%, and interactive/collaborative workflows in 3\%. This gap is striking: many surveyed systems automate tasks that can affect real systems, generate exploits, patch code, or interpret incidents, yet only a small fraction structurally encode human review or approval.


Finally, self-improvement remains sparse. Table~\ref{tab:approach-agent-self-improvement-learning} shows that learning from feedback within a session appears in 32\% of papers, but example-guided correction appears in only 1\%, storing successful trajectories in 3\%, updating fine-tuning data in 2\%, reward-signal design in 9\%, and environment-based RL in 9\%. This means that most agents adapt locally through tool feedback or iterative prompting, but few accumulate durable experience across tasks or improve through controlled training loops. For security applications, this is a serious limitation: repeated vulnerability classes, recurring tool failures, common exploit chains, and SOC triage patterns should become reusable experience rather than disappearing after each run.

\subsubsection{Future Research Directions}\label{sssec:approach-future}
Future work should make agent architecture more auditable by treating verification, policy control, and role separation as first-class design elements. Modular pipelines should specify which component is responsible for planning, acting, evidence collection, validation, reporting, and risk control. Critic/verifier modules should be assessed separately from generation modules, because a system that generates many plausible findings but validates few of them is different from one that produces fewer but more reliable outputs. Similarly, guardrail layers should be connected to concrete enforcement points, such as tool-call authorization, scope checking, retrieval filtering, command execution, memory writes, and report release, extending the capability-bounded tool invocation of \textsc{ZT-ICAS}~\cite{101}.

Memory should evolve from passive context accumulation into managed security state. Future agents should distinguish working memory, long-term memory, retrieved knowledge, executable traces, and analyst-approved facts. They should also record provenance: where a fact came from, which tool produced it, when it was observed, and whether it has been validated. This is particularly important for repository auditing, penetration testing, incident response, and threat intelligence, where stale or poisoned context can mislead later actions, as \textsc{AgentPoison}~\cite{14chen2024agentpoison} demonstrates for retrieval-backed memory. Memory systems should also support controlled updates, conflict resolution, deletion, and compartmentalization so that one compromised input does not contaminate the entire agent state.

Perception pipelines should become more adversarially robust. Since security agents ingest untrusted code, logs, websites, binaries, prompts, and tool outputs, future work should separate data from instructions and treat external artifacts as potentially hostile. Retrieval-augmented perception should include source ranking, provenance, freshness checks, and prompt-injection filtering. Summarization should preserve evidence links rather than merely compressing text. For binary, malware, and protocol tasks, perception should also integrate structured representations from disassemblers, sandboxes, dynamic traces, and protocol parsers rather than relying primarily on natural-language descriptions, generalizing the analyzer-backed interface of \textsc{ClearAgent}~\cite{29}.

Reasoning and planning should explicitly represent uncertainty. Future agents should attach model confidence and evidence. Confidence-based branching, voting, verifier checks, and human checkpoints should be used when uncertainty affects safety or cost, and papers should state their reasoning paradigm explicitly rather than leaving it unspecified. Planning should also include recovery behavior: how the agent detects a failed assumption, rolls back a bad action, requests clarification, or performs explicit backtracking. This is especially important for penetration testing, repair, and autonomous cyber defense, where early mistakes can cascade.

The action space should deepen toward security-native tool integration. Future agents should not only call tools, but also understand tool preconditions, parse outputs reliably, validate results, and decide when a tool is inappropriate. Static-analysis engines, symbolic execution, taint analysis, formal verification, fuzzers, sandboxes, SIEM/SOAR systems, cloud/IaC platforms, and cyber ranges should be integrated through typed interfaces rather than free-form command strings whenever possible, following the sound-tool-in-the-loop design of \textsc{RepoAudit}~\cite{89guo2025repoauditautonomousllmagentrepositorylevel}. Tool-call accuracy, permission boundaries, and side effects should be logged as part of the agent trajectory. This would make agent behavior easier to reproduce, audit, and constrain.

Human involvement should be designed around risk rather than added as an afterthought. Fully autonomous operation is reasonable for sandboxed benchmarks, low-risk analysis, or offline report generation, but real-world security workflows often require approval gates. Future work should define when a human must be in the loop, when a human can remain on the loop, and when autonomy is acceptable; \textsc{PentestGPT}~\cite{79} is instructive because it occupies both positions at different points in the same workflow. High-risk actions should trigger review policies. Human interfaces should expose evidence, uncertainty, planned actions, and alternatives, not merely final answers.

Finally, future agents should learn across tasks in a controlled and measurable way. In-session feedback is useful, but security work contains reusable structure: successful exploit chains, failed commands, false-positive patterns, patch templates, alert-triage rationales, and tool-specific recovery strategies. Future systems should store successful and failed trajectories with metadata, retrieve them for similar tasks, and use them to build fine-tuning or preference data when appropriate, extending the trajectory-synthesis pipeline of \textsc{Cyber-Zero}~\cite{40zhuo2026cyberzero} beyond CTF writeups to operational security tasks. Reinforcement learning and environment-based improvement are promising for API testing, cyber ranges, autonomous defense, and attack simulation, but reward design must avoid unsafe shortcuts. The long-term goal should be security agents that improve from experience while preserving auditability, scope control, and reproducibility.

\subsection{Applications of Agentic Approaches in Software/Systems Security}\label{subsec:application-gaps-future}
The survey results on Application ($\S$\ref{ssec:resultapplication}) show that LLM-based agents have already been applied to a wide range of software and systems security tasks. 
This breadth indicates that agentic systems are not confined to one narrow security subfield. However, the application landscape is uneven: current work is concentrated around vulnerability-centered and offensive workflows, while several operationally important defensive, low-level, and governance-heavy security tasks remain much less mature.

\subsubsection{Limitations and Challenges of Existing Works}\label{sssec:application-limitations}
Existing applications are strongly concentrated around the vulnerability lifecycle, especially vulnerability detection, penetration testing, exploitability assessment, and repair. This concentration is natural because these tasks are text-, code-, and tool-rich: agents can inspect code, read documentation, call scanners, run commands, and generate reports. It has also produced some of the most concrete progress in the surveyed corpus, including repository-level auditing in \textsc{RepoAudit}~\cite{89guo2025repoauditautonomousllmagentrepositorylevel}, executable CVE exploitation in CVE-Bench~\cite{38zhu2025cvebenchbenchmarkaiagents}, bug-bounty-style detect/exploit/patch tasks in BountyBench~\cite{25zhang2026bountybench}, and patch generation in \textsc{PatchAgent}~\cite{75}. At the same time, this concentration means that the field's view of ``security agent'' is still shaped heavily by finding, exploiting, or fixing vulnerabilities. Other important security tasks, such as access-control analysis, supply-chain security, asset inventory, security policy governance, long-running compliance checks, and analyst knowledge management, receive much less systematic treatment.

The surveyed applications are also fragmented across the security lifecycle. Many systems address one stage well, such as detecting a vulnerable function, solving a CTF challenge, triaging an alert, or generating a patch, but fewer connect these stages into a coherent operational workflow. Detection-oriented agents often stop at candidate findings; exploitation-oriented agents often stop at proof of compromise; repair-oriented agents often assume a localized bug and available tests; SOC-oriented agents often focus on evidence synthesis rather than closed-loop containment and recovery. This separation makes it hard to assess whether an agent can support the full path from discovery to verification, prioritization, remediation, validation, reporting, and follow-up monitoring. BountyBench~\cite{25zhang2026bountybench} is notable because it explicitly separates detect, exploit, and patch tasks, but such lifecycle-spanning designs are still uncommon.

Offensive applications have advanced faster than many defensive and governance-oriented applications. Penetration testing, red-teaming, CTF solving, exploit generation, and web vulnerability scanning are well represented, and systems such as \textsc{PentestGPT}~\cite{79}, \textsc{AutoPentester}~\cite{21}, \textsc{EnIGMA}~\cite{46abramovich2025enigma}, and \textsc{PTFusion}~\cite{84WANG2026103731} illustrate substantial progress in long-horizon offensive workflows. However, post-exploitation tasks such as privilege escalation, lateral movement, persistence analysis, and cleanup are much less developed. On the defensive side, autonomous cyber-defense and SOC systems are emerging, but many works remain in simulation or controlled-alert settings. In real environments, defensive agents must reason about business impact, false-positive cost, service availability, escalation policies, and human accountability, which are harder to encode than a CTF flag or exploit success condition.

Several security domains remain underexplored despite their practical importance. Malware analysis and Android-app security appear in only a small part of the taxonomy, even though agents could help combine static features, dynamic traces, threat intelligence, family attribution, and analyst-facing reports, as suggested by \textsc{MAD-Agent}~\cite{67}. Binary analysis and reverse engineering are also comparatively sparse, partly because they require agents to interpret disassembly, decompiled code, debugger state, calling conventions, and low-level execution artifacts. Protocol fuzzing and software testing have promising examples such as \textsc{MultiFuzz}~\cite{72} and AutoRestTest~\cite{2}, but they are not yet as broadly explored as penetration testing or source-code vulnerability detection. This imbalance suggests that current agents are strongest where security evidence can be represented as natural language, source code, or familiar command output, and weaker where the application requires specialized representations or stateful execution semantics.

Application realism is another persistent challenge. Many surveyed systems are assessed on CTF platforms, reproduced CVEs, synthetic networks, curated repositories, or benchmark tasks. These settings are valuable because they provide control, safety, and measurable outcomes, but they do not fully capture the constraints of production security work. Real deployments require authentication boundaries, incomplete asset inventories, noisy telemetry, legacy systems, change-management policies, legal scope limits, disclosure procedures, and human approval chains. Even when agents produce correct technical outputs, the application may still fail if it does not fit the workflow of a SOC analyst, cloud engineer, application-security reviewer, or incident commander. Thus, application success depends not only on agent capability, but also on integration with the surrounding security process.

Finally, many applications expose dual-use and operational-safety concerns. Agents that automate vulnerability discovery, exploit generation, web scanning, red teaming, or jailbreak generation can be useful for defense, but the same capabilities can cause harm if released or deployed without clear boundaries. Some surveyed works discuss sandboxing, scoped environments, or guardrails, yet application-level risk management is not consistently treated as part of the task definition. This is a limitation of the application literature itself: a security application should specify not only what the agent is trying to accomplish, but also where it is authorized to act, what evidence it may collect, what actions require approval, and what outputs should be withheld, sanitized, or disclosed responsibly.

\subsubsection{Future Research Directions}\label{sssec:application-future}
Future work should broaden the application portfolio beyond vulnerability-centered and offensive workflows. Vulnerability detection and penetration testing will remain important, but the field also needs stronger agentic support for software supply-chain security, dependency risk assessment, access-control review, cloud and IaC governance, identity and permission analysis, asset discovery, compliance evidence collection, secure configuration management, and security-policy maintenance. These tasks are less glamorous than exploitation, but they represent recurring work in real security programs and are well suited to agents that can gather evidence, reconcile documentation, reason over policies, and produce auditable recommendations.

Future applications should also connect isolated tasks into lifecycle-aware workflows. A mature application should be able to carry evidence across stages: a suspected vulnerability should be verified, mapped to exploitability and impact, prioritized against asset context, repaired or mitigated, validated against proof-of-concept triggers and regression tests, and summarized for human review. For example, the direction suggested by BountyBench~\cite{25zhang2026bountybench}, \textsc{RepoAudit}~\cite{89guo2025repoauditautonomousllmagentrepositorylevel}, and \textsc{PatchAgent}~\cite{75} could be extended into end-to-end application-security agents that preserve traceability from discovery to patch validation. Similar lifecycle integration is needed for SOC workflows, where alert triage should connect to enrichment, hypothesis testing, containment planning, response execution, and post-incident reporting.

The field should invest in security-domain-specific agents rather than relying mainly on generic tool-using assistants. Malware agents should integrate disassemblers, sandboxes, YARA/Sigma rules, behavioral traces, and threat-intelligence feeds. Binary-analysis agents should work with structured intermediate representations, debugger sessions, symbolic execution, and decompiler output. Protocol and fuzzing agents should reason over RFCs, message grammars, state machines, coverage feedback, and crash triage. Cloud and IaC agents should understand provider-specific APIs, identity policies, deployment graphs, and blast-radius analysis. In each case, the application should define the agent's domain model, tool semantics, evidence format, and acceptance criteria, not merely expose a shell and ask the LLM to proceed.

Future work should deepen defensive and human-centered applications. SOC, SOAR, incident-response, and cyber-defense agents should be designed around analyst workflows, including evidence provenance, escalation paths, uncertainty communication, and approval gates. \textsc{Cognitive SOC}~\cite{32} illustrates the value of evidence-backed narratives, but future systems should also assess whether agents reduce analyst workload, preserve situational awareness, and support accountable decisions under time pressure. For autonomous cyber defense, agents should reason not only about attack mitigation, but also about service continuity, recovery cost, mission impact, and operator trust. These concerns make defensive applications different from offensive benchmarks where success is often a single exploit or flag.

Application-level safety should become part of task design. Future papers should specify authorization scope, environment assumptions, allowed and forbidden actions, human approval requirements, logging obligations, and release constraints. Offensive applications should default to sandboxed or consent-based targets, and exploit-generation systems should report how dangerous artifacts are controlled. Defensive applications should define when the agent may recommend an action versus execute it. This risk-aware framing is especially important for web scanning, exploit development, agent-security red teaming, and autonomous response, where an apparently successful application can still be unsafe if it crosses operational or legal boundaries.

Finally, future research should study deployment fit and transfer across application contexts. A security agent that works on CTF tasks may not transfer to enterprise penetration testing; a repository auditor may not transfer from curated open-source projects to proprietary monorepos; a SOC summarizer may not transfer across organizations with different logging schemas and playbooks. Future applications should therefore report environmental assumptions, integration requirements, failure modes, and adaptation mechanisms. Cross-domain studies are also needed: for example, techniques from penetration-testing agents may help asset-discovery agents, while evidence-linking methods from SOC agents may improve vulnerability-report generation. The long-term goal is not only to build agents for many isolated security tasks, but to understand which application patterns generalize and which require domain-specific design.

\subsection{Assessment of Agentic Approaches in Security Applications}\label{subsec:assessment-gaps-future}
The survey results on Assessment ($\S$\ref{ssec:resultassessment}) show that the community has started to move beyond one-dimensional accuracy reporting, but the assessment practice for LLM-based security agents is still immature. The surveyed papers collectively assess task effectiveness, agent trajectories, safety and reliability, baselines, and datasets; however, these dimensions are used unevenly. Task-level success, classification, time, and resource metrics are relatively common, while tool-call quality, safety, reproducibility, and human-utility measurements remain sparse.

\subsubsection{Limitations and Challenges of Existing Works}\label{sssec:eval-limitations}
Existing assessments still emphasize final task outcomes more than the agentic process that produces them. In Table~\ref{tab:assessment-task-effectiveness-metrics}, the most common task-effectiveness family is success/completion, used by 51\% of the surveyed papers, followed by precision/recall/F1 at 34\%, time/responsiveness at 25\%, and accuracy/classification at 23\%. These metrics are necessary for vulnerability detection, exploit execution, patch success, malware classification, and cyber-defense tasks, but they are not sufficient for assessing agentic approaches. An agent may reach the right answer through brittle tool use, hallucinated reasoning, excessive cost, unsafe side effects, or a trajectory that analysts cannot audit. Conversely, an agent may fail the final benchmark task but still make useful intermediate progress, such as correctly identifying an attack surface or generating a partially valid patch.

The process-level evidence reported by existing papers is uneven. Table~\ref{tab:assessment-agentic-process-trajectory} shows that cost/token efficiency appears in 32\% of papers and steps to completion in 21\%, but hallucination rate and report/explanation metrics appear in only 11\% each, iteration count in 10\%, artifact correctness in 3\%, economic/bounty value in 2\%, and tool-call accuracy in only 1\%. This is a major limitation because long-horizon security agents plan, invoke tools, observe outputs, update memory, and produce intermediate artifacts. Resource use is now reported by a meaningful minority, but tool-call correctness is almost never measured directly, and hallucination and other trajectory-quality properties remain much less visible than final outcomes.

Safety and reliability are also underrepresented as assessment targets. Table~\ref{tab:assessment-safety-reliability} records consistency in 18\% of papers, attack success rate in 10\%, defense success in 8\%, prompt-injection robustness in 6\%, scope containment in 5\%, utility-security tradeoff in 4\%, and overhead in 3\%. Even the most common family therefore appears in fewer than one fifth of the corpus. This is concerning because many LLM-based security agents are deployed in adversarial contexts by design: they inspect untrusted code, parse attacker-controlled logs, browse web applications, retrieve external knowledge, execute commands, or interact with sandboxes. Studies such as AgentDojo~\cite{10debenedetti2024agentdojo}, AgentPoison~\cite{14chen2024agentpoison}, RedAgent~\cite{86}, and ZT-ICAS~\cite{101} show that the agent itself can become an attack surface, but adversarial assessment is not yet routine.

The Dataset results further reveal a tension between realism and reproducibility. Table~\ref{tab:assessment-dataset-origin} shows substantial use of purpose-built programs and environments (50\%), production systems (28\%), curated public artifacts (23\%), and threat feeds (17\%). Table~\ref{tab:assessment-dataset-characterisation-axes} shows that 36\% of papers use fully real-world datasets and 47\% use semi-synthetic datasets. Fully public datasets appear in 50\% of papers, while some datasets require registration, are restricted, proprietary/internal, or leave access unspecified. Ground truth also remains difficult for open-ended tasks: only 16\% of papers use explicitly expert/reference-verified labels, vulnerability datasets may contain noisy labels, and exploit, repair, and SOC assessments require stronger oracles than surface fluency.

Finally, current assessments are often hard to compare. Table~\ref{tab:assessment-comparative-baseline} shows that 36\% of papers compare with non-AI tools, 28\% with other agent systems, 27\% with prior ML/DL methods, 26\% include ablations, 12\% compare with humans, and 9\% compare with plain LLMs. These numbers show that tool, model, agent, and component-level comparisons are increasingly common, but direct human and plain-LLM baselines remain uncommon. Assessment coverage is also uneven across application areas: Table~\ref{tab:assessment-dataset-by-security-task} shows stronger dataset support for cyber-defense simulation, penetration testing and CTF tasks, SOC alerting, vulnerability detection, smart-contract analysis, and AI-agent security, while asset discovery, malware analysis, web-threat detection, security testing and fuzzing, and repair datasets are represented by fewer papers. Resource reporting has improved---cost/token efficiency appears in 32\% of papers---but safety overhead appears in only 3\%, and measurement conventions remain inconsistent despite being decisive for repository-scale auditing, SOC-scale alert streams, and continuous security pipelines.

\vspace{-4pt}
\subsubsection{Future Research Directions}\label{sssec:eval-future}
Future work should report assessment as a multi-dimensional profile rather than a single score. A mature assessment protocol for agentic security systems should include at least four dimensions: \textit{task effectiveness}, which captures whether the intended security outcome is achieved; \textit{trajectory quality}, which captures how the agent reasons, uses tools, and revises its state; \textit{safety and reliability}, which captures whether the agent remains bounded and robust under adversarial inputs; and \textit{cost and reproducibility}, which captures whether the result can be practically rerun and deployed. Benchmarks such as BountyBench~\cite{25zhang2026bountybench}, CVE-Bench~\cite{38zhu2025cvebenchbenchmarkaiagents}, and AgentDojo~\cite{10debenedetti2024agentdojo} point in this direction because they assess agents in executable or stateful environments, but the field still lacks a shared reporting template that makes results comparable across security applications.

Future assessments should treat the agent trajectory as a first-class object. For penetration testing and exploitation, this means measuring not only whether the target was compromised, but also how many steps were needed, whether reconnaissance actions were relevant, whether tool outputs were interpreted correctly, and whether unsafe or out-of-scope actions occurred. For repair, this means measuring the sequence of localization, patch generation, validation, and revision, as in PatchAgent~\cite{75} and RepairAgent~\cite{88}. For SOC and incident-response systems, this means assessing evidence linkage, citation coverage, hallucination rate, and explanation quality, as illustrated by Cognitive SOC~\cite{32}. More generally, future benchmarks should release execution traces or logs where possible, so researchers can analyze not only end states but also failure modes inside agent trajectories.

Safety assessment should also become a default component whenever an agent ingests untrusted artifacts or interacts with external tools. Future studies should test prompt-injection robustness, memory and retrieval poisoning, scope containment, consistency across repeated runs, and utility-security tradeoffs. Such assessments should report both security and utility: a defense is not satisfactory if it prevents compromise only by preventing the agent from completing useful work. Responsible assessment is equally important because many surveyed systems automate exploitation, red teaming, jailbreak generation, or vulnerability discovery. Future benchmarks should use sandboxed environments, clear scope boundaries, safe release policies, and artifact sanitization, while also reporting what is withheld and why.

Dataset construction is another important future direction. Future datasets should combine real-world structure, executable oracles, and public reproducibility. CVE-Bench~\cite{38zhu2025cvebenchbenchmarkaiagents} is a strong example because it reproduces real-world CVEs in sandboxed web applications and provides automated assessment against attack targets. BountyBench~\cite{25zhang2026bountybench} is another example because it uses real bug-bounty systems and defines detect, exploit, and patch tasks with localized checks. For repository-level auditing, RepoAudit~\cite{89guo2025repoauditautonomousllmagentrepositorylevel} highlights the importance of real codebases and validation against confirmed or fixed bugs. Future benchmarks should release environment setup scripts, task specifications, expected outputs, traces, validation oracles, and partial-credit rubrics whenever release is ethically and legally possible.

Comparative assessment should be strengthened through stronger baselines, controlled agent-to-agent comparisons, and ablations. Agent systems should be compared against strong non-agent baselines when those baselines exist, such as traditional scanners, fuzzers, static analyzers, APR tools, SOC runbooks, or reinforcement-learning defenders. AutoRestTest~\cite{2} and MultiFuzz~\cite{72} are useful examples because they compare against established testing and fuzzing tools. Agent-to-agent comparisons should control for model, budget, tool access, and environment; otherwise, it is difficult to know whether the improvement comes from the agent architecture, the underlying LLM, a larger token budget, or privileged tool access. Ablation studies should also become standard, as illustrated by PatchAgent~\cite{75}, AutoRestTest~\cite{2}, and PropertyGPT~\cite{83}, because the term ``agent'' hides many components, including planning, memory, retrieval, tool wrappers, validators, critics, and feedback loops.

Finally, future research should broaden assessment coverage and make cost explicit. Malware analysis needs datasets that combine static evidence, dynamic behavior, family attribution, and analyst-facing reports, as suggested by MAD-Agent~\cite{67}. Asset discovery and OSINT need assessments that capture both retrieval quality and time-to-discovery, not just final asset lists. SOC/SOAR systems need more alert streams with evidence provenance and analyst-review criteria. Repair and hardening need datasets that connect real vulnerabilities, proof-of-concept triggers, functional tests, and patch validity. Across all such tasks, future papers should report token usage, wall-clock time, monetary cost, infrastructure assumptions, and variance across runs. In this sense, assessment of LLM-based security agents is itself a security problem: the benchmark must be realistic enough to measure capability, controlled enough to prevent harm, and transparent enough to support scientific progress.

\subsection{Cross-Cutting Future Research Directions}\label{subsec:frd}\label{subsec:crosscutting-frd}
The preceding subsections discussed future work separately from the perspectives of approach, application, and assessment. These three perspectives, however, should not evolve independently. For LLM-based agents in software and systems security, technical architecture determines what applications are feasible; application context determines which risks and evidence matter; and assessment determines whether the resulting system is trustworthy, useful, and reproducible. We therefore summarize several cross-cutting research directions that span all three survey aspects.

\subsubsection{Co-Designing Agent Architectures, Security Tasks, and Assessment Protocols}
Future work should co-design the agent approach, target application, and assessment protocol from the beginning. Many current papers introduce an agent architecture, apply it to a task, and then assess the final outcome with task-specific metrics. This pipeline is understandable for early-stage research, but it can hide mismatches between design and use. A penetration-testing agent, a repository-auditing agent, a SOC triage agent, and a patch-generation agent should not differ only in prompts and tools; they require different authority boundaries, evidence models, success conditions, and failure-handling mechanisms.

A cross-cutting design methodology should therefore begin with the security task's operational requirements. If the application is vulnerability repair, the architecture should include localization, patch generation, validation, and regression-safety components, and the assessment should measure both security fixes and functional preservation. If the application is SOC investigation, the architecture should preserve evidence provenance and analyst review points, and the assessment should measure evidence linkage, hallucination, timeliness, and human usefulness. If the application is exploit generation or red teaming, the architecture should encode scope control and sandboxing, and the assessment should report both capability and containment. This co-design would make the term agent more precise: the agent would be defined by the task it can safely perform, the evidence it can use, and the assessment standard it must satisfy.

\subsubsection{Evidence-Centered and Auditable Agentic Security}
Across all three survey aspects, evidence is the central missing abstraction. Approach-wise, agents need memory, perception, reasoning, and tool-use mechanisms that distinguish observed facts from hypotheses, generated text, retrieved knowledge, and validated findings. Application-wise, security decisions must be traceable to concrete artifacts such as code locations, logs, packets, exploit outputs, test results, SBOM entries, CVE records, or analyst annotations. Assessment-wise, final-answer correctness is insufficient unless the supporting evidence can be inspected.

Thus, future agents should maintain explicit evidence graphs or evidence ledgers. Each security-relevant claim should record its source, time, tool or model that produced it, confidence, validation status, and downstream uses. This would support repository auditing, incident response, threat intelligence, repair, and penetration testing alike. For example, \textsc{RepoAudit}~\cite{89guo2025repoauditautonomousllmagentrepositorylevel} points toward evidence-constrained repository exploration, while \textsc{Cognitive SOC}~\cite{32} highlights evidence-linked incident narratives. Future work should generalize these ideas into reusable evidence-management layers for security agents. Such layers would also improve assessment: benchmarks could check not only whether the final answer is correct, but whether the supporting chain of evidence is complete, faithful, and minimally sufficient.

\subsubsection{Risk-Aware Autonomy and Governance}
The survey shows that many security agents are designed for autonomous or near-autonomous operation, but autonomy has different implications across applications. A fully autonomous agent that summarizes alerts is different from one that executes exploits, modifies production code, changes firewall rules, or updates cloud policies. Approach research must therefore expose control points for authorization, rollback, human approval, and policy enforcement. Application research must define what actions are allowed in each security setting. Assessment research must measure whether agents stay within those boundaries under normal, erroneous, and adversarial conditions.

Future work should develop risk-aware autonomy levels for security agents. Low-risk tasks, such as offline summarization or documentation retrieval, may permit fully autonomous execution. Medium-risk tasks, such as vulnerability triage or patch recommendation, may require human-on-the-loop review. High-risk tasks, such as exploit execution, production remediation, disclosure decisions, and active response, should require explicit human-in-the-loop approval or policy-verified execution. These autonomy levels should be reported as part of the agent design and benchmark setup. They should also be stress-tested against prompt injection, retrieval poisoning, malicious tool output, and scope-confusion attacks, following the concerns raised by benchmarks and systems such as AgentDojo~\cite{10debenedetti2024agentdojo}, AgentPoison~\cite{14chen2024agentpoison}, RedAgent~\cite{86}, and ZT-ICAS~\cite{101}.

\subsubsection{Realistic, Reproducible, and Safe Benchmark Ecosystems}
A recurring issue across the approach, application, and assessment results is the lack of benchmark ecosystems that are simultaneously realistic, reproducible, and safe. Realistic applications need live services, repositories, attack surfaces, logs, binaries, cloud configurations, or cyber-range states. Reproducible assessment needs stable setup scripts, task definitions, labels, oracles, and comparable baselines. Safe release needs scope boundaries, artifact sanitization, and clear policies for exploit or vulnerability disclosure. Most existing studies optimize only part of this triangle.

Future benchmark ecosystems should package tasks as executable, instrumented environments rather than static datasets whenever the application requires interaction. CVE-Bench~\cite{38zhu2025cvebenchbenchmarkaiagents} and BountyBench~\cite{25zhang2026bountybench} illustrate this direction for vulnerability exploitation and bug-bounty workflows; similar benchmark ecosystems are needed for SOC investigation, cloud/IaC security, malware analysis, protocol fuzzing, repair, access-control review, and supply-chain security. These benchmarks should include task metadata, allowed actions, expected evidence, validation oracles, baseline tools, resource budgets, and safety policies. They should also preserve trajectories where possible, because agent behavior is defined by the path from observation to action, not only by the final output.

\subsubsection{Cost, Scalability, and Resource Governance}
Cost and scalability cut across all three aspects. Approach choices such as multi-agent orchestration, long-context retrieval, repeated tool calls, verifier loops, and persistent memory can improve capability but also increase token usage, latency, and infrastructure cost. Application settings such as repository-scale auditing, SOC alert streams, cloud posture management, and internet-scale asset discovery require sustained operation rather than one-off benchmark runs. Assessment must therefore account for resource use as a first-class outcome.

Future research should report and optimize token cost, wall-clock time, tool-call count, infrastructure requirements, memory growth, and variance across runs. More importantly, agents should make resource-aware decisions: when to retrieve more context, when to call an expensive model, when to invoke a heavyweight analyzer, when to stop exploring, and when to ask for human input. This suggests a need for budget-aware planning and adaptive execution policies. Such policies should be assessed against both security effectiveness and resource consumption, because an agent that is accurate but too expensive or slow may be impractical for continuous security workflows.

\subsubsection{Human-Agent Collaboration as a Security Interface}
Human involvement should be studied as a cross-cutting design problem rather than a minor workflow option. Approach research should design interfaces that expose uncertainty, evidence, alternatives, and planned actions. Application research should identify where humans add judgment, authority, or domain context: approving exploit attempts, prioritizing vulnerabilities, interpreting business impact, validating incident narratives, or deciding whether to deploy a patch. Assessment research should measure whether agent outputs actually improve human decisions, reduce workload, and preserve accountability.

Future studies should therefore move beyond fully automated benchmark scores and include human-centered assessment where the application requires professional judgment. For SOC, incident-response, vulnerability-management, and cloud-security settings, useful measures include analyst time saved, review accuracy, trust calibration, escalation quality, and error recovery. Human-agent collaboration should also be bidirectional: humans should be able to correct agent state, mark evidence as trusted or untrusted, constrain future actions, and inspect why an agent made a recommendation. In high-impact security settings, a usable agent is not merely one that produces the right answer; it is one whose reasoning and authority can be governed by the people responsible for the system.

\subsubsection{Toward Generalizable Security-Agent Engineering Principles}
Finally, the field needs reusable engineering principles that explain which agent patterns generalize across security tasks and which must remain domain-specific. Some capabilities, such as evidence tracking, tool-use validation, memory provenance, risk-based approval, and cost-aware planning, appear broadly useful across vulnerability detection, pentesting, incident response, repair, and SecOps. Other capabilities, such as exploit-chain construction, malware sandbox interpretation, protocol-state modeling, or cloud-policy reasoning, require domain-specific representations and tools.

Future work should compare agent designs across multiple security applications under controlled conditions. This would help identify when a multi-agent architecture is genuinely useful, when retrieval improves robustness, when verifier modules reduce false positives, when fine-tuning is preferable to prompting, and when human checkpoints improve safety without eliminating utility. The longer-term goal is to move from isolated prototypes to an engineering discipline for LLM-based security agents: a body of patterns, benchmarks, reporting standards, and safety practices that connects approach, application, and assessment into one coherent research agenda.

\section{Threats to Validity}\label{sec:threats}
Our survey results are subject to validity threats because the studied area is recent, fast-moving, and terminologically unstable. LLM-based agents for software and systems security are described using overlapping terms such as autonomous agents, agentic workflows, multi-agent systems, tool-using LLMs, and LLM-based security assistants. Moreover, many surveyed works target dual-use security tasks, which affects dataset release, benchmark design, reproducibility, and the kinds of assessments that authors can report. We therefore discuss the main threats associated with our literature search, taxonomy derivation, paper attribution, and result analysis.

\vspace{3pt}
\noindent
{\bf Literature search.}
The first threat concerns coverage of the literature. We used keyword-based search and forward/backward snowballing to identify papers published between January 2023 and March 2026, but relevant papers may still have been missed. This risk is heightened because the terminology around LLM agents is not yet standardized: some papers may describe similar systems as assistants, copilots, planners, tool-augmented LLMs, or multi-step workflows without using the word ``agent.'' Conversely, papers that use the term ``agent'' may not implement an agentic system in the sense used by this survey. We mitigated this threat by combining multiple agent-related and security-task-related keywords, manually screening candidate papers, and using snowballing from the seed set to recover papers that keyword search alone may not find.

A second search-related threat comes from our scope decisions. We focused on peer-reviewed papers and excluded studies exclusively available on arXiv, short papers, posters, non-English papers, and papers outside the January 2023 to March 2026 window. These criteria improve the consistency and inspectability of the corpus, but they may exclude emerging systems, industry reports, technical blogs, benchmark repositories, or very recent preprints that influence practice. This is a tradeoff of the survey design. Our results should therefore be interpreted as a structured view of the peer-reviewed literature in the selected time window, not as an exhaustive inventory of all deployed or publicly released LLM-based security-agent systems.

\vspace{3pt}
\noindent
{\bf Survey taxonomy derivation.}
Our taxonomy was derived through manual coding of the surveyed papers into the three aspects of Approach, Application, and Assessment. Although this process was grounded in recurring concepts observed in the corpus, alternative researchers could choose different abstraction levels. For example, one taxonomy might separate web penetration testing from CTF solving, while another might group both under offensive assessment. Similarly, one taxonomy might treat retrieval, memory, and context management as one design attribute, while ours separates them when they imply different agent capabilities.

We mitigated this threat by using the taxonomy for paper attribution and result analysis, rather than treating it as a standalone conceptual proposal. Items were retained when they captured recurring distinctions in the surveyed papers and helped explain differences in agent design, security task coverage, or assessment practice. We also organized the taxonomy around three high-level questions that are stable across the literature: how the agent is built, what security task it performs, and how its behavior is assessed. Even so, the taxonomy should be viewed as an evolving synthesis. As the field matures, new attributes may be needed for issues such as long-term memory governance, agent supply-chain risk, tool-permission models, and deployment-time monitoring.

\vspace{3pt}
\noindent
{\bf Paper attribution.}
The third threat concerns the mapping of papers to taxonomy items. Many LLM-based security agents are multi-component systems, and a single paper may legitimately belong to multiple categories. For instance, a system may combine retrieval, planning, tool execution, verification, and reporting; it may address both vulnerability detection and repair; and it may assess both final task success and trajectory properties. Such multi-label attribution is necessary, but it introduces judgment calls about how much evidence is sufficient to assign an item.

We mitigated this threat by coding papers according to explicit descriptions in the paper text, including architecture descriptions, workflows, datasets, metrics, and experimental protocols. For the Assessment metric tables, we revisited the full text of every paper with an initially blank row, examining the methods, experimental setup, results, and appendices where available rather than relying on the abstract or introduction. We added every mapping supported by a reported measure and retained a blank only when the relevant metric family remained unspecified. When a paper described multiple relevant capabilities, we allowed multiple attributions rather than forcing a single primary category. We also considered duplicate or substantially overlapping publications to avoid counting the same contribution multiple times. Nevertheless, some ambiguity remains. Authors do not always use consistent terminology, and papers vary greatly in how much detail they provide about prompts, tools, memory, orchestration, datasets, baselines, and release artifacts. Our attribution should therefore be read as a best-effort characterization of the available evidence rather than as a definitive implementation audit.

\vspace{3pt}
\noindent
{\bf Survey result analysis.}
The fourth threat concerns interpretation of the attributed results. Because LLM-based security agents depend heavily on foundation models, tool access, prompts, execution environments, and resource budgets, results reported by different papers are often not directly comparable. A performance difference may come from the agent architecture, the underlying model, the benchmark, the available tools, the amount of context, the number of attempts, or the assessment budget. The field is also changing quickly: models, APIs, cyber ranges, and benchmark versions may change after publication, and some reported results may become difficult to reproduce.

We mitigated this threat by emphasizing patterns across papers rather than ranking individual systems. Our analysis therefore focuses on recurring gaps, such as limited safety assessment, weak trajectory-level reporting, uneven baseline comparisons, concentration around vulnerability-centric applications, and insufficient human-governance mechanisms. These conclusions are less sensitive to any single paper's result. Still, the survey may understate industrial practice, overstate published benchmark coverage, or miss negative results that were not published. The future research directions we identify should be interpreted as grounded recommendations from the available literature, not as claims that every individual system suffers from every limitation.

\section{Conclusion}\label{sec:con}
This paper presented a systematic survey of LLM-based agents for software
and systems security, covering 100 peer-reviewed papers published between
January 2023 and March 2026 and organized around three aspects:
\emph{Approach}, \emph{Application}, and \emph{Assessment}.

Our survey shows that LLM-based security agents have rapidly moved from
simple prompt-response systems toward modular, tool-mediated, and often
autonomous workflows, and that the resulting field is unevenly developed
along a consistent axis. Agents are built to act: 91\% adopt a modular
pipeline, 88\% place the backbone LLM in a planner-and-actor role, and 95\%
support fully autonomous operation. They are far less often built to be
bounded or inspected: explicit critic or verifier roles appear in 15\% of
systems, guardrail layers in 13\%, and any human review point in 13\%.
Assessment mirrors this asymmetry. Outcome measures are routine---success
or completion in 51\% of papers, precision/recall/F1 in 34\%---while the
measures that would expose \emph{how} an outcome was reached are not:
hallucination rate appears in 11\%, scope containment in 5\%, and tool-call
accuracy in 1\%, despite nearly every surveyed system being tool-using.
Applications concentrate around the vulnerability lifecycle and offensive
assessment, leaving post-access phases, low-level artifact analysis, and
governance-oriented tasks such as access-control review and supply-chain
security comparatively undeveloped.

The central message of this survey is that security agents should be
studied as operational systems, not merely as prompts wrapped around LLMs.
Their value depends on how they gather evidence, use tools, maintain state,
respect authority boundaries, recover from errors, communicate with humans,
and behave under adversarial conditions. Future progress in this field will require co-design
across the three aspects: architectures shaped by the security tasks they
perform, applications that encode evidence and risk requirements, and
assessments that measure process quality, safety, cost, and human oversight
alongside final outcomes. The field has built agents able to act; building
agents whose authority is bounded and whose behavior is auditable remains
the open problem.

\bibliographystyle{plain}
\bibliography{surveyed-papers-normalized,reference}


\end{document}